\documentclass{IEEEoj}
\usepackage{cite}
\usepackage{amsmath,amssymb,amsfonts}
\usepackage{algorithmic}
\usepackage{graphicx,color}
\usepackage{textcomp}
\def\BibTeX{{\rm B\kern-.05em{\sc i\kern-.025em b}\kern-.08em
    T\kern-.1667em\lower.7ex\hbox{E}\kern-.125emX}}
\AtBeginDocument{\definecolor{ojcolor}{cmyk}{0.93,0.59,0.15,0.02}}
\def\OJlogo{\vspace{-14pt}\includegraphics[height=20pt]{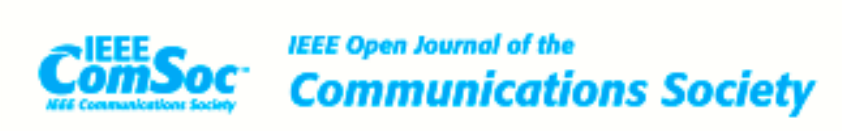}}

\usepackage{cite}
\usepackage{amsmath,amssymb,amsfonts}
\usepackage{algorithmic}
\usepackage{graphicx, color}
\usepackage{textcomp}
\def\BibTeX{{\rm B\kern-.05em{\sc i\kern-.025em b}\kern-.08em
    T\kern-.1667em\lower.7ex\hbox{E}\kern-.125emX}}

\usepackage{adjustbox}     
\usepackage{cite}
\usepackage[hyphens]{url}
\usepackage{amsmath}
\usepackage{amsthm}
\usepackage{amssymb}
\usepackage{array}
\usepackage{url}
\usepackage{graphicx}
\usepackage{lipsum}
\usepackage{listings}
\usepackage{xcolor}
\usepackage{nomencl}
\usepackage[caption=false,font=footnotesize]{subfig}
\usepackage[ruled,vlined]{algorithm2e}
\usepackage{colortbl}
\usepackage[acronym]{glossaries}
\usepackage{longtable}
\usepackage{soul}

\definecolor{backcolour}{RGB}{255, 247, 210}
\definecolor{codeorange}{RGB}{226, 121, 0}
\definecolor{codepink}{RGB}{249, 38, 114}
\definecolor{codegreen}{RGB}{166, 226, 46}
\definecolor{codeblue}{RGB}{0, 171, 205}
\definecolor{numbercolor}{RGB}{117, 113, 94}

\makeatletter
\def\lst@makecaption{%
  \def\@captype{table}%
  \@makecaption
}
\makeatother

\newlength{\xfigwd}
\newcolumntype{P}[1]{>{\centering\arraybackslash}p{#1}}
\newcolumntype{M}[1]{>{\centering\arraybackslash}m{#1}}
\renewcommand{\arraystretch}{1.5}
\arrayrulecolor[HTML]{45526C}
\definecolor{table_white}{RGB}{248, 245, 241}
\definecolor{table_red}{RGB}{248, 164, 136}
\definecolor{table_green}{RGB}{90, 168, 151}
\definecolor{table_blue}{RGB}{69, 82, 108}
\definecolor{table_banana}{RGB}{200, 187, 120}
\definecolor{table_x}{RGB}{150, 200, 200}

\DeclareMathOperator*{\argmin}{arg\,min}

\makeglossaries

\newacronym{3gpp}{3GPP}{3rd Generation Partnership Project}
\newacronym{5g}{5G}{5th Generation}
\newacronym{ioe}{IoE}{Internet-of-Everything}
\newacronym{mmwave}{mmWave}{Millimeter-Wave}
\newacronym{ofdm}{OFDM}{Orthogonal Frequency Division Multiplexing}
\newacronym{6g}{6G}{6th Generation}
\newacronym{phy}{PHY}{Physical Layer}
\newacronym{ai}{AI}{Artificial Intelligence}
\newacronym{im}{IM}{Index Modulation}
\newacronym{mimo}{MIMO}{Multiple-Input Multiple-Output}
\newacronym{ris}{RIS}{Reconfigurable Intelligent Surfaces}
\newacronym{nn}{NN}{Neural Network}
\newacronym{ml}{ML}{Machine Learning}
\newacronym{dl}{DL}{Deep Learning}
\newacronym{mcs}{MCS}{Modulation and Coding Scheme}
\newacronym{e2e}{E2E}{End-to-End}
\newacronym{dnn}{DNN}{Deep Neural Network}
\newacronym{mc}{MC}{Multicarrier}
\newacronym{noma}{NOMA}{Non-Orthogonal Multiple Access}

\newacronym{cnn}{CNN}{Convolutional Neural Networks}
\newacronym{rnn}{RNN}{Recurrent Neural Networks}
\newacronym{relu}{ReLU}{Rectified Linear Unit}
\newacronym{sgd}{SGD}{Stochastic Gradient Descent}
\newacronym{adam}{ADAM}{Adaptive Moment Estimation}
\newacronym{bptt}{BPTT}{Backpropagation Through Time}
\newacronym{lstm}{LSTM}{Long Short-Term Memory}
\newacronym{papr}{PAPR}{Peak-to-Average Power Ratio}
\newacronym{cp}{CP}{Cyclic Prefix}

\newacronym{mu}{MU}{Multi-User}
\newacronym{csi}{CSI}{Channel State Information}
\newacronym{bs}{BS}{Base Station}
\newacronym{rf}{RF}{Radio Frequency}
\newacronym{mld}{MLD}{Maximum Likelihood Detector}
\newacronym{zf}{ZF}{Zero-Forcing}
\newacronym{mmse}{MMSE}{Minimum-Mean Squared Error}
\newacronym{ber}{BER}{Bit Error Rate}
\newacronym{dac}{DAC}{Digital-to-Analog Converter}
\newacronym{qam}{QAM}{Quadrature Amplitude Modulation}
\newacronym{psk}{PSK}{Phase-Shift Keying}
\newacronym{ae}{AE}{Autoencoder}
\newacronym{amp}{AMP}{Approximate Message Passing}
\newacronym{sdrx}{SDRX}{Semidefinite Relaxation}
\newacronym{oamp}{OAMP}{Orthogonal Approximate Message Passing}
\newacronym{apm}{APM}{Amplitude-Phase Modulation}
\newacronym{gsm}{GSM}{Generalised Spatial Modulation}
\newacronym{iid}{i.i.d.}{Independent and Identically Distributed}
\newacronym{spd}{SPD}{Sphere Decoding}
\newacronym{ts}{TS}{Tabu Search}
\newacronym{vblast}{V-BLAST}{Vertical Bell Laboratories Layered Space-Time}
\newacronym{sic}{SIC}{Successive Interference Cancellation}
\newacronym{ldamp}{LDAMP}{Learned Denoising-Based Approximate Message Passing}
\newacronym{doa}{DOA}{Direction-of-Arrival}
\newacronym{mse}{MSE}{Mean Squared Error}
\newacronym{nmse}{NMSE}{Normalized Mean Squared Error}
\newacronym{lamp}{LAMP}{Learned Approximate Message Passing}
\newacronym{adc}{ADC}{Analog-to-Digital Converter}
\newacronym{gamp}{GAMP}{Generalised Approximate Message Passing}
\newacronym{gan}{GAN}{Generative Adversarial Network}
\newacronym{snr}{SNR}{Signal-to-Noise Ratio}
\newacronym{sm}{SM}{Spatial Modulation}
\newacronym{awgn}{AWGN}{Additive White Gaussian Noise}
\newacronym{ldpc}{LDPC}{Low-Density Parity-Check}
\newacronym{bpsk}{BPSK}{Binary Phase Shift Keying}
\newacronym{qpsk}{QPSK}{Quadrature Phase Shift Keying}
\newacronym{slp}{SLP}{Symbol-Level Precoding}
\newacronym{rtn}{RTN}{Radio Transformer Network}
\newacronym{c2po}{C2PO}{Biconvex 1-Bit Precoding}
\newacronym{nno-c2po}{NNO-C2PO}{Neural Network Optimized Biconvex 1-Bit Precoding}
\newacronym{ide2}{IDE2}{Iterative Discrete Estimation}
\newacronym{wmmse}{WMMSE}{Weighted Minimum Mean Squared Error}
\newacronym{admm}{ADMM}{Alternating Direction Method of Multipliers}
\newacronym{siso}{SISO}{Single-Input Single-Output}
\newacronym{sdr}{SDR}{Software-Defined Radio}
\newacronym{tas}{TAS}{Transmit Antenna Selection}
\newacronym{knn}{KNN}{$K$-Nearest Neighbors}
\newacronym{svm}{SVM}{Support Vector Machine}
\newacronym{fvg}{FVG}{Feature Vector Generation}
\newacronym{edas}{EDAS}{Euclidean Distance-Optimized Antenna Selection}
\newacronym{ser}{SER}{Symbol Error Rate}
\newacronym{tac}{TAC}{Transmit Antenna Combination}

\newacronym{ici}{ICI}{Intercarrier Interference}
\newacronym{sc}{SC}{Single-carrier}
\newacronym{lmmse}{LMMSE}{Linear Minimum Mean Error Square}
\newacronym{ifft}{IFFT}{Inverse Fast Fourier Transform}
\newacronym{oobe}{OOBE}{Out-of-Band Emission}
\newacronym{fbmc}{FBMC}{Filter Bank Multicarrier}
\newacronym{gfdm}{GFDM}{Generalized Frequency Division Multiplexing}
\newacronym{w-ofdm}{W-OFDM}{Windowed-OFDM}
\newacronym{ufmc}{UFMC}{Universal Filtered Multicarrier}
\newacronym{ofdm-im}{OFDM-IM}{Orthogonal Frequency Division Multiplexing with Index Modulation}
\newacronym{otfs}{OTFS}{Orhogonal Time Frequency Space}
\newacronym{usrp}{USRP}{Universal Software Radio Peripherals}

\newacronym{em}{EM}{Electromagnetic}
\newacronym{los}{LOS}{Line-of-Sight Path}
\newacronym{qos}{QoS}{Quality-of-Service}
\newacronym{iot}{IoT}{Internet-of-Things}
\newacronym{cs}{CS}{Compressed Sensing}
\newacronym{sl}{SL}{Supervised Learning}
\newacronym{miso}{MISO}{Multiple-Input Single-Output}
\newacronym{oma}{OMA}{Orthogonal Multiple Access}
\newacronym{mui}{MUI}{Multiuser Interference}
\newacronym{usl}{USL}{Unsupervised Learning}
\newacronym{mlp}{MLP}{Multi-Layer Perceptron}
\newacronym{ddpg}{DDPG}{Deep Deterministic Policy Gradient}
\newacronym{d2d}{D2D}{Device-to-Device}

\newacronym{rssi}{RSSI}{Received Signal Strength Indicator}
\newacronym{dqn}{DQN}{Deep Q-learning Network}

\begin{document}
\receiveddate{XX Month, XXXX}
\reviseddate{XX Month, XXXX}
\accepteddate{XX Month, XXXX}
\publisheddate{XX Month, XXXX}
\currentdate{XX Month, XXXX}
\doiinfo{OJCOMS.2022.1234567}

\title{Deep Learning-Aided 6G Wireless Networks: A Comprehensive Survey of \\ Revolutionary PHY Architectures}

\author{\uppercase{Burak~Ozpoyraz}\authorrefmark{1}, \IEEEmembership{Graduate Student Member,~IEEE},
\uppercase{A. Tugberk~Dogukan}\authorrefmark{1}, \IEEEmembership{Graduate Student Member,~IEEE},
\uppercase{Yarkin~Gevez}\authorrefmark{1}, \IEEEmembership{Graduate Student Member,~IEEE},
\uppercase{Ufuk~Altun}\authorrefmark{1}, \IEEEmembership{Graduate Student Member,~IEEE},
\uppercase{Ertugrul~Basar}\authorrefmark{1}, \IEEEmembership{Senior Member,~IEEE}}

\affil{Communications Research and Innovation Laboratory (CoreLab), Department of Electrical and Electronics Engineering, Koç University, Sariyer 34450, Istanbul, Turkey (e-mail: bozpoyraz20@ku.edu.tr; adogukan18@ku.edu.tr; ygevez21@ku.edu.tr; ualtun20@ku.edu.tr; ebasar@ku.edu.tr.}

\corresp{Corresponding author: Ertugrul Basar (e-mail: ebasar@ku.edu.tr).}
\authornote{This work is supported by TUBITAK under Grant Number 121C254.}
\markboth{DEEP LEARNING-AIDED 6G WIRELESS NETWORKS: A COMPREHENSIVE SURVEY OF REVOLUTIONARY PHY ARCHITECTURES}{OZPOYRAZ \textit{et al.}}

\begin{abstract}
Deep learning (DL) has proven its unprecedented success in diverse fields such as computer vision, natural language processing, and speech recognition by its strong representation ability and ease of computation. As we move forward to a thoroughly intelligent society with 6G wireless networks, new applications and use cases have been emerging with stringent requirements for next-generation wireless communications. Therefore, recent studies have focused on the potential of DL approaches in satisfying these rigorous needs and overcoming the deficiencies of existing model-based techniques. The main objective of this article is to unveil the state-of-the-art advancements in the field of DL-based physical layer methods to pave the way for fascinating applications of 6G. In particular, we have focused our attention on four promising physical layer concepts foreseen to dominate next-generation communications, namely massive multiple-input multiple-output systems, sophisticated multi-carrier waveform designs, reconfigurable intelligent surface-empowered communications, and physical layer security. We examine up-to-date developments in DL-based techniques, provide comparisons with state-of-the-art methods, and introduce a comprehensive guide for future directions. We also present an overview of the underlying concepts of DL, along with the theoretical background of well-known DL techniques. Furthermore, this article provides programming examples for a number of DL techniques and the implementation of a DL-based multiple-input multiple-output by sharing user-friendly code snippets, which might be useful for interested readers.
\end{abstract}



\maketitle

\printglossary[style=tree, type=\acronymtype, title=Nomenclature, nogroupskip, nonumberlist]

\begin{figure}
   \centering
   \includegraphics[width=\linewidth]{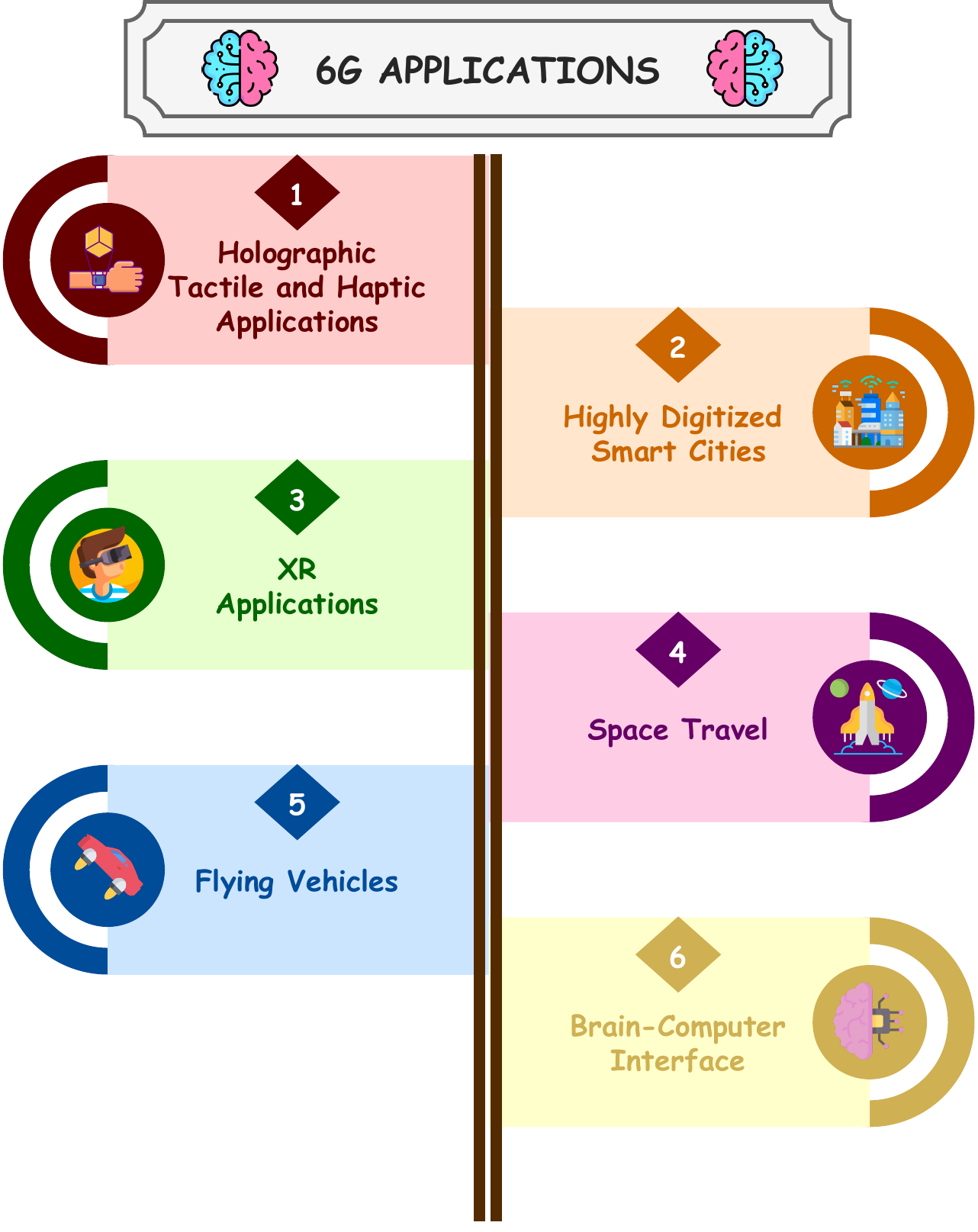}
   \caption{Emerging applications of 6G wireless networks.}
   \vspace{-0.5cm}
   \label{fig:6G_Applications}
\end{figure}


\vspace{-2ex}
\section{Introduction}  \label{sec:intro}
\begin{figure}
   \centering
   \includegraphics[width=\linewidth]{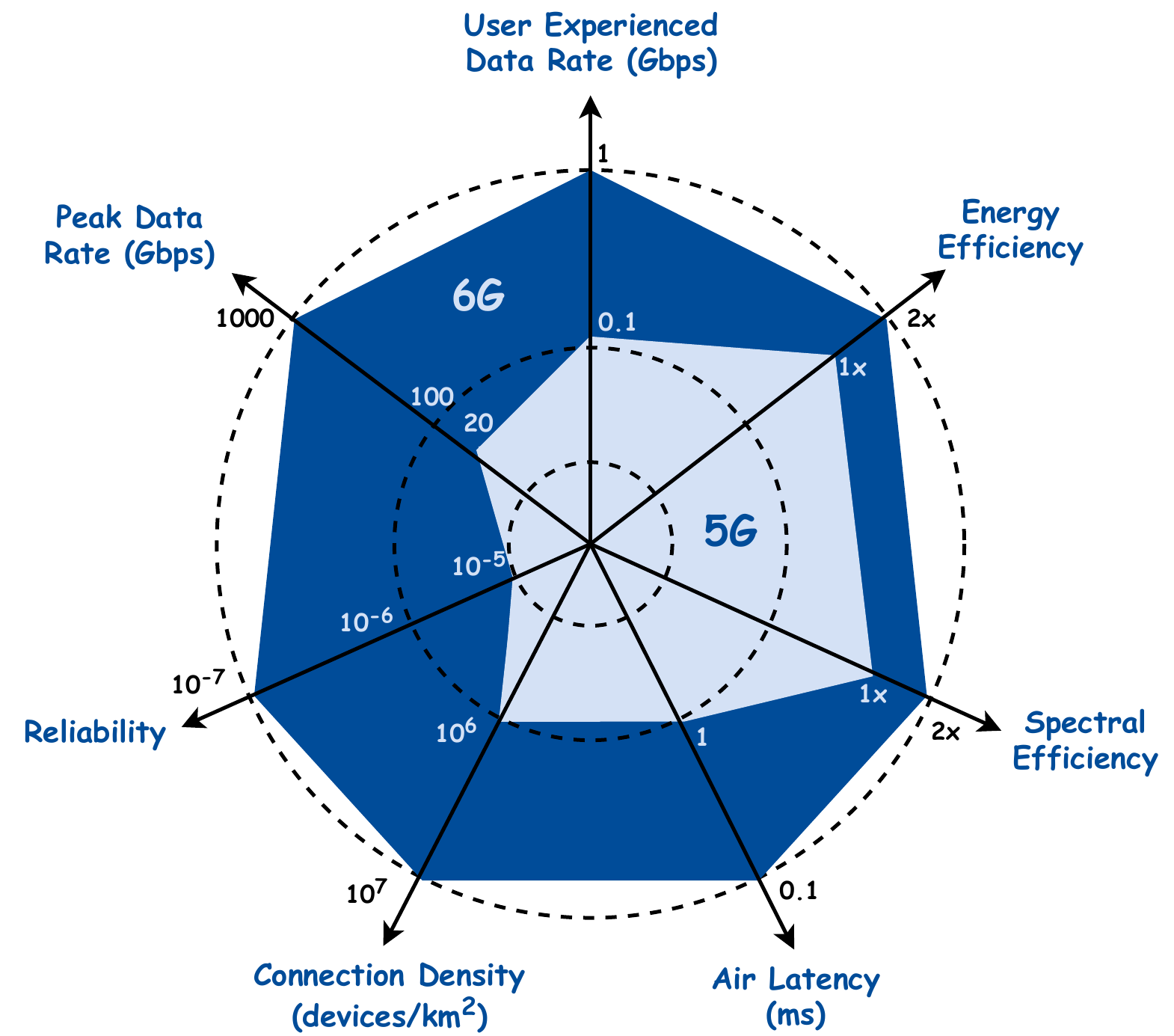}
   \caption{Comparison of 5G and 6G specifications~\cite{Burak_8_Samsung5Gvs6GChart}.}
   \label{fig:5Gvs6G}
\end{figure}

\PARstart{S}{tarting} with the \acrshort{3gpp} (3rd Generation Partnership Project) Release-15 standard in March 2017, the brand new 5th generation (\acrshort{5g}) wireless communication technology has made significant progress worldwide~\cite{Burak_1_Release15},~\cite{Burak_2_Statista}. 3GPP completed the Release-17 Stage-2 freeze in June 2021 and is planning the next freeze in March 2022~\cite{Burak_3_Release17}. As these releases have been proceeding, the deployment process of 5G has attained a high level of maturity. According to a recent GSA report, 176 operators from 72 countries had launched commercial 5G networks as of August 2021~\cite{Burak_4_GSA}. However, the recent advancements in {the} implementation of 5G have exposed the drawbacks and limitations of this technology despite being nominated as the primary enabler of Internet-of-Everything (\acrshort{ioe}) systems~\cite{Burak_5_2019_Vision6G}. Even though 5G has introduced more flexibility and efficiency to wireless networks by utilizing new technologies such as millimeter-wave (\acrshort{mmwave}) communication and multi-numerology orthogonal frequency division multiplexing (\acrshort{ofdm}), the pledged revolutionary mobile communication perspective has remained a pipe dream thus far~\cite{Burak_5_2019_Vision6G},~\cite{Burak_6_2020_RISParadigm6G}. Moreover, although 5G systems are ready to support data rate-hungry enhanced mobile broadband services and ultra-reliable low-latency communications, it is questionable whether they can support future IoE applications~\cite{Burak_6_2020_RISParadigm6G}. Therefore, researchers have been investigating sophisticated communication technologies and developing intriguing concepts for 6th generation (\acrshort{6g}) wireless networks, which are projected to come into life after 2030. Even if cellular communication systems have advanced to a new level with the development of 5G, expectations of the intelligent information society of 2030 and beyond will be remarkably compelling. 6G will be the essential provider of a highly digital community by connecting everything, enabling almost unlimited wireless coverage, establishing full-vertical networks, and supporting holographic and high-precision communications for tactile and haptic applications~\cite{Burak_7_2019_6GKeyTechnol}. In addition, new IoE services ranging from extended reality applications and flying vehicles to space travel and brain-computer interfaces will come to reality with 6G networks, making 6G the true master of IoE~\cite{Burak_5_2019_Vision6G}, as illustrated in Fig.~\ref{fig:6G_Applications}. However, these new and fascinating applications will bring strict and very challenging requirements such as ultra-high reliability, low latency, substantially high data rate, high energy and spectral efficiency, and dense connectivity. Fig.~\ref{fig:5Gvs6G} compares the specifications of 5G and 6G, elucidating the 6G expectations~\cite{Burak_8_Samsung5Gvs6GChart}. The developments and expectations on the way toward these 6G technologies in the literature have also been comprehensively reviewed and discussed by some other research groups \mbox{\cite{ikram2022road,alsabah20216g,barakat20216g}}.

\begin{figure}
   \centering
   \includegraphics[width=\linewidth]{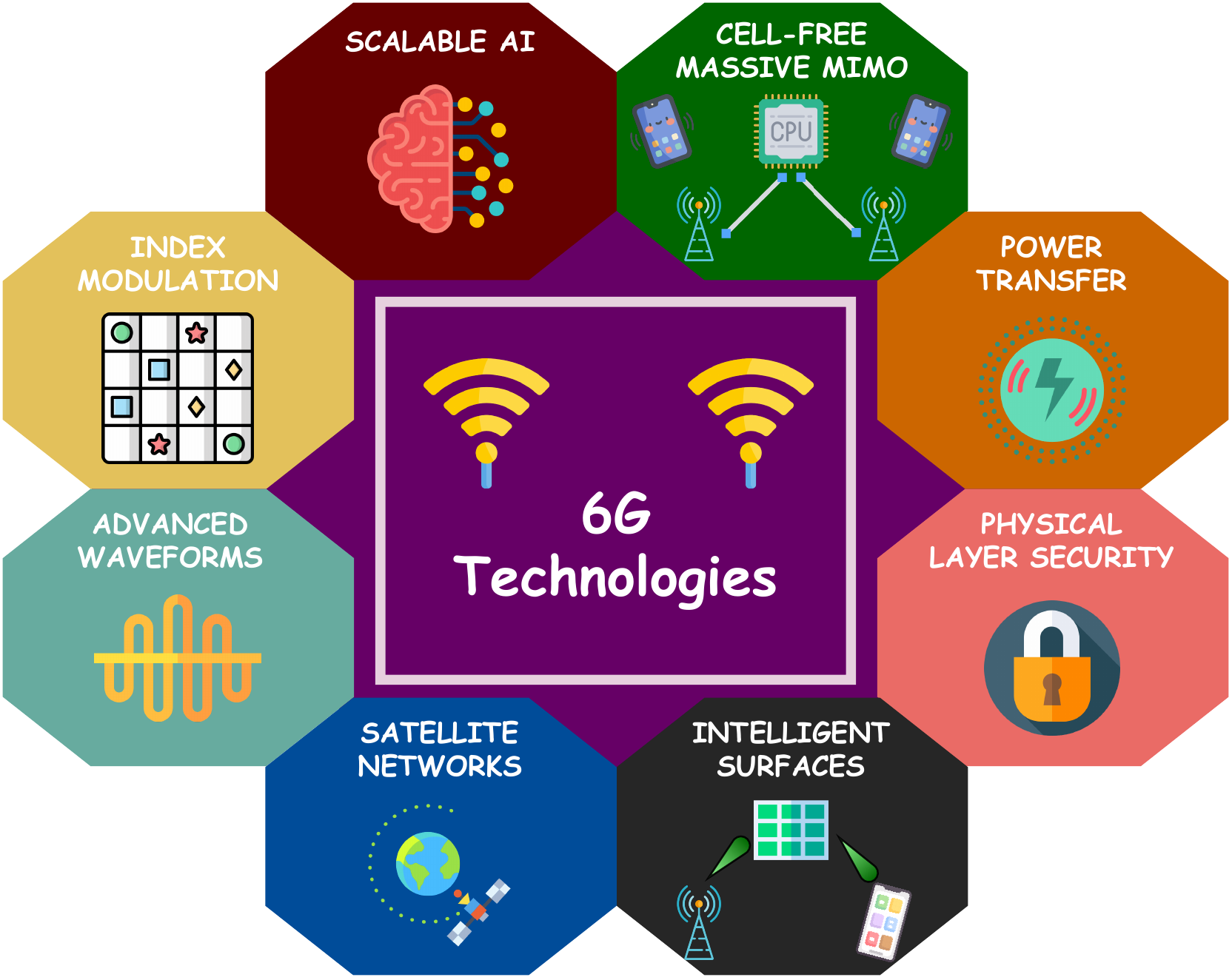}
   \caption{An overview of emerging PHY techniques that can dominate the 6G era.}
   \label{fig:6G_Technologies}
\end{figure}

The stringent demands of 6G have driven researchers to look for sophisticated physical layer (\acrshort{phy}) techniques. Beyond using more spectrum in upper 100 GHz bands, such as high-frequency mmWave and terahertz bands, 6G might incorporate numerous emerging technologies such as scalable artificial intelligence (\acrshort{ai}), index modulation (\acrshort{im})~\cite{Burak_9_2017_IMNextGen},~\cite{Burak_10_2021_MultidimIM}, cell-free massive multiple-input multiple-output (\acrshort{mimo}) systems~\cite{Burak_11_2019_CellFreeParadigm},~\cite{Burak_12_2020_ScalableCellFree}, reconfigurable intelligent surfaces (\acrshort{ris})~\cite{Burak_6_2020_RISParadigm6G},~\cite{Burak_13_2019_WirelessCommRIS},~\cite{Burak_14_2021_RISFutureNetworks}, PHY security~\cite{Wu2018},~\cite{Benzaid2020}, advanced waveforms~\cite{yardimci1},~\cite{yardimci8}, satellite and non-terrestrial networks~\cite{Burak_15_2019_LEOSmallSatelliteCons5GBeyond},~\cite{Burak_16_2021_SurveyOnAerialRAN}, and wireless power transfer~\cite{Burak_17_2019_WPTOverview},~\cite{Burak_5_2019_Vision6G}, as shown in Fig.~\ref{fig:6G_Technologies}. Even though all these technologies will significantly impact 6G, AI might be the leading innovator of 6G networks, supporting a completely new outlook for wireless networks, particularly in the PHY.

{The aforementioned current technologies optimize several network functions by using model-based techniques that provide characteristics of the policy involved. These techniques, yet, can be too sophisticated to be implemented physically in terms of run time, or they might contain too much abstraction to work in a broad environment. Conversely, AI-based solutions may adjust to dynamically changing scenarios and localized characteristics by absorbing knowledge of the target communication environment. AI can be implemented in a highly configurable infrastructure with a wide range of network flexibility. AI has the ability to forecast certain limits and handle vast amounts of data so that AI-integrated 6G will assist in processing the final volume of metadata with less resources and computational burden.}

AI has a long and successful history, dating back to approximately 1940s when neural networks (\acrshort{nn}) first appeared in intelligent systems. Machine learning (\acrshort{ml}) and deep learning (\acrshort{dl}) approaches have been proven to be competitive frameworks over the years for challenging tasks such as computer vision, robotics, and natural language processing, in which building a concrete mathematical model is relatively difficult. For example, although it is almost mission impossible to develop an analytical model or robust algorithm for detecting handwritten digits (digit recognition) or different objects in an image (image segmentation), DL techniques can accomplish these tasks with a performance exceeding human level. On the other hand, communication technologies hinge on numerous mathematical models and theories requiring expert knowledge, such as information theory and channel modeling. Information signals stream from optimal transmitter designs with modulation, coding, and signaling schemes over a range of mathematically defined channel models to be reliably detected at the receiver, in which each block is optimized individually. In addition to this well-defined PHY architecture, the ability of secure transmission along with eliminating various hardware imperfections makes communications a complicated and mature field. Therefore, ML and DL techniques must reach lofty goals to outperform existing technologies and provide discernable advantages. Applying DL approaches to optimize the building blocks separately, ranging from modulation and coding schemes (\acrshort{mcs}) to symbol detection and channel estimation algorithms, would not produce much of a change~\cite{Burak_18_2017_AnIntroDLPHY}. End-to-end (\acrshort{e2e}) optimization of communication systems as a whole in complex scenarios, such as unknown channel models and ultra-high mobility conditions, is the most fascinating concept that we expect to see in 6G, and we call this \textit{PHY revolution}. Switching from the block structure of current PHY communications, where each block is responsible for a different signal processing task, to E2E optimization using DL methods represents a radical paradigm shift that will entirely transform 6G and beyond wireless communications.

\begin{figure*}
   \centering
   \includegraphics[width=14cm]{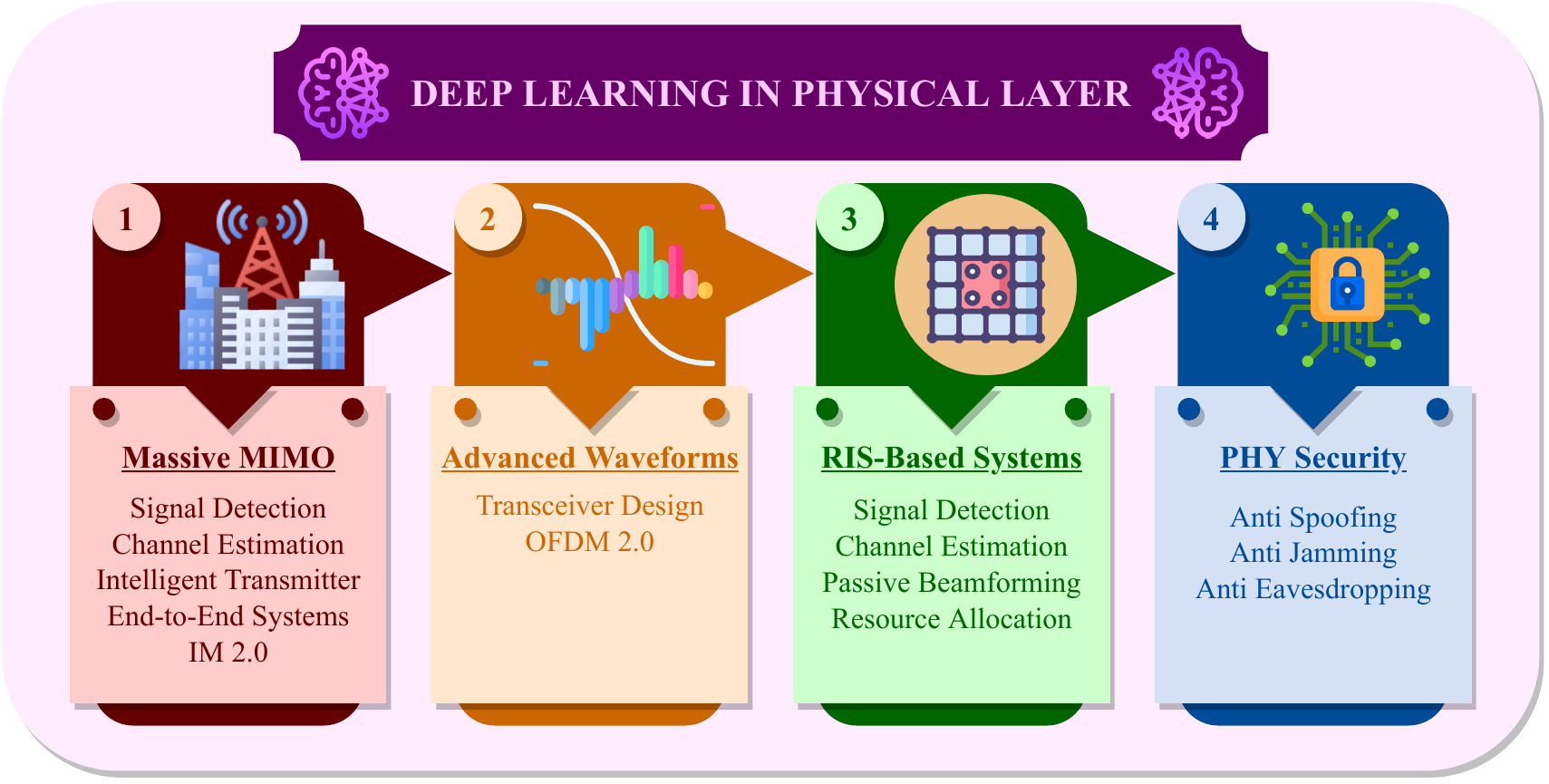}
   \caption{Four leading research directions in 6G on which this comprehensive survey article investigates the DL solutions.}
   \label{fig:FourMainConcepts}
\end{figure*}


Researchers have been studying the use of ML and DL techniques to overcome the drawbacks of current systems{,} particularly in the past couple of years. However, despite being exceptionally promising, DL methods experience some difficulties that restrict their viability in challenging communication scenarios. The highly-parameterized architecture of deep NNs (\acrshort{dnn}s) necessitates massive datasets to learn an appropriate mapping from features to desired outcomes, which increases the computing complexity and complicates the training process. Thus, utilizing solely data-based DNNs as a black box and leaving all predictions to model weights is questionable, at least for now, since datasets regarding communications have not reached a satisfactory level yet. Due to the limits of model-based and data-based methods, model-driven DNNs have emerged, which combine DNNs' powerful learning and mapping abilities with expert knowledge to get the most potential advantage. The two most common model-driven DNN strategies are deep unfolded networks, where DNN layers replicate iterations of an existing iterative algorithm, and hybrid networks, where DNNs help conventional models and enhance efficiency~\cite{Burak_19_2021_ModelBasedML}. As the DL literature on communication technologies has progressed, the data scarcity problem has also become a thing of the past, paving the way for entirely data-based DL systems.

\subsection{{Contributions}}
In this comprehensive survey article, we investigate DL applications from the perspective of four prominent and innovative 6G concepts in the PHY, including massive MIMO systems, advanced multicarrier (\acrshort{mc}) waveforms, RIS-empowered systems, and PHY security, as illustrated in Fig.~\ref{fig:FourMainConcepts}. We thoroughly present the overview of the up-to-date literature on DL for PHY design, reveal the progress made so far, and compare the considered models in terms of their key characteristics. Moreover, considering the limited coverage of and overwhelming need for the implementation and programming steps of DL concepts in the literature, we exhibit example implementations by providing simple code snippets to enlighten readers interested in this field. In this regard, we also hope that our article might be helpful for wireless researchers who {want} to have hands-on experience on DL-based wireless system design. {To our knowledge, this is the only survey in the literature that focuses on 6G from the perspective of PHY architectures and presents a complete guideline to PHY researchers.}

In particular, we first shed light on DL approaches that address the challenges in massive MIMO detection and channel estimation in order to overcome the drawbacks of traditional methodologies. Combining DL-based detection and channel estimation schemes, we present entirely intelligent large-scale MIMO receiver frameworks. Furthermore, we analyze how to replace the conventional signal processing blocks in a classical massive MIMO transmitter with DL structures to provide optimal transmission schemes with low complexity. Our long-term objective for intelligent massive MIMO systems is to integrate DL-based transmitters and receivers to develop innovative E2E communication systems even in unknown channel models. As a glaring enhancement for MIMO systems, IM techniques have attracted huge interest recently. Therefore, we also examine how to improve the efficiency of the existing IM techniques and present DL-empowered IM 2.0 solutions. We also provide an implementation example for a DL-based IM system with simplified code snippets to guide interested readers on programming a DL-based MIMO system in Python.

For MC communication systems, we primarily focus on the DL-based transceiver design to improve various performance metrics. In most studies, NNs have been designed to be implemented in the receiver side of various MC communication systems with the purpose of outperforming conventional algorithms. Data-driven DL-based receiver designs are introduced to perform channel estimation and signal detection tasks. Furthermore, by using expert knowledge, model-driven NNs have been presented to further improve the performance and decrease the computational complexity. Recent studies have focused on transceiver designs optimizing the transmitter and receiver pipeline as a whole by benefiting from the E2E approach, which is put forward as a promising direction. Moreover, an extensive literature review is provided to amend the weaknesses of OFDM, which is by far the most popular waveform in current standards. Particularly, DL techniques utilizing deep unfolding stand out as promising solutions to combat major problems of OFDM.

On the side of RIS-assisted communication systems, we investigate DL-based studies that present novel frameworks to overcome the existing drawbacks of earlier RIS-based systems. We primarily introduce DL approaches constituting intelligent channel estimation and signal detection processes. We conclude that employment of DL approaches requires less pilots and training overhead than conventional techniques. Furthermore, we focus on emerging DL-based NNs eliminating model dependency for phase configuration in RIS-assisted schemes. The presented passive beamforming designs prove that DL-based processes, which unveil a direct mapping between phase shifts and received signal strength, provide robustness in different channel environments with reduced computational complexity. We also review the recent studies that present standalone operations employing fully passive elements to suit the passive nature of RISs and reduce the training overhead. Besides, due to the growing interest in non-orthogonal multiple access (\acrshort{noma}) solutions for next-generation communication systems, we examine DL-based systems improving clustering and power allocation performance of RIS-assisted NOMA designs. Thereby, we discuss that the utilization of DL approaches provides enhanced flexibility and robust performance under dynamic states and {the} number of users compared to traditional methodologies.

\begin{figure}
   \centering
   \includegraphics[width=\linewidth]{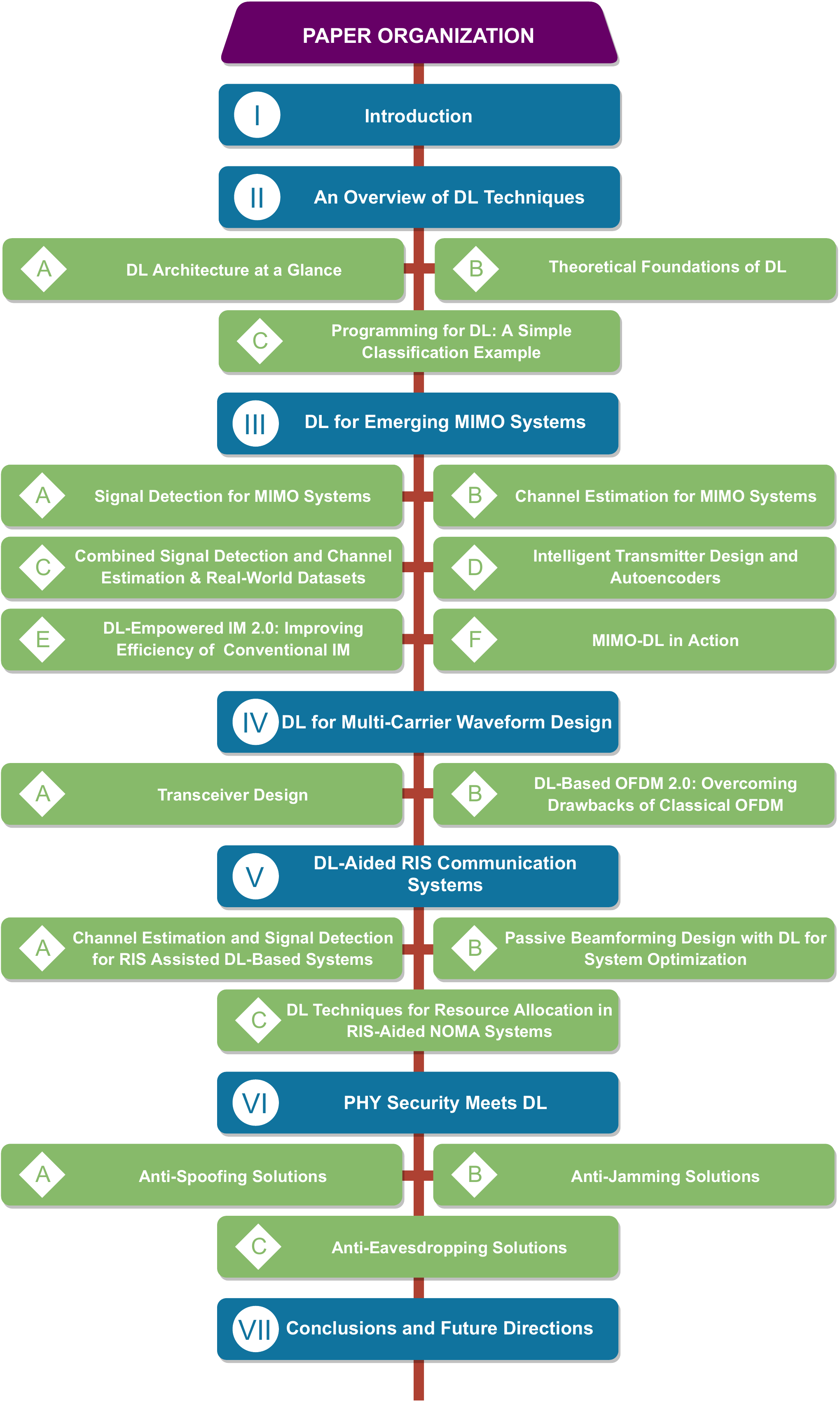}
   \caption{Organization of the article.}
   \label{fig:PaperOrganization}
\end{figure}


Finally, we focus on DL-assisted PHY security approaches and investigate the existing studies in the literature. DL is implemented to PHY security schemes from three main perspectives based on the attack type: spoofing, jamming, and eavesdropping. Against spoofers, DL is mostly used at receivers to improve authentication accuracy. Especially, DL improves the legitimate/illegitimate classification problem of the authentication test and brings robustness against unknown and varying environments. DL is also implemented for anti-jamming purposes. However, most of the anti-jamming methods employ reinforcement learning to avoid jammed frequencies and apply DL for the purpose of better Q-functions approximation. Moreover, various secure communication systems implemented DL algorithms in their system models to improve the secrecy rate against eavesdroppers. Particularly, we reveal that these studies employ data-driven models to generate coding or beamforming schemes which {minimize} the leaked information to eavesdropper.

\subsection{{Organization}}

The organization of this article is given in Fig.~\ref{fig:PaperOrganization}. In Section II, we briefly discuss the structure of DL and introduce basic DL techniques from both theoretical and programming perspectives. We begin our extensive literature review with emerging MIMO technologies in Section III and provide an implementation example for a DL-based MIMO system model. Section IV presents  DL solutions for sophisticated MC waveforms. In Section V, we delve into DL-aided RIS communication systems. We discuss how DL methods can enable secure communications in Section VI. Finally, we conclude the article with Section VII by providing our future perspectives.

\section{An Overview of DL Techniques} \label{sec:overview}
In this section, we will discuss the general structure of DL models and explain the training procedures of NNs to provide a quick overview of DL techniques. We will emphasize fundamental concepts and parameters that form DL models to facilitate understanding {of} proposed methods in the literature. We will also dig into the details of commonly used DL techniques, namely DNN, convolutional NNs (\acrshort{cnn}), and recurrent NNs (\acrshort{rnn}), by providing their theoretical background and motivations. In addition, we will discuss the applications of each technique in wireless networks. We will conclude this section with an image classification task using a simple CNN architecture and the well-known CIFAR10 dataset~\cite{Burak_20_2009_CIFAR10Paper},~\cite{Burak_21_CIFAR10Website}.

\subsection{DL Architecture at a Glance}
\begin{figure}
   \centering
   \includegraphics[width=7cm]{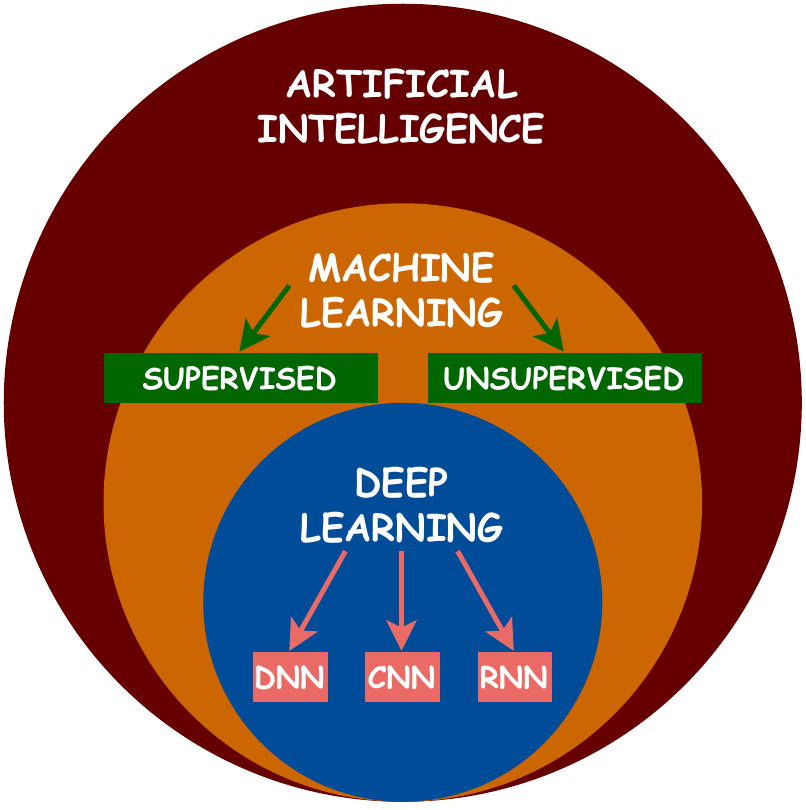}
   \caption{Demonstration of the relationship between DL, ML, and AI.}
   \label{fig:AI_ML_DL}
\end{figure}

DL is a sub-branch of ML, with multiple stacked processing layers, that {enable} predictions, classifications, or other decisions by learning data representations. Given raw data, in contrast to classical ML methods that strongly depend on domain-expert features, DL models generate non-linear input-output mappings to execute actions on a goal objective. Fig.~\ref{fig:AI_ML_DL} illustrates the relationship between DL, ML, and AI. Each layer in an NN architecture transforms a representation to a higher level, where each transformation unlocks the learning of a more sophisticated characteristic of raw data. A model may be shallow or deep depending on the number of layers, the depth of an NN, imitating either a simplistic or complicated function of raw input data by being sensitive to vital features and insensitive to irrelevant changes such as environment or nearby objects. Aside from the input and output layers, each layer is a hidden layer consisting of hidden units known as neurons. During {the} representation learning process, hidden neurons in a layer calculate a weighted sum of their preceding layer's inputs and transfer the result to the next layer through a non-linear activation function, corresponding to one iteration of forward-propagation. Starting with the input layer, which passes the weighted sum of raw input data to the first hidden layer, each hidden layer handles its computation. In the end, the output layer produces the desired outcomes, which might be probabilities, category scores, or any other metric. DL researchers have identified several activation functions to introduce a non-linearity effect into forward-propagation, such as the rectified linear unit (\acrshort{relu}), leaky ReLU, tanh, sigmoid, and more. The most popular one is the ReLU function and defined as $f(a) = \text{max}(a, 0)$~\cite{Burak_22_2015_DL},~\cite{Burak_23_2019_DLMobileNetwSurvey}.

\begin{figure}
   \centering
   \includegraphics[width=\linewidth]{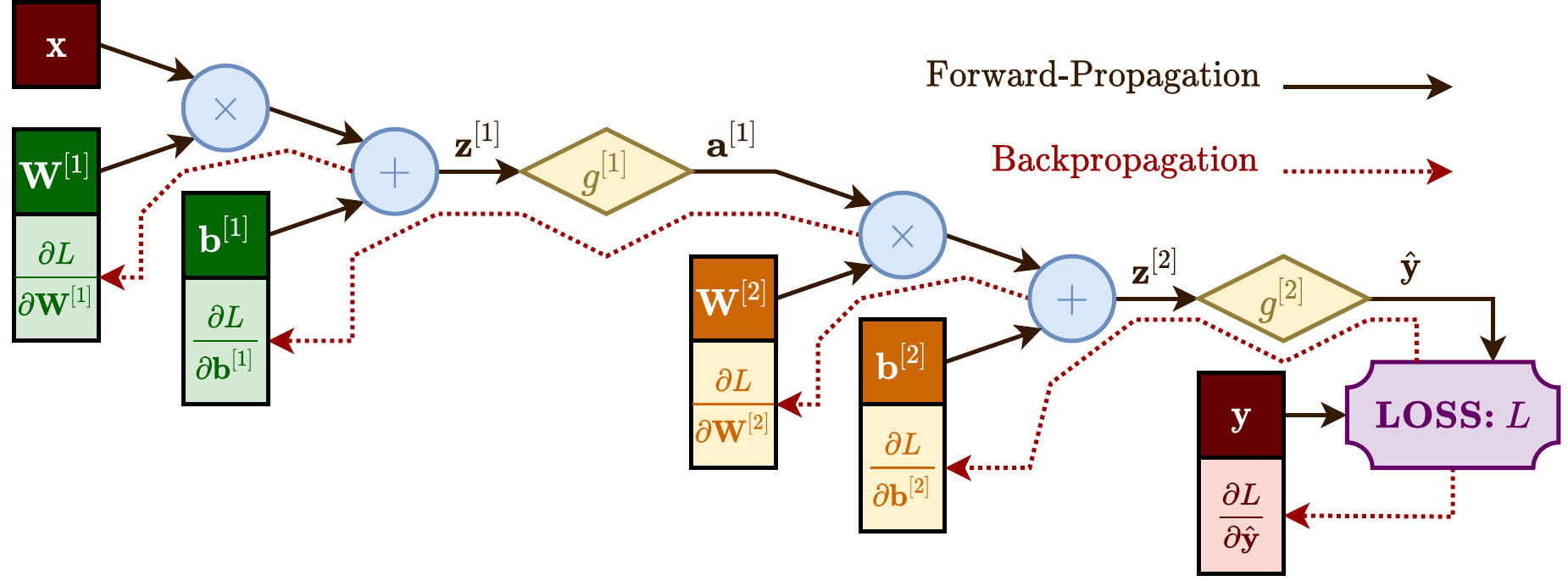}
   \caption{Computation graph of a single hidden layer DNN including forward-propagation, loss calculation, and backpropagation.}
   \label{fig:DL_ComputationGraph}
\end{figure}

When an NN completes a forward-propagation iteration, the algorithm computes a loss function by comparing the output scores with desired labels. This loss function represents {a} model error, which is subsequently utilized to update internal parameters, also known as model weights. {The} update of these adjustable weights corresponds to {the} training of {the} DL model. During training, the algorithm calculates the derivative of {the} loss function with respect to each weight to observe the amount of resultant error when weight was increased or decreased. In the opposite direction of forward-propagation, the algorithm computes weight gradients using the chain rule~\cite{Burak_24_1986_LearningBackPropErrors}, which obtains derivatives going backward, called backpropagation. DL model updates its weights by changing them in the reverse direction of calculated gradients to decrease {the} error value after backpropagation~\cite{Burak_22_2015_DL}. {The} update rule is one of the most critical aspects of DL applications since it highly impacts overall learning performance and training time. Thus, researchers have developed different optimizers, including momentum~\cite{Burak_25_1999_MomentumGD}, root mean square propagation, stochastic gradient descent (\acrshort{sgd})~\cite{Burak_26_2008_StochasticApproxMethod}, adaptive moment estimation (\acrshort{adam})~\cite{Burak_27_2015_ADAM}, and many more. Recently, Facebook Research has published a new update algorithm, called MADGRAD~\cite{Burak_28_MADGRADGitHub}, and the DL world is debating whether this new technique can replace the ADAM optimizer. Fig.~\ref{fig:DL_ComputationGraph} demonstrates a single run of a training procedure including forward-propagation, loss calculation, and backpropagation for a single hidden layer DNN, where the given variables will be defined shortly. A typical DL model takes hundreds of millions of these iterations to ensure that the model observes a good deal of instances and that the model error converges to a reasonable value depending on the application. A testing procedure follows training to examine whether the model can generalize to unseen occurrences, also known as the compositionality of DL models. The difference between training and test loss values represents model variance and indicates that {the} model is overfitting, in which DL model memorizes rather than learning. On the other hand, the difference between training and the Bayes error values corresponds to model bias and reveals that {the} model is underfitting, which necessitates more training. It is a common mistake to investigate model bias by comparing training error with zero error. Indeed, the Bayes error, the known optimal value that can be human-level performance or other values, should be the benchmark for bias, which is not necessarily zero.

We have provided a brief introduction to DL structures and training procedures, including insights of forward-propagation and backpropagation so far. In the following, we will go over the theoretical background of three frequently used DL architectures: DNN, CNN, and RNN, and present the implementation of a simple image classification task.

\begin{figure*}
   \centering
   \includegraphics[width=\textwidth]{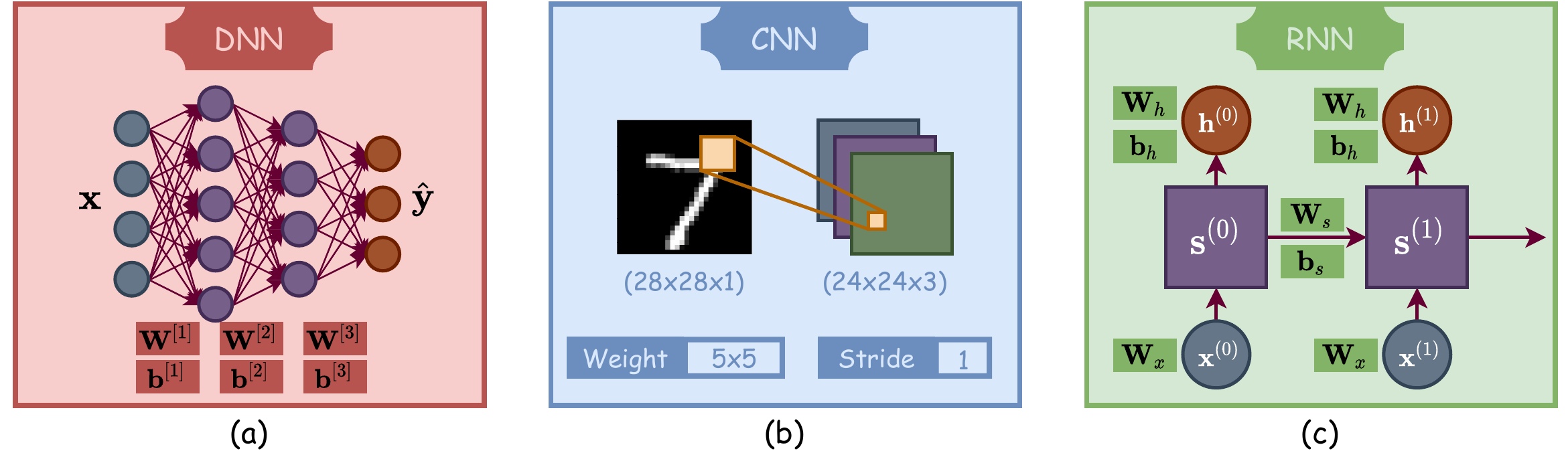}
   \caption{Three common DL architectures: (a) DNN, (b) CNN, (c) RNN.}
   \label{fig:DNN_CNN_RNN}
\end{figure*}


\subsection{Theoretical Foundations of DL}
We will begin by analyzing DNNs, which are by far the most common DL structures used in DL applications. A DNN consists of multiple stacked layers, with each neuron connected to all neurons in the preceding and the following layers creating a fully connected structure. The number of neurons in the input and output layers equal{s} the number of features in the raw input data and the number of categories, classes, or other targets according to the application, respectively. Fig.~\ref{fig:DNN_CNN_RNN}(a) depicts the architecture of a DNN with two hidden layers and three classes. Each link between two neurons represents a real-valued weight that forms the basis of the learning mechanism. It is critical to carefully initialize model weights to optimize learning performance and avoid vanishing or exploding gradients. A widespread initialization method is {the} random generation of weight matrices with uniform or Gaussian distributed elements. Another popular approach is Xavier initialization~\cite{Burak_29_2010_DeepFNN}, developed by Xavier Glorot and Yoshua Bengio, in which randomly generated weights between layers $[l-1]$ and $[l]$ are scaled by $\sqrt{2 / (n^{[l-1]} + n^{[l])}}$, where $n^{[l]}$ is the number of neurons in the $l^{\text{th}}$ layer. The forward-propagation of the DNN given in Fig.~\ref{fig:DNN_CNN_RNN}(a) for a single instance is given as follows:
\begin{equation} \label{DNN_FP}
    \begin{aligned}
        \text{Hidden-1: } \mathbf{z}^{[1]} &= \mathbf{W}^{[1]} \mathbf{x} + \mathbf{b}^{[1]}, \quad \mathbf{a}^{[1]} = g^{[1]}(\mathbf{z}^{[1]}),
        \\
        \text{Hidden-2: } \mathbf{z}^{[2]} &= \mathbf{W}^{[2]} \mathbf{a}^{[1]} + \mathbf{b}^{[2]}, \quad \mathbf{a}^{[2]} = g^{[2]}(\mathbf{z}^{[2]}),
        \\
        \text{Output: } \mathbf{z}^{[3]} &= \mathbf{W}^{[3]} \mathbf{a}^{[2]} + \mathbf{b}^{[3]}, \quad \hat{\mathbf{y}} = g^{[3]}(\mathbf{z}^{[3]}).
    \end{aligned}
\end{equation}
In (\ref{DNN_FP}), $\mathbf{x}$ and $\hat{\mathbf{y}}$ represent $(4 \times 1)$-dimensional raw input data and $(3 \times 1)$-dimensional class scores, respectively, while $\mathbf{W}^{[l]}$, $\mathbf{b}^{[l]}$, $\mathbf{z}^{[l]}$, $g^{[l]}$, and $\mathbf{a}^{[l]}$ correspond to $(n^{[l]} \times n^{[l-1]})$-dimensional weight matrix, $(n^{[l]} \times 1)$-dimensional bias vector, $(n^{[l]} \times 1)$-dimensional weighted sum, the activation function, and $(n^{[l]} \times 1)$-dimensional layer output of the $l^{\text{th}}$ layer, respectively. The activation function $g$ is non-linear since using linear activation functions eliminates the effect of stacking multiple layers and results in the same performance as the model with a single output layer, as indicated in (\ref{DNN_FP_LinearActivation}) where the activation function is $g(x) = x$:
\begin{equation} \label{DNN_FP_LinearActivation}
    \begin{split}
        \hat{\mathbf{y}} &= \mathbf{W}^{[3]}(\mathbf{W}^{[2]}(\mathbf{W}^{[1]} \mathbf{x} + \mathbf{b}^{[1]}) + \mathbf{b}^{[2]}) + \mathbf{b}^{[3]}
        \\
        &= \bar{\mathbf{W}} \mathbf{x} + \bar{\mathbf{b}},
    \end{split}
\end{equation}
where $\bar{\mathbf{W}}=\mathbf{W}^{[3]} \mathbf{W}^{[2]} \mathbf{W}^{[1]}$ and $\bar{\mathbf{b}}=\mathbf{W}^{[3]} \mathbf{W}^{[2]} \mathbf{b}^{[1]} + \mathbf{W}^{[3]} \mathbf{b}^{[2]} + \mathbf{b}^{[3]}$. As we discussed earlier, ReLU is a suitable option for hidden layers, while the output layer employs the softmax activation function, given by (\ref{softmax}), to get output scores:
\begin{equation} \label{softmax}
    \mathbf{softmax}(\hat{y}_{i}) = \frac{e^{\hat{y}_{i}}}{\sum_{c=1}^{3} e^{\hat{y}_{c}}}.
\end{equation}
Here, $\hat{y}_{i}$ is the score of $i^{\text{th}}$ class. {Yet, a DL network's output layer need not always employ a softmax activation function. This is dependent on the work that should be handled using the DL technique, in fact. The addressed problem that is being handled shapes the selection of the output's layer activation function. While any regression problem best fits with linear activation function, classification problems require different types of activation functions at the output layer, such as sigmoid and softmax.}

 When the algorithm obtains the output scores by forward-propagation, it calculates the loss value depending on these scores. The loss function might vary according to the application, and we will use the categorical cross-entropy loss function for this classification model:
\begin{equation} \label{categorical_cross_entropy_loss}
    L(\mathbf{y},\hat{\mathbf{y}}) = - \sum_{c=1}^{3} y_{c} \text{log}(\hat{y}_{c}),
\end{equation}
where $\mathbf{y}$ is the one-hot vector of the ground truth labels consisting of all zeros except one in the index of the correct class. The next step is backpropagation, in which the gradients are computed by taking the derivative of the loss function with respect to the weights and biases. (\ref{chain_rule}) provides the calculation of the gradients of the first hidden layer's weight matrix and bias vector, using the well-known chain rule:
\begin{equation} \label{chain_rule}
    \begin{aligned}
        \frac{\partial L}{\partial \mathbf{W}^{[1]}} &= 
        \frac{\partial L}{\partial \hat{\mathbf{y}}} 
        \frac{\partial \hat{\mathbf{y}}}{\partial \mathbf{z}^{[3]}} 
        \frac{\partial \mathbf{z}^{[3]}}{\partial \mathbf{a}^{[2]}}
        \frac{\partial \mathbf{a}^{[2]}}{\partial \mathbf{z}^{[2]}}
        \frac{\partial \mathbf{z}^{[2]}}{\partial \mathbf{a}^{[1]}}
        \frac{\partial \mathbf{a}^{[1]}}{\partial \mathbf{z}^{[1]}}
        \frac{\partial \mathbf{z}^{[1]}}{\partial \mathbf{W}^{[1]}},
        \\[5pt]
        \frac{\partial L}{\partial \mathbf{b}^{[1]}} &= 
        \frac{\partial L}{\partial \hat{\mathbf{y}}} 
        \frac{\partial \hat{\mathbf{y}}}{\partial \mathbf{z}^{[3]}} 
        \frac{\partial \mathbf{z}^{[3]}}{\partial \mathbf{a}^{[2]}}
        \frac{\partial \mathbf{a}^{[2]}}{\partial \mathbf{z}^{[2]}}
        \frac{\partial \mathbf{z}^{[2]}}{\partial \mathbf{a}^{[1]}}
        \frac{\partial \mathbf{a}^{[1]}}{\partial \mathbf{z}^{[1]}}
        \frac{\partial \mathbf{z}^{[1]}}{\partial \mathbf{b}^{[1]}}.
    \end{aligned}
\end{equation}
The final stage of the single training pass of the DNN model performs the update of the model weights and biases. Using the SGD optimization method, the algorithm updates the weight matrix and the bias vector of the $l^{\text{th}}$ layer by changing them in the opposite direction of their gradients by their gradients scaled by a learning constant $\alpha$:
\begin{equation} \label{SGD_update}
    \begin{aligned}
        \mathbf{W}^{[l]} &:= \mathbf{W}^{[l]} - \alpha \frac{\partial L}{\partial \mathbf{W}^{[l]}}, \\
        \mathbf{b}^{[l]} &:= \mathbf{b}^{[l]} - \alpha \frac{\partial L}{\partial \mathbf{b}^{[l]}}.
    \end{aligned}
\end{equation}

The second well-known DL architecture is CNN, also known as ConvNets. A CNN architecture presumes similarity between various areas of a 2D feature vector and aims at capturing correlations between these parts, as shown in Fig.~\ref{fig:DNN_CNN_RNN}(b)~\cite{Burak_22_2015_DL},~\cite{Burak_23_2019_DLMobileNetwSurvey},~\cite{Burak_129_KnetCNN}. The first key idea underlying the motivation of the CNN architecture is local connections. In contrast to DNNs, where each hidden unit links with all hidden units of the former and latter layers, CNNs connect a single unit in a convolutional layer to a local patch in the previous layer. Raw input data of a CNN model consists of multiple 2D feature vectors corresponding to channels. For example, a $28 \times 28$ RGB image comprises three 2D feature vectors representing the pixels in the RGB color channels. It is most likely that local regions of an RGB image are correlated, and statistics are comparable, which means a pattern appearing in an area might also come out of another region, such as two ears of a cat in different parts of an image~\cite{Burak_22_2015_DL},~\cite{Burak_129_KnetCNN}. Therefore, it is logical to utilize these spatial relationships and link a patch of input data to a hidden unit rather than a fully connected structure~\cite{Burak_129_KnetCNN}.

The second major characteristic that forms the CNN baseline is weight sharing. Each local patch in the 2D feature vector connects to a distinct hidden unit in the following convolutional layer, where all links use the same weight matrix. The motivation behind weight sharing is to decrease the number of learnable parameters. The example given in~\cite{Burak_129_KnetCNN} describes this motivation perfectly. Let us assume that we have $482 \times 415$ input data with three channels, where the input size is $600090$. Considering the dense connection of DNNs, the parameter size is approximately $600$K for a single hidden unit in the next layer, increasing to more than $600$M for $1000$ hidden units. Many parameters increase the training time substantially, crack the available memory, and necessitate powerful regularizers to prevent overfitting. On the other hand, a $40 \times 40$ shared weight matrix spanning local patches provides $4800$ learnable parameters, allowing $125$ hidden layers with $1000$ units each against a single hidden layer in the dense connection. As a result, the number of parameters and model size reduce significantly along with {the} enhancement of the generalization capacity, preventing memorizing data rather than learning.

\begin{figure}
   \centering
   \includegraphics[width=\linewidth]{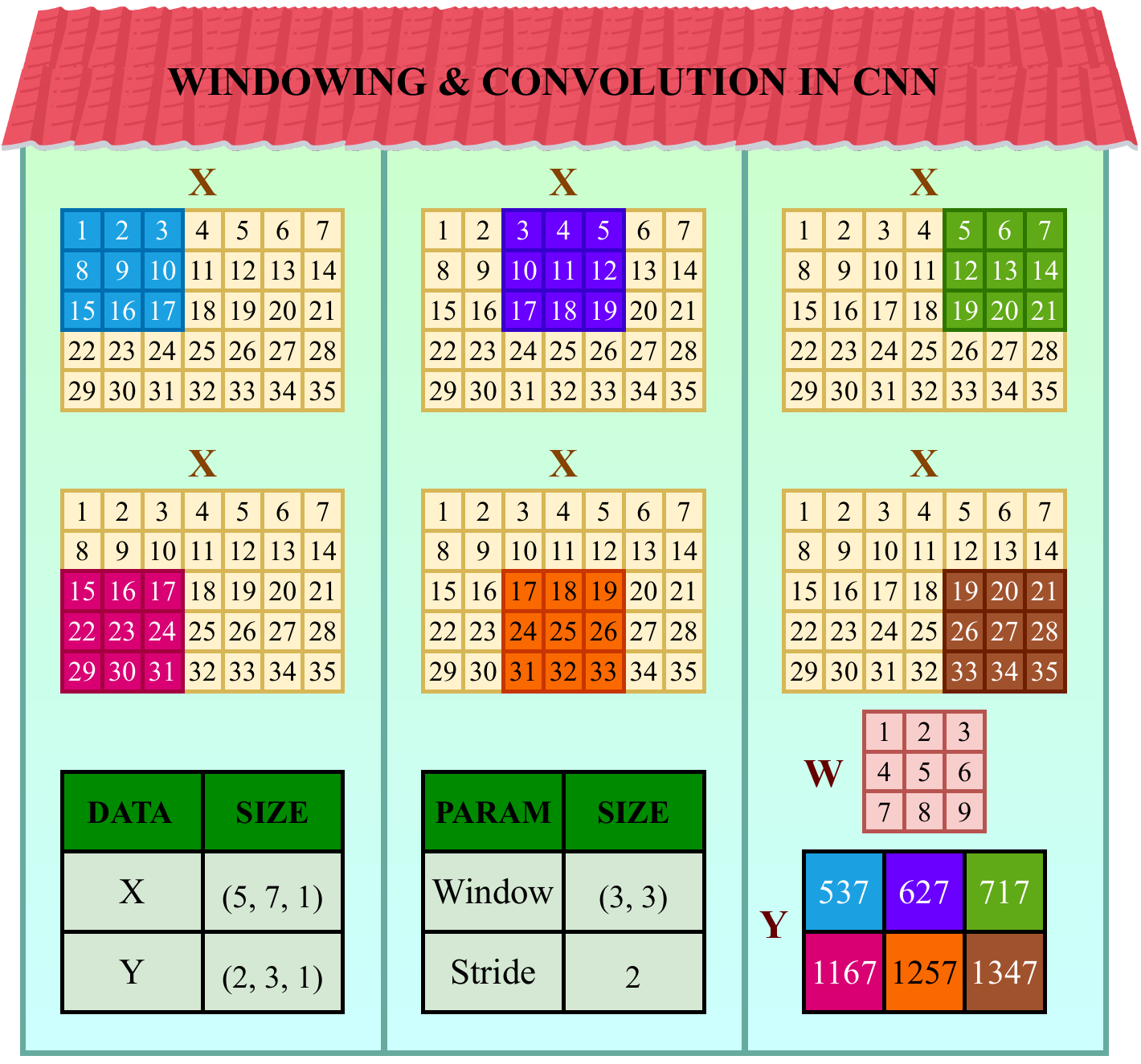}
   \caption{Windowing procedure and the convolution operation of a CNN model.}
   \label{fig:CNN_Windowing}
\end{figure}

The primary operation of CNN forward-propagation is convolution, which links local regions to hidden units. Specifically, a CNN model calculates the weighted sum of a patch in the 2D feature vector by the convolution operation to pass the patch's features to a single hidden unit of the next layer. A single patch represents the window of the CNN forward-propagation whose size determines the size of the shared weight matrix, and CNN travels across all windows to move on to the following layer. These windows might, and most probably, intersect, in which the shifting size corresponds to the stride of the current convolutional layer. A simple example of this windowing procedure and the convolution operation is given in Fig.~\ref{fig:CNN_Windowing}. In this example, the $5 \times 7$ input data $\mathbf{X}$ and the $2 \times 3$ output data $\mathbf{Y}$ both include a single channel; however, they might differ depending on the model architecture. When there are more channels in the input data, the $3 \times 3$ shared weight spans other channels in the same way as shown in Fig.~\ref{fig:CNN_Windowing}. The dimensions of the input and output data may dispute according to window size and stride value. Therefore, amending a padding layer composed of all zeros to the input data equalizes the dimensions, if necessary.


A pooling layer follows the convolutional layer in a typical CNN model architecture, which computes a single statistic, such as the average or the maximum, of the input patches~\cite{Burak_129_KnetCNN}. Pooling layers enable the CNN model to become robust to minor distortions without requiring additional learnable parameters~\cite{Burak_22_2015_DL},~\cite{Burak_129_KnetCNN}. The stride value of a pooling layer is generally as large as the window size. The initial stages of a CNN model are typically composed of stacked convolution, non-linear activation function, and pooling, followed by further convolutional and dense layers~\cite{Burak_22_2015_DL}. The loss calculation and backpropagation procedures are the same as that of DNNs, where the algorithm computes the gradients of the weights by taking the derivative of the loss with respect to the weights.

The DL techniques that we have examined so far suffer from a common shortcoming of being memoryless, where each layer's output depends only on its input and the corresponding weights without considering the past. This is where RNNs come into play for applications including sequential data, such as speech recognition and natural language processing, thanks to their ability to utilize previous parts. RNNs process an input sequence with correlations between the elements one sample at a time, and the outputs not just depend on the input and weights but also the past information, as given in Fig.~\ref{fig:DNN_CNN_RNN}(c)~\cite{Burak_22_2015_DL},~\cite{Burak_23_2019_DLMobileNetwSurvey},~\cite{Burak_130_KnetRNN}. The architecture of an RNN might vary according to the application. For example, each layer might yield an output, like in Fig.~\ref{fig:DNN_CNN_RNN}(c), for a sequence generation application such as image captioning or a single outcome for a sequence classification such as assessing the sentiment of a product review if it is positive or negative~\cite{Burak_130_KnetRNN}. For a sequential input data $\mathbf{X}$, such as a sentence or a speech, a typical RNN's hidden layer performs the following forward-propagation:
\begin{equation} \label{RNN_forward_propagation}
    \begin{aligned}
        \mathbf{s}^{(t)} &= \text{tanh}(\mathbf{W}_{x} \mathbf{x}^{(t)} + \mathbf{W}_{s} \mathbf{s}^{(t-1)} + \mathbf{b}_{s}),
        \\
        \mathbf{h}^{(t)} &= \mathbf{W}_{h} \mathbf{s}^{(t)} + \mathbf{b}_{h}.
    \end{aligned}
\end{equation}
Here, $\mathbf{s}^{(t)}$ is the internal state of the model at time $t$, which forms the memory unit by containing the information about the past elements, $\mathbf{W}_{x}$ and $\mathbf{W}_{s}$ are the weights that connect the $t^{\text{th}}$ input sample $\mathbf{x}^{(t)}$ and the previous state $\mathbf{s}^{(t-1)}$ to the current state, respectively, $\mathbf{b}_{s}$ is the bias of $\mathbf{s}^{(t)}$, and $\mathbf{b}_{h}$ is the bias of the layer output $\mathbf{h}^{(t)}$. Similar to the weight sharing procedure in a CNN, the hidden layers of an RNN share the weights and biases, which prevent overfitting and decrease the model complexity~\cite{Burak_23_2019_DLMobileNetwSurvey}.

The training procedure of an RNN resembles that of DNNs and CNNs, where the RNN model serves as a multi-layer feed-forward NN with each layer corresponding to a time step, shared weights across these time steps, and single or multiple outputs according to the application. Following the loss calculation, the RNN model is trained via the backpropagation through time (\acrshort{bptt}) algorithm, which is the SGD unfolded in time~\cite{Burak_23_2019_DLMobileNetwSurvey},~\cite{Burak_130_KnetRNN}. However, the BPTT algorithm is challenging since it causes gradient vanishing that inhibits information delivery or exploding that paralyzes the training. The reason for this is repeated multiplication with the weight matrix during the backward pass. Researchers have suggested several solutions for this problem, such as initializing weights from a previously trained model, gradient clipping that prevents gradients from exceeding a threshold, and sophisticated optimization techniques like ADAM~\cite{Burak_130_KnetRNN}. As a powerful alternative to these solutions, the Long Short-Term Memory (\acrshort{lstm}) networks are designed to address gradient vanishing or exploding issues along with providing longer memorization than a classical RNN by enhancing the network with explicit memory. Let us consider a language model aiming at predicting the upcoming word based on the formers~\cite{Burak_131_ColahLSTM}. RNNs perform well in anticipation when the gap between relevant information and the prediction point is narrow:
\begin{align*}
    \textit{The }
    \underbrace{\textit{\fcolorbox{orange}{table_red}{clouds} are in the \underline{\textcolor{red}{\textbf{sky}}}.}}_{\textbf{{\normalsize \textcolor{table_green}{Small Gap}}}}
\end{align*}
Unfortunately, RNNs can not find the relevant information as the distance increases, and it is possible for the distance becomes considerably large. In such cases, it is beneficial to employ LSTM networks to learn long-term dependencies:
\begin{align*}
    \textit{I'm from }
    \underbrace{\textit{\fcolorbox{orange}{table_red}{France} and I'm good at speaking fluent \underline{\textcolor{red}{\textbf{French}}}.}}_{\textbf{{\normalsize \textcolor{table_red}{Large Gap}}}}
\end{align*}

According to our comprehensive research, DNNs are by far the most popular DL architectures used in the PHY design of wireless communications. Specifically, DNNs are utilized for signal detection, channel estimation, peak-to-average power ratio (\acrshort{papr}) reduction in MC waveform designs, and passive beamforming designs in RIS-based systems. CNN-based models also have an immense role in the literature of DL-based PHY designs. There are use-cases like modulation recognition, signal classification, removing cyclic prefix (\acrshort{cp}) in MC waveform designs, and anti-spoofing \& anti-jamming in PHY security schemes.
{It is also worth mentioning that a CNN does not have to utulize 2D feature vectors.  There are several studies using 1D CNNs in the literature. Particularly, the literature in wireless communication widely benefits from 1D CNNs in numerous tasks such as network management, traffic analysis so on \cite{lotfollahi2020deep},\cite{montieri2021packet}, \cite{lopez2017network}.}

RNNs find a moderate pace in DL-based PHY methods, with certain applications for MC waveform designs under high mobility and for anti-eavesdropping PHY security schemes. Considering future wireless networks, DNNs, CNNs, and RNNs have {the} potential to be utilized for modeling multi-attribute mobile data, spatial mobile data analysis, and temporal data modeling, respectively~\cite{Burak_23_2019_DLMobileNetwSurvey}. Once we understand the theoretical background and major applications of the most popular DL architectures, we may now delve into DL programming with Python. Despite presenting an image classification application is not the primary focus of this work, we provide a simple example here to illustrate the basics of DL programming.

\subsection{Autoencoders and Generative Adversarial Networks }

\begin{figure}
   \centering
   \includegraphics[width=\linewidth]{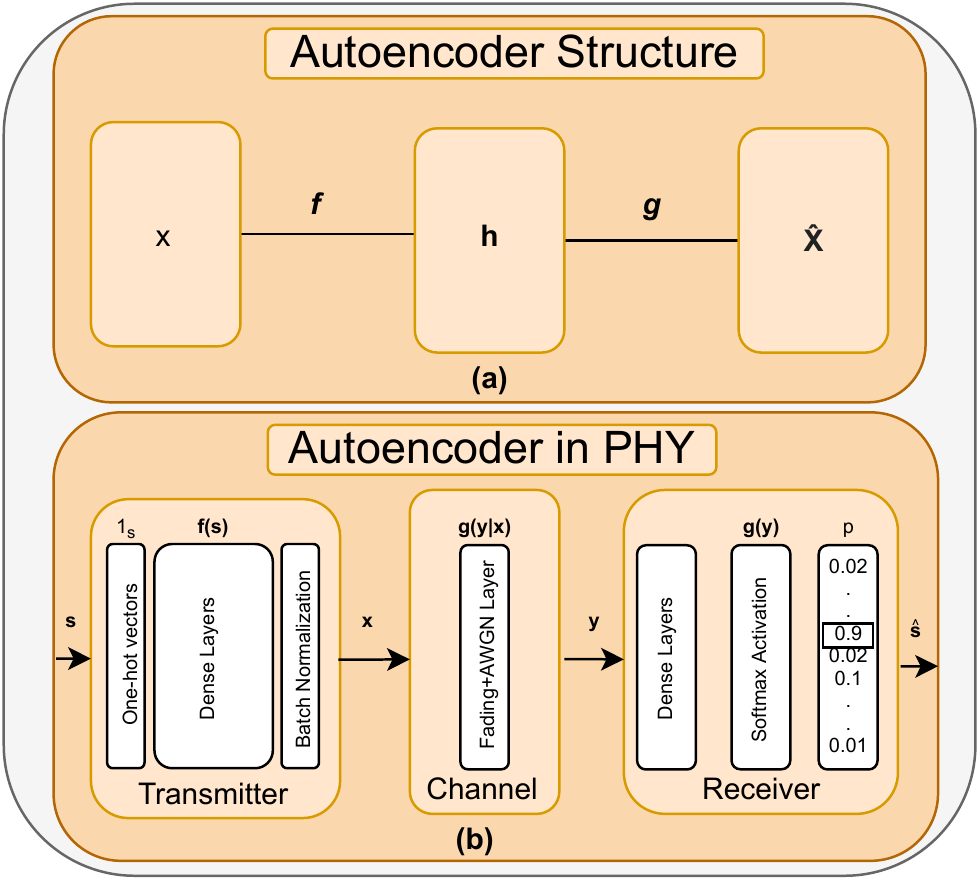}
   \caption{An autoencoder scheme (a) mapping  the input x to the output x where function f and g  stands for encoder and decoder, respectively (b) from wireless communication perspective}
   \label{fig:AE_Scheme}
\end{figure}

We also elaborate and present the studies pursuing end-to-end learning schemes in this survey. Since the end-to-end learning concept shows a significant contribution to performance enhancement in wireless communication schemes, several studies have been published using this still maturing autoencoder approach. 
An autoencoder is a kind of neural network that is trained to pass its input to the output the same. The network can be divided into two components:
\begin{equation} \label{encoder}
    \begin{aligned}
\mathbf{h}=\mathbf{f(x),}
    \end{aligned}
\end{equation}
\begin{equation} \label{decoder}
    \begin{aligned}
\mathbf{\hat{x}}=\mathbf{g(h).}
    \end{aligned}
\end{equation}
In (\ref{encoder}), f(x) function defines the encoding process where x is the input data to be converted to h code. Also, g(h) function in (\ref{decoder}) defines the decoding process converts this h code to the output $\mathbf{\hat{x}}$ with the g function. The simple architecture of an autoencoder is shown in Fig.  \ref{fig:AE_Scheme}(a). An autoencoder is not particularly effective if it simply learns the set $\mathbf{g(f(x))=x}$ all the time. Conversely, autoencoders are made so that they cannot be perfect at copying. They are typically regulated so that they can only duplicate input that closely resembles the training data and only with some degree of accuracy. The model frequently learns beneficial features of the data because it must decide which aspects of the input should be duplicated. The concept of an encoder and a decoder has been expanded by modern autoencoders to include stochastic mappings in addition to deterministic functions.

{Through a sequence of effective parametric linear algebra operations, autoencoders in wireless communication, as shown in Fig.  \ref{fig:AE_Scheme}(b), enable the physical layer transmitter and receiver to take the form of essentially unconstrained mappings. By attempting to reduce BER corresponding reconstruction loss, deep learning allows us to arrive at a solution to the whole communications system design challenge. This approach is capable of being applied over a wide range of channels and impairment models to provide better-suited solutions. Tailor-made waveforms may accomplish innovative and unparagoned performance under challenging channel conditions by learning physical layer information (encoder, decoder) and representation solutions in this end-to-end manner.}

{A generative adversarial network (GAN) resembles an inside-out autoencoder in appearance. Yet, unlike autoencoders compressing input data, GANs convert their low-dimensional input into high-dimensional data at the inner network. Using two neural networks in competition with one another (adversarial), GANs are computational structures that produce new, artificial instances of data that can be misinterpreted for genuine data. They are extensively utilized in the creation of images, videos, and voices. 
The neural network named "Generator" creates new data instances while they are authenticated by another neural network named "Discriminator" as shown in Fig. \ref{fig:GAN}. Each sample of data that is reviewed by the discriminator is evaluated to determine if it is a part of the training dataset or not.
The primary distinction between Autoencoders and GANs is the way that learn. Autoencoders may be thought of as addressing a semi-supervised learning issue since they aim to reproduce a picture with the least amount of loss possible. In other respects, GAN handles problems involving unsupervised learning. The training period for the two approaches could be treated as the most significant distinction. The required training time for GAN is longer; nevertheless, autoencoders provide less credible results (such as decreased image quality). GANs were thus considered and found to be far more stable.}

\begin{table}
    \caption{Labels and descriptions of the categories in the CIFAR10 dataset.}

    \centering
    \begin{tabular}{|M{1.7cm}|M{1.7cm}||M{1.7cm}|M{1.7cm}|}
        \hline
        
        \cellcolor {table_blue}
        \textbf{\textcolor{table_white}{Label}} & 
        \cellcolor{table_blue}
        \textbf{\textcolor{table_white}{Description}} &
        \cellcolor{table_blue}
        \textbf{\textcolor{table_white}{Label}} & 
        \cellcolor{table_blue}
        \textbf{\textcolor{table_white}{Description}} \\
        \hline
        
        \cellcolor{table_white} \textcolor{table_blue}{$0$} &
        \cellcolor{table_white} \textcolor{table_blue}{Airplane} &
        \cellcolor{table_white} \textcolor{table_blue}{$5$} &
        \cellcolor{table_white} \textcolor{table_blue}{Dog} \\
        \hline
        
        \cellcolor{table_white} \textcolor{table_blue}{$1$} &
        \cellcolor{table_white} \textcolor{table_blue}{Automobile} &
        \cellcolor{table_white} \textcolor{table_blue}{$6$} &
        \cellcolor{table_white} \textcolor{table_blue}{Frog} \\
        \hline
        
        \cellcolor{table_white} \textcolor{table_blue}{$2$} &
        \cellcolor{table_white} \textcolor{table_blue}{Bird} &
        \cellcolor{table_white} \textcolor{table_blue}{$7$} &
        \cellcolor{table_white} \textcolor{table_blue}{Horse} \\
        \hline
        
        \cellcolor{table_white} \textcolor{table_blue}{$3$} &
        \cellcolor{table_white} \textcolor{table_blue}{Cat} &
        \cellcolor{table_white} \textcolor{table_blue}{$8$} &
        \cellcolor{table_white} \textcolor{table_blue}{Ship} \\
        \hline
        
        \cellcolor{table_white} \textcolor{table_blue}{$4$} &
        \cellcolor{table_white} \textcolor{table_blue}{Deer} &
        \cellcolor{table_white} \textcolor{table_blue}{$9$} &
        \cellcolor{table_white} \textcolor{table_blue}{Truck} \\
        \hline
        
    \end{tabular}
    \label{tab:CIFAR10Categories}
\end{table}

\subsection{Programming for DL: A Simple Classification Example}
We have analyzed the theoretical foundations of DNNs, CNNs, and RNNs so far. The programming stage of a DL application is as important as building the DL-based system since the feasibility of the considered method requires thorough examination via extensive computer simulations. Thus, we will provide the coding stage of a basic CNN architecture using Python and Keras~\cite{Burak_132_Keras}. We will build a CNN-based image classifier model with training using the CIFAR10 dataset~\cite{Burak_20_2009_CIFAR10Paper},~\cite{Burak_21_CIFAR10Website}. CIFAR10 consists of $60000$ $(32 \times 32)$ RGB images including $10$ categories as given in Table~\ref{tab:CIFAR10Categories}. $50000$ images correspond to the training data while the rest $10000$ images construct the test data. We will begin with calling the necessary modules: \vspace{-1ex}

\begin{figure}
   \centering
   \includegraphics[width=\linewidth]{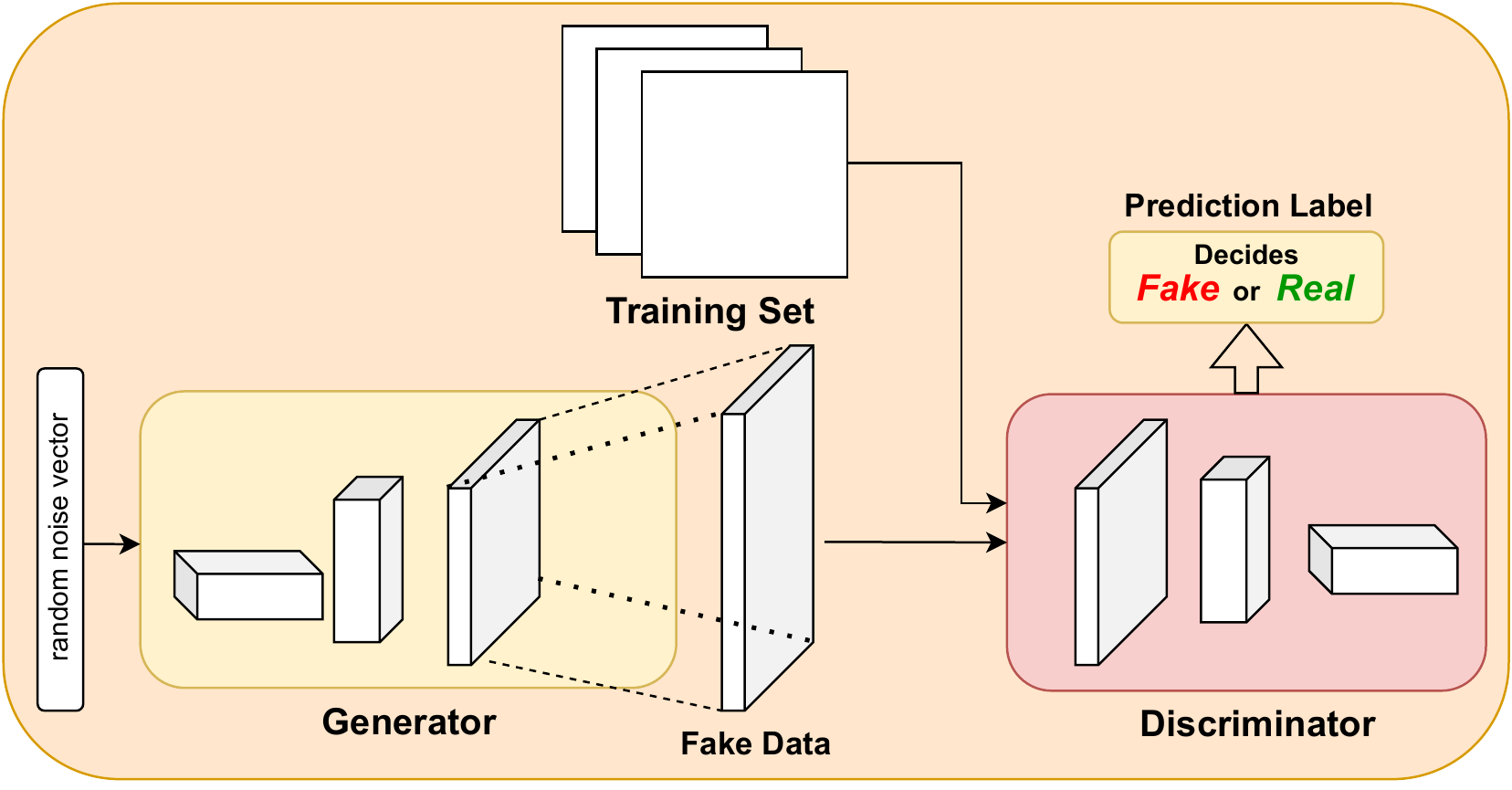}
   \caption{{A generative adversarial network scheme.}}
   \label{fig:GAN}
\end{figure}
\begin{lstlisting}
    from tensorflow.keras.preprocessing.image import img_to_array
    from tensorflow.keras.datasets.cifar10 import load_data
    from tensorflow.keras.layers import Conv2D, MaxPooling2D
    from tensorflow.keras.layers import Dense, Flatten
    from tensorflow.keras.models import Sequential
    from PIL import Image
    import numpy as np
\end{lstlisting}
Here, the \textcolor{codeblue}{\textbf{img\_to\_array}} module converts an image to the corresponding 3D pixel array, \textcolor{codeblue}{\textbf{load\_data}} module fetches the CIFAR10 dataset, \textcolor{codeblue}{\textbf{Conv2D}}, \textcolor{codeblue}{\textbf{MaxPooling2D}}, \textcolor{codeblue}{\textbf{Dense}}, and \textcolor{codeblue}{\textbf{Flatten}} represent the convolutional and pooling layers of a CNN, dense layer of a DNN, and flattening layer which squashes its input, respectively. The \textcolor{codeblue}{\textbf{Sequential}} module creates a DL model by adding the layers consecutively. The \textcolor{codeblue}{\textbf{Image}} module is useful to process an image and \textbf{numpy} is the well-known scientific computing module. Now, we can import the CIFAR10 data and preprocess it by normalizing each pixel to (0-255) range: 

\begin{lstlisting}
    (x_train, y_train), (x_test, y_test) = load_data()
    x_train = x_train.astype("float32") / 255
    x_test = x_test.astype("float32") / 255
\end{lstlisting}

The \textcolor{codeorange}{\textit{\textbf{astype}}} function converts data to "float32" data type. The first ten images in the training data and their labels with corresponding descriptions are given in Fig.~\ref{fig:CIFAR10Data}. After preprocessing and manually observing the data, we can now build the CNN model. We use a simple architecture to keep the model complexity and training time moderate, which can be created as follows: 

\begin{figure}
   \centering
   \includegraphics[width=\linewidth]{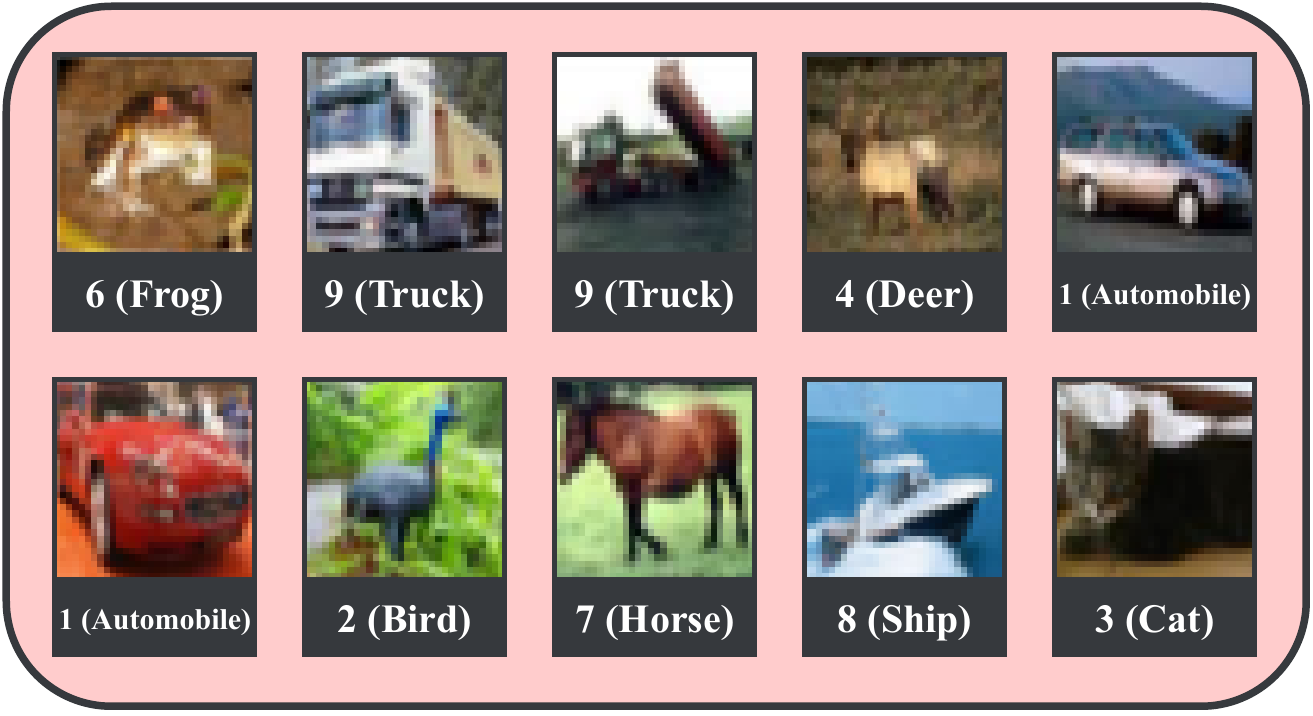}
   \caption{First ten images in the CIFAR10 training data with their labels and corresponding descriptions.}
   \label{fig:CIFAR10Data}
\end{figure}


\begin{lstlisting}
    model = Sequential([
        Conv2D(32, 3, input_shape=(32, 32, 3), activation="relu"),
        Conv2D(32, 3, activation="relu"),
        MaxPooling2D(2, 2),
        Conv2D(64, 3, activation="relu"),
        Conv2D(64, 3, activation="relu"),
        MaxPooling2D(2, 2),
        Flatten(),
        Dense(64, activation="relu"),
        Dense(10, activation="softmax")
    ])
    model.compile(loss="sparse_categorical_crossentropy", optimizer="adam", metrics=["accuracy"])
\end{lstlisting}

We use \textbf{sparse\_categorical\_crossentropy} loss function to calculate the loss value which performs exactly the same as the typical categorical cross-entropy loss function given in (\ref{categorical_cross_entropy_loss}). However, in this case the ground truth label is a single digit from 0 to 9 representing the correct label instead of a one-hot vector. Subsequently, we train our CNN model to observe train/test loss and accuracy values: 

\begin{lstlisting}
    history = model.fit(x_train, y_train, validation_split=0.1, batch_size=128, epochs=50)
\end{lstlisting}

Here, the \textcolor{codeorange}{\textit{\textbf{validation\_split}}} represents the percentage of the train data to be utilized as validation data to observe the loss and accuracy changes of the unseen data during the training. The \textcolor{codeorange}{\textit{\textbf{batch\_size}}} is the number of images that train the model in a single iteration, where a single \textcolor{codeorange}{\textit{\textbf{epoch}}} is composed of training all batches. The \textbf{history} variable includes the loss and accuracy values corresponding to the train and validation data for each epoch, which is shown in Fig.~\ref{fig:LossAccuracy}. The graph indicates that the train loss and accuracy {decrease} and {increase} smoothly, where the train accuracy converges to nearly $97\%$. Therefore there is almost no bias and the model does not underfit. However, the validation loss {rises} after almost the tenth iteration and the validation accuracy converges to approximately $72\%$. Thus, there is a considerable variance and the model strongly overfits to the training data. This can also be observed {by} evaluating the trained model on the test data: 

\begin{figure}
   \centering
   \includegraphics[width=\linewidth]{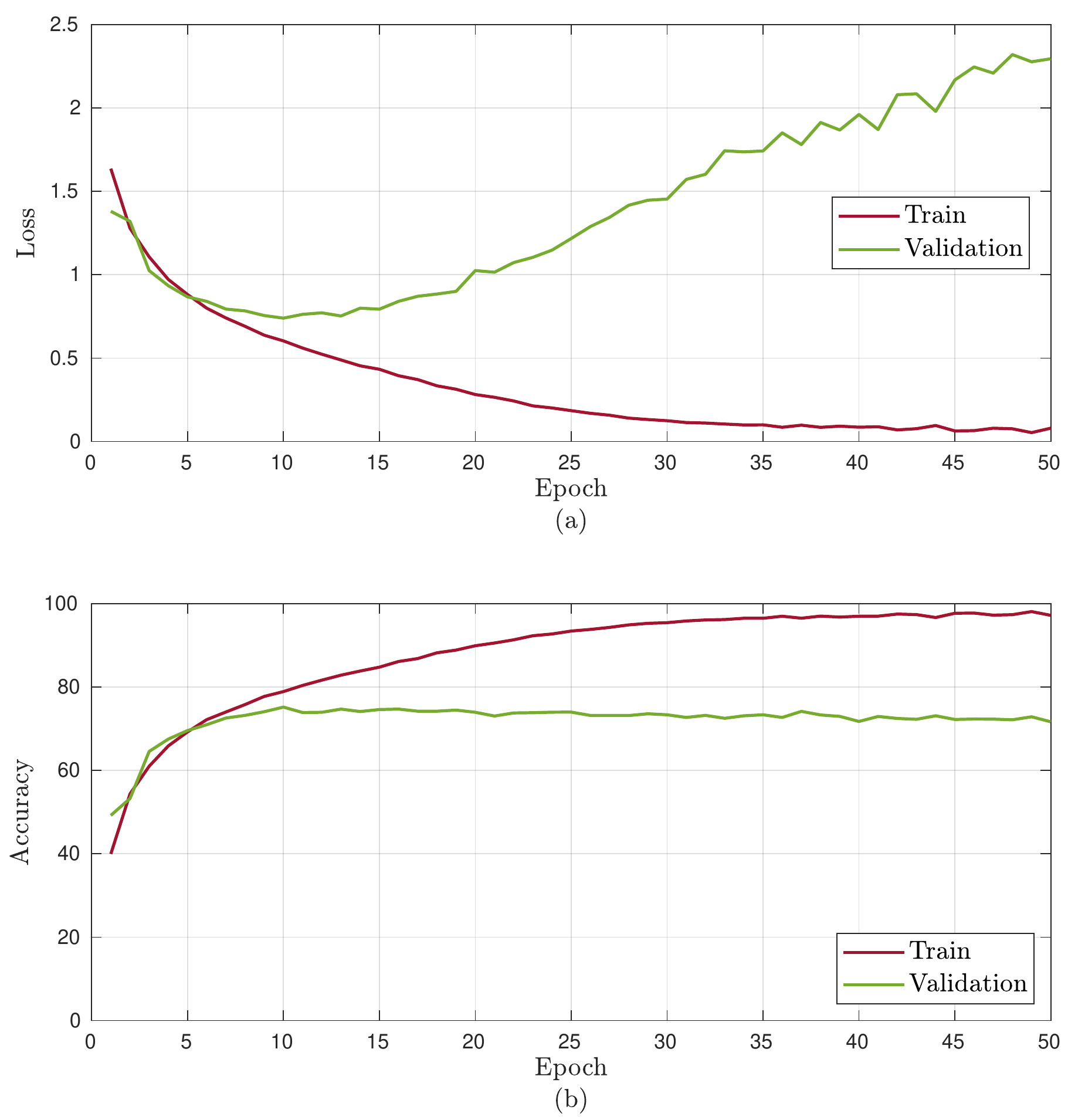}
   \caption{The loss and accuracy graphs of the train and validation data.}
   \label{fig:LossAccuracy}
\end{figure}

\begin{figure}
   \centering
   \includegraphics[width=\linewidth]{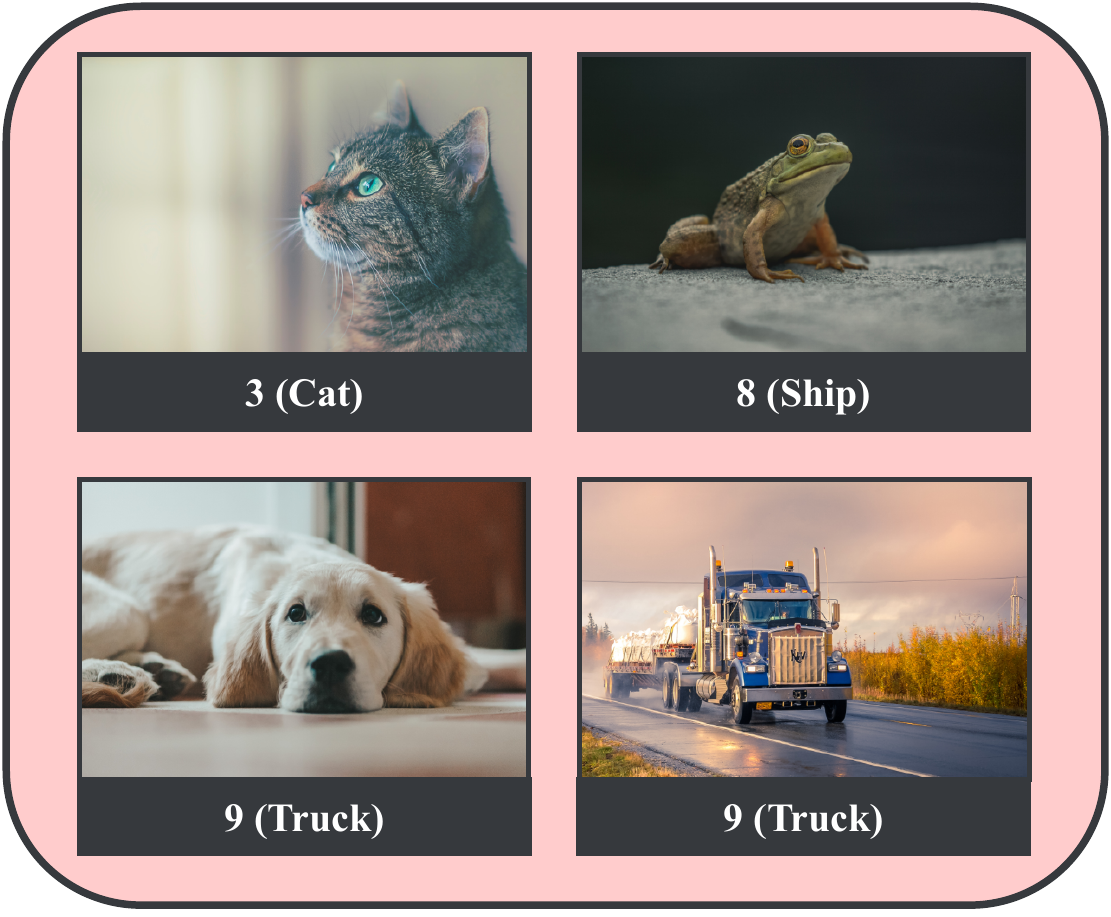}
   \caption{Predictions of the trained CNN model on the unseen images.}
   \label{fig:Predictions}
\end{figure}

\begin{lstlisting}
    _, accuracy = model.evaluate(x_test, y_test)
    print("Accuracy: %.2f"%(accuracy * 100))
\end{lstlisting}
\begin{itemize}
    \item[$\gg$] Accuracy: 71.10
\end{itemize}
As a result, this model requires a hyperparameter tuning to improve its generalization ability, which is out of our scope. As the last operation, we can try explicit images to predict their categories: 
\begin{lstlisting}
    image = Image.open('/../image')
    display(image)
    image = image.resize((32, 32))
    img_array = img_to_array(image)
    img_array = img_array.reshape(-1, 32, 32, 3)
    c = np.argmax(model.predict(img_array))
\end{lstlisting}

Here, we first import the image and resize {it} to make {it} suitable for our model. Thereafter, we convert the image to the corresponding 3D pixel array. Since we feed the model with a train data including 50000 instances, the input of the model has the dimension of (N, 32, 32, 3). Thus, we reshape the 3D pixel array to have 1 as the first dimension and predict the class of the image by taking the maximum index of the output array. Fig.~\ref{fig:Predictions} shows the predictions of our model on four explicit images that the model has never seen before. The model correctly recognizes the cat and the truck, while it classifies the given dog and frog as a truck and a ship, which supports the low validation accuracy in Fig.~\ref{fig:LossAccuracy}.

We have covered the basics of DL architectures, introduced an overview of the three mostly used DL techniques: DNN, CNN, and RNN, and provided the programming stage of an image classification task. In the following, we will begin our investigations of DL applications on PHY technologies with MIMO systems.

\section{DL for Emerging MIMO Systems}  \label{sec:mimo}
\begin{figure}
   \centering
   \includegraphics[width=\linewidth]{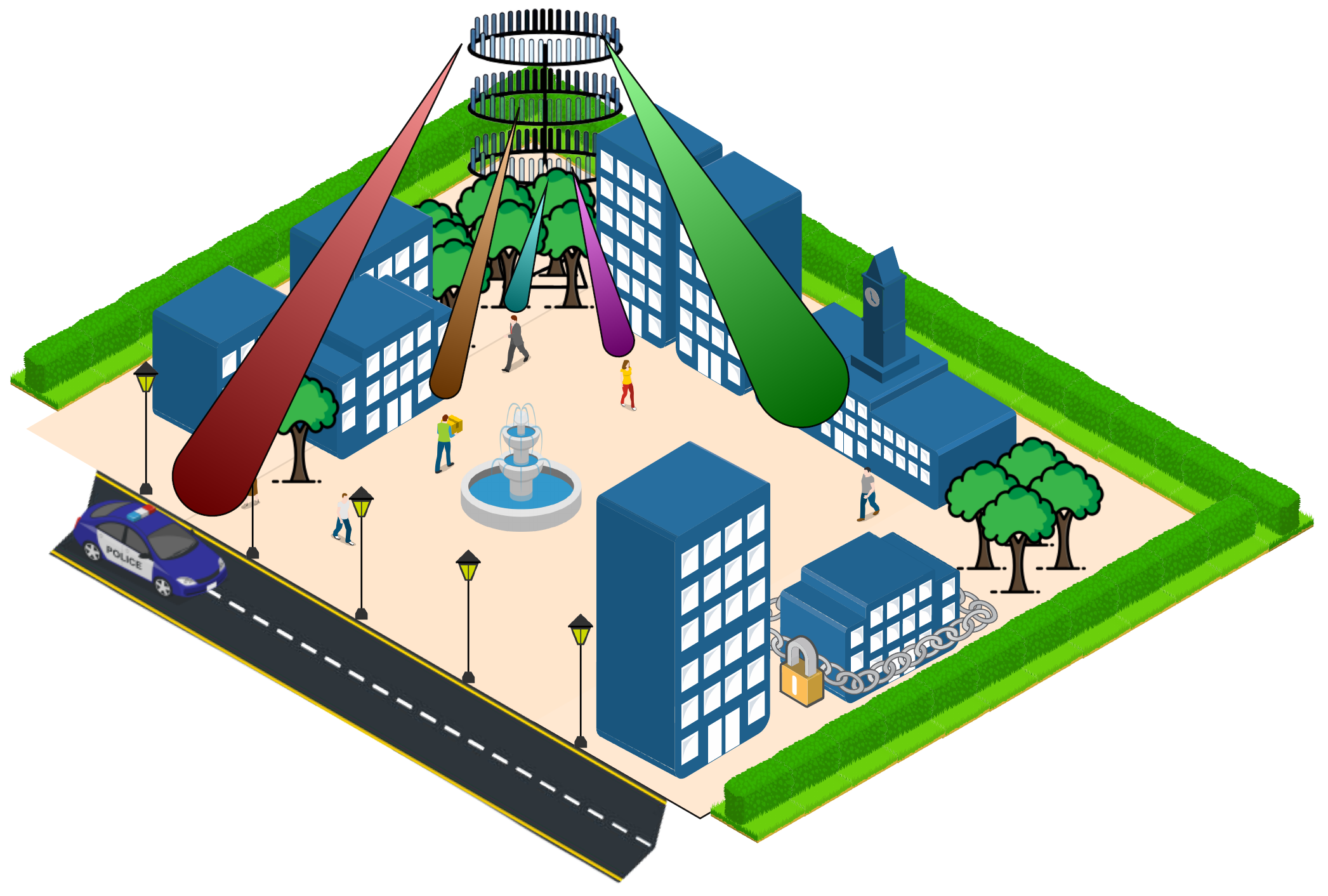}
   \caption{An example massive MU-MIMO scenario with a large antenna array supporting multiple users.}
   \label{fig:MassiveMIMO}
\end{figure}

Next-generation wireless communication technologies are expected to bring forth a variety of improvements such as extremely high data rate and spectral efficiency. The brand new mobile communication technology, 5G, has attracted the attention of both academic and industrial circles with the promise of a maximum of 20 Gb/s peak data rate along with other major improvements in latency and energy/spectrum efficiency. The key enabler of a fully intelligent world, 6G, will introduce even higher specs such as a peak data rate of at least 1 Tb/s and a user-experienced data rate of 1 Gb/s~\cite{Burak_5_2019_Vision6G}. These significant improvements are possible with the use of multiple antennas at both transmitter and receiver sides as in early standards, which is known as the MIMO technology. MIMO systems can provide spatial diversity gain where the transmitted symbols are received from multiple paths and the fading effect is mitigated. Alternatively, spatial multiplexing gain is obtained by MIMO systems where the capacity of the channel increases linearly with the minimum of transmit and receive antennas~\cite{Burak_30_2004_KeyToGigabitWireless}. Scaling up the conventional MIMO systems with a few hundred antennas that simultaneously serve many users, massive multi-user (\acrshort{mu}) MIMO technology has become a breakthrough. Massive MU-MIMO utilizes all the benefits of the conventional MIMO while eliminating the MU interference to further enhance the transmission performance, which can be seen in Fig.~\ref{fig:MassiveMIMO}. Massive MIMO provides 10 times or more capacity compared to the conventional MIMO via its powerful spatial multiplexing ability. In addition, it enhances the energy efficiency on the order of 100 times via the ability of sharp beamforming to the small regions with a large number of antennas~\cite{Burak_31_2014_MassiveMIMONextGen}.

Although massive MIMO technology brings immense benefits, it comes at some prices. The most shining feature of massive MIMO, spatial multiplexing, requires reliable channel state information (\acrshort{csi}). During the downlink transmission of conventional MIMO systems such as in the Long Term Evolution (4G) standard, the base station (\acrshort{bs}) transmits a pilot sequence so that the users estimate the channel responses and feed them back to the BS. Nevertheless, this method is not suitable {for} massive MIMO systems since each user would have to estimate a few hundred channel responses which give{s} rise to a huge feedback overhead. Another reason is that pilots should be orthogonal between the antennas and this requires a considerable amount of orthogonal resources~\cite{Burak_31_2014_MassiveMIMONextGen}. Another challenge of massive MIMO technology is the number of required radio frequency (\acrshort{rf}) chains. When each transmit antenna at the BS is activated to transmit a different symbol, the number of required RF chains equals the number of transmit antennas and this increases the hardware cost. As a promising solution to {the} inherent problems of MIMO mentioned above, IM appears as a strong candidate for next-generation wireless communication technologies. Along with decreasing the need for {the} massive number of RF chain requirements, IM techniques provide high energy and spectral efficiency by conveying additional information bits utilizing the indices of the building blocks of a communication system~\cite{Burak_9_2017_IMNextGen}. One of the most critical shortcomings of IM techniques is the detection complexity. The more bits transmitted via IM, the more complex the detector becomes. The optimum detector is the maximum likelihood detector (\acrshort{mld}) and in some cases, it is impossible to use it because of its unacceptable level of complexity. At this point, linear detectors, such as zero-forcing (\acrshort{zf}) and minimum-mean squared error (\acrshort{mmse}) detectors, take place with the cost of {a} worse bit error rate (\acrshort{ber}). Thus, there is a trade-off between the complexity and the BER performance of currently used model-based detectors.

\begin{figure}
   \centering
   \includegraphics[width=\linewidth]{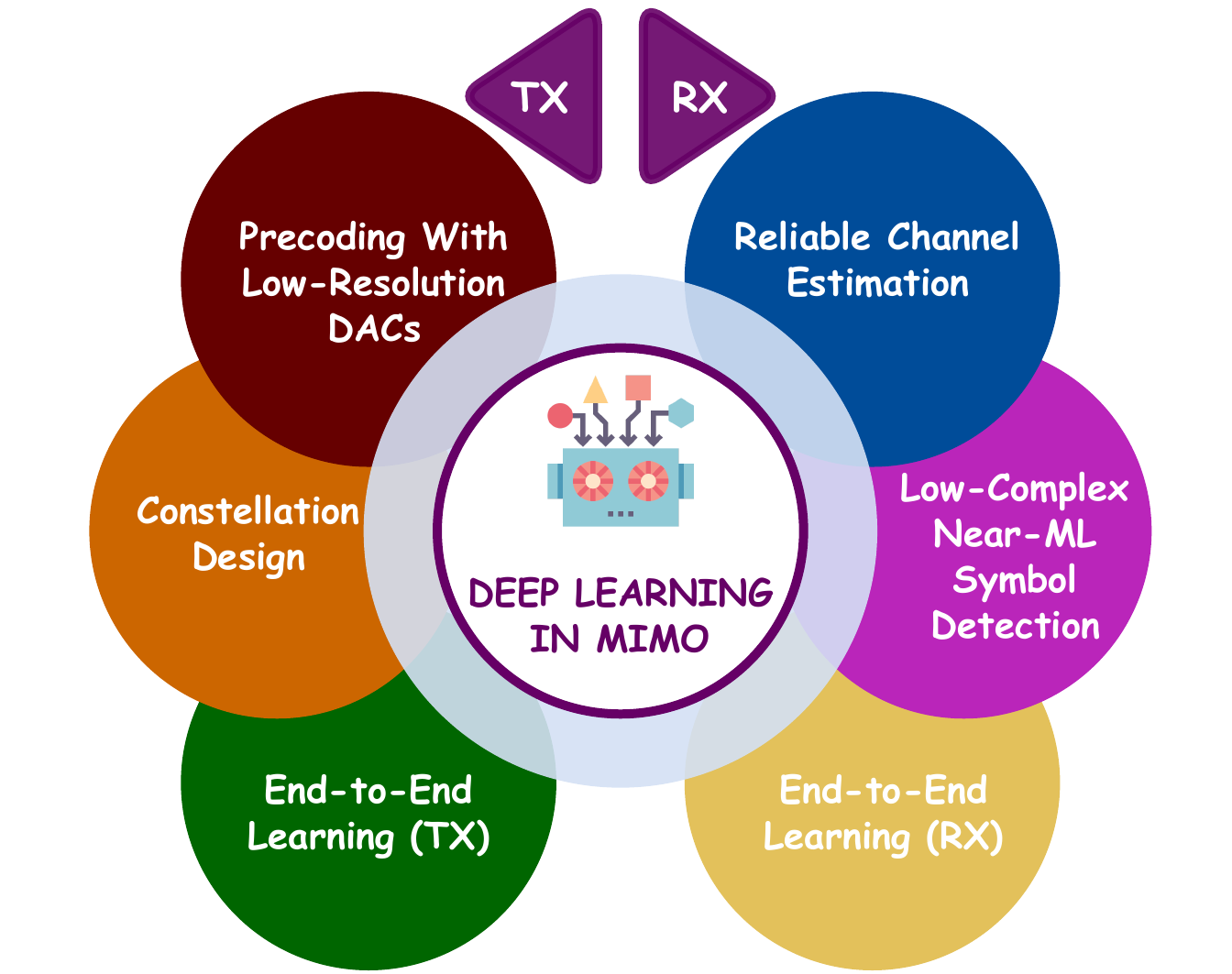}
   \caption{DL applications for the transmitter and receiver sides of a MIMO system.}
   \label{fig:DL_ApplicationsOnMIMO}
\end{figure}

Preprocessing of transmitting signals, often known as precoding, is another vital topic in efficient massive MU-MIMO systems. The BS uses precoding before transmission to mitigate the fading channel impact and cancel MU interference. However, precoding techniques that currently exist in the standards require high-resolution digital-to-analog converters (\acrshort{dac}s) which bring high complexity to the system. Therefore, the design of precoders with low-resolution DACs is of interest among the issues of intelligent massive MIMO transmitter design. Within the context of intelligent MIMO transmitters, another blooming topic is constellation design since the performance of communication systems is highly affected by the shape of the constellation. From their infancy to the cutting-edge technology, information bits are modulated by a limited number of constellations such as quadrature amplitude modulation (\acrshort{qam}) and phase-shift keying (\acrshort{psk}). Next-generation intelligent MIMO transmitters are expected to modulate information bits by using complex-shaped constellations that are optimized for the transmission environments. It is the mission of the intelligent black box to find the best locations for the constellation points and to optimize the decision boundaries for these points~\cite{Burak_32_2020_SurveyDLConsDesign}. However, to obtain a performance improvement concerning the conventional constellation designs, the receiver must be aware of the decision boundaries of the used complex-shaped constellations. Therefore, it is mission impossible for the conventional detectors to decode information bits without the initial knowledge of the constellation design. This is where E2E learning appears as one of the remarkably charming concepts for the next-generation wireless communication systems, where the aim is to fully optimize both the transmitter and receiver by representing the whole system as an autoencoder (\acrshort{ae})~\cite{Burak_33_2018_E2EWithoutACHannelModel}.

DL techniques gleam as a bright solution to those aforementioned drawbacks and further needs of next-generation massive MU-MIMO systems thanks to their ability to learn a suitable mapping from input to output. Model-based blocks of a conventional communication system that depend on optimized mathematical models might be replaced by intelligent data-based blocks that are robust to dynamic changes of the transmission environment. \textit{The ultimate goal is to achieve E2E intelligent communication systems, which will, in our opinion, revolutionize next-generation technology}. In Fig.~\ref{fig:DL_ApplicationsOnMIMO}, the applications of DL techniques for MIMO systems are given from a high-level perspective with a categorization under transmitter and receiver parts. The following subsections will shed light on the state-of-the-art research in the literature of DL applications for MIMO systems as well as future directions and open issues in this area. Specifically, in Subsection III.A, we will demonstrate the use of DL in detection for MIMO systems. Subsection III.B investigates DL-based MIMO channel estimation schemes. By combining DL-aided MIMO detectors and channel estimators, we discuss intelligent MIMO receivers and introduce the real-world datasets that might be {a} benchmark for future studies in Subsection III.C. Subsection III.D describes the intelligent MIMO transmitter designs and AEs. DL-based cutting-edge IM technologies are described in Subsection III.E. Finally, Subsection III.F demonstrates an example implementation of a DL-based IM detector for interested readers.


\subsection{Signal Detection for MIMO Systems}

The primary motivation behind using DL techniques in MIMO detectors is to \textit{overcome the trade-off between computational complexity and BER performance}. The MLD operates depending on a strong mathematical model and {it is proved that MLD provides the optimum BER performance under ideal conditions~\cite{Burak_39_2020_ADNNArchitectureGSM, Burak_40_2020_BDNN}.} Since DL techniques, especially DNNs, are general function approximators, the model of the MLD can be represented by NNs with significantly low complexity. {These approximations are not expected to outperform MLD. Instead, it is expected to reach the performance of MLD without requiring the impractical complexity of MLD.} In what follows, we reveal and summarize a brief overview of the literature in this direction.

The authors of~\cite{Burak_34_2017_DeepMIMODetection} and~\cite{Burak_35_2019_LearningToDetect} study a DL-based detector called DetNet for a MIMO system. The structure of DetNet is slightly different than a well-known DNN, where it is designed based on the iterative projected gradient descent algorithm. The BER performance of the DetNet is compared to the state-of-the-art detectors, approximate message passing (\acrshort{amp}), and semidefinite relaxation (\acrshort{sdrx}) under the assumption of perfect CSI at the receiver. As a similar approach to~\cite{Burak_34_2017_DeepMIMODetection} and~\cite{Burak_35_2019_LearningToDetect}, the study in~\cite{Burak_36_2018_ModelDrivenDLNetwork_MIMODetection} aims to create a model-driven DL-based detector for a MIMO system based on the existing iterative detection algorithm, orthogonal AMP (\acrshort{oamp}). The proposed network, OAMP-Net, wields some additional trainable parameters to improve the BER performance of the OAMP detector under perfect CSI.

Another member of model-driven DL-based networks appears in~\cite{Burak_37_2019_PartialLearning}, where the authors combine model-based and data-based detectors to create a partial learning scheme. An NN detects a portion of the amplitude-phase modulation (\acrshort{apm}) symbols in a MIMO system as the non-linear data-based part of the detector. The rest of the APM symbols are detected by a conventional ZF detector as the linear model-based part of the detector. This scheme also requires perfect CSI at the receiver.

In contrast to the former solutions, a pure data-driven DNN architecture is proposed to detect symbols transmitted over a multipath fading channel with bandwidth constraints in~\cite{Burak_38_2019_LearningforDetection}. Both flat fading and frequency-selective fading channels are implemented to analyze the BER performance of the detector. For a range of pilot overhead percentages in the transmitted signal, the proposed detector is compared to the theoretical BER of the Rayleigh fading channel and the decision feedback equalizer without requiring CSI at the receiver.

As discussed earlier, DL-based detectors are expected to decrease the complexity in MIMO system models with IM techniques. Within this context, data-driven and model-driven DNNs are investigated in~\cite{Burak_39_2020_ADNNArchitectureGSM} and~\cite{Burak_40_2020_BDNN}, respectively. Instead of a single giant DNN, the work in~\cite{Burak_39_2020_ADNNArchitectureGSM} suggests utilizing data-driven sub-DNNs for the detection process of a MIMO system with generalized spatial modulation (\acrshort{gsm}). The proposed architecture performs similarly to the MLD and the single DNN detector with lower complexity under the assumption that the fading channel is fixed and the noise is independent and identically distributed (\acrshort{iid}) Gaussian. However, when the noise is correlated or has a deviation in its distribution, the sub-DNN architecture outperforms the MLD and the single DNN detector. CSI at the receiver is not necessary for fixed fading channels, but it is essential for time-varying channels. On the other hand, in~\cite{Burak_40_2020_BDNN}, the authors propose a model-driven DL-based block detector for GSM inspired by linear block detectors such as Block-ZF and Block-MMSE detectors. The APM symbol of each active transmit antenna is detected by a DNN, assuming that perfect CSI is available at the receiver. The B-DNN model outperforms the linear detectors and performs similarly to the MLD with lower complexity.

The treatises examined so far follow the fundamental principle of DNNs, which is stacking the hidden layers to build a deep network and to increase the learning ability. However, the authors in~\cite{Burak_41_2020_ParallelDeepDetection} realize that deepening the network after a certain number of hidden layers does not considerably improve the BER performance. Therefore, they propose a parallel detection network, where several DL detection networks perform in parallel to introduce a diversity effect. In order to obtain diversity, a special loss function is designed. Detection is carried out based on the perfect knowledge of the CSI. Likewise, in~\cite{Burak_42_2020_DLBasedParallelDetector}, a DL-based detector composed of many parallel modified DNNs introduces a diversity gain in a MIMO system model. The modified DNNs employ the residual learning principle and differ in several ways and introduce diversity to the system. Assuming perfect CSI, computer simulations unveil that the proposed detector outperforms the ZF, MMSE, and single DNN-based detectors.

The kind of NN employed in DL-based detectors is another key factor that draws attention. Within this perspective, the work in~\cite{Burak_43_2019_EfficientMIMO} compares DNN- and CNN-based detectors to address the issues of robust detection with imperfect CSI and efficient DL framework for low-complex detection. This comparison reveals that the DNN-based detector performs significantly better in terms of the BER performance. In addition, it outperforms the ZF, MMSE, and DetNet~\cite{Burak_34_2017_DeepMIMODetection},~\cite{Burak_35_2019_LearningToDetect} detectors with both perfect and imperfect CSI at the receiver. However, the MLD detector has a considerably better BER performance than the proposed detector. With a similar approach, DNN-, CNN-, and RNN-based detectors are designed without channel dependency and compared to the MLD in~\cite{Burak_44_2019_ImplementationMethodologies}. Computer simulations in this study show that the RNN-based approach performs the best, whereas the DNN-based {approach is} the poorest. The fact that DNN-based architecture outperforms CNN-based architecture in~\cite{Burak_43_2019_EfficientMIMO} but degrades in~\cite{Burak_44_2019_ImplementationMethodologies} demonstrates that the optimum DL architecture is dependent on the system model and environment.

So far, the suggested detectors have relied on channel assumptions without considering real-world conditions. However, according to~\cite{Burak_45_2020_AdaptiveNeuralDetection} and~\cite{Burak_46_2020_Hypernetwork}, the main issue with the existing DL-based detectors is either unpleasant performance over practical spatially correlated channels or a computational burden because of retraining for each channel realization. Thus, the authors of~\cite{Burak_45_2020_AdaptiveNeuralDetection} focus on a robust DL-based MIMO detector that can optimize itself via online training during the transmission over realistic and spatially correlated channel models. The proposed detector, called MMNet, is built upon iterative soft-thresholding algorithms. Experiments conducted by using a dataset of channel realizations from the 3GPP 3D MIMO channel indicate that MMNet outperforms the classical approaches such as AMP and SDRX,  as well as recent DL techniques like DetNet~\cite{Burak_34_2017_DeepMIMODetection},~\cite{Burak_35_2019_LearningToDetect}, and OAMPNet~\cite{Burak_36_2018_ModelDrivenDLNetwork_MIMODetection}, under the assumption of perfect CSI. The work in~\cite{Burak_46_2020_Hypernetwork} introduces an additional NN, a HyperNetwork, that eliminates the need for online retraining on top of a modified version of MMNet to create the HyperMIMO detector. Computer simulation results reveal that the HyperMIMO detector outperforms MMSE and OAMPNet~\cite{Burak_36_2018_ModelDrivenDLNetwork_MIMODetection} and performs similarly to the MMNet and the MLD with less computing cost under the perfect CSI.

\begin{table*}
    \caption{An overview of the literature on DL-based MIMO detectors considering five key factors (Red hues indicate CSI requirement or without real-world dataset and green hues indicate no CSI requirement or with real-world dataset).}

    \centering
    \begin{tabular}{|M{3cm}|M{2cm}|M{2cm}|M{2cm}|M{3cm}|M{2cm}|}
        \hline
        
        \cellcolor{table_blue} \textbf{\textcolor{table_white}{Paper}} & 
        \cellcolor{table_blue} \textbf{\textcolor{table_white}{CSI Requirement}} & 
        \cellcolor{table_blue} \textbf{\textcolor{table_white}{DL Structure}} & 
        \cellcolor{table_blue} \textbf{\textcolor{table_white}{Real-World Dataset}} & 
        \cellcolor{table_blue} \textbf{\textcolor{table_white}{Benchmarks}} & 
        \cellcolor{table_blue} \textbf{\textcolor{table_white}{Data- or Model-Driven}} \\
        \hline
        
        \cellcolor{table_white} \textcolor{table_blue}{Deep MIMO Detection~\cite{Burak_34_2017_DeepMIMODetection}} & 
        \cellcolor{table_red} &
        \cellcolor{table_white} \textcolor{table_blue}{DetNet} &
        \cellcolor{table_red} &
        \cellcolor{table_white} \textcolor{table_blue}{AMP, SDRX} &
        \cellcolor{table_white} \textcolor{table_blue}{Model} \\
        \hline
        
        \cellcolor{table_white} \textcolor{table_blue}{Learning to Detect~\cite{Burak_35_2019_LearningToDetect}} & 
        \cellcolor{table_red} &
        \cellcolor{table_white} \textcolor{table_blue}{DetNet} &
        \cellcolor{table_red} &
        \cellcolor{table_white} \textcolor{table_blue}{AMP, SDRX} &
        \cellcolor{table_white} \textcolor{table_blue}{Model} \\
        \hline
        
        \cellcolor{table_white} \textcolor{table_blue}{Model-Driven MIMO Detection~\cite{Burak_36_2018_ModelDrivenDLNetwork_MIMODetection}} & 
        \cellcolor{table_red} &
        \cellcolor{table_white} \textcolor{table_blue}{OAMP-Net} &
        \cellcolor{table_red} &
        \cellcolor{table_white} \textcolor{table_blue}{OAMP} &
        \cellcolor{table_white} \textcolor{table_blue}{Model} \\
        \hline
        
        \cellcolor{table_white} \textcolor{table_blue}{Partial Learning~\cite{Burak_37_2019_PartialLearning}} & 
        \cellcolor{table_red} &
        \cellcolor{table_white} \textcolor{table_blue}{DNN} &
        \cellcolor{table_red} &
        \cellcolor{table_white} \textcolor{table_blue}{ZF, MLD, Single DNN} &
        \cellcolor{table_white} \textcolor{table_blue}{Model} \\
        \hline
        
        \cellcolor{table_white} \textcolor{table_blue}{Learning for Detection~\cite{Burak_38_2019_LearningforDetection}} & 
        \cellcolor{table_green} &
        \cellcolor{table_white} \textcolor{table_blue}{DNN} &
        \cellcolor{table_red} &
        \cellcolor{table_white} \textcolor{table_blue}{Decision Feedback Equalizer, Theoretical Rayleigh} &
        \cellcolor{table_white} \textcolor{table_blue}{Data} \\
        \hline
        
        \cellcolor{table_white} \textcolor{table_blue}{Sub-DNNs for GSM~\cite{Burak_39_2020_ADNNArchitectureGSM}} & 
        \cellcolor{table_white} \textcolor{table_blue}{Flexible} &
        \cellcolor{table_white} \textcolor{table_blue}{Sub-DNNs} &
        \cellcolor{table_red} &
        \cellcolor{table_white} \textcolor{table_blue}{MLD, Single DNN} &
        \cellcolor{table_white} \textcolor{table_blue}{Data} \\
        \hline
        
        \cellcolor{table_white} \textcolor{table_blue}{B-DNN~\cite{Burak_40_2020_BDNN}} & 
        \cellcolor{table_red} &
        \cellcolor{table_white} \textcolor{table_blue}{Block-DNN} &
        \cellcolor{table_red} &
        \cellcolor{table_white} \textcolor{table_blue}{B-ZF, B-MMSE, MLD} &
        \cellcolor{table_white} \textcolor{table_blue}{Model} \\
        \hline
        
        \cellcolor{table_white} \textcolor{table_blue}{Parallel DL~\cite{Burak_41_2020_ParallelDeepDetection}} & 
        \cellcolor{table_red} &
        \cellcolor{table_white} \textcolor{table_blue}{PDN} &
        \cellcolor{table_red} &
        \cellcolor{table_white} \textcolor{table_blue}{SPD, DetNet} &
        \cellcolor{table_white} \textcolor{table_blue}{Data} \\
        \hline
        
        \cellcolor{table_white} \textcolor{table_blue}{DL-Based Parallel Detector~\cite{Burak_42_2020_DLBasedParallelDetector}} & 
        \cellcolor{table_red} &
        \cellcolor{table_white} \textcolor{table_blue}{Parallel Modified DNN} &
        \cellcolor{table_red} &
        \cellcolor{table_white} \textcolor{table_blue}{ZF, MMSE, Single DNN} &
        \cellcolor{table_white} \textcolor{table_blue}{Data} \\
        \hline
        
        \cellcolor{table_white} \textcolor{table_blue}{Efficient MIMO Detection~\cite{Burak_43_2019_EfficientMIMO}} & 
        \cellcolor{table_red} &
        \cellcolor{table_white} \textcolor{table_blue}{DNN, CNN} &
        \cellcolor{table_red} &
        \cellcolor{table_white} \textcolor{table_blue}{ZF, MMSE, MLD, DetNet} &
        \cellcolor{table_white} \textcolor{table_blue}{Data} \\
        \hline
        
        \cellcolor{table_white} \textcolor{table_blue}{Implementation Methodologies~\cite{Burak_44_2019_ImplementationMethodologies}} & 
        \cellcolor{table_green} &
        \cellcolor{table_white} \textcolor{table_blue}{DNN, CNN, RNN} &
        \cellcolor{table_red} &
        \cellcolor{table_white} \textcolor{table_blue}{MLD} &
        \cellcolor{table_white} \textcolor{table_blue}{Data} \\
        \hline
        
        \cellcolor{table_white} \textcolor{table_blue}{Adaptive Neural Detection~\cite{Burak_45_2020_AdaptiveNeuralDetection}} & 
        \cellcolor{table_red} &
        \cellcolor{table_white} \textcolor{table_blue}{MMNet} &
        \cellcolor{table_green} &
        \cellcolor{table_white} \textcolor{table_blue}{MMSE, MLD, AMP, SDRX, \acrshort{vblast}, DetNet, OAMPNet} &
        \cellcolor{table_white} \textcolor{table_blue}{Data} \\
        \hline
        
        \cellcolor{table_white} \textcolor{table_blue}{HyperMIMO~\cite{Burak_46_2020_Hypernetwork}} & 
        \cellcolor{table_red} &
        \cellcolor{table_white} \textcolor{table_blue}{Modified MMNet} &
        \cellcolor{table_green} &
        \cellcolor{table_white} \textcolor{table_blue}{MMSE, MLD, OAMPNet, MMNet} &
        \cellcolor{table_white} \textcolor{table_blue}{Data} \\
        \hline
        
        \cellcolor{table_white} \textcolor{table_blue}{Model-Driven Massive MIMO~\cite{Burak_47_2020_AModelDrivenDLMethodMassiveMIMO}} & 
        \cellcolor{table_red} &
        \cellcolor{table_white} \textcolor{table_blue}{DNN} &
        \cellcolor{table_red} &
        \cellcolor{table_white} \textcolor{table_blue}{Iterative~\cite{Burak_48_2017_LowComplexityNearOptimal}, MMSE, DetNet} &
        \cellcolor{table_white} \textcolor{table_blue}{Model} \\
        \hline
        
        \cellcolor{table_white} \textcolor{table_blue}{DL-Based Viterbi~\cite{Burak_49_2020_ViterbiNet}} & 
        \cellcolor{table_green} &
        \cellcolor{table_white} \textcolor{table_blue}{ViterbiNet} &
        \cellcolor{table_red} &
        \cellcolor{table_white} \textcolor{table_blue}{Viterbi, Sliding Bidirectional RNN} &
        \cellcolor{table_white} \textcolor{table_blue}{Data} \\
        \hline
        
        \cellcolor{table_white} \textcolor{table_blue}{DL-Based Sphere Decoding~\cite{Burak_50_2019_DLBasedSphereDecoding}} & 
        \cellcolor{table_red} &
        \cellcolor{table_white} \textcolor{table_blue}{DNN} &
        \cellcolor{table_red} &
        \cellcolor{table_white} \textcolor{table_blue}{SPD Variants} &
        \cellcolor{table_white} \textcolor{table_blue}{Data} \\
        \hline
        
        \cellcolor{table_white} \textcolor{table_blue}{DL-Aided Tabu Search~\cite{Burak_51_2020_DLAidedTabuSearch}} & 
        \cellcolor{table_red} &
        \cellcolor{table_white} \textcolor{table_blue}{FS-Net, DL-Aided TS} &
        \cellcolor{table_red} &
        \cellcolor{table_white} \textcolor{table_blue}{ZF, MMSE, SPD, Ordered SIC, DetNet} &
        \cellcolor{table_white} \textcolor{table_blue}{Data} \\
        \hline
        
        \cellcolor{table_white} \textcolor{table_blue}{RE-MIMO~\cite{Burak_52_2021_REMIMO}} & 
        \cellcolor{table_red} &
        \cellcolor{table_white} \textcolor{table_blue}{RE-MIMO} &
        \cellcolor{table_red} &
        \cellcolor{table_white} \textcolor{table_blue}{MMSE, MLD, AMP, SDRX, V-BLAST, DetNet, OAMPNet, OAMPNet-2} &
        \cellcolor{table_white} \textcolor{table_blue}{Data} \\
        \hline
    \end{tabular}
    \label{tab:MIMO_DetectorsTable}
\end{table*}

The development of DL-based detectors by recreating mathematical models using DNNs is popular among researchers.~\cite{Burak_34_2017_DeepMIMODetection},~\cite{Burak_35_2019_LearningToDetect}, and~\cite{Burak_36_2018_ModelDrivenDLNetwork_MIMODetection} are examples with model-driven DL architectures. In addition, the work in~\cite{Burak_47_2020_AModelDrivenDLMethodMassiveMIMO} presents a DNN-based detector for massive MIMO systems. The DNN architecture is model-driven since it depends on an existing iterative detection method given in~\cite{Burak_48_2017_LowComplexityNearOptimal}. There are additional parameters to be learned for MU interference cancellation. The suggested detector outperforms the iterative~\cite{Burak_48_2017_LowComplexityNearOptimal}, DetNet~\cite{Burak_34_2017_DeepMIMODetection},~\cite{Burak_35_2019_LearningToDetect}, and MMSE detectors, assuming perfect CSI at the receiver. However, there are also data-driven architectures utilizing the same concept. The authors of~\cite{Burak_49_2020_ViterbiNet} offer a MIMO detector dubbed ViterbiNet that eliminates the CSI reliance by integrating DNN architecture into the Viterbi algorithm. Their simulation results indicate that the ViterbiNet detector has a superior performance compared to the conventional Viterbi decoder. Another classical MIMO detection approach, sphere decoding (\acrshort{spd}), meets with DNNs in the study of~\cite{Burak_50_2019_DLBasedSphereDecoding}. The proposed method learns the radius of the decoding hypersphere, outperforming previous SPD versions and coming close to the MLD with much less complexity assuming the perfect CSI. The study in~\cite{Burak_51_2020_DLAidedTabuSearch} explores the use of DL within tabu search (\acrshort{ts}) detection. Two models are proposed: FS-Net and DL-Aided TS, with the latter building on the former. Assuming perfect CSI is available at the receiver, the proposed DL-Aided TS method decreases complexity by roughly 90\% compared to existing TS algorithms while keeping almost the same performance.

As opposed to the literature inspected thus far, the authors of~\cite{Burak_52_2021_REMIMO} consider the MU case and present an NN-based MU-MIMO detector called RE-MIMO. They are concerned with resilience against channel misspecification, the capacity to manage a varying number of users, and invariance to the sequence in which users interact with the system. The proposed model-driven DL detector depends on the neural augmentation strategy, which combines the learning procedure with the inductive biases.

We classify the investigated DL-based detectors by five key specifications in Table~\ref{tab:MIMO_DetectorsTable}. It should be noted that red and green hues in the CSI column indicate CSI requirement and no CSI requirement, while they represent computer simulations without and with real-world dataset in the Real-World Dataset column, respectively. There are two noteworthy outcomes deduced from Table~\ref{tab:MIMO_DetectorsTable}. The first is that, to function as intended, most system models require the perfect CSI. The second issue is that most DL-based detectors have never been tested in a real-world setting. In the follow-up, we will scrutinize the literature of DL-based channel estimators in MIMO systems.

\subsection{Channel Estimation for MIMO Systems}

The fundamental rationale for employing DL methods in massive MIMO channel estimation is to \textit{enable reliable channel estimation with considerably fewer pilots than the number of transmit antennas and low-resolution ADCs at the BS}. This leads to a robust detection process with a reasonable amount of resources for channel estimation.

A DL-based channel estimator for downlink massive MIMO systems is presented in~\cite{Burak_53_2019_CEforMassiveMIMO}, where the pilot length is less than the number of transmit antennas at the BS. The proposed channel estimator consists of two stages, the first of which involves two DNNs working together to perform pilot design and pilot-aided channel estimation. In the second stage, the channel estimation and symbol detection are implemented iteratively by another DNN to enhance the performance of the estimation, which is called data-aided channel estimation. Computer simulations reveal that the two-stage estimator outperforms the conventional MMSE-based data-aided channel estimator given in~\cite{Burak_54_2014_DataAidedLargeAntennaSystems}. The authors of~\cite{Burak_55_2019_JointPDandCE} extend this study to uplink MU-MIMO systems. The recommended channel estimator, unlike~\cite{Burak_53_2019_CEforMassiveMIMO}, skips the data-aided estimation stage. Separate DNNs design the pilot signals for the users at the initial part of the pilot-aided estimation stage. Subsequently, separate DNNs estimate the channels of the users using successive interference cancellation (\acrshort{sic}). The suggested estimator outperforms both the traditional MMSE estimator and the DL approach without SIC. In addition, both~\cite{Burak_53_2019_CEforMassiveMIMO} and~\cite{Burak_55_2019_JointPDandCE} investigates the optimal pilot length to obtain higher capacity.

Channel estimation in mmWave massive MIMO systems is complicated since the number of RF chains is limited. The authors of~\cite{Burak_56_2018_CEforBeamspacemmWaveMassiveMIMO} propose a learned denoising-based AMP (\acrshort{ldamp}) network to solve this problem. The LDAMP network is an image recovery network that treats the channel matrix as a 2D image, with a single pilot for all channels in the antenna array. The LDAMP network outperforms the conventional AMP variants. The study in~\cite{Burak_57_2018_SuperResolChannelandDOAEstimation} focuses on DNN-based direction-of-arrival (\acrshort{doa}) estimation in addition to DNN-based channel estimation. Specifically, two DNN-based algorithms operate DOA and channel estimation processes without the need for pilot symbols. Extensive computer simulations verify that the suggested DOA and channel estimation algorithms exceed the traditional techniques in terms of mean squared error (\acrshort{mse}) of DOA and BER performance. Motivated by the drawbacks of~\cite{Burak_56_2018_CEforBeamspacemmWaveMassiveMIMO}, the authors of~\cite{Burak_58_2020_DLBasedBeamspaceCEformmWaveMassiveMIMO} develop a better denoising network based on CNNs along with the LDAMP network, resulting in the fully convolutional denoising AMP algorithm. The required pilot length is less than the number of channel components to be estimated. The fully convolutional denoising AMP algorithm has a successive performance over LDAMP~\cite{Burak_56_2018_CEforBeamspacemmWaveMassiveMIMO} and other variants of AMP in terms of both normalized MSE (\acrshort{nmse}) and achievable sum rate. Another channel estimator built upon the classical learned AMP (\acrshort{lamp}) method is presented in~\cite{Burak_59_2021_DLforBeamspaceCEinMillimeterWaveMassiveMIMO}, which is prior-aided Gaussian mixture LAMP. In the first stage of the Gaussian mixture LAMP algorithm, a new shrinkage function is developed based on the Gaussian mixture distribution of beamspace channel elements. Then, the Gaussian mixture LAMP{-}based beamspace channel estimation is performed depending on the derived shrinkage function in the second stage. The proposed algorithm outperforms the OMP, AMP, and LAMP algorithms in terms of NMSE. The methods suggested in~\cite{Burak_56_2018_CEforBeamspacemmWaveMassiveMIMO},~\cite{Burak_58_2020_DLBasedBeamspaceCEformmWaveMassiveMIMO}, and~\cite{Burak_59_2021_DLforBeamspaceCEinMillimeterWaveMassiveMIMO} are simulated using the well-known Saleh-Valenzuela channel model. The study~\cite{Burak_59_2021_DLforBeamspaceCEinMillimeterWaveMassiveMIMO} also ran further computer simulations with the practical ray-tracing channel dataset provided by~\cite{Burak_60_2019_DeepMIMOPaper} and~\cite{Burak_61_DeepMIMOWebsite}.

Along with less pilot-aided channel estimation, DL-based estimators pave the way for reliable prediction even with low-resolution analog-to-digital converters (\acrshort{adc}s). In~\cite{Burak_62_2019_DLBasedCEforMixedResolADCs}, a massive MIMO BS equipped with mixed-resolution ADCs implements channel estimation with DNNs. A direct-input DNN estimates channels using the received signals of all antennas, whereas a selective-input prediction DNN estimates channels of the antennas with low-resolution ADCs via received signals of the antennas with high-resolution ADCs. The proposed algorithm has a superior performance over the classical MMSE and the expectation{-}maximization Gaussian-mixture generalized AMP methods. The idea of using mixed-resolution ADCs is improved in~\cite{Burak_63_2021_JointCEandMixedResolADCsAllocation}, where the authors propose an optimization technique for mixed-resolution ADCs allocation along with the channel estimation and pilot design. Specifically, the CENet estimates channels, the SELNet assigns one-bit and high-resolution ADCs, and the PDNet constructs the pilot matrix. The suggested framework outperforms the conventional generalized AMP (\acrshort{gamp}) and gridless GAMP in terms of NMSE. Besides, the complete framework is compared with the partial scenarios in which all antennas at the BS have one-bit ADCs, and only the CENet exists. Another partial scheme includes all one-bit ADC antennas with both PDNet and CENet. The partial design with non-optimized high-resolution ADCs, PDNet, and CENet performs similarly to the complete framework.

Using DL methods, it is feasible to reduce ADC resolution even further, in which accurate channel estimation is possible despite complete one-bit ADCs. The study in~\cite{Burak_64_2019_SegmentAverage} presents a DNN-based channel estimator for one-bit massive MIMO systems. The pilot signal is divided into segments in the first stage, where each one forms the input of a DNN. The estimated channel is the average of all DNNs' outputs. Next, detected data symbols using the initial estimation are interpreted as the remaining pilot signals and utilized to update the initial channel estimation, similar to the procedure in~\cite{Burak_53_2019_CEforMassiveMIMO}. The proposed algorithm surpasses the traditional LS and Bussgang linear MMSE algorithms in terms of MSE under both i.i.d. and spatially correlated channels. Similarly, the work in~\cite{Burak_65_2020_DLforMassiveMIMO1BitADCs} proposes a DNN-based channel estimator for one-bit massive MIMO systems. The authors perform investigations to determine the length and architecture of the pilot sequence required to ensure that the quantized received signals are mapped to the channel matrix. According to these analyses, the more BS antennas, the better the channel estimation performance for the same pilot length. Equivalently, the more BS antennas, the less required pilot length for the same channel estimation performance. The suggested network outclasses the GAMP algorithm under the indoor massive MIMO channel scenario offered by~\cite{Burak_60_2019_DeepMIMOPaper} and~\cite{Burak_61_DeepMIMOWebsite}. As opposed to the former one-bit channel estimators,~\cite{Burak_66_2021_1BitCEUsingConditionalGAN} introduces a conditional generative adversarial network (\acrshort{gan}) that can learn an adaptive loss function to optimize the training procedure and predict the channels. The proposed conditional GAN algorithm is compared with expectation maximization Gaussian-mixture GAMP algorithm, two-stage algorithm~\cite{Burak_53_2019_CEforMassiveMIMO}, LDAMP, channel and DOA estimation algorithm~\cite{Burak_57_2018_SuperResolChannelandDOAEstimation}, and one-bit channel estimator~\cite{Burak_64_2019_SegmentAverage} using the ray-tracing channel dataset given by~\cite{Burak_60_2019_DeepMIMOPaper} and~\cite{Burak_61_DeepMIMOWebsite}. Computer simulations reveal that the conditional GAN algorithm has a {more} successive performance than the benchmarks in terms of average NMSE.

Except for~\cite{Burak_53_2019_CEforMassiveMIMO}, all of the channel estimators we have looked at thus far have only considered uplink scenarios. However, downlink channel estimation is also crucial in terms of reliable symbol detection at the user end. It is common to assume that the downlink and uplink channels are reciprocal, and the BS feedbacks the estimated uplink channels to the users. Nevertheless, the detection performance degrades when the channel is not reciprocal. Thus, the authors of~\cite{Burak_67_2019_ULDLChannelCalibration} introduce a channel calibration network, called CaliNet, based on DNNs to estimate downlink channels accurately from the uplink channels when there is no reciprocality. Computer simulation results indicate that CaliNet performs better than the Argos method, and similar to the Cram\'er-Rao Bound.

In addition to the prior DL-based channel estimators, the study in~\cite{Burak_68_2020_DoubleDeepQLearning} investigates reinforcement learning in channel estimation of industrial wireless networks. The proposed algorithm is double deep Q-learning, where the pilot length equals the number of channel components to be estimated. The recommended approach beats the estimated MMSE in an indoor industrial manufacturing setting with 4.7 dB path-loss, 8.6 dB shadowing, and 29 dB obstacles. However, the ideal MMSE surpasses the suggested technique in the same scenario. Similar to how~\cite{Burak_68_2020_DoubleDeepQLearning} examines a realistic indoor environment,~\cite{Burak_69_2021MLAssistedCEinMassiveMIMO} suggests an RNN-based channel estimator for massive MIMO systems in the non-line-of-sight 5G Quadriga 2.0 channel model~\cite{Burak_70_2014_QuaDRiGaPaper},~\cite{Burak_71_QuaDRiGaWebsite}. The algorithm iteratively searches the channel taps and denoises their amplitudes in the time domain. Considering frame error rate, simulation results for a user speed of 5 km/h show that the RNN-based iterative channel estimator outperforms the traditional MMSE technique.

Considering seven key factors, we categorize the reviewed DL-based channel estimators in Table~\ref{tab:MIMO_ChannelEstimatorsTable}. The red and green colors in the Pilot Design column represent if pilot design exists in the study, while they indicate system model with and without low-resolution ADCs in the last column. FCDAMP, EM-GM-GAMP, and GM-LAMP represent fully convolutional denoising AMP, expectation maximization Gaussian-mixture GAMP, and Gaussian-mixture learned AMP, respectively. The most eye-catching observation from Table~\ref{tab:MIMO_ChannelEstimatorsTable} is that almost all of the DL-based estimators work for uplink massive MIMO scenarios. It is a favorable consequence that the majority of the approaches fulfill the motivation of using fewer pilots for DL-based channel estimators. It is also encouraging to note that realistic channel scenarios are becoming increasingly common, albeit still insufficiently so, within the framework of DL-based channel estimators. The next subsection combines the DL-based detectors and channel estimators to build an intelligent receiver framework, and provides an insight to the real-world datasets.

\begin{table*}
    \caption{An overview of the literature on DL-based MIMO channel estimators considering seven key factors (Red hues indicate no pilot design, no low-resolution ADCs or without real-world dataset and green hues indicate pilot design, low-resolution ADCs or with real-world dataset).}

    \centering
    \begin{tabular}{|M{3cm}|M{1.5cm}|M{1cm}|M{1cm}|M{3cm}|M{1.5cm}|M{1cm}|M{1.5cm}|}
        \hline
        
        \cellcolor{table_blue} \textbf{\textcolor{table_white}{Paper}} & 
        \cellcolor{table_blue} \textbf{\textcolor{table_white}{DL Structure}} & 
        \cellcolor{table_blue} \textbf{\textcolor{table_white}{Real-World Dataset}} & 
        \cellcolor{table_blue} \textbf{\textcolor{table_white}{Pilot Length}} &
        \cellcolor{table_blue} \textbf{\textcolor{table_white}{Benchmarks}} & 
        \cellcolor{table_blue} \textbf{\textcolor{table_white}{Downlink or Uplink}} &
        \cellcolor{table_blue} \textbf{\textcolor{table_white}{Pilot Design}} &
        \cellcolor{table_blue} \textbf{\textcolor{table_white}{Low-Resolution ADCs}} \\
        \hline
        
        \cellcolor{table_white} \textcolor{table_blue}{DL-Based Two-Stage~\cite{Burak_53_2019_CEforMassiveMIMO}} &
        \cellcolor{table_white} \textcolor{table_blue}{DNN} & 
        \cellcolor{table_red} &
        \cellcolor{table_green} \textcolor{table_white}{Low} &
        \cellcolor{table_white} \textcolor{table_blue}{Data-Aided MMSE~\cite{Burak_54_2014_DataAidedLargeAntennaSystems}} &
        \cellcolor{table_white} \textcolor{table_blue}{Downlink} &
        \cellcolor{table_green} &
        \cellcolor{table_red} \\
        \hline
        
        \cellcolor{table_white} \textcolor{table_blue}{Joint Pilot Design \& Channel Estimation for MU~\cite{Burak_55_2019_JointPDandCE}} &
        \cellcolor{table_white} \textcolor{table_blue}{DNN} & 
        \cellcolor{table_red} &
        \cellcolor{table_green} \textcolor{table_white}{Low} &
        \cellcolor{table_white} \textcolor{table_blue}{MMSE} &
        \cellcolor{table_white} \textcolor{table_blue}{Uplink} &
        \cellcolor{table_green} &
        \cellcolor{table_red} \\
        \hline
        
        \cellcolor{table_white} \textcolor{table_blue}{DL-Based Beamspace mmWave Massive MIMO (2018)~\cite{Burak_56_2018_CEforBeamspacemmWaveMassiveMIMO}} &
        \cellcolor{table_white} \textcolor{table_blue}{LDAMP} & 
        \cellcolor{table_red} &
        \cellcolor{table_green} \textcolor{table_white}{Low} &
        \cellcolor{table_white} \textcolor{table_blue}{AMP Variants} &
        \cellcolor{table_white} \textcolor{table_blue}{Uplink} &
        \cellcolor{table_red} &
        \cellcolor{table_red} \\
        \hline
        
        \cellcolor{table_white} \textcolor{table_blue}{Super Resolution Channel \& DOA Estimation~\cite{Burak_57_2018_SuperResolChannelandDOAEstimation}} &
        \cellcolor{table_white} \textcolor{table_blue}{DNN} & 
        \cellcolor{table_red} &
        \cellcolor{table_green} \textcolor{table_white}{No Pilot} &
        \cellcolor{table_white} \textcolor{table_blue}{Various Conventional Compressed Sensing Methods} &
        \cellcolor{table_white} \textcolor{table_blue}{Uplink} &
        \cellcolor{table_red} &
        \cellcolor{table_red} \\
        \hline
        
        \cellcolor{table_white} \textcolor{table_blue}{DL-Based Beamspace mmWave Massive MIMO (2020)~\cite{Burak_58_2020_DLBasedBeamspaceCEformmWaveMassiveMIMO}} &
        \cellcolor{table_white} \textcolor{table_blue}{FCDAMP} & 
        \cellcolor{table_red} &
        \cellcolor{table_green} \textcolor{table_white}{Low} &
        \cellcolor{table_white} \textcolor{table_blue}{LDAMP, AMP Variants} &
        \cellcolor{table_white} \textcolor{table_blue}{Uplink} &
        \cellcolor{table_red} &
        \cellcolor{table_red} \\
        \hline
        
        \cellcolor{table_white} \textcolor{table_blue}{DL-Based Beamspace mmWave Massive MIMO (2021)~\cite{Burak_59_2021_DLforBeamspaceCEinMillimeterWaveMassiveMIMO}} &
        \cellcolor{table_white} \textcolor{table_blue}{GM-LAMP} & 
        \cellcolor{table_green} &
        \cellcolor{table_green} \textcolor{table_white}{Low} &
        \cellcolor{table_white} \textcolor{table_blue}{AMP, LAMP} &
        \cellcolor{table_white} \textcolor{table_blue}{Uplink} &
        \cellcolor{table_red} &
        \cellcolor{table_red} \\
        \hline
        
        \cellcolor{table_white} \textcolor{table_blue}{Mixed-Resolution ADCs~\cite{Burak_62_2019_DLBasedCEforMixedResolADCs}} &
        \cellcolor{table_white} \textcolor{table_blue}{DNN, CNN} & 
        \cellcolor{table_red} &
        \cellcolor{table_green} \textcolor{table_white}{Low} &
        \cellcolor{table_white} \textcolor{table_blue}{MMSE, EM-GM-GAMP} &
        \cellcolor{table_white} \textcolor{table_blue}{Uplink} &
        \cellcolor{table_red} &
        \cellcolor{table_green} \textcolor{table_white}{Mixed} \\
        \hline
        
        \cellcolor{table_white} \textcolor{table_blue}{Joint Channel Estimation \& Mixed ADCs Allocation~\cite{Burak_63_2021_JointCEandMixedResolADCsAllocation}} &
        \cellcolor{table_white} \textcolor{table_blue}{PDNet, SELNet, CENet} & 
        \cellcolor{table_red} &
        \cellcolor{table_white} \textcolor{table_blue}{Varying} &
        \cellcolor{table_white} \textcolor{table_blue}{GAMP, Gridless GAMP} &
        \cellcolor{table_white} \textcolor{table_blue}{Uplink} &
        \cellcolor{table_green} &
        \cellcolor{table_green} \textcolor{table_white}{Mixed} \\
        \hline
        
        \cellcolor{table_white} \textcolor{table_blue}{Segment Average Based Channel Estimation~\cite{Burak_64_2019_SegmentAverage}} &
        \cellcolor{table_white} \textcolor{table_blue}{DNN} & 
        \cellcolor{table_green} &
        \cellcolor{table_red} \textcolor{table_white}{High} &
        \cellcolor{table_white} \textcolor{table_blue}{LS, Bussgang Linear MMSE} &
        \cellcolor{table_white} \textcolor{table_blue}{Uplink} &
        \cellcolor{table_red} &
        \cellcolor{table_green} \textcolor{table_white}{1-Bit} \\
        \hline
        
        \cellcolor{table_white} \textcolor{table_blue}{DL for 1-Bit ADCs \& Fewer Pilots~\cite{Burak_65_2020_DLforMassiveMIMO1BitADCs}} &
        \cellcolor{table_white} \textcolor{table_blue}{DNN} & 
        \cellcolor{table_green} &
        \cellcolor{table_green} \textcolor{table_white}{Low} &
        \cellcolor{table_white} \textcolor{table_blue}{GAMP} &
        \cellcolor{table_white} \textcolor{table_blue}{Uplink} &
        \cellcolor{table_red} &
        \cellcolor{table_green} \textcolor{table_white}{1-Bit} \\
        \hline
        
        \cellcolor{table_white} \textcolor{table_blue}{1-Bit MU Conditional GAN~\cite{Burak_66_2021_1BitCEUsingConditionalGAN}} &
        \cellcolor{table_white} \textcolor{table_blue}{conditional GAN} & 
        \cellcolor{table_green} &
        \cellcolor{table_green} \textcolor{table_white}{Low} &
        \cellcolor{table_white} \textcolor{table_blue}{EM-GM-GAMP,~\cite{Burak_53_2019_CEforMassiveMIMO}, LDAMP,~\cite{Burak_57_2018_SuperResolChannelandDOAEstimation},~\cite{Burak_64_2019_SegmentAverage}} &
        \cellcolor{table_white} \textcolor{table_blue}{Uplink} &
        \cellcolor{table_red} &
        \cellcolor{table_green} \textcolor{table_white}{1-Bit} \\
        \hline
        
        \cellcolor{table_white} \textcolor{table_blue}{UL-DL Channel Calibration~\cite{Burak_67_2019_ULDLChannelCalibration}} &
        \cellcolor{table_white} \textcolor{table_blue}{CaliNet (DNN)} & 
        \cellcolor{table_red} &
        \cellcolor{table_green} \textcolor{table_white}{Low} &
        \cellcolor{table_white} \textcolor{table_blue}{Argos, Cram\'er-Rao Bound} &
        \cellcolor{table_white} \textcolor{table_blue}{Both} &
        \cellcolor{table_red} &
        \cellcolor{table_red} \\
        \hline
        
        \cellcolor{table_white} \textcolor{table_blue}{Double Deep Q Learning~\cite{Burak_68_2020_DoubleDeepQLearning}} &
        \cellcolor{table_white} \textcolor{table_blue}{DDQL} & 
        \cellcolor{table_green} &
        \cellcolor{table_red} \textcolor{table_white}{High} &
        \cellcolor{table_white} \textcolor{table_blue}{MMSE} &
        \cellcolor{table_white} \textcolor{table_blue}{No Info} &
        \cellcolor{table_red} &
        \cellcolor{table_red} \\
        \hline
        
        \cellcolor{table_white} \textcolor{table_blue}{Iterative RNN~\cite{Burak_69_2021MLAssistedCEinMassiveMIMO}} &
        \cellcolor{table_white} \textcolor{table_blue}{RNN} & 
        \cellcolor{table_green} &
        \cellcolor{table_red} \textcolor{table_white}{No Info} &
        \cellcolor{table_white} \textcolor{table_blue}{MMSE} &
        \cellcolor{table_white} \textcolor{table_blue}{No Info} &
        \cellcolor{table_red} &
        \cellcolor{table_red} \\
        \hline
    \end{tabular}
    \label{tab:MIMO_ChannelEstimatorsTable}
\end{table*}

\subsection{Combined Signal Detection and Channel Estimation \& Real-World Datasets}

After examining DL-based detectors and channel estimators, we can now integrate them in a single algorithm to get a completely intelligent receiver architecture, which will lead us to our ultimate objective of E2E intelligent communication systems. In~\cite{Burak_72_2019_JointCHandSDMIMOSTBC}, a single DNN jointly executes channel estimation and symbol detection for a MIMO-Alamouti system, a unique case of space-time block codes with two transmit antennas. The four-layer DNN takes the real and imaginary parts of the received signal as the input and generates the probabilities of the possible transmit symbol combinations, without explicitly estimating the channel. The recommended network has 1 dB signal-to-noise ratio (\acrshort{snr}) loss compared to MLD detection with perfect CSI and 3 dB SNR gain compared to MLD detection with imperfect CSI, at the same BER.

The authors of~\cite{Burak_73_2020_DeepSM} direct their research towards intelligent receiver for spatial modulation (\acrshort{sm}) considering time-variant dynamic channels. The suggested network, called DeepSM, consists of two parallel DNNs, with the upper DNN updating the CSI and the lower one detecting the transmitted symbols at each time slot. The initial CSI is computed using the LS technique utilizing a pilot signal with a length equal to the number of transmit antennas. DeepSM is compared to the conventional model-based and data-driven DNN-based receivers over both time-variant and time-invariant channel models. Both benchmarks implement LS channel estimation and assume that the channel is invariant during the whole simulation. Subsequently, the conventional model-based receiver employs MLD detection, whereas the data-driven DNN-based receiver detects the transmitted symbols by a DNN. Extensive computer simulations reveal that DeepSM outperforms the benchmark schemes in all scenarios. Similar to the iterative architecture of DeepSM, the work in~\cite{Burak_74_2020_ModelDrivenDLforMIMODetection} proposes a model-driven DL-based joint channel estimation and symbol detection network, OAMP-Net2, by unfolding the existing iterative algorithm OAMP. OAMP-Net2 performs initial pilot-based channel estimation and detects information symbols afterward. Following that, data-aided channel estimation and symbol detection processes take place iteratively. The recommended approach outperforms the AMP, OAMP, MMSE, MMSE-SIC, DetNet~\cite{Burak_34_2017_DeepMIMODetection},~\cite{Burak_35_2019_LearningToDetect}, and OAMP-Net~\cite{Burak_36_2018_ModelDrivenDLNetwork_MIMODetection} in different MIMO setups and modulation orders. MLD and SPD detectors, on the other hand, outclass OAMP-Net2 with their significantly larger complexity. In addition, OAMP-Net2 is specifically compared with OAMP-Net~\cite{Burak_36_2018_ModelDrivenDLNetwork_MIMODetection} in various MIMO setups and modulation orders considering both i.i.d. and correlated channel scenarios, in which OAMP-Net2 beats the predecessor in all situations. Furthermore, the authors examine OAMP-Net2 under the practical Quadriga 3GPP 3D MIMO channel model~\cite{Burak_70_2014_QuaDRiGaPaper},~\cite{Burak_71_QuaDRiGaWebsite}. 

The authors of~\cite{Burak_75_2020_PilotAssitedCEandSDinULMUMIMO} propose two DL-based receiver structures for uplink MU-MIMO systems. The first technique, FullCon, is a data-driven DNN-based algorithm and detects the information bits by the received signal directly without explicitly executing channel estimation. MdNet, on the other hand, separates the channel estimation and symbol detection phases. DNN-based channel estimation follows the LS estimation in the first stage, while the DL-based detector comes after ZF detection in the second stage. Both receiver algorithms are tested using various MIMO configurations and the number of hidden layers and neurons inside the layers. In addition, they are compared to two conventional receiver structures first of which consists of LS channel estimation and MMSE detection, while the second one includes LS channel estimation and projected gradient iterative detection. According to computer simulation results, both FullCon and MdNet outperform traditional techniques in $2 \times 2$ and $4 \times 4$ MIMO setups. One notable outcome of computer simulations is that FullCon has superior performance than MdNet in $2 \times 2$ MIMO, whereas MdNet takes the lead in $4 \times 4$ MIMO.

\begin{figure}
   \centering
   \includegraphics[width=\linewidth]{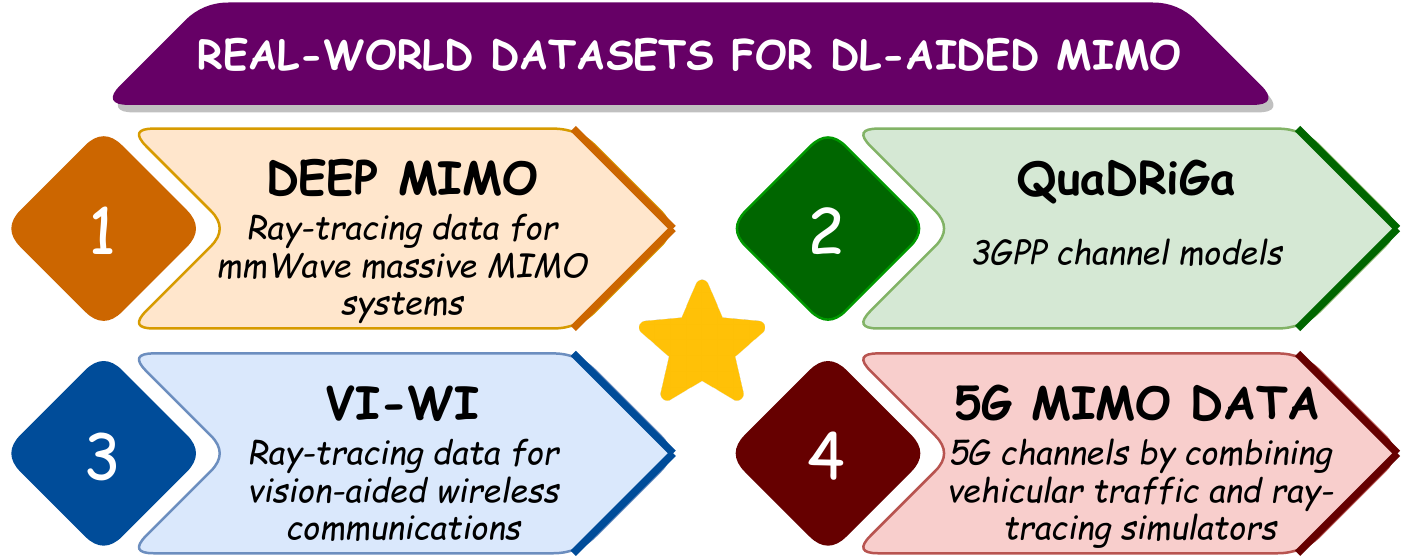}
   \caption{Major real-world datasets that are considered for DL-aided MIMO techniques.}
   \label{fig:RealWorldDatasets}
\end{figure}

As investigated so far, the literature of DL-based receiver architectures has grown to maturity. However, to incorporate them into real-world systems and standards, it is necessary to examine the feasibility of these designs using real-world datasets. It is also critical to have a sufficiently big dataset to reproduce the results of existing algorithms in the literature, create benchmarks, and compare different algorithms based on universal data~\cite{Burak_60_2019_DeepMIMOPaper}. As seen from Tables~\ref{tab:MIMO_DetectorsTable} and~\ref{tab:MIMO_ChannelEstimatorsTable}, the popularity of practical datasets is not satisfying, which motivates researchers to focus on generating datasets using real-world scenarios. In~\cite{Burak_60_2019_DeepMIMOPaper} and~\cite{Burak_61_DeepMIMOWebsite}, a generic dataset, dubbed DeepMIMO, for mmWave and massive MIMO systems is proposed. The DeepMIMO channels are generated based on accurate ray-tracing data collected from Remcom Wireless InSite software~\cite{Burak_76_RemcomWirelessInsiteWebsite}. Therefore, this dataset includes the geometry of the surroundings as well as the transmitter and receiver positions. In addition, the data creation process is parametric, allowing researchers to fine-tune parameters based on the desired scenario. Similarly,~\cite{Burak_77_2020_ViWiPaper} and~\cite{Burak_78_WiViDataset} present a parametric and scalable data production framework for vision-aided wireless communications based on the 3D modeling and ray-tracing provided by~\cite{Burak_76_RemcomWirelessInsiteWebsite}. The objective is the co-existence of visual and wireless data, with visual data, such as that collected from LiDAR (light detection and ranging) sensors or cameras, assisting wireless communication systems that function in the same device or area. Quadriga, which stands for quasi deterministic radio channel generator, is a tool for simulating practical MIMO radio channels for networks such as indoor environments and satellite communications~\cite{Burak_70_2014_QuaDRiGaPaper},~\cite{Burak_71_QuaDRiGaWebsite}. As stated in~\cite{Burak_71_QuaDRiGaWebsite}, Quadriga facilitates {the creation of} channel models given by entities such as 3GPP. It can also be considered as an application of the well-known WINNER model with modifications for satellite communications via its geometry-based stochastic approach. It extends the terrestrial snapshot-based simulation system to produce complex-valued baseband time series. The work in~\cite{Burak_79_2018_5GMIMOData} provides a procedure for creating realistic channel environments for various 5G scenarios by combining a vehicle traffic simulator and a ray-tracing simulator. By producing propagation channel data, the goal is to aid DL-based solutions to challenges linked to the PHY of mmWave MIMO in 5G. To sum up, the analyzed studies that propose real-world datasets for MIMO communications are summarized in Fig.~\ref{fig:RealWorldDatasets}. In the next subsection, we will discuss DL-aided algorithms within the context of smart MIMO transmitters and E2E communications.

\subsection{Intelligent Transmitter Design and Autoencoders}

We analyzed how to optimize individual blocks of a MIMO receiver separately and jointly in the previous subsections. Thus, we show that DL networks can learn and optimize a complete MIMO receiver framework for a range of optimization criteria. However, we have yet to achieve our ultimate goal of communication technology revolution with fully intelligent frameworks. Therefore, we will investigate DL approaches that optimize individual blocks of a MIMO transmitter and eventually learn a whole transmitter framework tuned with the corresponding receiver architecture. The basic idea underlying E2E communications is to \textit{replace the separate blocks of a MIMO transmitter, channel, and receiver with a single DL network that can be trained as an AE to address the conventional block structure's sub-optimization issue}~\cite{Burak_18_2017_AnIntroDLPHY},~\cite{Burak_80_2021_ChannelAutoencoderSOTAChallengesTrends}. As a departure, we will first delve into the details of DL-based constellation design in the following part.

As previously mentioned, constructing constellations with the aid of DL networks is an E2E process since the network must optimize both the positions and decision regions of the constellation points. The objective is to \textit{maximize the mutual information between the input and output of the channel, even when the channel model is unknown}~\cite{Burak_32_2020_SurveyDLConsDesign}. The study in~\cite{Burak_81_2019_DLConsDesignfoAWGNRadarInterference} provides an AE for designing optimal constellations and receiver architectures for additive white Gaussian noise (\acrshort{awgn}) channels with additive radar interference. The optimization metric is mutual information. Under various channel model assumptions for different SNR regimes, several demapping procedures are suggested. In addition, the proposed network enables using powerful coding methods such as turbo codes or low-density parity-check (\acrshort{ldpc}) codes. Computer simulations reveal that the AE-based network outperforms the architectures with the traditional constellations. The network in~\cite{Burak_82_2019_GeometricProbabilisticConsShaping} conducts probabilistic shaping to improve the constellations along with the geometric shaping. The probabilistic shaping optimizes the probability of occurrence of the constellation points. Even when the geometry of the constellation is fixed, such as QAM, the probabilistic shaping improves the achievable information rate. This network also maximizes mutual information using AEs, as in~\cite{Burak_81_2019_DLConsDesignfoAWGNRadarInterference}. The proposed joint shaping method achieves near-capacity performance and outperforms both non-shaped QAM and just geometric-shaped QAM across AWGN and Rayleigh fading channels.~\cite{Burak_83_2020_DLAEformUserWirelessInterferenceChannel} analyzes AE-based constellation design over a MU interference channel to address the dynamic interference problem. Specifically, the adaptive DL-based AE network can learn the degree of interference and enhance BER performance by adjusting each user's constellation accordingly. According to computer simulations, users' learned constellations are well-decomposed, and they concentrate on their clusters in the face of heavy interference. It is, therefore, much easier for the receiver to decode the information in the strong-interference environment. Furthermore, the AE-based network performs very similar to the typical uncoded binary PSK (\acrshort{bpsk}) and quadrature PSK (\acrshort{qpsk}), with around 2 dB improvement over 8-PSK and 16-QAM. 

In contrast to the formers,~\cite{Burak_84_2021_GrassmannianConsDesignNoncoherentMIMO} investigates AE-based Grassmannian constellation design in non-coherent MIMO systems, where neither transmitter nor the receiver requires CSI. Each point on the Grassmann manifold is a unitary matrix, and the distance between each point pair equals the chordal Frobenius norm. The recommended method is compared to Grassmannian constellations designed by traditional techniques and a non-Grassmannian constellation. Simulation results indicate that the proposed network exceeds all state-of-the-art approaches and attains the highest diversity order.

Along with intelligent constellation designs that are more resilient to dynamic propagation environments than the traditional fixed constellations, DL-based approaches have also been applied to precoding in massive MIMO transmitters. The goal is to use DL networks to \textit{offer reliable precoding with low-resolution DACs while substantially decreasing complexity}. The study in~\cite{Burak_85_2019_1BitPrecodingConsDesignviaAEBasedDL} investigates a one-bit precoding design with the optimum constellation design utilizing AEs. In a multicast massive MU-MIMO system where the BS broadcasts the information signal to all users, the transmitter composes a DNN followed by a binary layer to satisfy the one-bit precoding constraint. Because the gradients of the binary layer are always zero, implementing the standard backpropagation approach is complicated. Therefore, the authors employ an approximation which is a variant of the straight-through estimator. Experiments over fixed and varying channels show that the AE-based one-bit precoding and constellation design network outperforms the PSK and QAM transmissions with the conventional one-bit precoder~\cite{Burak_86_2018_1BitPrecodingandConsDesignMassiveMIMOQAM}. It is also worth noting that the performance of the AE-based constellation design with the standard one-bit precoder~\cite{Burak_86_2018_1BitPrecodingandConsDesignMassiveMIMOQAM} is the best, demonstrating the potential of model-driven DL networks. Likewise,~\cite{Burak_87_2020_RobustSLPviaAEBasedDL} provides AE-based symbol-level precoding (\acrshort{slp}) and constellation design for unicast massive MU-MIMO systems in which the BS transmits a unique information signal to each user. Moreover, the authors examine robust SLP design for classic QAM constellations in instances when AE-based constellations are challenging to execute in reality. According to the numerical results, the proposed AE-based SLP and constellation design network outperforms the AE-based SLP with QAM constellation and non-robust SLP with QAM constellation. Another DL-based SLP design is provided in~\cite{Burak_88_2021_DLBasedEfficientSLPforMUMISO}, where the authors propose an efficient precoding NN to mitigate MU interference in an MU-MISO system model. The optimization metric is the minimum quality-of-service of all users. Simulation results illustrate that the suggested network has superior performance than the conventional block-level precoding. However, the traditional convex optimized SLP scheme beats the recommended network with considerably greater complexity, indicating that the DL-based SLP network provides robust performance with significantly reduced complexity. The work in~\cite{Burak_89_2020_SupervisedDLforMIMOPrecoding} is similar to~\cite{Burak_85_2019_1BitPrecodingConsDesignviaAEBasedDL} and~\cite{Burak_87_2020_RobustSLPviaAEBasedDL} in that it offers an AE-based precoding network for MIMO systems. The transmitter consists of two DNNs that encode bits to symbols and precode these symbols with power normalization at the end. As seen from computer simulations, the AE-based precoding network surpasses the conventional linear precoding techniques such as ZF and MMSE and the non-linear Tomlinson-Harashima precoding assuming perfect CSI. The authors of~\cite{Burak_90_2021_DLBasedMIMOTransmissionPrecodingRTN} present a similar separate DNN structure of~\cite{Burak_89_2020_SupervisedDLforMIMOPrecoding} in the MIMO transmitter. The first of the two DNNs encodes bits to symbols, while the second does precoding. The receiver of this AE-based approach includes a radio transformer network (\acrshort{rtn}). The precoding network at the transmitter and the RTN at the receiver are trained in an E2E fashion to outclass the classical DL-based MIMO system. Simulations with different MIMO configurations reveal that the proposed combined precoding and RTN architecture outperforms the traditional linear precoding methods ZF and MMSE and the DL structures with only precoding or RTN. The authors also thoroughly analyze the precoding network's learned constellations. 

The substantial research in~\cite{Burak_91_2021_DLBasedRobustPrecodingForMassiveMIMO},~\cite{Burak_92_2020_RobustPrecodingADLApproach}, and~\cite{Burak_93_2021_DLBasedRobustPrecoderDownlink} results in a DL-aided precoding framework for downlink massive MU-MIMO systems with a uniform planar array at the BS. The objective of this framework is to implement precoding that maximizes the ergodic rate while limiting overall transmit power under a constraint by using both instantaneous and statistical CSI. The authors, however, convert this ergodic rate maximization issue to an enhanced quality-of-service problem to make it tractable, and the structure of optimum precoding emerges as a result. Thus, the proposed structure employs a DNN to successfully obtain the optimum precoding architecture and to mitigate complexity problem. Consequently, the proposed framework reduces complexity significantly while maintaining almost the same performance as the traditional iterative technique. Furthermore, the authors minimize complexity by splitting the optimization problem into two sections that evaluate instantaneous and statistical CSI, respectively. Experiments with the Quadriga channel model~\cite{Burak_70_2014_QuaDRiGaPaper},~\cite{Burak_71_QuaDRiGaWebsite} prove that both general and low-complexity frameworks produce near-optimal performance with considerably lower complexity than the classical iterative approach. 

The study in~\cite{Burak_94_2020_DLBasedPrecoderDesignMIMOFiniteAlphabetInputs} proposes another data-driven DNN-based precoding architecture in which the network fully learns the input-output relationship of a nearly optimum precoder to maximize mutual information. Simulation results show that the suggested network produces almost the same performance as the optimum precoder with significantly lower complexity than the traditional iterative precoders, just like in~\cite{Burak_91_2021_DLBasedRobustPrecodingForMassiveMIMO},~\cite{Burak_92_2020_RobustPrecodingADLApproach}, and~\cite{Burak_93_2021_DLBasedRobustPrecoderDownlink}. The work in~\cite{Burak_95_2021_DLBasedLinearPrecodingforMIMOFiniteAlphabetSignaling} takes a similar approach to~\cite{Burak_94_2020_DLBasedPrecoderDesignMIMOFiniteAlphabetInputs}, in which a data-driven DNN learns the behavior of an ideal precoder to maximize the mutual information with considerably reduced complexity. The suggested DNN accepts vectorized water-filling precoding matrix as input and yields the optimum one. Concerning various MIMO setups, simulations indicate that the recommended network achieves the optimum performance while substantially decreasing the execution time.

Apart from the AE-based and DNN-based precoding designs discussed so far, model-driven DL-based networks are also popular within the context of precoding design by combining the conventional algorithms with DL techniques. The work in~\cite{Burak_96_2019_NNOptimized1BitPrecodingMassiveMUMIMO} optimizes the conventional non-linear biconvex 1-bit precoding (\acrshort{c2po}) method by automatically tuning the algorithm parameters in a massive MU-MIMO system model. The NN optimized C2PO (\acrshort{nno-c2po}) unfolds the iterations of C2PO and utilizes backpropagation to adjust parameters. Simulations conducted using the Quadriga channel model~\cite{Burak_70_2014_QuaDRiGaPaper},~\cite{Burak_71_QuaDRiGaWebsite}  illustrate that the proposed NNO-C2PO algorithm requires about 50\% fewer iterations than C2PO for a similar BER performance. In addition, NNO-C2PO may employ the same parameters learned for a given channel model for various channel models with a moderate performance loss, indicating robustness to dynamic channel circumstances. In like manner, the study in~\cite{Burak_97_2020_ModelDrivenDLforMassiveMUMIMOFiniteAlphabetPrecoding} unfolds another conventional iterative algorithm named iterative discrete estimation (\acrshort{ide2}) to develop a model-driven DL network for massive MU-MIMO with finite-alphabet precoding. The proposed network, IDE2-Net, precodes the transmit signals using low-resolution DACs by modifying the current IDE2 algorithm with configurable parameters. Simulations reveal that IDE2-Net has substantially better performance than the IDE2 algorithm and slightly outperforms NNO-C2PO~\cite{Burak_96_2019_NNOptimized1BitPrecodingMassiveMUMIMO} for the same number of iterations. Following in the footsteps of the formers, the authors unfold the classical iterative conjugate gradient method for constant envelope precoding in~\cite{Burak_98_2020_ModelDrivenDLforMassiveMultiuserConstantEnvelope}. The suggested network, CEPNet, optimizes the traditional approach by introducing trainable parameters to decrease MU interference and computational cost. Simulations show that CEPNet significantly outperforms the conventional algorithm in terms of both BER and average achievable rate performances with less complexity. CEPNet is also resistant to channel estimation errors and channel model mismatch as in~\cite{Burak_96_2019_NNOptimized1BitPrecodingMassiveMUMIMO}. The study in~\cite{Burak_99_2020_IterativeAlgorithmInducedDeepUnfolding} unfolds the iterative weighted MMSE (\acrshort{wmmse}) algorithm into a layer-wise structure to maximize the sum rate. The proposed network embeds trainable parameters to the WMMSE algorithm to eliminate matrix inversions and reduce complexity. Unlike previous studies, the authors develop a CNN-based precoding scheme to compare their model-driven method to a data-driven network. Extensive simulation results in a massive MU-MIMO system model with varying numbers of transmit antennas and users show that the proposed method reaches the sum rate of the WMMSE algorithm and outperforms the CNN-based scheme with much less complexity. By unfolding, another conventional iterative technique, the alternate direction method of multipliers (\acrshort{admm}), is mapped to a DL framework in~\cite{Burak_100_2021_DLBasedSLPforLargeScaleAntenna}. The authors provide ADMM-Net for large-scale mmWave communications in which the DL-based SLP selects the optimum subset of RF chains to decrease power consumption. Simulation results illustrate that ADMM-Net outperforms conventional ADMM, orthogonal matching pursuit, and coordinated descent algorithms. 

Considering a different approach than prior research of model-driven DL networks that unfold existing iterative algorithms,~\cite{Burak_101_2021_CINN} presents a DL-based precoding architecture driven by the constructive interference communication model. The suggested CI-NN model includes a customized loss function implemented through a customized layer following the output. CI-NN attains the same performance as the conventional constructive interference model and outperforms the linear ZF precoder, as indicated in simulations for a different number of users. Besides, CI-NN is a user-adaptive method in which a trained model with a specific number of users operates in scenarios with a variable number of users, which is practical for dynamic environments.

\begin{figure}
   \centering
   \includegraphics[width=\linewidth]{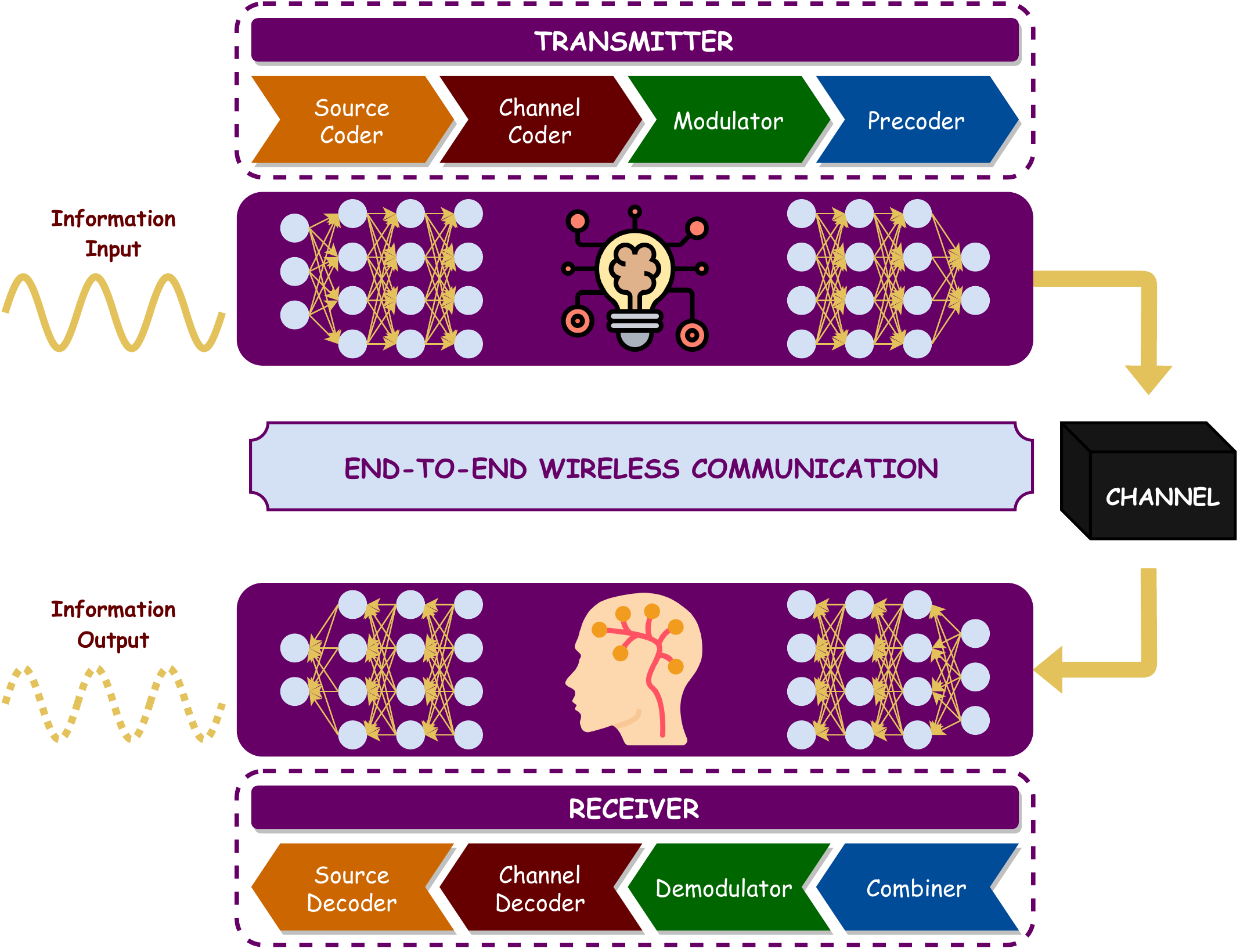}
   \caption{An E2E PHY communication system model, where individual signal processing blocks are replaced with a single adaptive DL network.}
   \label{fig:E2E}
\end{figure}

After thoroughly investigating the intelligent massive MU-MIMO transmitter architectures such as AE-based and DNN-based complex-shaped constellations and precoding designs, it is the perfect time to focus on E2E communication systems in which a single giant DL network substitutes for all individual blocks. This approach, illustrated by an example system model in Fig.~\ref{fig:E2E}, \textit{can enable E2E optimization without requiring a mathematical channel model}, which is highly practical since the mathematical definition of the channel might be challenging to obtain in some complicated environments. The main issue with the unknown channel model is that the conventional backpropagation technique, on which most of the DL networks rely, necessitates complete knowledge of the gradients in each layer. When the channel model is unknown, it is impossible to obtain gradients which complicate{s} the backpropagation. Researchers in this field have been proposing various solutions to this problem. The study in~\cite{Burak_102_2018_BackpropagatingThroughTheAir} presents an optimization technique called simultaneous perturbation stochastic optimization to approximate channel gradients so that the E2E model becomes trainable using the standard backpropagation method. The E2E network with the suggested approximation achieves the theoretical BER performance in AWGN and Rayleigh fading channels.

The authors of~\cite{Burak_33_2018_E2EWithoutACHannelModel} and~\cite{Burak_104_2019_ModelFreeTrainingE2ECommSysts} introduce a novel aspect to the unknown gradient challenge in which the E2E network iterates between the reinforcement learning-based transmitter training and supervised receiver training. The transmitter acts as an agent in an reinforcement learning problem, learning the environment using the loss function from the receiver as the reward. On the other hand, the receiver is trained in a supervised manner to obtain the loss function. Simulation results indicate that the proposed alternating method performs similarly to the E2E supervised DNN network with a known differentiable channel model. On top of this alternating training technique, the work in~\cite{Burak_105_2019_DeepRLAEWithNoisyFeedback} proposes a reliable feedback system in which the receiver can convey loss value to the transmitter without assuming a flawless feedback link. The suggested feedback system trains both ends of the communication as transmitter and receiver, removing the necessity for a perfect feedback link. When the E2E alternating network~\cite{Burak_33_2018_E2EWithoutACHannelModel},~\cite{Burak_104_2019_ModelFreeTrainingE2ECommSysts} equipped with the recommended feedback system is trained in a noisy feedback environment, it performs almost the same as the E2E alternating network with perfect feedback link, as simulations indicate. 

The studies in~\cite{Burak_106_2018_ChannelAgnosticE2ELearningConditionalGAN} and~\cite{Burak_107_2020_DLBasedE2EConditionalGANsUnknownChannel} provide a channel-agnostic E2E network with a conditional GAN modeling the channel to produce gradients. Using the encoded signals from the transmitter and the received pilot signals as the conditioning information, the conditional GAN generates the channel model and paves the way through supervised training. Extensive simulations yield that the suggested network achieves the performance of an E2E-DNN network with a known channel model over AWGN and Rayleigh fading channels. However, the authors of~\cite{Burak_108_2021_ResidualAidedE2ELearningWithoutKnownChannel} and~\cite{Burak_109_2021_E2ELearningCommSystWithoutKnownChannel} point out the gradient vanishing and overfitting issues of the GAN-based E2E network. They offer a residual-aided GAN that generates the difference between the transmitted and received signal. {R}esidual-aided GAN introduces extra gradients to the network to overcome {the} gradient vanishing problem. In addition, the loss function is modified by adding a regularizer to prevent overfitting. Simulations using theoretical AWGN and Rayleigh fading channels and practical ray-tracing channel dataset DeepMIMO~\cite{Burak_60_2019_DeepMIMOPaper},~\cite{Burak_61_DeepMIMOWebsite} illustrate that the proposed residual-aided GAN outperforms the E2E alternating network~\cite{Burak_33_2018_E2EWithoutACHannelModel},~\cite{Burak_104_2019_ModelFreeTrainingE2ECommSysts} and conventional GAN-based E2E network~\cite{Burak_106_2018_ChannelAgnosticE2ELearningConditionalGAN},~\cite{Burak_107_2020_DLBasedE2EConditionalGANsUnknownChannel} in all cases. 

Instead of a GAN imitating the channel, the studies of~\cite{Burak_110_2021_DLBasedE2EWithoutPilots} and~\cite{Burak_111_2020_BilinearConvolutionalAE_PilotFreeE2E} suggests a stochastic convolutional layer for representing channel in an E2E network without pilot signals. The receiver DNN consists of two DNNs for channel information extraction and data recovery, with the extracted channel information merged with the received signal via a bilinear production operation used for data recovery. Extensive simulations in frequency-selective and flat-fading MIMO channels yield that the recommended approach outperforms the LS and MMSE channel estimations while trailing the GAN-based E2E network. In addition, the E2E method is exceptionally resistant to channel correlations, whereas the conventional techniques deteriorate even with perfect CSI knowledge. Another intriguing experiment is wireless image transmission in which the authors test their E2E network with the recovered image quality, which performs better than conventional compression techniques.

E2E networks are easier to use when the channel model is accessible since channel gradients exist and the network is trainable in an E2E fashion. The pioneering study in this area, which is given in~\cite{Burak_18_2017_AnIntroDLPHY}, analyzes an E2E network for a given channel model and a loss function. This comprehensive research includes various investigations. First, the AE-based E2E network is compared against an uncoded BPSK scheme at several coding rates, with the proposed network outperforming the latter. To observe expert knowledge in E2E networks, the authors amend an RTN on top of the AE-based network, which improves performance. Extending the AE concept to multiple transmitters and receivers, the authors examine the performance of an AE-based E2E network in an interference channel. Comparison against a time-sharing scheme reveals substantial gains. In addition, a CNN-based DL network yields promising performance within the context of modulation classification and outperforms the expert feature approaches. This exhaustive treatise opens the door for our ultimate goal of PHY revolution. The authors of~\cite{Burak_112_2017_PHYDLEncodingsMIMOFadingChannel} apply this innovative E2E technique to MIMO system models and investigate the performance of AEs across Rayleigh fading channels for various MIMO configurations. In the first case, an AE optimizes its encoding scheme during training to introduce spatial diversity in a $2 \times 1$ MIMO setup. Compared to the well-known Alamouti coding, the AE creates a superior encoding scheme without CSI at the transmitter, as demonstrated in simulations, where the AE achieves slightly higher diversity in two time slots. This improvement is the consequence of the uneven distribution of power across antennas. In the second case, another AE is trained to obtain spatial multiplexing with lower BER than a traditional MIMO system in a $2 \times 2$ MIMO setting. Similar to the situation of spatial diversity, the AE generates better spatial multiplexing code and yields substantially higher BER performance. 

\cite{Burak_113_2018_LearningPHYLayerSchemeMIMOInterferenceChannel} extends the single-input single-output (\acrshort{siso}) IC system over AWGN channel investigated in~\cite{Burak_18_2017_AnIntroDLPHY} to SISO and MIMO interference channel systems over Rayleigh fading channel. The proposed AE~\cite{Burak_18_2017_AnIntroDLPHY} with a modulation order of four in a SISO interference channel system performs slightly better than the standard single-user no interference MIMO system using the QPSK modulation scheme. The demonstration of the received signals reveals well-separated four regions corresponding to the constellation points, which indicates how effectively the AE learns how to transmit with a modulation order of four. With the AE trained in a MIMO interference channel, performance improvement becomes considerable compared to the standard single-user no interference MIMO system. Assuming the channel model is available as in~\cite{Burak_18_2017_AnIntroDLPHY}, the study in~\cite{Burak_114_2019_E2EBlockAEPHYLayerBasedOnNN} proposes a block AE-based E2E network to deal with the dynamic input lengths, which is notable in terms of efficiency. The suggested network employs RNN to build a memory mechanism. Simulation results show that the RNN-based AE network has superior performance over the CNN-based networks in AWGN and custom fading channels. Furthermore, the authors examine the learned 64-QAM constellation diagram. 

The work in~\cite{Burak_115_2020_TrainableCommunicationSystemsConceptsProto} provides another thorough research that includes numerous examinations for E2E communication systems. The principal motivation is to switch from symbol-wise AE to bit-wise AE for the first time in the literature to show that optimizing bit metrics improves the overall performance. Bit-wise AE with IEEE 802.11n (Wi-Fi 4) LDPC code outperforms the symbol-wise AE with IEEE 802.11n LDPC code and the conventional communication using PSK or QAM. Also, the learned constellation analysis proves that bit-wise optimization leads to a better separation in the constellation, which is a solid indicator of performance improvement. The authors further enhance the BER performance by replacing the demapper of the model with the iterative demapping and decoding scheme, which results in a superior BER performance compared to the bit-wise AE without iterative demapping and decoding at the receiver. In addition, they optimize the LDPC codes used in iterative demapping and decoding-aided bit-wise AE and obtain enhancement in the BER performance. Although the AWGN channel is assumed in computer simulations, the strength of the proposed technique is its flexibility to be applied to any channel model without modification, as demonstrated by simulations utilizing software-defined radio (\acrshort{sdr}). The study in~\cite{Burak_116_2020_AEBasedRobustTransceiversFadingChannelsDNNs} investigates a solution to the problem of finding $(n,k)$ block codes to maximize the minimum Hamming distance between the codewords by using AEs. Along with a custom loss function, the authors suggest two receiver structures depending on DNN and bidirectional RNN. The proposed model performs almost the same as the optimal block codes in the AWGN channel, and the minimum Hamming distance learned by the AE achieves its optimal value specified in theory. In addition, AE-based block code outperforms the theoretical optimum under channel model mismatches in which the performance difference increases as the correlation level grows.

According to the reviewed major studies in the literature, we note that there is noticeable progress toward entirely intelligent and E2E optimized wireless communication systems by replacing all separate blocks in conventional communication systems with a single massive DL network. Without a doubt, there will be more developments soon to incorporate these innovative approaches into real life. In the next subsection, we will explore the emerging applications of DL in transforming classical IM to make it a compelling technology for 6G.

\subsection{DL-Empowered IM 2.0: Improving Efficiency of Conventional IM}

\begin{figure}
   \centering
   \includegraphics[width=8cm]{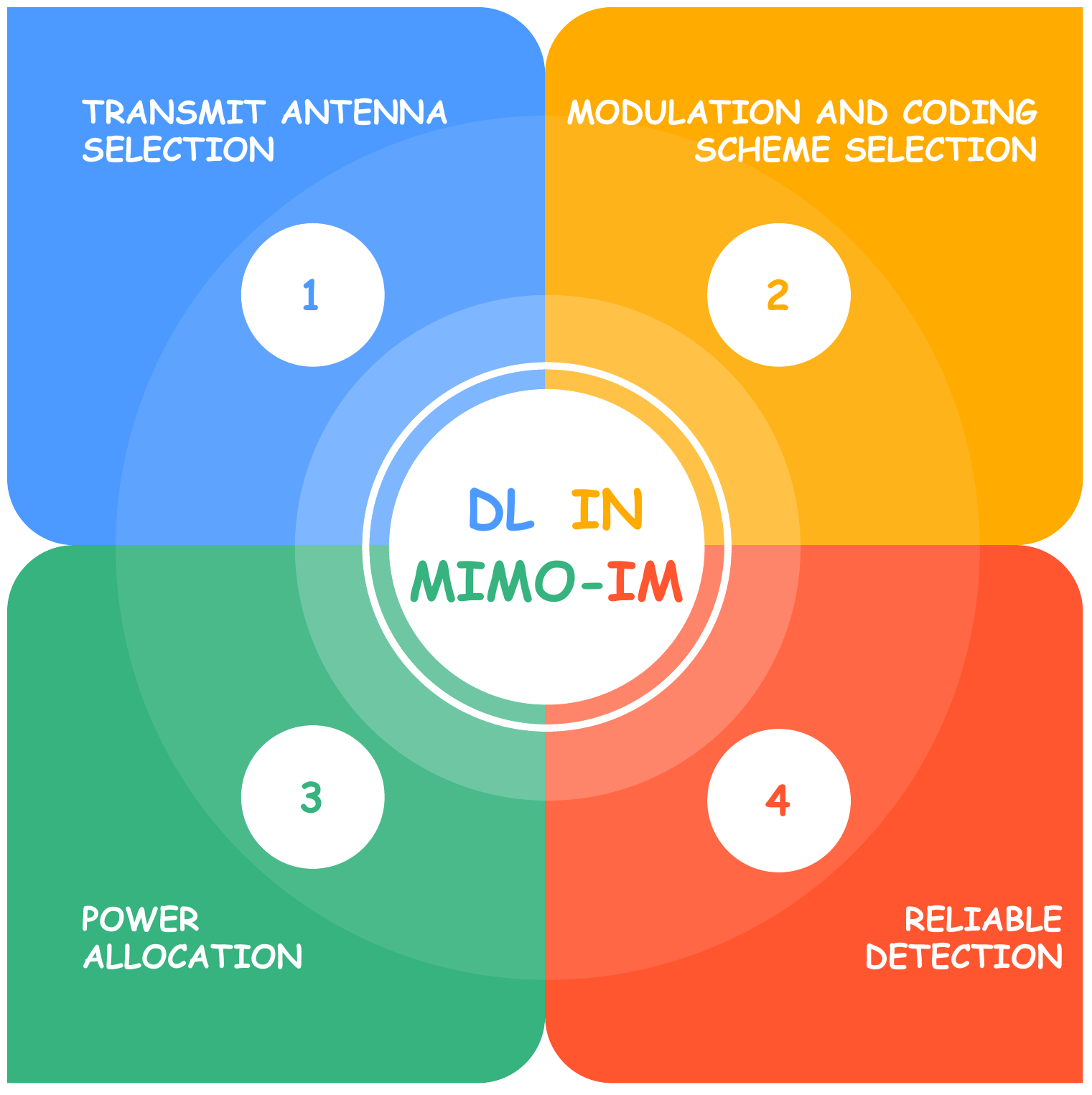}
   \caption{Four main applications of DL in MIMO-IM.}
   \label{fig:DL_ApplicationsOnIM}
\end{figure}

We have covered the main developments in DL applications for massive MIMO systems thus far, beginning with symbol detection at the receiver and progressing to E2E intelligent communication systems. Another promising aspect within the 6G vision is MIMO-IM, which, as previously stated, enables high energy and spectral efficiency together while also decreasing the number of required RF chains by leveraging the building blocks of a communication system as additional information sources. However, the trade-off between detector complexity and BER performance is an issue for IM systems. In addition, employing several advanced algorithms such as transmit antenna selection (\acrshort{tas}), power allocation, and MCS selection can further improve the BER performance of IM systems. Because of the immense complexity of traditional methods to implement these algorithms, DL approaches seem appealing for the efficiency improvement of IM systems. Fig.~\ref{fig:DL_ApplicationsOnIM} categorizes the primary contributions of DL to IM systems.

The study in~\cite{Burak_117_2019_AdaptiveSMBasedOnML} proposes a novel data-driven framework for TAS and power allocation issues in SM-MIMO systems by considering two different ML techniques. As the initial contribution, the framework achieves feasible solutions with decreased complexity comparing conventional optimization approaches by utilizing supervised-learning classifiers such as the $K$-nearest neighbors (\acrshort{knn}) and support vector machine (\acrshort{svm}). The authors examine a DNN-based framework for optimization tasks in the second part. Furthermore, they explore the impact of feature vector design on BER performance for both methods. Specifically, they suggest three distinct feature vector generation (\acrshort{fvg}) methods: conventional FVG with the modulus of the channel elements, separate FVG with the modulus of the real and imaginary parts of the channel elements, and joint FVG with the correlation of the channel matrix's column pairs. Extensive simulations over various MIMO configurations and modulation schemes show that SVM-based TAS-SM outperforms KNN-based TAS-SM, while DNN-based TAS-SM exceeds both. Joint FVG has superior performance than the other two FVG approaches with near-optimal performance. Similarly, the DNN-based power allocation in SM with joint FVG achieves the best BER performance compared to other power allocation scheme and FVG combinations. On top of that,~\cite{Burak_118_2021_NovelDNNBasedASArchitectureForSM} minimizes the computing overhead by removing the repeated elements of the antenna selection process. The proposed DNN outperforms the proposed method in~\cite{Burak_117_2019_AdaptiveSMBasedOnML} with fewer hidden neurons in which both schemes employ the joint FVG technique. However, Euclidean distance-optimized antenna selection (\acrshort{edas}) has substantial performance than these DNN-based TAS-SM techniques.

In~\cite{Burak_119_2020_EfficientSelectiononSMAntennasLearningBoosting}, the authors suggest a DNN-based and a gradient boosting decision tree based TAS-SM technique. Considering the importance of features obtained from the channel matrix, they employ channel matrix correlation, same with the joint FVG in~\cite{Burak_117_2019_AdaptiveSMBasedOnML}, as the feature vector. Simulation results indicate that both DNN- and gradient boosting decision tree based TAS-SM methods perform near-optimal performance while outperforming conventional SM and random TAS-SM schemes. In addition, gradient boosting decision tree attains lower complexity than DNN with slightly worse BER performance. Similar to~\cite{Burak_119_2020_EfficientSelectiononSMAntennasLearningBoosting}, the work in~\cite{Burak_120_2020_TASforLSMIMOGSMwithML} provides a DNN-based and a decision tree based TAS-GSM scheme under practical channel scenarios. The authors create a $16 \times 4$ test-bed using SDRs for tests and estimation of the impairments of the channel for theoretical optimization benchmark. Experiments reveal that the suggested techniques outperform EDAS while DNN-based TAS-GSM exceeds decision tree based TAS-GSM under the real-life scenario. The authors of~\cite{Burak_121_2021_SLCBasedTASforSMMIMO} offer another TAS-SM scheme depending on DNNs combining the modulus of the channel matrix and the channel matrix correlation as the feature vector, which corresponds to the combination of conventional FVG and joint FVG in~\cite{Burak_117_2019_AdaptiveSMBasedOnML}. DNN-based TAS-SM with the recommended FVG type performs similar{ly} to EDAS while outperforming random forest decision{-}based TAS-SM and various conventional TAS-SM schemes and achieve{s} lower complexity. 

The TAS problem is addressed in~\cite{Burak_122_2021_TASforFullDuplexSMBasedOnML} for full-duplex SM systems in which self-interference is a crucial issue using DNN and SVM methods. The authors first derive the upper and lower bounds of the full-duplex SM's channel capacity and subsequently utilize these derivations in EDAS and capacity-optimized antenna selection to generate a training label set. In contrast to previous TAS methods majority of which use joint FVG of~\cite{Burak_117_2019_AdaptiveSMBasedOnML}, they propose eigenvector FVG based on principal component analysis to extract vector features of the channel matrix and enhance the BER performance. Extensive simulations illustrate that the proposed eigenvector FVG leads to a better performance than joint FVG of~\cite{Burak_117_2019_AdaptiveSMBasedOnML} in various transmit antenna number and modulation scheme settings. The proposed DNN and SVM methods function almost the same and are comparable to EDAS. They outperform capacity-optimized antenna selection, KNN of~\cite{Burak_117_2019_AdaptiveSMBasedOnML}, and other traditional antenna selection algorithms.

Apart from TAS and power allocation, MCS selection is another essential concept for enhancing the efficiency of IM systems. The study in~\cite{Burak_123_2018_LearningApproachforOptimalCodebookinSM} describes a DNN-based codebook selection method for SM-MIMO systems. The receiver predicts the symbol error rate (\acrshort{ser}) corresponding to each codebook using the instantaneous channel state and sends the optimal codebook back to the transmitter. Each codebook assigns different constellation combinations to transmit antennas, where the total number of bits per channel use remains constant. Dynamically switching between these codebooks according to DNN-based SER predictions yields significantly better results than using any codebook throughout the transmission. The work in~\cite{Burak_124_2020_NNAidedComputationMutualInfoAdaptationSM} offers a DNN-based approach to {calculating} mutual information of an SM-MIMO system on the fly, allowing for dynamic MCS adaptation. The authors explore different input feature vectors, where they contain some features from channel matrix and SNR. The proposed DNN-based method obtains mutual information almost perfectly with significantly lower complexity than traditional Taylor and Jensen approximations, paving the door for MCS adaptation. Furthermore, the authors expand their analyses in~\cite{Burak_125_2019_NNAidedComputationofGSMCapacity} to calculate the channel capacity of a GSM-MIMO system model, for the same purpose with~\cite{Burak_124_2020_NNAidedComputationMutualInfoAdaptationSM} as allowing dynamic adaptation. The suggested DNN-based model computes the capacity with almost no error, as provided by simulations. 

\cite{Burak_126_2019_DLAssistedRateAdaptationinSMLinks} presents a DNN-based code rate selection technique in SM-MIMO systems. The proposed DNN extracts the norms of channel matrix columns and the angles between each pair of channel matrix columns and combines them with SNR as input features, as described in~\cite{Burak_124_2020_NNAidedComputationMutualInfoAdaptationSM}. The DNN output is the maximum coding rate that is suitable under a BER constraint. Simulation results show that DNN-based dynamic code rate selection delivers near-maximum throughput and substantially outperforms the fixed-rate scenarios under a BER constraint. The authors extend their model in~\cite{Burak_127_2019_SMLinkAdaptationDLApproach} to predict the optimal MCS rather than only code rate depending on the same input features. The DNN-based MCS selection approach attains near-maximum performance, as does the sole coding rate optimization scenario in~\cite{Burak_126_2019_DLAssistedRateAdaptationinSMLinks}, indicating enhanced efficiency in SM-MIMO systems.

In contrast to the former studies,~\cite{Burak_128_2021_ANNBasedAdaptiveSM} provides a DNN-based modulation order selection scheme without considering the coding rate of the system. The recommended DNN model predicts the optimal modulation order by maximizing the minimum Euclidean distance, similar to the methods implemented in TAS-IM algorithms. Simulation results indicate that DNN-based modulation order selection aided SM-MIMO significantly outperforms the classical SM with substantially lower complexity than conventional selection algorithms.

Studies discovered thus far unlock the potential of IM systems for future 6G and beyond communication technologies, owing to DL approaches that further enhance system efficiency by employing sophisticated algorithms such as TAS, power allocation, and MCS selection with remarkably low complexity. We may now conclude the examination of the literature on DL-aided MIMO systems and move on to the implementation of the generic MIMO-IM model presented in~\cite{Burak_40_2020_BDNN}, which might give a sense of MIMO-DL programming to interested readers.

\subsection{MIMO-DL in Action}
GSM is one of the most promising techniques for the PHY of future wireless communication systems by combining the benefits of SM and MIMO. Reducing the number of required RF chains by activating only a subset of the transmit antennas, GSM has lower complexity than conventional MIMO. It also improves SM efficiency by activating more transmit antennas and conveying more information bits. However, when the number of transmit antenna combinations and information bits increases, the complexity of MLD detection escalates to intolerable levels, as discussed earlier. On the other hand, linear detectors, such as ZF and MMSE detectors, produce significantly worse BER performance than MLD, introducing a trade-off into GSM systems. To address this BER-complexity trade-off, the authors of~\cite{Burak_40_2020_BDNN} suggest a DNN-based detector for GSM systems. They consider a GSM system model with $N_{t}$ transmit and $N_{r}$ receive antennas $(N_{r} < N_{t})$, where $N_{p}$ $(2 \leq N_{p} \ll N_{t})$ out of $N_{t}$ transmit antennas are selected by index bits to be activated in any time slot. There are $N = 2^{\lfloor \text{log}_{2} \binom{N_{t}}{N_{p}} \rfloor}$ legitimate transmit antenna combinations (\acrshort{tac}), where each active antenna transmits a symbol from normalized $M$-QAM constellation $\mathcal{S}$. Thus, the total number of bits transmitted in each time slot equals $B = \text{log}_{2}N + N_{p}\text{log}_{2}M$.

Assuming a quasi-static flat fading MIMO channel matrix, $\mathbf{H} \in \mathbb{C}^{N_{r} \times N_{t}}$, the entries of which follow complex Gaussian distribution $\mathcal{C} \mathcal{N}(0, 1)$, the received signal vector $\mathbf{y} \in \mathbb{C}^{N_{r} \times 1}$ can be expressed as follows:
\begin{equation} \label{received_signal_vector}
    \mathbf{y} = \mathbf{H}\mathbf{x} + \mathbf{n} =
    \mathbf{H}_{I} \mathbf{s} + \mathbf{n},
\end{equation}
where $\mathbf{x}$ is the vector of transmit symbols, $\mathbf{n} \in \mathbb{C}^{N_{r} \times 1}$ is the additive white Gaussian noise samples vector with complex Gaussian distribution $\mathcal{C} \mathcal{N}(\mathbf{0}, \sigma^{2} \mathbf{\mathbf{I}})$, $\mathbf{s}$ is the symbol vector of the active transmit antennas, and $\mathbf{H}_{I}$ is the channel matrix corresponding to the active transmit antennas.

\begin{figure*}[t]
   \centering
   \includegraphics[width=\textwidth]{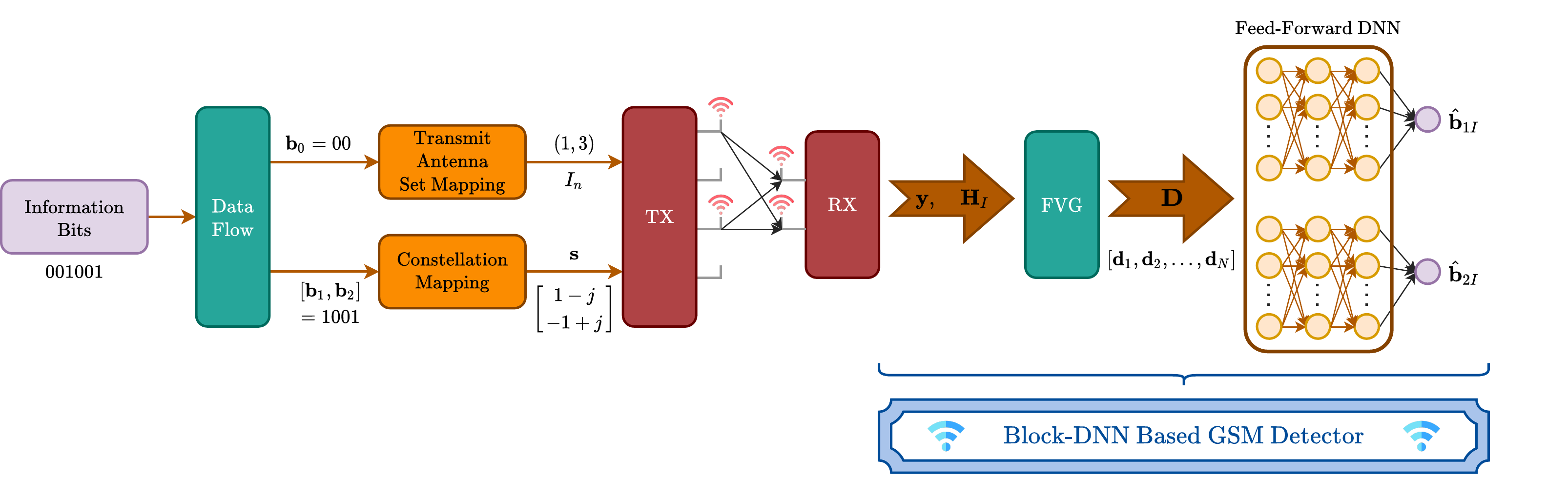}
   \caption{System model of the B-DNN detector that integrates a DNN into classical block detection for GSM systems~\cite{Burak_40_2020_BDNN}.}
   \label{fig:BDNN}
\end{figure*}


The optimal MLD detector for this GSM transmission scheme performs an exhaustive search over all possible TACs and information symbols, as shown below
\begin{equation} \label{ML_Detection}
    (\hat{I}, \hat{\mathbf{s}}) = \argmin_{I \in \mathbb{I}, \mathbf{s} \in \mathbb{S}} ||\mathbf{y} - \mathbf{H}_{I} \mathbf{s}||^{2}_{F},
\end{equation}
where $\mathbb{I} = \{I_{1},I_{2}, \dots, I_{N}\}$ is the set of legitimate TACs and $\mathbb{S}$ is the set of possible QAM symbol vectors of active transmit antennas. The linear detectors deal with solving an inverse operation including the channel matrix, $\mathbf{H}$, which can be given as follows:
\begin{equation} \label{ZF_MMSE_Detectors}
    \begin{aligned}
        &\hat{\mathbf{x}}_{ZF} = (\mathbf{H}^{H}\mathbf{H})^{-1} \mathbf{H}^{H} \mathbf{y},
        \\
        &\hat{\mathbf{x}}_{MMSE} = (\mathbf{H}^{H}\mathbf{H} + \sigma^{2}\mathbf{I})^{-1} \mathbf{H}^{H} \mathbf{y}.
    \end{aligned}
\end{equation}
Here, $\hat{\mathbf{x}}_{ZF}$ and $\hat{\mathbf{x}}_{MMSE}$ represent detected symbol vectors by ZF and MMSE detectors, respectively. To perform a successful inverse operation, $N_{r}$ should at least equal to $N_{t}$ which ensures that $\mathbf{H}^{H}\mathbf{H}$ has a full rank. However, this requirement reduces the practical feasibility of GSM in massive MIMO systems with hundreds of transmit antennas at the BS since it is almost impossible for a user equipment to install that many antennas. Therefore, block linear detectors slightly change the equations in (\ref{ZF_MMSE_Detectors}) and employ $\mathbf{H}_{I}$ instead of $\mathbf{H}$, which results in $N$ solutions of $\hat{\mathbf{s}}_{I}$ for each legitimate TAC. In the final stage, block linear detectors compare the Euclidean distance between the received signal vector $\mathbf{y}$ and each $\mathbf{H}_{I} \hat{\mathbf{s}}_{I}$ product to estimate selected TAC and QAM symbols, as given below
\begin{equation} \label{block_linear_detection_final_stage}
    \begin{aligned}
        (\hat{I}) &= \argmin_{I \in \mathbb{I}} ||\mathbf{y} - \mathbf{H}_{I} \hat{\mathbf{s}}_{I}||^{2}_{F},
        \\
        (\hat{\mathbf{s}}) &= \hat{\mathbf{s}}_{\hat{I}}.
    \end{aligned}   
\end{equation}

This block-structured detection scheme achieves better BER performance than the classical linear detectors since it incorporates the MLD approach to the linear detectors. However, complexity increases with the growing number of transmit antennas, as expected, and the block detection still performs worse than the optimal MLD due to the suboptimal linear part. Thus, the authors of~\cite{Burak_40_2020_BDNN} replace the linear component of the block detection with a DNN to enhance the accuracy of $\hat{\mathbf{s}}_{I}$ estimation. Fig.~\ref{fig:BDNN} demonstrates the GSM transmitter and the considered block-DNN detector.

\begin{table}
    \caption{Network and training parameters~\cite{Burak_40_2020_BDNN}.}

    \centering
    \begin{tabular}{|M{1.7cm}|M{1.7cm}||M{1.7cm}|M{1.7cm}|}
        \hline
        
        \cellcolor{table_blue} \textbf{\textcolor{table_white}{Parameters}} & 
        \cellcolor{table_blue} \textbf{\textcolor{table_white}{Value}} &
        \cellcolor{table_blue} \textbf{\textcolor{table_white}{Parameters}} & 
        \cellcolor{table_blue} \textbf{\textcolor{table_white}{Value}} \\
        \hline
        
        \cellcolor{table_white} \textcolor{table_blue}{Input Nodes} &
        \cellcolor{table_white} \textcolor{table_blue}{$2(N_{r} + N_{r} N_{p})$} &
        \cellcolor{table_white} \textcolor{table_blue}{Learning Rate} &
        \cellcolor{table_white} \textcolor{table_blue}{$0.005$} \\
        \hline
        
        \cellcolor{table_white} \textcolor{table_blue}{Hidden Layer} &
        \cellcolor{table_white} \textcolor{table_blue}{$3$} &
        \cellcolor{table_white} \textcolor{table_blue}{Number of Training Set} &
        \cellcolor{table_white} \textcolor{table_blue}{$15.000.000$} \\
        \hline
        
        \cellcolor{table_white} \textcolor{table_blue}{Output Nodes} &
        \cellcolor{table_white} \textcolor{table_blue}{$M$} &
        \cellcolor{table_white} \textcolor{table_blue}{Number of Validation Set} &
        \cellcolor{table_white} \textcolor{table_blue}{$5.000.000$} \\
        \hline
        
        \cellcolor{table_white} \textcolor{table_blue}{Hidden Layer Activation} &
        \cellcolor{table_white} \textcolor{table_blue}{ReLU} &
        \cellcolor{table_white} \textcolor{table_blue}{Epoch} &
        \cellcolor{table_white} \textcolor{table_blue}{$50$} \\
        \hline
        
        \cellcolor{table_white} \textcolor{table_blue}{Output Layer Activation} &
        \cellcolor{table_white} \textcolor{table_blue}{Softmax} &
        \cellcolor{table_white} \textcolor{table_blue}{BPSK Hidden Nodes} &
        \cellcolor{table_white} \textcolor{table_blue}{$128$-$64$-$32$} \\
        \hline
        
        \cellcolor{table_white} \textcolor{table_blue}{Loss Function} &
        \cellcolor{table_white} \textcolor{table_blue}{Cross-Entropy} &
        \cellcolor{table_white} \textcolor{table_blue}{QPSK Hidden Nodes} &
        \cellcolor{table_white} \textcolor{table_blue}{$256$-$128$-$64$} \\
        \hline
        
        \cellcolor{table_white} \textcolor{table_blue}{Optimizer} &
        \cellcolor{table_white} \textcolor{table_blue}{SGD} &
        \cellcolor{table_white} \textcolor{table_blue}{$16$-QAM Hidden Nodes} &
        \cellcolor{table_white} \textcolor{table_blue}{$512$-$256$-$128$} \\
        \hline
        
    \end{tabular}
    \label{tab:BDNN_NetworkTrainingParameters}
\end{table}

As shown in Fig.~\ref{fig:BDNN}, the block-DNN detector consists of two stages: FVG and feed-forward DNN. In the first stage, the FVG generates the input feature vector $\mathbf{d}_{i}$ of the feed-forward DNN by processing the raw data composed of the received signal vector $\mathbf{y}$ and the channel matrix $\mathbf{H}_{I}$ corresponding to the $i^{\text{th}}$ legitimate TAC. As described in~\cite{Burak_117_2019_AdaptiveSMBasedOnML}, the selected FVG type significantly impacts model performance, and the joint FVG results in the best performance among the other FVG types for the model suggested in~\cite{Burak_117_2019_AdaptiveSMBasedOnML}. Hence, the authors of~\cite{Burak_40_2020_BDNN} explore the influence of FVG type on detection reliability and evaluate the BER performance of the block-DNN detector for three FVG kinds described in~\cite{Burak_117_2019_AdaptiveSMBasedOnML}: separate FVG, joint FVG, and conventional FVG. Combining the feature vectors of all legitimate TACs, the FVG creates the feature matrix $\mathbf{D}$ and feeds it to the feed-forward DNN. In the second stage, the DNN yields a $\hat{\mathbf{s}}_{I}$ for each $\mathbf{d}_{i}$. The feed-forward DNN contains $N_{p}$ sub-DNNs to estimate the QAM symbol corresponding to each active transmit antenna. Finally, the block-DNN detector employs (\ref{block_linear_detection_final_stage}) to obtain the selected TAC and QAM symbols. We reproduce some of the figures in~\cite{Burak_40_2020_BDNN} using the network and training parameters given in Table~\ref{tab:BDNN_NetworkTrainingParameters} and share some portion of example codes. The training procedure of any block-DNN detector ignores the GSM scheme at the transmitter, assuming all transmit antennas are active, which makes sense since DNN is only responsible for detecting QAM symbols given a TAC. In addition, the received signal does not experience AWGN during the training. Here is an example model creation of a DNN with separate FVG for QPSK transmission, which is another version of model definition in Keras different than what we provided in Section II-B: \vspace{-1ex}

\begin{lstlisting}
    from tensorflow.keras import Input, Model
    from tensorflow.keras.layers import  Dense, BatchNormalization
    from tensorflow.keras.optimizers import SGD
    from tensorflow.keras.regularizers import l2
    y_lst, l_lst, m_lst = [], [], []
    M, Nt, Np, Nr = 4, 2, 2, 2
    n_x, n_y = 2 * Nr + 2 * Nr * Np, M
    x = Input(shape=(n_x,))
    for i in range(Np):
        h1 = Dense(256, kernel_regularizer=l2(l=0.001), activation="relu", name="D" + str(i) + "1")(x)
        b1 = BatchNormalization(name="BN" + str(i) + "1")(h1)
        h2 = Dense(128, kernel_regularizer=l2(l=0.001), activation="relu", name="D" + str(i) + "2")(b1)
        b2 = BatchNormalization(name="BN" + str(i) + "2")(h2)
        h3 = Dense(64, kernel_regularizer=l2(l=0.001), activation="relu", name="D" + str(i) + "3")(b2)
        b3 = BatchNormalization(name="BN" + str(i) + "3")(h3)
        y = Dense(n_y, kernel_regularizer=l2(l=0.001), activation="softmax", name="O" + str(i))(b3)
        y_lst.append(y)
        l_lst.append("categorical_crossentropy")
        m_lst.append("accuracy")
    model = Model(inputs=x, outputs=y_lst)
    SGD_opt = SGD(lr=0.005, nesterov=True)
    model.compile(optimizer=SGD_opt, loss=l_lst, metrics=m_lst)
\end{lstlisting}

\begin{figure*}
   \centering
   \includegraphics[width=\textwidth]{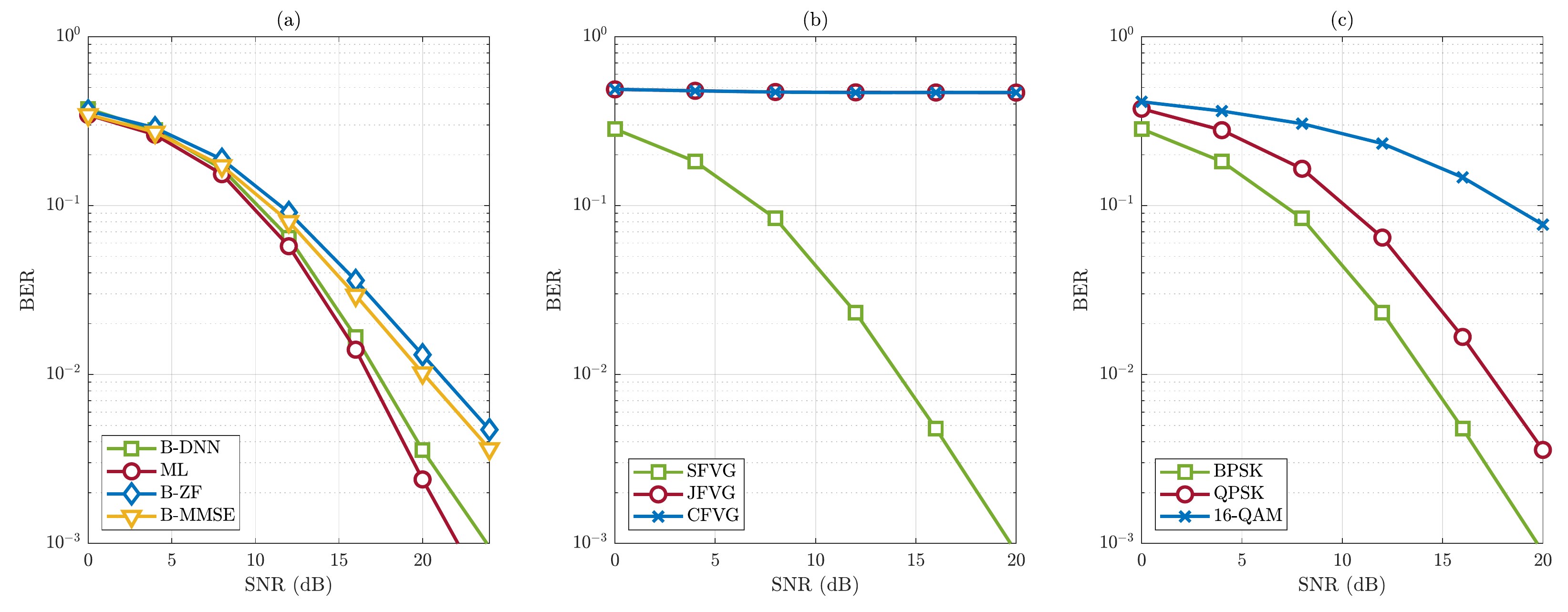}
   \caption{BER comparisons for the B-DNN detector (a) Conventional detectors vs B-DNN detector, (b) B-DNN detector with various FVG types, (c) B-DNN detector under various modulation levels.}
   \label{fig:BER_Figures}
\end{figure*}


Here, we create a chain of layers instead of using the \textcolor{codeblue}{\textbf{Sequential}} module for each active transmit antenna by a \textcolor{codepink}{\textbf{for}} loop, and combine them to a single \textcolor{codeblue}{\textbf{Model}}. The \textcolor{codeblue}{\textbf{BatchNormalization}} and \textcolor{codeblue}{\textbf{l2}} modules represent the batch normalization layer and l2 regularizer, respectively. The next step is defining PHY parameters prior to constructing the training data: 
\begin{lstlisting}
    import WirelessCommLib as wcl
    N_tot = wcl.Combination(Nt, Np)
    ns = int(np.floor(np.log2(N_tot)))
    m = int(np.log2(M))
    ni = Np * m
    n_tot = ns + ni
    N = 2 ** ns
    is_normalized = True
    ss = wcl.Constellation(M, mod_type, is_normalized)
    TAC_set = wcl.OptimalTAC_Set(Nt, Np, N)
\end{lstlisting}
The \textbf{WirelessCommLib} is our package for PHY-related tasks, in which \textbf{Combination(a, b)} function corresponds to the number of picking unordered b items out of a items, \textbf{Constellation} returns a symbol set given a modulation level $M$ and modulation type, like PSK or QAM. \textbf{OptimalTAC\_Set} creates the set of legitimate TACs from all possible TACs. Optimality indicates that the TAC set is not created by randomly picking some of the TACs, instead TAC set is created so that the number of each transmit antenna in the set is either equal or close to each other. Now, we can construct the input training data: 
\begin{lstlisting}
    Ns = 15000000
    FVG_type = "SFVG"
    bit_matrix = np.random.randint(2, size=(n_tot, Ns))
    for j in range(Ns):
        bit_array = bit_matrix[:, j]
        x = wcl.EncodeBits(bit_array, ss, TAC_set, ns, m, Nt, Np)
        H = wcl.Channel([Nr, Nt])
        y = np.matmul(H, x)
        train_input_data[j, :] = np.concatenate((wcl.FVG(y, FVG_type), wcl.FVG(H, FVG_type)))[:, 0]
\end{lstlisting}
The number of training instances is equal to $N_{s}$. For input of each training data, we first randomly generate a bit array and encode these bits into a transmit vector $\mathbf{x}$ by using \textbf{EncodeBits} function. Subsequently, we randomly generate a i.i.d. channel matrix $\mathbf{H}$ whose elements follow the complex Gaussian distribution $\mathcal{CN}(0, 1)$ via \textbf{Channel} function, and multiply it with $\mathbf{x}$. The input data is the concatenation of the feature vectors of the received signal vector $\mathbf{y}$ and the channel matrix $\mathbf{H}$. The labels for this training data are the correct symbols transmitted from all antennas: \vspace{-1ex}
\begin{lstlisting}
    train_output_data = []
    for i in range(Np):
        labels = np.zeros((Ns, M))
        for j in range(Ns):
            start = ns + i * m
            stop = ns + (i + 1) * m
            bits = bit_matrix[start : stop, j]
            labels[j, wcl.Bin2Dec(bits)] = 1
        train_output_data.append(labels)
\end{lstlisting}
Here, \textbf{Bin2Dec} function converts the given bit array to the corresponding symbol. Once the training data and the model is ready, we can train the model: \vspace{-1ex}
\begin{lstlisting}
    model.fit(train_input_data, train_output_data, validation_split=0.25, batch_size=512, epochs=50, shuffle=True)
\end{lstlisting}
The codes, trained models, simulations for testing the B-DNN model and comparisons with the conventional linear detectors and MLD, as well as the generated figures are available in our online database\footnote{https://github.com/burakozpoyraz/Block-DNN}. Fig.~\ref{fig:BER_Figures}(a) illustrates that the proposed block-DNN detector outperforms the conventional block-linear detectors and performs similar to the optimal MLD detection assuming $N_{t} = 4$, $N_{r} = 2$, $N_{p} = 2$, and QPSK modulation. It should be noted that, compared to MLD, the proposed B-DNN model can achieve similar results with less time complexity. Fig~\ref{fig:BER_Figures}(b) shows the impact of FVG type on the BER performance of B-DNN detector under the same setting with the former except BPSK modulation. In contrast to the model of~\cite{Burak_117_2019_AdaptiveSMBasedOnML}, where joint FVG produces the best performance, the block-DNN detector that employs separate FVG delivers the best BER performance, while the other FVG types cause unreliable communication with a BER value of almost 0.5 across all SNR levels. Thus, the appropriate FVG type might vary depending on the application, necessitating thorough examination during the validation step. Finally, Fig.~\ref{fig:BER_Figures}(c) provides the BER performance of the block-DNN detector under various modulation levels, assuming the same antenna settings as before. The block-DNN with BPSK obtains the best BER performance, while 16-QAM performs the worst, as expected. As a trade-off, increasing the modulation order provides higher bit rates.

With this striking implementation, we may now end this part and go on to the following section's examination of DL methods for MC waveforms.

\section{DL for Multicarrier Waveform Design} \label{sec:waveform}
 Waveform, which characterizes the physical shape of information{-}carrying signals, is one of the key components of wireless communication systems~\cite{yardimci1}. Existing waveforms can be categorized into two groups as single-carrier (SC) and MC waveforms. In order to increase the data rate in a SC or MC system, symbol duration needs to be decreased or equivalently bandwidth occupied for data transmission should be increased.  However, {with} the adoption of wider bandwidth, SC waveforms suffer from frequency-selective fading that causes inter-symbol interference (ISI) since symbol duration is less than the delay spread of the wireless channel.  Since complex equalization techniques should be applied to reduce ISI, it is very challenging to provide high data rates with the existing SC waveforms. In {an} MC system, the frequency band is divided into many sub-bands which are also known as subcarriers~\cite{yardimci5}. These subcarriers are employed in parallel to convey information bits simultaneously. The spacing between two consecutive subcarriers is selected such that each of them undergoes flat fading in the frequency domain. Hence, a simple single-tap equalizer may be utilized to remove the channel effect in MC systems. Consequently, MC techniques enable wideband transmission to convey information; therefore, they yield higher data rates than SC techniques.
 
 OFDM appears as the most popular MC waveform and has been used in numerous standards such as LTE and the IEEE 802.11 family due to its simple and effective structure. Owing to the overlapped orthogonal subcarriers, OFDM is capable of using the spectrum efficiently. Moreover, the time frequency grid of OFDM allows the flexible use of resource elements. In conventional OFDM systems, modulated data symbols for each data subcarrier are determined in the frequency domain by mapping information bits to the PSK/QAM constellation. A certain number of subcarriers are allocated for the transmission of pilot symbols in order to perform channel estimation at the receiver side. These pilot symbols can be inserted into the OFDM symbol with block or comb type methods~\cite{yardimci2}. Channel coefficients can be estimated in the time-domain or frequency-domain and numerous estimation techniques, which provide different performance and complexity, have been proposed for these two domains~\cite{yardimci6, yardimci7}. In the frequency domain, the channel frequency response is estimated by exploiting pilot symbols and is interpolated to obtain the channel frequency response over data symbols. For example, the least squares (LS) and the linear minimum mean error square (\acrshort{lmmse}) are two well-known frequency domain channel estimation methods. After inserting pilot symbols, the time domain OFDM signal is obtained by employing the inverse fast Fourier transform (\acrshort{ifft}). Furthermore, a CP is embedded in the beginning or end of the OFDM symbol to eliminate ISI. Additionally, the CP allows us to model the frequency selective channel as circular convolution. After adding a CP, the OFDM signal is transmitted through the wireless communication channel. At the receiver side, the CP is removed and the frequency domain OFDM signal is acquired by taking FFT. Channel estimation is performed and equalization is applied to remove the effect of the wireless channel. Finally, the signal is demodulated to obtain information.
 
 Although OFDM has several advantages as discussed above, it also has many drawbacks such as high PAPR, sensitivity to frequency and timing errors, high out-of-band emissions (\acrshort{oobe}), and CP\&pilot overhead. After IFFT operation, the subcarriers are randomly summed up in the time domain and this may cause high peaks due to the overlapping of the peak amplitudes of different signals. The power amplifier at the transmitter works in the nonlinear region due to these high peaks, generating distortion and spectral dispersion. These peaks give rise to high PAPR that reduces the efficiency of the power amplifier, analog-to-digital, and digital-to-analog converters. Therefore, researchers have designed various PAPR reduction techniques for OFDM transmission systems~\cite{yardimci4}. Another drawback of OFDM is sensitivity to intercarrier interference (\acrshort{ici}) caused by high mobility, phase noise, timing offset, and carrier frequency offset (CFO). The channel in a mobile wireless environment rapidly changes with respect to time. This places a strain on the receiver since it must precisely estimate the channel prior to coherent detection. The most critical consequence of this time variation is the distortion of the orthogonality between subcarriers resulting in ICI, whose intensity is determined by the amount of channel time variation. Furthermore, ICI may also occur due to the frequency difference between the local oscillators of the transmitter and receiver. Since ICI disrupts OFDM subcarriers, data detection becomes less accurate, and a single tap equalizer can not provide satisfactory performance. Thus, more complicated methods are required to decode OFDM symbols. To combat this difficult interference mitigation problem under high mobility environments or {in} the presence of CFO, several decoding algorithms have been proposed~\cite{yardimci3}. Another important concern with OFDM systems is their excessive out-of-band emissions (OOBE) that should be diminished to prevent adjacent channel interference. The OFDM signal has a rectangular pulse in the time domain, which generates a sinc signal in the frequency domain. The sidelobes of the sinc signals at the edge carriers bring tremendous interference; therefore, they should be minimized. OOBE decreases through various windowing/filtering methods together with guard band allocation to fulfill the spectrum mask criteria of the different standards. Nonetheless, spectral efficiency degrades due to the fixed guard band allocation. Lastly, in OFDM systems, inserting CP and pilot symbols is highly indispensable. However, the spectral efficiency reduces because of reserving a considerable number of resources for CP and pilot overhead. 
 
 In recent years, many different and attractive OFDM-based MC waveforms have been designed to alleviate the drawbacks of OFDM. Windowed-OFDM (\acrshort{w-ofdm}) has been proposed to reduce OOBE by applying windowing operation to classical OFDM. Filter bank MC (\acrshort{fbmc}) \cite{alitugberk1}, which implements subcarrier-wise filtering operation, has been introduced. FBMC can use the spectrum more efficiently and is more resilient under high mobility compared to OFDM in the expense of additional signal processing. Generalized frequency division multiplexing (\acrshort{gfdm}) \cite{alitugberk2}, which also performs filtering at {the} subcarrier level like FBMC, appears as a remarkably flexible waveform since it does not have to meet the orthogonality requirement. Another engaging waveform is universal filtered MC (\acrshort{ufmc}) that applies filtering to subbands instead of subcarriers \cite{alitugberk3}. UFMC needs less redundancy in comparison to FBMC; however, it is not applicable for very high data rates. As mentioned previously, since the orthogonality of its subcarriers is disrupted, OFDM is highly susceptible to high Doppler which leads to ICI. Orthogonal time frequency space (\acrshort{otfs}) modulation appears as a promising waveform that brings a clever solution to this disadvantage of OFDM \cite{alitugberk5}. Even in high-Doppler channels, OTFS assures that each transmitted symbol has a near-constant channel gain by converting the time-varying multipath channel into a two-dimensional delay-Doppler channel. Finally, OFDM with IM (\acrshort{ofdm-im}) emerges as a promising OFDM-based waveform {that} changes only the modulation/demodulation processes of plain OFDM system and offers an additional degree of freedom for waveform design thanks to the {flexibility} of IM \cite{alitugberk4}. It is clear that all waveforms mentioned above {have both some} advantages and disadvantages over OFDM. Because of its simple structure, OFDM was chosen for also 5G networks, and we believe that it will maintain its importance for future wireless systems.
 
\begin{figure}
   \centering
   \includegraphics[width=\linewidth]{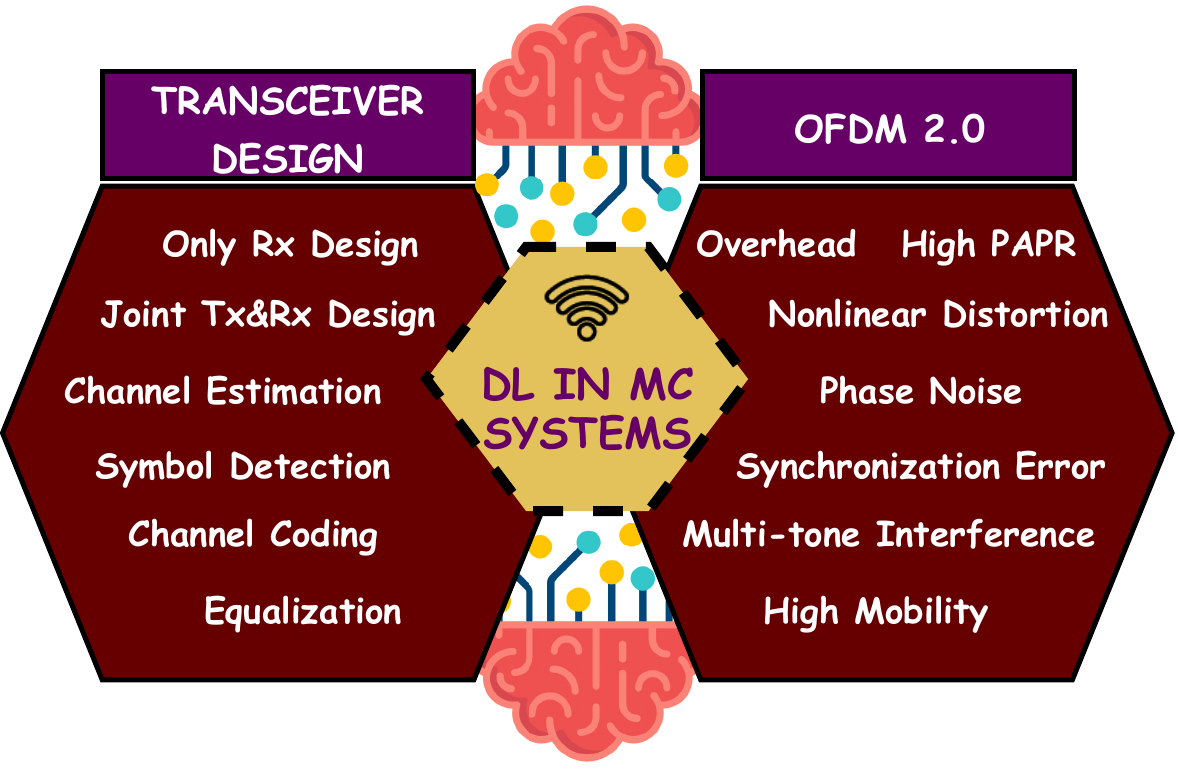}
   \caption{DL applications in MC systems.}
   \label{fig:DLMC}
\end{figure}

Against this background, DL emerges as an appealing tool to solve challenging problems in designing MC transmission systems. For instance, DL can be used to improve the subblocks of MC systems such as channel estimation and symbol detection or to optimize the whole transmission and reception pipeline jointly. In literature, a vast number of interesting DL-based methods are developed to enhance the performance {of} existing MC schemes. Under the subsection of transceiver design, as seen from Fig.~\ref{fig:DLMC}, we present both DL-based receiver architectures and also joint designs of transmitter and receiver. Then, in the second subsection, DL-based techniques, which address the shortcomings of OFDM, are investigated. These techniques provide significant improvements that encourage us to believe in the feasibility of DL-based solutions for future MC systems.

\subsection{Transceiver Design}

In this subsection, an overview of the studies in the literature review on DL-based receiver and transceiver designs, that is the studies that focus on the joint optimization of both transmitter and receiver structures, is provided. For the receiver design, most studies mainly focus on the implementation of DL-based techniques for channel estimation and symbol detection. For the transceiver design, the processing blocks at the transmitter and receiver are devised together by using AE-based approaches to optimize the system as a whole.

DL has been implemented for channel estimation and symbol detection in an OFDM system for the first time in~\cite{tugberk1}. In this study, DNNs have been shown to be capable of learning and evaluating wireless channel properties. The proposed DNN architecture, which is trained with the simulated OFDM symbols, has been designed as data-driven and fully-connected. When the number of pilots is reduced or CP is completely removed, the proposed DNN-based receiver outperforms the traditional algorithms such as LS and MMSE in terms of error performance. Moreover, the DNN-based receiver provides better error performance than LS and MMSE in the presence of clipping noise. In~\cite{tugberk28}, a low complexity DNN called SimNet with simple architecture and {{shorter} training has been designed to perform channel estimation. A meta{-}learning{-}based NN called robust channel estimation with meta NNs, RoemNet, is also introduced~\cite{tugberk51}.  Since it leverages meta-learner, RoemNet can tackle new channel learning problems with a minimal number of pilots. DeepRx, which improves the channel estimation and symbol detection {performance} significantly by adjusting the inputs of NN in a clever way, is another DL-based data{-}driven approach~\cite{tugberk3}. Since it supports 5G frame structure including different modulation schemes or pilot configurations, DeepRx is compatible with 5G communication systems. The output of DeepRx has been determined such that various QAM schemes can be supported by training only a single NN. The authors investigated both coded and uncoded error performance of DeepRx and compared it to the well-known LMMSE receiver. It is indicated that DeepRx outperforms LMMSE for different pilot configurations. Additionally, DeepRx has superiority over LMMSE in the presence of Doppler shift and inter-cell interference. The aforementioned NNs work with real-valued tensors and this can cause DNN to have more complexity and a decrease in {their} performance. In~\cite{tugberk5}, a deep complex-valued convolutional network, that works in complex field{s} and decodes bits from time domain OFDM signals without requiring any IFFT/FFT operation, has been proposed. 

A model{-}driven approach called ComNet~\cite{tugberk11}, which designs two different DNNs for channel estimation and symbol detection blocks, has been proposed instead of performing channel estimation and symbol detection jointly in~\cite{tugberk1}. These two DNNs are initialized by conventional wireless communication solutions. Thanks to the benefit of expert knowledge, it becomes possible to reduce the need for training data and converge faster with {the} model{-}driven method compared to data driven one~\cite{tugberk1}. Additionally, simulation results {have} shown that ComNet outperforms rival channel estimation and symbol detection algorithms such as LMMSE channel estimation, fully-connected DNN~\cite{tugberk1}, MMSE symbol detection. The mismatch between the channel model for offline training and the real environment, {which} causes a performance gap between the simulation and the over-the-air test, is detected in~\cite{tugberk35}. To overcome this problem, a novel online training system called SwitchNet receiver is designed to catch channel characteristics that were neglected during offline training. SwitchNet pretrains multiple channel estimation RefineNet of ComNet with diverse channel conditions and selects the network by online learning. Another model{-}driven approach is DeepWiPHY~\cite{tugberk14}, a DL-based OFDM receiver that is completely compatible with the newest Wi-Fi standard IEEE 802.11ax. In contrast to above mentioned designs, DeepWiPHY receiver is trained with not only a synthetic data set but also a real-world data set created by using universal software radio peripheral (\acrshort{usrp}) modules. Finally, a model{-}driven technique, that divides channel estimation into three CNN-based networks, is also proposed~\cite{tugberk43}. 

\begin{figure}
   \centering
   \includegraphics[width=\linewidth]{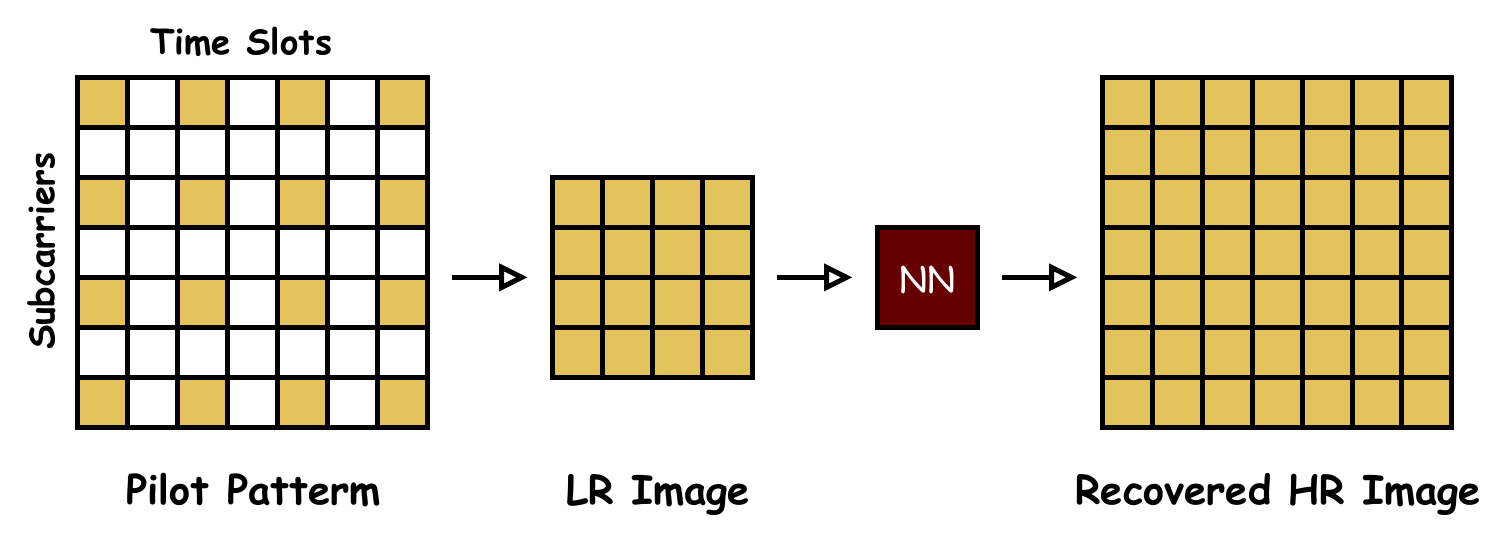}
   \caption{Representation of the time-frequency grid as a 2D image~\cite{tugberk8}.}
   \label{fig:Image2D}
\end{figure}


Authors of~\cite{tugberk54, tugberk8, tugberk59, tugberk32, tugberk27, tugberk56, tugberk39, tugberk73, tugberk47} have considered the time frequency response of wireless channel as a 2D image as in Fig.~\ref{fig:Image2D} where LR and HR stand for low resolution and high resolution, respectively. In~\cite{tugberk54}, the channel response at the pilot positions is regarded as low resolution image and the estimated channel is a high resolution image. A DL-based receiver called ChannelNet including two different CNNs, super-resolution CNN~\cite{tugberk54_10} and denoising CNN~\cite{tugberk54_11}, which improves the resolution of low resolution image and eliminates the effect of noise, respectively, is designed. In~\cite{tugberk8}, inspired from the image super
resolution technique~\cite{tugberk8_9} and conditioned
image synthesis technology~\cite{tugberk8_10}, two cascaded NNs, channel estimation network and channel conditional
recovery network, have been used to estimate the channel and detect the signal. In~\cite{tugberk59}, a DL-based channel estimation method, that does not need for training, has been introduced for multidimensional OFDM signals. Another channel estimation technique, that combines the conventional LS estimator with fast super resolution CNN, FSRCNN,~\cite{tugberk27_13} to increase the estimation performance, has been presented~\cite{tugberk27}. A deep
residual channel estimation network, ReEsNet, whose architecture is optimized by taking advantage of residual learning,~\cite{tugberk32} is introduced to outperform ChannelNet~\cite{tugberk54}. A CNN-based NN architecture called FreqTimeNet~\cite{tugberk39,tugberk73}, that performs learning in both time and frequency domain, is introduced for channel estimation. FreqTimeNet decreases the complexity by facilitating the orthogonality between two domains. Comprehensive simulation results have indicated that FreqTimeNet outperforms ChannelNet and ReEsNet. Furthermore, another NN design called AttenFreqTimeNet, that includes a widely used DL technique called attention mechanism, has been presented. Since it employs SNR information, AttenFreqTimeNet outperforms FreqTimeNet. A CNN-based DL structure called dual CNN, that takes advantage of both angle-delay domain and spatial-frequency domain, is introduced. Moreover, another novel network called HyperNet is proposed to further enhance the robustness. In~\cite{tugberk47}, a GAN-based NN called SRGAN is exploited for super resolution to obtain the whole channel response. Due to {the} employment of discriminator, the proposed channel estimator has superiority over ReEsNet and traditional techniques such as LMMSE.

It is also possible to implement DL-based techniques at both transmitter and receiver side{s}. Thus, the whole OFDM transceiver pipeline can be optimized jointly by utilizing the AE-based structure. In~\cite{tugberk6}, authors have shown that it is possible to integrate the E2E system into an OFDM system. {Furthermore}, for wireless picture transmission across multipath fading channels with nonlinear dis{s}}tortion, a DL-based joint source{-}channel coding technique is given~\cite{tugberk34}. Lastly, in~\cite{tugberk2}, a DL-based MC system called MC-AE, that optimizes modulation and demodulation operations jointly to maximize coding and diversity gain, is presented. {This MC-AE system is explained in the following subsection in detail.}

On the other hand, various NNs have been designed for channel estimation and symbol detection in OFDM-IM{-}based schemes. In~\cite{tugberk4}, a fully-connected DNN, called DeepIM, is used for symbol detection of OFDM-IM. The efficiency of the DNN in signal detection of OFDM-IM is demonstrated in terms of both error performance and decoding complexity. A preprocessing based on domain knowledge is applied to the received signal before inputting it to DeepIM. CNN-based receiver has also been proposed where the received symbols are converted to polar coordinates to ease the decoding of information bits transmitted by active subcarriers~\cite{tugberk13}. To enhance the performance of channel estimation and symbol detection for pilot-assisted OFDM-IM, complex DNN and complex CNN have been exploited~\cite{tugberk55}. Additionally, a detector called DeepDM that includes both a DNN and a CNN is introduced for an OFDM-IM{-}based scheme, dual-mode OFDM-IM (DM-OFDM-IM)~\cite{tugberk12}. In DeepDM, CNN and DNN are used to detect index bits and QAM/PSK modulated bits, respectively. Another receiver called IMNet is also proposed to detect IM{-}aided MIMO-OFDM, IM-MIMO-OFDM, signals~\cite{tugberk22}. Finally, the performance of DNN-based receiver is investigated for sparse vector coding OFDM which can also be considered as a member of OFDM-IM family~\cite{tugberk23}.   

DNN-based receivers have been designed for also other waveforms such as UFMC, FBMC, GFDM, and OTFS. In~\cite{tugberk62}, for two-stage index modulated UFMC system, a DL-based signal detector called TSIMNet, is proposed over multipath underwater acoustic channels. DNN-based and ResNet-DNN based receivers have been presented for channel estimation and symbol detection of FBMC systems in~\cite{tugberk46, tugberk57}, respectively. Moreover, in \cite{alitugberk6}, a damped generalized approximation message passing method is suggested to minimize receiver complexity in OTFS systems, with the damping factors tuned using DL approaches. A data-driven DNN-based method \cite{alitugberk7}, another low-complexity DNN-based technique \cite{alitugberk8} that equalizes received signal at the symbol level, and 2D CNN-based design \cite{alitugberk9} have been proposed for the equalization task at the OTFS receiver. Additionally, in \cite{alitugberk10}, authors have presented a novel DNN-based framework for the equalization of both SISO and MIMO-OTFS systems. DNN-based transceiver architectures have also been introduced for the detection of OTFS signals under IQ imbalance \cite{alitugberk11} and PAPR reduction \cite{alitugberk12}. Lastly, DL-aided receivers are presented for GFDM and GFDM-IM systems~\cite{tugberk67,tugberk66}.

In Table~\ref{tab:table_tugberk_1} and \ref{tab:table_tugberk_OTFS}, we demonstrate DL-based channel estimation and symbol detection methods with their respective NN techniques for OFDM and OFDM-based alternative waveforms such as FBMC and GFDM. Here, we note that SD and CE represent symbol detection and channel estimation, respectively. It can be seen from Table~\ref{tab:table_tugberk_1} that CNN is more frequently exploited particularly for channel estimation due to its ability {to learn} the correlation in a time-frequency grid.

\begin{table}[ht!]

\caption{An overview of DL-based transceiver designs for MC systems.}

\centering

\setlength{\tabcolsep}{3pt} 
\renewcommand{\arraystretch}{1.0}

\begin{tabular}
{|M{1.1cm}|M{2.1cm}|M{0.3cm}|M{0.3cm}|M{1.1cm}|M{1.2cm}|}
\hline
 \cellcolor{table_blue} \textbf{\textcolor{table_white}{Reference}} &  \cellcolor{table_blue} \textbf{\textcolor{table_white}{Name}}                                                          &  \cellcolor{table_blue} \textbf{\textcolor{table_white}{CE}} &  \cellcolor{table_blue} \textbf{\textcolor{table_white}{SD}} &  \cellcolor{table_blue} \textbf{\textcolor{table_white}{NN Type}} &  \cellcolor{table_blue} \textbf{\textcolor{table_white}{Waveform}} \\ 
 
\hline
\cellcolor{table_white} \textcolor{table_blue}{\cite{tugberk1}}               & 
\cellcolor{table_white} \textcolor{table_blue}{FC-DNN}               & 
\cellcolor{table_green}           & 
\cellcolor{table_green}           & 
\cellcolor{table_white} \textcolor{table_blue}{DNN}             & 
\cellcolor{table_white} \textcolor{table_blue}{OFDM}              \\ 

\hline
\cellcolor{table_white} \textcolor{table_blue}{\cite{tugberk3}}               & 
\cellcolor{table_white} \textcolor{table_blue}{DeepRx}          & 
\cellcolor{table_green}           & 
\cellcolor{table_green}           & 
\cellcolor{table_white} \textcolor{table_blue}{CNN}         & 
\cellcolor{table_white} \textcolor{table_blue}{OFDM}              \\ 

\hline
\cellcolor{table_white} \textcolor{table_blue}{\cite{tugberk5}}               & 
\cellcolor{table_white} \textcolor{table_blue}{DCCN}            & 
\cellcolor{table_green}           & 
\cellcolor{table_green}           & 
\cellcolor{table_white} \textcolor{table_blue}{CNN}        & 
\cellcolor{table_white} \textcolor{table_blue}{OFDM}              \\ 

\hline
\cellcolor{table_white} \textcolor{table_blue}{\cite{tugberk8}}               & 
\cellcolor{table_white} \textcolor{table_blue}{CENet}           &
\cellcolor{table_green}           & 
\cellcolor{table_green}           & 
\cellcolor{table_white} \textcolor{table_blue}{SAN, GAN}    & 
\cellcolor{table_white} \textcolor{table_blue}{OFDM}              \\ 

\hline
\cellcolor{table_white} \textcolor{table_blue}{\cite{tugberk11}}          &
\cellcolor{table_white} \textcolor{table_blue}{ComNet}      &
\cellcolor{table_green}           &
\cellcolor{table_green}           &
\cellcolor{table_white} \textcolor{table_blue}{DNN, LSTM}   &
\cellcolor{table_white} \textcolor{table_blue}{OFDM}              \\ 

\hline
\cellcolor{table_white} \textcolor{table_blue}{\cite{tugberk14}}          & 
\cellcolor{table_white} \textcolor{table_blue}{DeepWiPHY}   & 
\cellcolor{table_green}           & 
\cellcolor{table_green}           & 
\cellcolor{table_white} \textcolor{table_blue}{DNN, CNN}         & 
\cellcolor{table_white} \textcolor{table_blue}{OFDM}              \\ 

\hline
\cellcolor{table_white} \textcolor{table_blue}{\cite{tugberk35}}                 &
\cellcolor{table_white} \textcolor{table_blue}{SwitchNet}          &
\cellcolor{table_green}           &
\cellcolor{table_green}           &
\cellcolor{table_white} \textcolor{table_blue}{DNN}         &
\cellcolor{table_white} \textcolor{table_blue}{OFDM}              \\

\hline
\cellcolor{table_white} \textcolor{table_blue}{\cite{tugberk6}}                 &
\cellcolor{table_white} \textcolor{table_blue}{N/A}          &
\cellcolor{table_green}           &
\cellcolor{table_green}           &
\cellcolor{table_white} \textcolor{table_blue}{AE, DNN}         &
\cellcolor{table_white} \textcolor{table_blue}{OFDM}              \\

\hline
\cellcolor{table_white} \textcolor{table_blue}{\cite{tugberk34}}                 &
\cellcolor{table_white} \textcolor{table_blue}{N/A}          &
\cellcolor{table_green}           &
\cellcolor{table_green}           &
\cellcolor{table_white} \textcolor{table_blue}{AE, CNN}         &
\cellcolor{table_white} \textcolor{table_blue}{OFDM}              \\ 

\hline
\cellcolor{table_white} \textcolor{table_blue}{\cite{tugberk6}}                 &
\cellcolor{table_white} \textcolor{table_blue}{MC-AE}          &
\cellcolor{table_red}           &
\cellcolor{table_green}           &
\cellcolor{table_white} \textcolor{table_blue}{AE, DNN}         &
\cellcolor{table_white} \textcolor{table_blue}{OFDM}              \\

\hline
\cellcolor{table_white} \textcolor{table_blue}{\cite{tugberk28}}                 & 
\cellcolor{table_white} \textcolor{table_blue}{SimNet}             & 
\cellcolor{table_green}           &
\cellcolor{table_red}            &
\cellcolor{table_white} \textcolor{table_blue}{DNN}         &
\cellcolor{table_white} \textcolor{table_blue}{OFDM}              \\ 

\hline
\cellcolor{table_white} \textcolor{table_blue}{\cite{tugberk27}}                 &
\cellcolor{table_white} \textcolor{table_blue}{FSRCE}              &
\cellcolor{table_green}           &
\cellcolor{table_red}            &
\cellcolor{table_white} \textcolor{table_blue}{CNN}         &
\cellcolor{table_white} \textcolor{table_blue}{OFDM}              \\ 

\hline
\cellcolor{table_white} \textcolor{table_blue}{\cite{tugberk32}}                 & 
\cellcolor{table_white} \textcolor{table_blue}{ReEsNet}            & 
\cellcolor{table_green}           &
\cellcolor{table_red}            & 
\cellcolor{table_white} \textcolor{table_blue}{CNN}         & 
\cellcolor{table_white} \textcolor{table_blue}{OFDM}              \\ 

\hline
\cellcolor{table_white} \textcolor{table_blue}{\cite{tugberk39}}                 &
\cellcolor{table_white} \textcolor{table_blue}{FreqTimeNet}        &
\cellcolor{table_green}           &
\cellcolor{table_red}            &
\cellcolor{table_white} \textcolor{table_blue}{DNN}         &
\cellcolor{table_white} \textcolor{table_blue}{OFDM}              \\ 

\hline
\cellcolor{table_white} \textcolor{table_blue}{\cite{tugberk43}}                 &
\cellcolor{table_white} \textcolor{table_blue}{N/A}                  &
\cellcolor{table_green}           &
\cellcolor{table_red}            &
\cellcolor{table_white} \textcolor{table_blue}{CNN}              &
\cellcolor{table_white} \textcolor{table_blue}{OFDM}              \\ 

\hline
\cellcolor{table_white} \textcolor{table_blue}{\cite{tugberk47}}                 &
\cellcolor{table_white} \textcolor{table_blue}{SRGAN}              &
\cellcolor{table_green}           &
\cellcolor{table_red}            &
\cellcolor{table_white} \textcolor{table_blue}{GAN}              &
\cellcolor{table_white} \textcolor{table_blue}{OFDM}              \\ 

\hline
\cellcolor{table_white} \textcolor{table_blue}{\cite{tugberk51}}                 &
\cellcolor{table_white} \textcolor{table_blue}{RoemNet}            &
\cellcolor{table_green}           &
\cellcolor{table_red}            &
\cellcolor{table_white} \textcolor{table_blue}{DNN}              &
\cellcolor{table_white} \textcolor{table_blue}{OFDM}              \\ 

\hline
\cellcolor{table_white} \textcolor{table_blue}{\cite{tugberk54}}                 &
\cellcolor{table_white} \textcolor{table_blue}{ChannelNet}         &
\cellcolor{table_green}           &
\cellcolor{table_red}            &
\cellcolor{table_white} \textcolor{table_blue}{CNN}         &
\cellcolor{table_white} \textcolor{table_blue}{OFDM}              \\ 

\hline
\cellcolor{table_white} \textcolor{table_blue}{\cite{tugberk56}}                 &
\cellcolor{table_white} \textcolor{table_blue}{DualCNN}            &
\cellcolor{table_green}           &
\cellcolor{table_red}            &
\cellcolor{table_white} \textcolor{table_blue}{CNN, RNN}    &
\cellcolor{table_white} \textcolor{table_blue}{OFDM}              \\ 

\hline
\cellcolor{table_white} \textcolor{table_blue}{\cite{tugberk59}}                 &
\cellcolor{table_white} \textcolor{table_blue}{DCE}                &
\cellcolor{table_green}           &
\cellcolor{table_red}            &
\cellcolor{table_white} \textcolor{table_blue}{DNN}              &
\cellcolor{table_white} \textcolor{table_blue}{OFDM}              \\ 

\hline
\cellcolor{table_white} \textcolor{table_blue}{\cite{tugberk73}}                 &
\cellcolor{table_white} \textcolor{table_blue}{AttenFreqTimeNet} &
\cellcolor{table_green}           &
\cellcolor{table_red}            &
\cellcolor{table_white} \textcolor{table_blue}{DNN}         &
\cellcolor{table_white} \textcolor{table_blue}{OFDM}              \\ 

\hline
\cellcolor{table_white} \textcolor{table_blue}{\cite{tugberk4}}               &
\cellcolor{table_white} \textcolor{table_blue}{DeepIM}          &
\cellcolor{table_red}            &
\cellcolor{table_green}           &
\cellcolor{table_white} \textcolor{table_blue}{DNN}         &
\cellcolor{table_white} \textcolor{table_blue}{OFDM-IM}           \\ 

\hline
\cellcolor{table_white} \textcolor{table_blue}{\cite{tugberk12}}              &
\cellcolor{table_white} \textcolor{table_blue}{DeepDM}          &
\cellcolor{table_red}            &
\cellcolor{table_green}           &
\cellcolor{table_white} \textcolor{table_blue}{DNN, CNN}         & \cellcolor{table_white} \textcolor{table_blue}{OFDM-IM}           \\ 

\hline
\cellcolor{table_white} \textcolor{table_blue}{\cite{tugberk13}}              &
\cellcolor{table_white} \textcolor{table_blue}{CNN-IM}          &
\cellcolor{table_red}            &
\cellcolor{table_green}           &
\cellcolor{table_white} \textcolor{table_blue}{CNN}         &
\cellcolor{table_white} \textcolor{table_blue}{OFDM-IM}           \\ 

\hline
\cellcolor{table_white} \textcolor{table_blue}{\cite{tugberk22}}              &
\cellcolor{table_white} \textcolor{table_blue}{IMNet}           &
\cellcolor{table_red}           &
\cellcolor{table_green}           &
\cellcolor{table_white} \textcolor{table_blue}{CNN}         &
\cellcolor{table_white} \textcolor{table_blue}{OFDM-IM}           \\ 

\hline
\cellcolor{table_white} \textcolor{table_blue}{\cite{tugberk23}}              &
\cellcolor{table_white} \textcolor{table_blue}{Deep SVC}        &
\cellcolor{table_red}            &
\cellcolor{table_green}           &
\cellcolor{table_white} \textcolor{table_blue}{CNN}         &
\cellcolor{table_white} \textcolor{table_blue}{OFDM-IM}           \\ 

\hline
\cellcolor{table_white} \textcolor{table_blue}{\cite{tugberk55}}               &
\cellcolor{table_white} \textcolor{table_blue}{N/A}                &
\cellcolor{table_green}           &
\cellcolor{table_green}           &
\cellcolor{table_white} \textcolor{table_blue}{DNN, CNN}         &
\cellcolor{table_white} \textcolor{table_blue}{OFDM-IM}           \\ 

\hline
\cellcolor{table_white} \textcolor{table_blue}{\cite{tugberk62}}              &
\cellcolor{table_white} \textcolor{table_blue}{TSIMNet}         &
\cellcolor{table_green}           &
\cellcolor{table_green}           &
\cellcolor{table_white} \textcolor{table_blue}{DNN}          &
\cellcolor{table_white} \textcolor{table_blue}{UFMC-IM}           \\ 

\hline
\cellcolor{table_white} \textcolor{table_blue}{\cite{tugberk46}}              &
\cellcolor{table_white} \textcolor{table_blue}{DL-CE}           &
\cellcolor{table_green}           &
\cellcolor{table_green}           &
\cellcolor{table_white} \textcolor{table_blue}{DNN}          &
\cellcolor{table_white} \textcolor{table_blue}{FBMC}              \\ 

\hline
\cellcolor{table_white} \textcolor{table_blue}{\cite{tugberk57}}              &
\cellcolor{table_white} \textcolor{table_blue}{Res-DNN}         &
\cellcolor{table_green}           &
\cellcolor{table_green}           &
\cellcolor{table_white} \textcolor{table_blue}{DNN}          &
\cellcolor{table_white} \textcolor{table_blue}{FBMC}              \\

\hline
\cellcolor{table_white} \textcolor{table_blue}{\cite{tugberk67}}              &
\cellcolor{table_white} \textcolor{table_blue}{N/A}               &
\cellcolor{table_red}            &
\cellcolor{table_green}           &
\cellcolor{table_white} \textcolor{table_blue}{DNN, CNN}     &
\cellcolor{table_white} \textcolor{table_blue}{GFDM}              \\ 

\hline
\cellcolor{table_white} \textcolor{table_blue}{\cite{tugberk66}}          &
\cellcolor{table_white} \textcolor{table_blue}{DeepConvIM}  &
\cellcolor{table_red}                &
\cellcolor{table_green}               &
\cellcolor{table_white} \textcolor{table_blue}{DNN, CNN}        &
\cellcolor{table_white} \textcolor{table_blue}{GFDM-IM}           \\ 

\hline
\end{tabular}

\label{tab:table_tugberk_1}

\end{table}

\begin{table}[ht!]

\caption{An overview of DL-based transceiver designs for OTFS systems (Red hues indicate no CE or no SD and green hues indicate CE or SD).}

\centering
\begin{tabular}{|M{1.1cm}|M{1.95cm}|M{0.3cm}|M{0.3cm}|M{1.1cm}|M{1.2cm}|}
\hline
 \cellcolor{table_blue} \textbf{\textcolor{table_white}{Reference}} &  \cellcolor{table_blue} \textbf{\textcolor{table_white}{Name}}                                                          &  \cellcolor{table_blue} \textbf{\textcolor{table_white}{CE}} &  \cellcolor{table_blue} \textbf{\textcolor{table_white}{SD}} &  \cellcolor{table_blue} \textbf{\textcolor{table_white}{NN Type}} &  \cellcolor{table_blue} \textbf{\textcolor{table_white}{Waveform}} \\ 
 
\hline
\cellcolor{table_white} \textcolor{table_blue}{\cite{alitugberk6}}               & 
\cellcolor{table_white} \textcolor{table_blue}{DL-Based GAMP}               & 
\cellcolor{table_red}           & 
\cellcolor{table_green}           & 
\cellcolor{table_white} \textcolor{table_blue}{Unfolding}             & 
\cellcolor{table_white} \textcolor{table_blue}{OTFS}              \\ 

\hline
\cellcolor{table_white} \textcolor{table_blue}{\cite{alitugberk7}}               & 
\cellcolor{table_white} \textcolor{table_blue}{N/A}          & 
\cellcolor{table_red}           & 
\cellcolor{table_green}           & 
\cellcolor{table_white} \textcolor{table_blue}{DNN}         & 
\cellcolor{table_white} \textcolor{table_blue}{OTFS}              \\ 

\hline
\cellcolor{table_white} \textcolor{table_blue}{\cite{alitugberk8}}               & 
\cellcolor{table_white} \textcolor{table_blue}{Symbol-DNN}               & 
\cellcolor{table_red}           & 
\cellcolor{table_green}           & 
\cellcolor{table_white} \textcolor{table_blue}{DNN}             & 
\cellcolor{table_white} \textcolor{table_blue}{OTFS}              \\ 

\hline
\cellcolor{table_white} \textcolor{table_blue}{\cite{alitugberk9}}               & 
\cellcolor{table_white} \textcolor{table_blue}{N/A}          & 
\cellcolor{table_red}           & 
\cellcolor{table_green}           & 
\cellcolor{table_white} \textcolor{table_blue}{CNN}         & 
\cellcolor{table_white} \textcolor{table_blue}{OTFS}              \\ 

\hline
\cellcolor{table_white} \textcolor{table_blue}{\cite{alitugberk10}}               & 
\cellcolor{table_white} \textcolor{table_blue}{N/A}          & 
\cellcolor{table_red}           & 
\cellcolor{table_green}           & 
\cellcolor{table_white} \textcolor{table_blue}{RNN}         & 
\cellcolor{table_white} \textcolor{table_blue}{OTFS}              \\ 

\hline
\cellcolor{table_white} \textcolor{table_blue}{\cite{alitugberk11}}               & 
\cellcolor{table_white} \textcolor{table_blue}{N/A}          & 
\cellcolor{table_green}           & 
\cellcolor{table_green}           & 
\cellcolor{table_white} \textcolor{table_blue}{DNN}         & 
\cellcolor{table_white} \textcolor{table_blue}{OTFS}              \\ 

\hline
\cellcolor{table_white} \textcolor{table_blue}{\cite{alitugberk12}}               & 
\cellcolor{table_white} \textcolor{table_blue}{N/A}          & 
\cellcolor{table_green}           & 
\cellcolor{table_green}           & 
\cellcolor{table_white} \textcolor{table_blue}{DNN}         & 
\cellcolor{table_white} \textcolor{table_blue}{OTFS}              \\ 

\hline

\end{tabular}

\label{tab:table_tugberk_OTFS}

\end{table}

\subsection{DL-Based OFDM 2.0: Overcoming Drawbacks of Classical OFDM}

In this subsection, an overview of the literature on DL-based techniques to deal with the main disadvantages of OFDM is given. By enjoying particularly deep unfolding approach, numerous DL-based schemes are proposed to improve the performance of classical OFDM systems.

In~\cite{tugberk7}, a novel PAPR reduction network, PRNet, is proposed where PAPR and BER performance are jointly optimized thanks to its AE-based architecture. Furthermore, a low-complex real-valued NN is presented to reduce PAPR and minimize BER at the same time~\cite{tugberk80}. Since the proposed method in~\cite{tugberk80} is implemented in the time domain unlike PRNet, it can provide a reduction in complexity. Nonetheless, optimizing different metrics increases the computational complexity considerably. Motivated from model-driven approach, A DL-based tone reservation network, namely TRNet, is proposed to improve only the PAPR performance to increase the training speed~\cite{tugberk29}. TRNet improves the performance of the classical tone reservation method and in contrast to PRNet, it is applied only at the transmitter side. Another tone reservation{-}based DL method called DL-TR is introduced in~\cite{tugberk65} where conventional tone reservation method is unfolded to design a DNN. Moreover, in \cite{tugberk44}, a model-driven DL-based tone reservation technique, which yields low-complexity due to its clever design, is proposed by unfolding an iterative tone reservation scheme as a layer of DNN. In~\cite{tugberk69}, the usage of residual NNs with soft-clipping is presented as a novel PAPR reduction approach{es}. Authors of~\cite{tugberk52} have proposed to employ a DNN at the transmitter to develop a high-dimensional modulation method that enables regulation of {both the} PAPR and adjacent channel leakage ratio. Another NN is implemented at the receiver side to decode information bits. Finally, an AE-based PAPR reduction method is introduced for pre-coded OFDM signals without reducing the OOBE performance~\cite{tugberk70}.
 
In order to remove the CP in OFDM systems, a model-driven DL technique based on orthogonal approximate message passing (DL-OAMP) is presented~\cite{tugberk76, tugberk10}. The channel estimation module of ComNet~\cite{tugberk11} and the OAMP detection NN, OAMP-Net, that merges the OAMP method and DL by incorporating a few trainable parameters, are included in the DL-OAMP receiver. Authors have shown that the complexity of DL-OAMP is lower than ComNet and DL-OAMP is capable of adapting time-varying channels. In~\cite{tugberk10}, the performance of OAMP-NET is investigated {not only by} extensive simulation and also {by} employing a real communication system. Another model-driven DL method based on OAMP algorithm has also been proposed for CP removal in MIMO-OFDM systems~\cite{tugberk71}. Although CP is inserted to time-domain OFDM signal to avoid ICI, its length may not be long enough in some practical scenarios. In case of insufficient CP, model-driven DL-based receivers are designed for the detection of SISO-OFDM~\cite{tugberk78} and MIMO-OFDM~\cite{tugberk77} signals. In addition to CP, pilot overhead needs to be considered in OFDM systems to increase spectral efficiency. In~\cite{tugberk15}, firstly, the number of pilot symbols is removed partially without a decrease in error performance. Secondly, by using an AE-based NN, pilots are completely removed along with a learned constellation or superimposed pilots. In order to increase the throughput considerably, authors of~\cite{tugberk18} addressed both CP and pilot reduction issues and demonstrated that it is possible to eliminate CP and pilot entirely by exploiting E2E learning. In~\cite{tugberk19}, for frequency division duplex massive MIMO-OFDM systems, a{n} NN-based combined downlink pilot design and channel estimation approach is presented. An efficient pilot reduction approach is also suggested for reducing pilot overhead and saving time-frequency resources for data transmission by progressively pruning less important neurons from dense layers. Lastly, to solve the channel estimation problem, a pilotless AE-based E2E learning technique is presented in~\cite{tugberk79} where a CNN and two DNN are implemented in the transmitter and receiver side, respectively.

 As mentioned earlier, high mobility causes ICI in OFDM signals and this ICI disrupts the orthogonality between subcarriers and makes the detection process considerably challenging. DL emerges as an interesting tool to be able to detect OFDM signals in fast time varying environment. In~\cite{tugberk48}, a channel estimation network, ChanEstNet, which exploits a CNN and a RNN, is introduced for channel estimation in the high speed mobile scenarios. Moreover, this idea is extended to MIMO systems~\cite{tugberk31}. DL-based cascaded structures called ICINet~\cite{tugberk40} and Cascade-Net~\cite{tugberk21} are also proposed for channel estimation and symbol detection for rapidly time varying channels. Lastly, AE-based DL technique is also developed to cope with complex and fast varying environments in marine communications~\cite{tugberk20}.
  \begin{table}

\caption{An overview of DL-based designs focusing on OFDM drawbacks. {(The colors are used to cluster similar problem types.)}}

\setlength{\tabcolsep}{3pt} 
\renewcommand{\arraystretch}{1.0}

\centering
\begin{tabular}{|M{1.1cm}|M{1.6cm}|M{2cm}|M{1.4cm}|}


\hline

 \cellcolor{table_blue} \textbf{\textcolor{table_white}{Reference}} &  \cellcolor{table_blue} \textbf{\textcolor{table_white}{Name}} &  \cellcolor{table_blue} \textbf{\textcolor{table_white}{Problem}} &  \cellcolor{table_blue} \textbf{\textcolor{table_white}{NN Type}}  \\ \hline
 
\cellcolor{table_white} \textcolor{table_blue}{\cite{tugberk7}}  & \cellcolor{table_white} \textcolor{table_blue}{PRNet}         
& \cellcolor{table_red} \textcolor{table_blue}{High PAPR}       & \cellcolor{table_white} \textcolor{table_blue}{AE, DNN}   \\ \hline

\cellcolor{table_white} \textcolor{table_blue}{\cite{tugberk80}} & \cellcolor{table_white} \textcolor{table_blue}{N/A}             
& \cellcolor{table_red} \textcolor{table_blue}{High PAPR}   & \cellcolor{table_white} \textcolor{table_blue}{AE, DNN}    \\ \hline

\cellcolor{table_white} \textcolor{table_blue}{\cite{tugberk29}} & \cellcolor{table_white} \textcolor{table_blue}{TRNet}         
& \cellcolor{table_red} \textcolor{table_blue}{High PAPR}          & \cellcolor{table_white}  \textcolor{table_blue}{AE, DNN}       \\ \hline

\cellcolor{table_white} \textcolor{table_blue}{\cite{tugberk65}} & \cellcolor{table_white} \textcolor{table_blue}{DL-TR}         
& \cellcolor{table_red} \textcolor{table_blue}{High PAPR}           & \cellcolor{table_white}  \textcolor{table_blue}{Unfolding}    \\ \hline

\cellcolor{table_white} \textcolor{table_blue}{\cite{tugberk44}} & \cellcolor{table_white} \textcolor{table_blue}{N/A}             
& \cellcolor{table_red} \textcolor{table_blue}{High PAPR}           &  \cellcolor{table_white}  \textcolor{table_blue}{Unfolding}  \\ \hline

\cellcolor{table_white} \textcolor{table_blue}{\cite{tugberk69}} & \cellcolor{table_white} \textcolor{table_blue}{RDNN}          
& \cellcolor{table_red} \textcolor{table_blue}{High PAPR}          &      \cellcolor{table_white} \textcolor{table_blue}{Residual DNN}       \\ \hline

\cellcolor{table_white} \textcolor{table_blue}{\cite{tugberk52}} & \cellcolor{table_white} \textcolor{table_blue}{N/A}             
& \cellcolor{table_red} \textcolor{table_blue}{High PAPR}   & \cellcolor{table_white} \textcolor{table_blue}{AE-Residual CNN} \\ \hline

\cellcolor{table_white} \textcolor{table_blue}{\cite{tugberk70}} & \cellcolor{table_white} \textcolor{table_blue}{N/A}             
& \cellcolor{table_red} \textcolor{table_blue}{High PAPR\&OOBE}     & \cellcolor{table_white}    \textcolor{table_blue}{AE, DNN}          \\ \hline

\cellcolor{table_white} \textcolor{table_blue}{\cite{tugberk76}} & \cellcolor{table_white} \textcolor{table_blue}{DL-OAMP}       
& \cellcolor{table_green} \textcolor{table_blue}{CP Overhead}    &  \cellcolor{table_white}   \textcolor{table_blue}{Unfolding}      \\ \hline

\cellcolor{table_white} \textcolor{table_blue}{\cite{tugberk10}} & \cellcolor{table_white} \textcolor{table_blue}{DL-OAMP}       
& \cellcolor{table_green} \textcolor{table_blue}{CP Overhead}   &    \cellcolor{table_white}  \textcolor{table_blue}{Unfolding}        \\ \hline

\cellcolor{table_white} \textcolor{table_blue}{\cite{tugberk71}} & \cellcolor{table_white} \textcolor{table_blue}{CG-OAMP-NET}   
& \cellcolor{table_green} \textcolor{table_blue}{CP Overhead}    & 
\cellcolor{table_white} \textcolor{table_blue}{Unfolding}  \\ 
\hline

\cellcolor{table_white} \textcolor{table_blue}{\cite{tugberk78}} & \cellcolor{table_white} \textcolor{table_blue}{DetNet-IG}     
& \cellcolor{table_green} \textcolor{table_blue}{CP Overhead}   &   \cellcolor{table_white}    \textcolor{table_blue}{Unfolding} \\ 
\hline

\cellcolor{table_white} \textcolor{table_blue}{\cite{tugberk77}} & \cellcolor{table_white} \textcolor{table_blue}{N/A}             
& \cellcolor{table_green} \textcolor{table_blue}{CP Overhead}   &   
\cellcolor{table_white} \textcolor{table_blue}{Unfolding} \\ 
\hline

\cellcolor{table_white} \textcolor{table_blue}{\cite{tugberk15}} & \cellcolor{table_white} \textcolor{table_blue}{N/A}             
& \cellcolor{table_green} \textcolor{table_blue}{Pilot Overhead}&   \cellcolor{table_white}    \textcolor{table_blue}{AE, Residual CNN}\\ 
\hline

\cellcolor{table_white} \textcolor{table_blue}{\cite{tugberk18}} & \cellcolor{table_white} \textcolor{table_blue}{N/A}             
& \cellcolor{table_green} \textcolor{table_blue}{CP\&Pilot Overhead}&   \cellcolor{table_white} \textcolor{table_blue}{AE, Residual CNN}         \\ 
\hline

\cellcolor{table_white} \textcolor{table_blue}{\cite{tugberk19}} & \cellcolor{table_white} \textcolor{table_blue}{N/A}             
& \cellcolor{table_green} \textcolor{table_blue}{Pilot Overhead}    &\cellcolor{table_white}   \textcolor{table_blue}{CNN}     \\ 
\hline

\cellcolor{table_white} \textcolor{table_blue}{\cite{tugberk79}} & \cellcolor{table_white} \textcolor{table_blue}{N/A}             
& \cellcolor{table_green} \textcolor{table_blue}{Pilot Overhead}    &\cellcolor{table_white}   \textcolor{table_blue}{AE, CNN, DNN}     \\ 
\hline

\cellcolor{table_white} \textcolor{table_blue}{\cite{tugberk48}} & \cellcolor{table_white} \textcolor{table_blue}{ChanEstNet}    
& \cellcolor{table_banana} \textcolor{table_blue}{High Mobility}&  \cellcolor{table_white}   \textcolor{table_blue}{CNN, BiLSTM}  \\ 
\hline

\cellcolor{table_white} \textcolor{table_blue}{\cite{tugberk31}} & \cellcolor{table_white} \textcolor{table_blue}{N/A}             
& \cellcolor{table_banana} \textcolor{table_blue}{High Mobility}&    \cellcolor{table_white}    \textcolor{table_blue}{CNN, BiLSTM} \\ 
\hline

\cellcolor{table_white} \textcolor{table_blue}{\cite{tugberk40}} & \cellcolor{table_white} \textcolor{table_blue}{ICINet}        
& \cellcolor{table_banana} \textcolor{table_blue}{High Mobility}&   \cellcolor{table_white}     \textcolor{table_blue}{DNN, Residual CNN}  \\ 
\hline

\cellcolor{table_white} \textcolor{table_blue}{\cite{tugberk21}} & \cellcolor{table_white} \textcolor{table_blue}{Cascade-Net}   
& \cellcolor{table_banana} \textcolor{table_blue}{High Mobility}&    \cellcolor{table_white}    \textcolor{table_blue}{Unfolding}   \\ 
\hline

\cellcolor{table_white} \textcolor{table_blue}{\cite{tugberk20}} & \cellcolor{table_white} \textcolor{table_blue}{N/A}             
& \cellcolor{table_banana} \textcolor{table_blue}{High Mobility}&   \cellcolor{table_white}     \textcolor{table_blue}{AE, CNN, LSTM}  \\ 
\hline

\cellcolor{table_white} \textcolor{table_blue}{\cite{tugberk37}} & \cellcolor{table_white} \textcolor{table_blue}{HybridDeepRx}  
& \cellcolor{table_x} \textcolor{table_blue}{Nonlinear Distortion}&
 \cellcolor{table_white} \textcolor{table_blue}{Residual CNN}  \\ 
\hline

\cellcolor{table_white} \textcolor{table_blue}{\cite{tugberk41}} & \cellcolor{table_white} \textcolor{table_blue}{N/A}             
& \cellcolor{table_x} \textcolor{table_blue}{Nonlinear Distortion} & \cellcolor{table_white} \textcolor{table_blue}{DNN}  \\ 
\hline

\cellcolor{table_white} \textcolor{table_blue}{\cite{tugberk38}} & \cellcolor{table_white} \textcolor{table_blue}{N/A}             
& \cellcolor{table_x} \textcolor{table_blue}{Phase Noise Compensation} &                  
\cellcolor{table_white} \textcolor{table_blue}{DNN, CNN}  \\ 
\hline

\cellcolor{table_white} \textcolor{table_blue}{\cite{tugberk42}} & \cellcolor{table_white} \textcolor{table_blue}{N/A}             
& \cellcolor{table_x} \textcolor{table_blue}{Interference Suppression} &                  
\cellcolor{table_white} \textcolor{table_blue}{LSTM, DNN}  \\ 
\hline

\cellcolor{table_white} \textcolor{table_blue}{\cite{tugberk60}} & \cellcolor{table_white} \textcolor{table_blue}{N/A}             
& \cellcolor{table_x} \textcolor{table_blue}{Synchronization Error}& \cellcolor{table_white} \textcolor{table_blue}{DNN} \\ 
\hline

\cellcolor{table_white} \textcolor{table_blue}{\cite{tugberk17}} & \cellcolor{table_white} \textcolor{table_blue}{N/A}             
& \cellcolor{table_x} \textcolor{table_blue}{One-bit Quantization}&   \cellcolor{table_white} \textcolor{table_blue}{AE, DNN}  \\ 
\hline

\end{tabular}

\label{tab:table_tugberk_2}
\vspace{0.5cm}
\end{table}

 In OFDM transmission systems, DL can be also used to combat other impairments such as phase noise and synchronization errors. For example, a DL-based receiver called HybridDeepRx is introduced to detect nonlinearly distorted OFDM signals~\cite{tugberk37}. Additionally, in order to alleviate nonlinear distortion in a MIMO-OFDM system, authors of~\cite{tugberk41} have designed both model and data-driven DL-based receivers. Furthermore, to mitigate the phase noise in channel estimation and symbol detection in both GFDM and OFDM systems, a DL-based method including DNN and CNN, has been devised~\cite{tugberk38}. In~\cite{tugberk42}, LSTM and DNN-based technique has been utilized to diminish multi-tone interference in a polar coded OFDM system. In~\cite{tugberk60}, an enhanced version of FC-DNN~\cite{tugberk1} is presented to perform channel estimation and symbol detection in the presence of timing synchronization error. Finally, under the restriction of one-bit complex quantization, authors of~\cite{tugberk17} explore innovative DL-based techniques for an OFDM receiver.

In Table~\ref{tab:table_tugberk_2}, we provide a summary of the reviewed studies that focus on DL-based designs to overcome some important drawbacks of classical OFDM, such as high PAPR and CP overhead. Particularly, deep unfolding comes into view as a significant concept to deal with the challenging impairments of OFDM by unfolding conventional algorithms. For PAPR reduction, AE-based NNs are advantageous since they can still improve the error performance while reducing PAPR.

{Although the DL-based methods show great promise, there are still limitations and drawbacks related to the application of DL-based solutions for MC waveforms. One of the most important limitations is the fact that most studies focus on optimizing only one performance metric such as BER and PAPR. Unfortunately, the performance of MC systems relies on multiple key performance indicators, which should be considered in the design process. Implementation of these indicators to DL model design is a challenging task and has not been addressed in the literature. Also, the design process requires the involvement of various 6G applications such as eMBB, URLLC, mMTC, and their possible combinations. These applications require specific challenges that need to be considered in the DL model design.}

\subsection{{An Example AE-Based OFDM System}}

\begin{figure*}
   \centering
   \includegraphics[width=\linewidth]{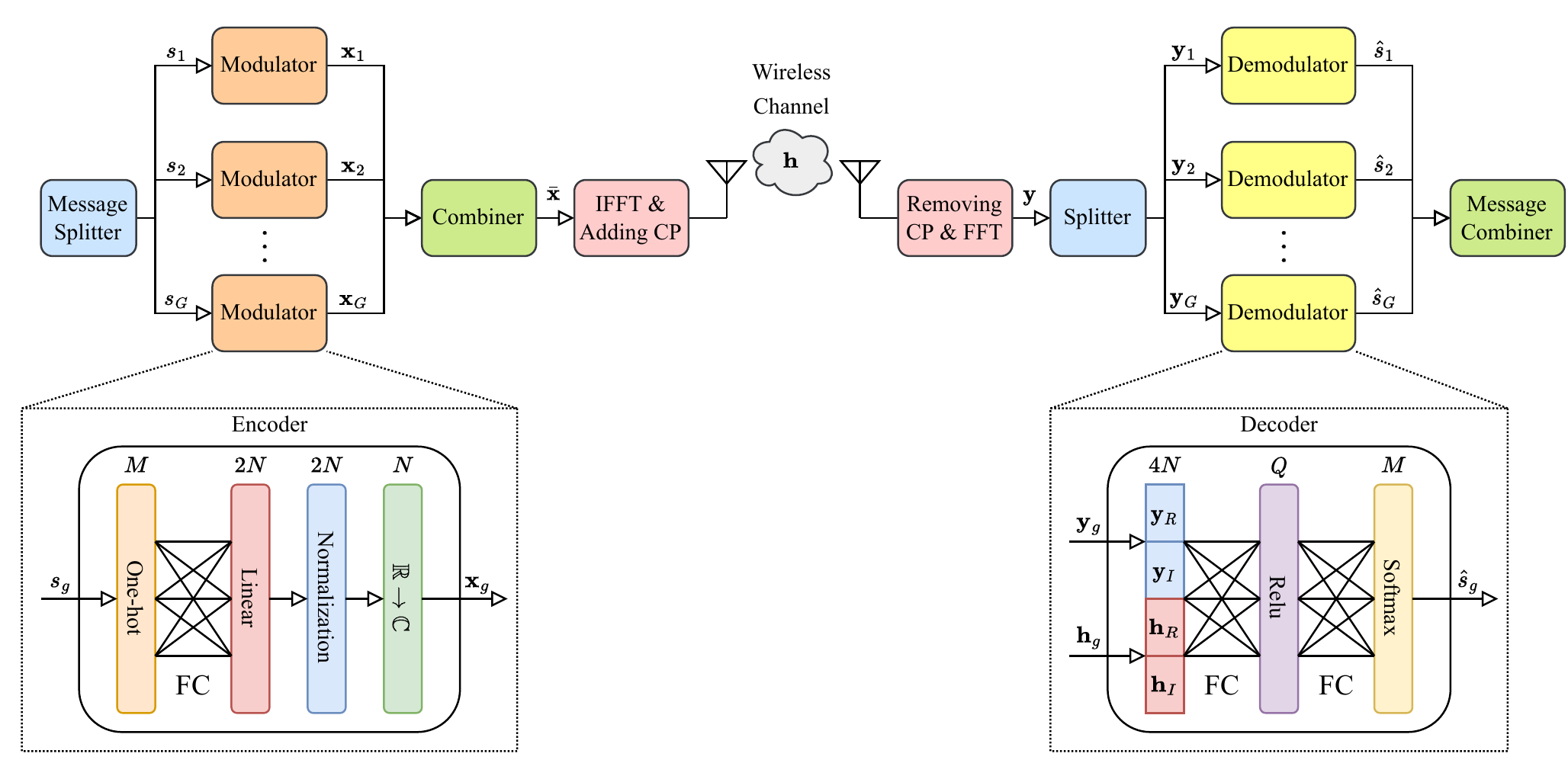}
   \caption{{The block diagram of MC-AE system ~\cite{tugberk2}.}}
   \label{fig:MCAE}
\end{figure*}

{
In this subsection, we introduce an AE-based MC (MC-AE) system proposed in \cite{tugberk2}. The block diagram of MC-AE is given in Fig. \ref{fig:MCAE}. In MC-AE, an OFDM system with $N_c$ number of subcarriers, is considered. These $N_c$ subcarriers are separated into $G$ blocks, each containing $N$ subcarriers, where $N=N_c/G$. For each subblock, AE method is performed independently. As seen in Fig. \ref{fig:MCAE}, the modulator and demodulator blocks of the overall OFDM system are modeled as encoder and decoder DNNs of an AE. At the transmitter side, for the $g$-th subblock, $g=1,\cdots,G$, the incoming message $s_g$  is mapped into a one-hot vector $\mathbf{s} \in \mathbb{R}^{M \times 1}$. Note that the incoming message $s_g$ can be a bit stream with length $m$. Then, $\mathbf{s}$ is passed through linear and normalization layers and converted into a complex-valued vector $\mathbf{x}_g$. After that, the overall OFDM symbol is obtained by concatenating all subblocks as $\Bar{\mathbf{x}}=[\mathbf{x}^{\mathrm{T}}_1,\mathbf{x}^{\mathrm{T}}_2,\cdots,\mathbf{x}^{\mathrm{T}}_G]^{\mathrm{T}}$. After performing OFDM transmission procedures, $\Bar{\mathbf{x}}$ is transmitted through the Rayleigh channel and the received signal is obtained. For the $g$th subblock, the frequency domain input-output relationship can be given as 
}
\begin{equation}
{
    \mathbf{y}_g = \mathbf{h}_g \odot \mathbf{x}_g + \mathbf{n}_g, 
    }
\end{equation}
{
where $\mathbf{y}_g$, $\mathbf{h}_g$, and $\mathbf{n}_g$ are the received signal, channel coefficient and AWGN vectors corresponding to the $g$th subblock, respectively. The elements of $\mathbf{h}_g$ and $\mathbf{n}_g$ follow the distributions $\mathcal{CN}(0,1)$ and $\mathcal{CN}(0,\sigma^2)$, respectively. The average received SNR is defined as $\Bar{\gamma} = E_s/\sigma^2$, where $E_s$ represents the average transmit power.
}
{
At the receiver side, the perfect CSI $\mathbf{h}$ is assumed to be known. The real and imaginary parts of $\mathbf{y}_g$ and $\mathbf{h}_g$ are concatenated and the resulting vector is given as an input to the decoder. As seen in Fig. \ref{fig:MCAE}, the inputted vector is passed through two different FC layers with activation functions ReLU and softmax. Finally, the estimated message $s_g$ is obtained. }

{
For the training of this AE-based structure, a set of random incoming messages and randomly generated channel and noise samples are used. The MSE loss function and SGD optimization method are exploited for training.}

\begin{figure}
   \centering
   \includegraphics[width=\linewidth]{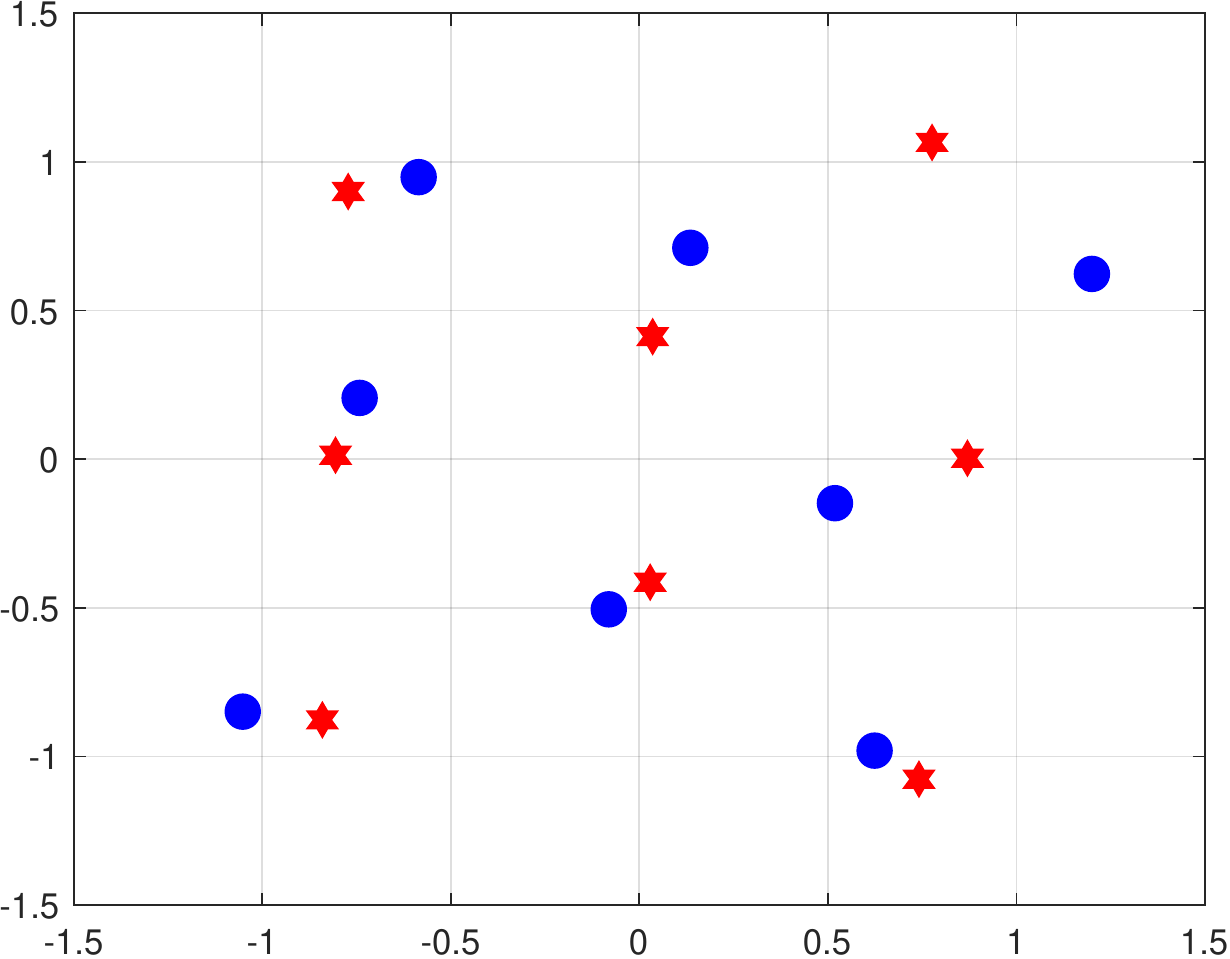}
   \caption{{The learned constellation of MC-AE for $N=2$ and $M=8$.}}
   \label{fig:MCAE_cons}
\end{figure}

{
The learned constellation for MC-AE system can be seen in Fig. \ref{fig:MCAE_cons}, where each marker represents the complex constellation point of a subcarrier. In \cite{tugberk2}, it is demonstrated that diversity and coding gains of MC-AE system are higher than benchmark schemes thanks to the learned constellation by the capability of AE structure. Therefore, the MC-AE system can provide outstanding error performance in fading channels. }

\section{DL-aided Communication Systems Through RIS} \label{sec:ris}

Communication through \acrshort{ris}s, emerges as a new and promising technology for next-generation communication systems due to its numerous advantages such as enabling smart radio environments with low cost and high efficiency. The RIS technology, which provides environmentally friendly solutions while increasing wireless communication performance, offer{s} attractive advantages at low cost and complexity due to its passive reflective architecture. Thus, the RIS technology has started to be seen as a strong candidate to be an indispensable part of next{-}generation systems, as an alternative to massive MIMO systems~\cite{Burak_6_2020_RISParadigm6G}. 

\begin{figure}[t]
   \centering
   \includegraphics[width=\linewidth]{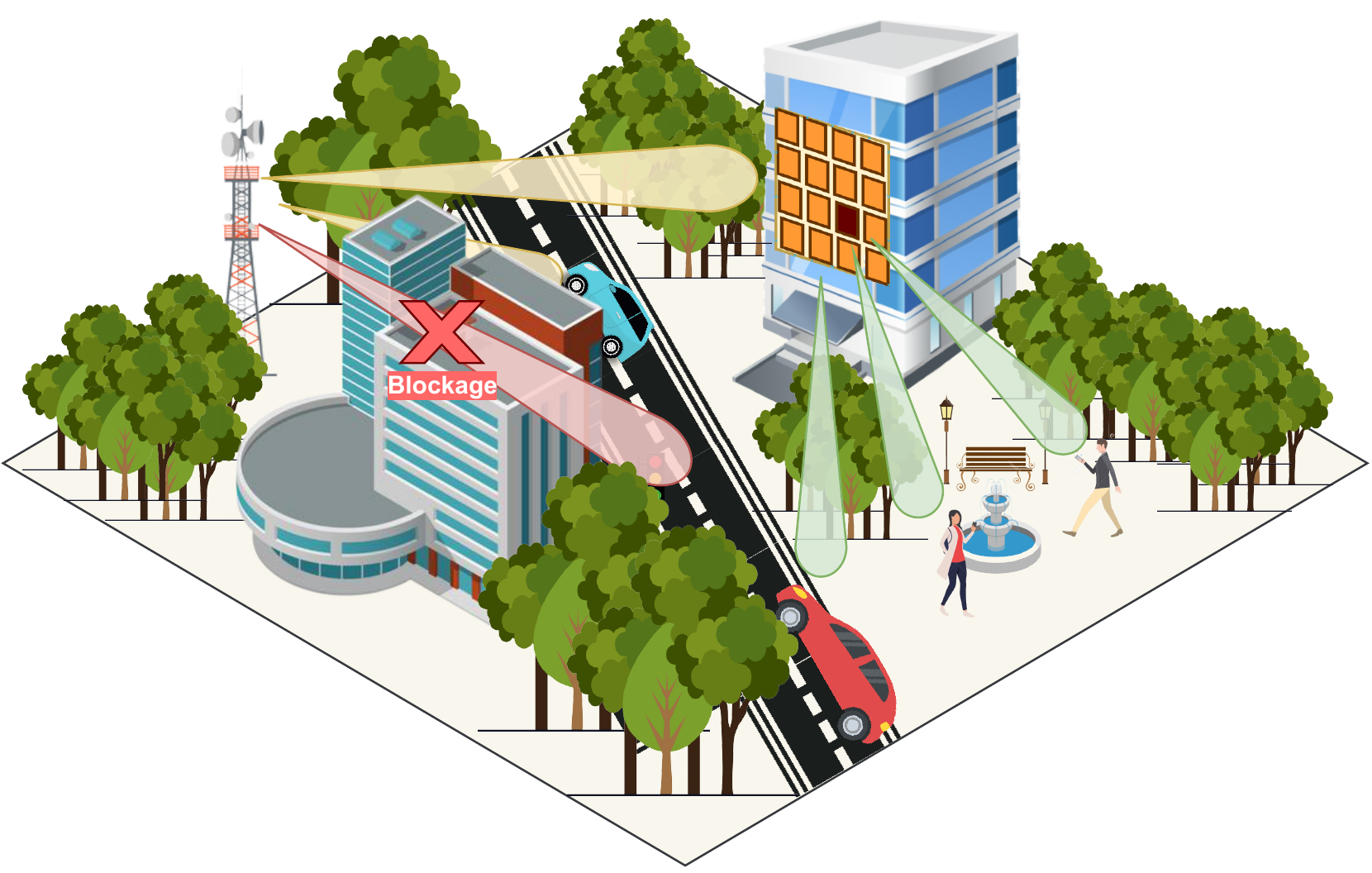}
   \caption{RIS-assisted wireless communication for multiple users.}
   \label{fig:RIS-assisted Wireless Communication Scenario for Multiple Users}
\end{figure}

An RIS interacts intelligently with incoming signals, aiming to expand energy efficiency and coverage in radio communication systems. The mentioned controlled interaction is realized with the adaptive meta elements on intelligent surfaces. It is aimed to obtain the optimum phase configuration and to optimize the wireless communication performance by manipulating the incoming signals thanks to these meta elements. Here, the incoming signals from the BS are controlled over-the-air in real-time and reflected to the receivers. Intelligent surfaces aim to increase the signal level at the receiver side as well as extend the signal coverage as illustrated in Fig.~\ref{fig:RIS-assisted Wireless Communication Scenario for Multiple Users}. Here, we consider the scenario of coverage expansion for multiple users having no line-of-sight path to the BS thanks to intelligent reflection phase configuration by an RIS coated on a building. To put it another way, RIS technologies enable intelligent wireless communication environments using software control methodologies. Thanks to their flexible mechanism, it is possible to work in different frequency bands such as terahertz communication with RISs~\cite{tarable2020meta} where it is crucial to establish mass communication and reduce interference between users.

An RIS, along with offering energy-efficient signal modulation at low cost in intelligent radio environments, takes an active role in secure and MU communication~\cite{tang2019wireless}. RIS-assisted beamforming techniques can ensure the secrecy of communication in the presence of a potential eavesdropper. The fact that it can be used for wireless energy transfer in low-energy applications such as wireless sensors, where continuous energy supply is required, induces us to see RISs frequently in areas  such as the IoT. Smart surfaces, which are capable of directing electromagnetic (\acrshort{em}) waves thanks to their unique structure and free-working nature, might be influential players in energy transfer~\cite{song2019smart}. Intelligent surfaces that act as reflective relays in the communication environment affected by poor environmental conditions are also used to improve the \acrshort{qos}.

All these advantages of RIS architectures and the success of passive  beamforming techniques show a solid commitment to channel knowledge~\cite{elbir2020deep}. The acquisition of multi-hop communication channel information in RIS-assisted communication schemes is more complicated than in traditional massive MIMO systems and can be addressed as an issue that needs to be studied
There beside, phase optimization algorithms show {a} strong commitment to the presence of the perfect channel information at the BS. Yet, acquiring this channel information can also become challenging due to hardware limitations, especially in RIS scenarios with passive reflective elements~\cite{Zhang2020DRLtaha2020deep}.

New scenarios brought by RISs for next-generation communication technologies will create new requirements that existing architectures will not be able to meet. Apart from the aforementioned challenges, the optimization of energy consumption of next-generation systems and providing storage options to meet the high training load of complex systems are some of the issues that need to be addressed by researchers. Adaptation performance of RISs to EM signals in radio environments  is one of the major issues that {are} open to improvements
For instance, in RIS-assisted \acrshort{iot} applications, the number of wireless devices connected to the intelligent surface can be vast. In such cases, the number of parameters required for system optimization will increase to the same extent~\cite{jiang2020learning}.

Advanced information processing techniques are essential in the response performance of smart surfaces to EM signals. This will require additional resources in terms of computation time and storage, as well as additional energy and bandwidth~\cite{leeJung2020deep}. The biggest obstacle in adapting existing systems to new communication technologies is that analytical models limit the working flexibility. Analytical {modeling} applies not only to network planning in the first stage in scenarios using RISs but also to resource management and network control where the adaptive approach is used. The complexity of RIS systems raises the possibility that existing traditional analytical models may become dysfunctional for them

Considering {the} situations mentioned above, uncertainties {arising} from RIS system configurations and channel dynamics complicate the system design~\cite{fengDRL2020deep}. Although signal processing techniques utilized by conventional RIS models provide solutions to many communication problems, their dependence on hypothetical mathematical models causes certain drawbacks. Model-independent solutions such as \acrshort{dl} stand out as a convenient solution to combat the aforementioned uncertainties. In RIS scenarios, the rapid change of the radio scattering and the hardware impairments result in certain mismatches, {such} that signal processing performance deteriorates. Thanks to its ability to process raw data and make it meaningful, DL can make a model-independent mapping by training with learnable parameters~\cite{elbir2020deep}. DL approaches yield more efficient and flexible results compared to traditional analytical model-dependent methods that are capable of working under only certain conditions. In other words, it would not be right to talk about the actual concept of "smart surfaces" in a communication environment where DL or \acrshort{ai} approaches are not implemented~\cite{Burak_14_2021_RISFutureNetworks}.  

\begin{figure}[t]
   \centering
   \includegraphics[width=\linewidth,height=5cm]{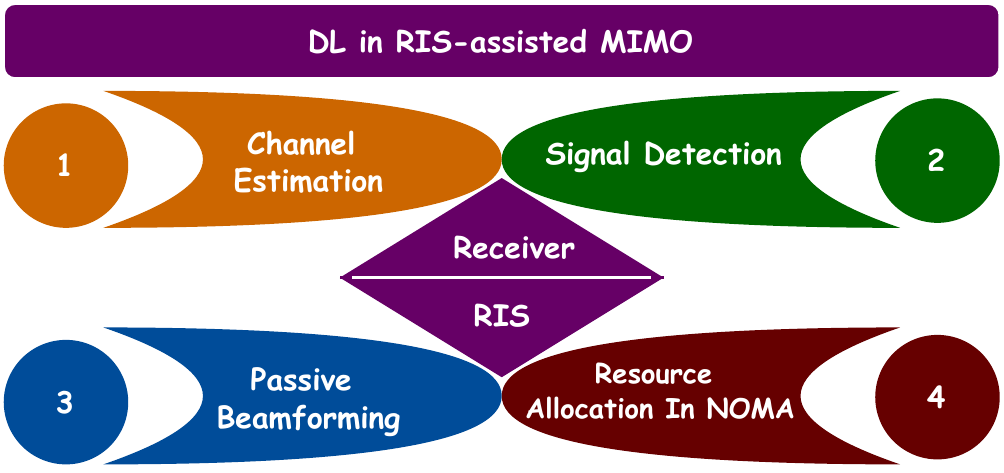}
   \caption{DL applications for RIS-assisted MIMO systems.}
   \label{fig:6G_Technologies-Page-2.pdf}
\end{figure}


In pursuit of the aforementioned advantages of DL-based approaches, many novel studies have been carried out in recent years using DL techniques in the field of RIS-assisted communication. This section is fictionalized on channel estimation and signal detection solutions, passive beamforming design scenarios, and resource allocation in NOMA techniques for RIS-aided DL-based systems. In Fig.~\ref{fig:6G_Technologies-Page-2.pdf}, the major applications of  DL  solutions in RIS-assisted systems are categorized from {the} perspective of receiver optimizations and RIS-side phase configurations. Finally, we discuss the actions {that} need to be taken to compensate {for} potential drawbacks of current applications.

\subsection{Channel Estimation and Signal Detection for RIS-assisted DL-based Systems}

Since the number of reflecting meta-elements increases in RIS-assisted massive MIMO systems, the channel acquisition process becomes more challenging due to {the} increased number of channels. In other respects, reliable and accurate channel acquisition carries critical importance for 6G and beyond technologies. The main inspiration behind DL-based solutions is balancing the trade-off between system complexity and achievable rate performance. Therefore, as can be seen in Table~\ref{tab:channel estimation and signal detection}, several studies are present in the literature {on} channel estimation and signal detection.

The biggest motivation for combining DL techniques with RIS-assisted wireless communication is the acquisition of mapping methods that can be unveiled without linear mathematical models. In this regard,~\cite{khan2019deep} performed a non-linear mapping between the sent and the received signals and optimized signal detection with a DL approach. A DL solution for signal detection proffered by~\cite{khan2019deep}, which contains individual learned models for each user. The foremost advantage of this scheme is that no channel estimation algorithm is required. Nonetheless, this system needs either beamforming optimization on the BS side or phase shift optimization directly by the RIS to improve the BER performance of users.

On the side of channel acquisition,~\cite{elbir2020deep} and~\cite{elbir2020federated} combined channel estimation scenarios with the DL approach and performed channel estimation for both reflected and direct paths. The proposed channel estimation technique, which employs the \acrshort{cnn} architecture, is highly dataset dependent. The diversity of data sets influences performance efficiency, which can sometimes limit the performance to local dimensions. This causes high training overhead during channel estimation learning. To overcome this problem,
\cite{liu2020deep} created the architecture of deep denoising NNs, that reduces the training load. In this model, the intelligent surface contains both active and passive elements concurrently. This architecture combines compressed sensing and DL approaches to achieve greater efficiency than either approach alone. However, the presence of active elements increases hardware complexity. Assuming that one of the main purposes is to reduce complexity in traditional mmWave massive MIMO architecture, this can be regarded as an issue that must be addressed.
 \begin{table*}
\caption{An overview of \acrshort{dl}-based channel estimation and signal detection studies for RIS systems.}
    \label{tab:channel estimation and signal detection}
\begin{tabular}{|p{1.3cm}|p{3cm}|p{12.385cm}|}
\hline
\cellcolor{table_blue} \textbf{\textcolor{table_white}{Reference}}&
\cellcolor{table_blue} \textbf{\textcolor{table_white}{Type of Algorithm}}& 
\cellcolor{table_blue} \textbf{\textcolor{table_white}{Application Scnearios}}
\\ \hline
\cellcolor{table_white} \textcolor{table_blue}{\cite{elbir2020deep} 
} 
&
\cellcolor{table_white} \textcolor{table_blue}{DL-CNN}&
\cellcolor{table_white} \textcolor{table_blue}{A novel CNN-based intelligent channel estimation framework without need of re-training for relocated users up to 4 degrees  }

\\ \hline
\cellcolor{table_white} \textcolor{table_blue}{
\cite{elbir2020federated } 
}& 
\cellcolor{table_white} \textcolor{table_blue}{Federated learning-CNN}&
\cellcolor{table_white} \textcolor{table_blue}{Channel   estimation for direct and cascaded paths via a novel DL approach in RIS-assisted mmWave MIMO systems}                                        
 
\\ \hline
\cellcolor{table_white} \textcolor{table_blue}{\cite{liu2020deep }  
}                                                           & 
\cellcolor{table_white} \textcolor{table_blue}{DL-Deep Denoising NN}                                        & 
\cellcolor{table_white} \textcolor{table_blue}{Combining the CS and DL to reduce training overhead by Deep Denoising NN architecture}     
\\ \hline
\cellcolor{table_white} \textcolor{table_blue}{\cite{he2021learning }
}                                                             & 
\cellcolor{table_white} \textcolor{table_blue}{DNN-Deep Unfolding}                                        & 
\cellcolor{table_white} \textcolor{table_blue}{ Deep unfolding with enhanced estimation performance at lower computational complexity and training overhead compared to the conventional least square estimator}   
\\ \hline
\cellcolor{table_white} \textcolor{table_blue}{\cite{taha2021enabling} }                                                             & 
\cellcolor{table_white} \textcolor{table_blue}{DL-CNN}                                        & 
\cellcolor{table_white} \textcolor{table_blue}{ CSI acquisition with the sparse channel sensors and optimization of the RIS phase configuration with reduced training overhead}  
\\ \hline
\cellcolor{table_white} \textcolor{table_blue}{\cite{aygul2021deep }  }                                                           & 
\cellcolor{table_white} \textcolor{table_blue}{MLP-DNN}                                        & 
\cellcolor{table_white} \textcolor{table_blue}{ Strengthen interaction between channel information and passive beamforming of conventional DL-based systems by practicing MLP architecture}  
\\ \hline
\cellcolor{table_white} \textcolor{table_blue}{\cite{shtaiwi2021ris }            }                                                 & 
\cellcolor{table_white} \textcolor{table_blue}{Unsupervised Learning-CNN}                                        & 
\cellcolor{table_white} \textcolor{table_blue}{ Two-stage novel channel estimation method developed to reduce the training load in mmWave communication scheme}  
\\ \hline
\cellcolor{table_white} \textcolor{table_blue}{\cite{kundu2021channel }                   }                                          & 
\cellcolor{table_white} \textcolor{table_blue}{DL-CNN}                                        & 
\cellcolor{table_white} \textcolor{table_blue}{A novel data-driven channel estimator in MISO architecture to achieve reduced system complexity}  
\\ \hline
\cellcolor{table_white} \textcolor{table_blue}{\cite{gaoDong2021deep }                }                                             & 
\cellcolor{table_white} \textcolor{table_blue}{DL-DNN}                                        & 
\cellcolor{table_white} \textcolor{table_blue}{Sequential trained three-stage synthetic DNNs to estimate cascaded channel by using fully passive elements}  
\\ \hline
\cellcolor{table_white} \textcolor{table_blue}{\cite{zhangGaogrouping2021deep }         }                                                    & 
\cellcolor{table_white} \textcolor{table_blue}{CENet-CNN }                                      & 
\cellcolor{table_white} \textcolor{table_blue}{Enhanced element-grouping method by investigating and eliminating channel interference to reduce pilot overhead with reliable channel estimation processes }  
\\ \hline
\end{tabular}
\end{table*}
Integration of DL-based techniques in RIS-aided systems has been challenging due to complicated training processes since the number of channels to be exploited is proportional with the number of reflecting elements. Therefore, the literature frequently focuses on this subject. DL-based channel estimation techniques reducing the training overhead are also presented in~\cite{taha2021enabling},~\cite{aygul2021deep}, and~\cite{shtaiwi2021ris} by different learning and network architectures.~\cite{aygul2021deep} aimed to strengthen the interaction between channel information and passive beamforming of conventional DL-based systems by practicing the multi-layer perceptron (MLP) architecture.~\cite{shtaiwi2021ris} presented a two-stage channel estimation method to reduce the training load in mmWave communication schemes. The considered DNN is trained with a reduced number of users in the first stage to learn effective channel parameters over active users. Then, the authors present a spatial-temporal–spectral framework {that} estimates deficient channel information in the second phase.~\cite{taha2021enabling} used a similar approach, in which the channel information obtained with only a few active elements was also used to estimate the remaining channels, resulting in almost no training overhead. In the first part of the offered two-tier solution, the authors performed CSI acquisition with the presented sparse channel sensors structure and then used the obtained information to optimize the RIS phase configuration. However, the presence of these controlled active elements used in channel reconstruction cannot practically meet the standalone operation principle and the passive nature of RISs. The presence of those active elements also increases the system. ~\cite{he2021learning} has been inspired by the high learning and prediction capability of DNNs and aimed low system complexity for mmWave communication. The proposed deep unfolding method gives higher estimation performance at lower computational complexity and learning load compared to the conventional LS estimator. The model-driven unfolding method, which is a sub-type of DNN and has the ability to learn from extensive synthetic data with lower iteration, follows an algorithm similar to traditional gradient descent optimizations.~\cite{kundu2021channel} proposes a novel data-driven channel estimator in MISO architecture to achieve reduced system complexity for a similar purpose of~\cite{he2021learning}. This study benefits from the fact that the CNN structure involves fewer parameters compared to prior architectures such as DNN. The proposed CNN-based channel estimator has superiority over traditional linear estimators in terms of implementation complexity. Nonetheless, extended scenarios, including mobility and plurality of users as well as multiple antennas assumption, will be vital to realize this architecture in practical applications. 

One of the fundamental motivations for exercising DL methods in channel estimation is to employ fewer pilots.~\cite{gaoDong2021deep} and~\cite{zhangGaogrouping2021deep} presents DL-based approaches achieving reliability in estimation accuracy while reducing the pilot overhead. Both studies perform entire channel extrapolation from sampled channels.~\cite{gaoDong2021deep} proposes sequential trained three synthetic DNNs. These DNNs first estimate the direct channel and then the cascaded channel. The last DNN predicts the cascaded channel by using fully passive elements at the final stage. 
\cite{zhangGaogrouping2021deep} also used multiple DNNs, which are cascaded and sequentially trained. The proposed study enhances the conventional element-grouping method by investigating and eliminating channel interference. Sampled channels refined by the first DNN as a result of interference elimination are used in the prediction of entire channels in the second stage. Both studies gain significant improvements in reducing the pilot overhead with reliable channel estimation processes.

Considering the current approaches in the literature, simpler wireless design schemes have been the focus of ongoing investigations on RIS-assisted systems. In order to measure the performance boundaries in wireless connections utilizing RISs, more accurate modeling approaches are necessary for the transmission of signals scattered by metasurfaces. Another challenge is that no amplifiers are present in an entirely passive RIS architecture. At that point, RISs might need a low-powered unit to be informed by current channel conditions and other communications blocks such as receiver{s} or transmitter{s} for the purpose of phase optimization. Benefiting these channel monitoring units embedded with RISs powered by novel energy harvesting methods might be necessary for energy-efficient approaches.

As concluded from the majority of recent studies, the presence of perfect CSI is crucial at the transmitter. Nevertheless, it is not a straightforward process to retrieve full channel knowledge in practice because of the passive nature of RIS-aided schemes. For this reason, the quick and accurate channel acquisition required for next-generation technologies causes a high training burden. 
For the present, supervised learning techniques are commonly employed to tackle channel issues in order to increase efficiency, and DL approaches investigate \acrshort{csi} structures of current designs. 
However, due to the limitations mentioned above, the research efforts can be expanded with different deployment scenarios. For instance, using RISs in a near-field setting might offer several exciting applications and benefits.

\begin{table*}
\caption{An overview of DL-based passive beamforming design studies.}
    \label{tab:DL-based passive beamforming}
\begin{tabular}{|p{1.3cm}|p{2.7cm}|p{2.8cm}|p{9.08cm}|}
\hline
\cellcolor{table_blue} \textbf{\textcolor{table_white}{Reference}}&
\cellcolor{table_blue} \textbf{\textcolor{table_white}{{Preliminary Summary }}}&
\cellcolor{table_blue} \textbf{\textcolor{table_white}{Type of   Algorithm}}& 
\cellcolor{table_blue} \textbf{\textcolor{table_white}{Application Scenarios}}


\\ \hline
\cellcolor{table_white} \textcolor{table_blue}{\cite{huang2019indoor}}                                                         &
\cellcolor{table_white} \textcolor{table_blue}{{Location-based Training}}                                                         &
\cellcolor{table_white} \textcolor{table_blue}{DNN}                                              & 
\cellcolor{table_white} \textcolor{table_blue}{The fingerprinting database to unveil mapping between measured user location and   RIS unit cells to maximize received signal strength}  
\\ \hline
\cellcolor{table_white} \textcolor{table_blue}{\cite{fengDRL2020deep}}                                                             & 
\cellcolor{table_white} \textcolor{table_blue}{{Phase Shifts Design}}                                                             & 
\cellcolor{table_white} \textcolor{table_blue}{Deep Reinforcement Learning-DDPG}                                          & 
\cellcolor{table_white} \textcolor{table_blue}{DDPG   algorithm that can quickly adapt the model for dynamic channel state and environmental   conditions  thanks to the continuous action space }
\\ \hline
\cellcolor{table_white} \textcolor{table_blue}{\cite{Lin2020gong2020optimization}}                                                           & 
\cellcolor{table_white} \textcolor{table_blue}{{DRL for Joint Beamforming}}                                                           &
\cellcolor{table_white} \textcolor{table_blue}{Deep Reinforcement Learning-DDPG }                                         & 
\cellcolor{table_white} \textcolor{table_blue}{An   optimization where the DDPG algorithm is responsible to search for the   optimum action at each decision during the network learning phase }   
\\ \hline
\cellcolor{table_white} \textcolor{table_blue}{\cite{ma2020distributed}}                                                            & 
\cellcolor{table_white} \textcolor{table_blue}{{Enhanced Data Rate and Protected Privacy}}                                                            & 
\cellcolor{table_white} \textcolor{table_blue}{Federated Learning-MLP }                                           & 
\cellcolor{table_white} \textcolor{table_blue}{The federated architecture to reduce transfer overhead offering beamforming optimization on the RIS   side    }    
\\ \hline
\cellcolor{table_white} \textcolor{table_blue}{\cite{gao2020resource}}                                                           & 
\cellcolor{table_white} \textcolor{table_blue}{{Power Allocation}}                                                           &
\cellcolor{table_white} \textcolor{table_blue}{Unsupervised Learning}                                           & 
\cellcolor{table_white} \textcolor{table_blue}{A novel usupervised learning architecture for beamforming design optimization}
\\ \hline
\cellcolor{table_white} \textcolor{table_blue}{\cite{Zhang2020DRLtaha2020deep}}                                                         & 
\cellcolor{table_white} \textcolor{table_blue}{{Standalone RIS Architecture}}                                                         &
\cellcolor{table_white} \textcolor{table_blue}{Deep Reinforcement Learning}                                              & 
\cellcolor{table_white} \textcolor{table_blue}{The architecture created an active learning scheme that quickly adapts to changing environmental conditions with high performance}       
\\ \hline
\cellcolor{table_white} \textcolor{table_blue}{\cite{ozdougan2020deep}}                                                       & 
\cellcolor{table_white} \textcolor{table_blue}{{Supervised Learning for Phase Shift Design}}                                                       & 
\cellcolor{table_white} \textcolor{table_blue}{Feed-forward NN}                                              & 
\cellcolor{table_white} \textcolor{table_blue}{The feed-forward NN fed with reflected symbols performing phase matrix optimization } 
\\ \hline 

\cellcolor{table_white} \textcolor{table_blue}{\cite{huang2020reconfigurable}}                                                         & 
\cellcolor{table_white} \textcolor{table_blue}{ {Historical Channel Knowledge}}                                                         & 
\cellcolor{table_white} \textcolor{table_blue}{ Deep Reinforcement Learning-DNN }     & 
\cellcolor{table_white} \textcolor{table_blue}{ DNN architecture evaluating historical line-of-sight path channels to interpret channel behaviour }  
\\ \hline
\cellcolor{table_white} \textcolor{table_blue}{\cite{song2021truly}}                                                           & 
\cellcolor{table_white} \textcolor{table_blue}{{Secrecy Rate Optimization}}                                                           &
\cellcolor{table_white} \textcolor{table_blue}{Unsupervised Learning-DNN}                    & 
\cellcolor{table_white} \textcolor{table_blue}{Supervised Learning approach for the secrecy rate optimization at  receiver and phase configuration optimization on the RIS side with reduced computational complexity }  
\\ \hline
\cellcolor{table_white} \textcolor{table_blue}{\cite{feriani2021robustness}}                                                     & 
\cellcolor{table_white} \textcolor{table_blue}{{Mobile User Scenario}}                                                     &
\cellcolor{table_white} \textcolor{table_blue}{Deep Reinforcement Learning}     & 
\cellcolor{table_white} \textcolor{table_blue}{Deep reinforcement learning approach achieving higher SNR for noisy channels and mobile scenarios comparing to conventional techniques }          
\\ \hline
\cellcolor{table_white} \textcolor{table_blue}{\cite{huangChen2021deep} }                                                         & 
\cellcolor{table_white} \textcolor{table_blue}{{Learning from Environment }}                                                         &
\cellcolor{table_white} \textcolor{table_blue}{Deep Reinforcement Learning-DNN }     & 
\cellcolor{table_white} \textcolor{table_blue}{ The relay selection optimization to reduce propagation loss over distance by the proposed Deep Reinforcement Learning model that can learn from the environment }                    
\\ \hline
\cellcolor{table_white} \textcolor{table_blue}{\cite{huangTerahertz2020hybrid}}                                                         & 
\cellcolor{table_white} \textcolor{table_blue}{{Terahertz Communication}}                                                         & 
\cellcolor{table_white} \textcolor{table_blue}{Deep Reinforcement Learning-DNN}     & 
\cellcolor{table_white} \textcolor{table_blue}{ A method that addresses the challenge in path loss optimization for RIS-aided terahertz communication having high molecular absorption and attenuation }   
\\ \hline
\cellcolor{table_white} \textcolor{table_blue}{\cite{nguyenD2D2021deep} }                                                         & 
\cellcolor{table_white} \textcolor{table_blue}{{Resource Allocation}}                                                         &
\cellcolor{table_white} \textcolor{table_blue}{Deep Reinforcement Learning-NN}     & 
\cellcolor{table_white} \textcolor{table_blue}{ Resource allocation  for D2D networks and phase shift configuration optimization in terms of achievable rate and computational time performance }   
\\ \hline
\cellcolor{table_white} \textcolor{table_blue}{\cite{leeJung2020deep}}                                                         & 
\cellcolor{table_white} \textcolor{table_blue}{{Energy-efficient Policy}}                                                         & 
\cellcolor{table_white} \textcolor{table_blue}{Deep Reinforcement Learning-DNN}     & 
\cellcolor{table_white} \textcolor{table_blue}{Deep Reinforcement Learning approach presenting a fully energy-efficient method by optimizing the ON/OFF state of RIS elements besides transmit power}     
\\ \hline
\cellcolor{table_white} \textcolor{table_blue}{\cite{alexandropoulos2020phase} }                                                         & 
\cellcolor{table_white} \textcolor{table_blue}{{Dynamic Positioning of RIS}}                                                        &
\cellcolor{table_white} \textcolor{table_blue}{MLP-Position Based}     & 
\cellcolor{table_white} \textcolor{table_blue}{The MLP-based NN addressing dynamic positioning of multiple RISs to overcome the storage and computational performance limitations }
\\ \hline
\cellcolor{table_white} \textcolor{table_blue}{\cite{mehmood2021throughput} }                                                         & 
\cellcolor{table_white} \textcolor{table_blue}{{Optimized Energy Expenditure }}                                                         &
\cellcolor{table_white} \textcolor{table_blue}{ Unsupervised Learning-NN  }     & 
\cellcolor{table_white} \textcolor{table_blue}{A NN trained by the approach of deep unsupervised learning optimizing both energy expenditure and phase configuration }   
\\ \hline
\cellcolor{table_white} \textcolor{table_blue}{\cite{sheen2021deep} }                                                         & 
\cellcolor{table_white} \textcolor{table_blue}{{Mapping without CSI}}                                                         & 
\cellcolor{table_white} \textcolor{table_blue}{ ML/DL-DNN}     & 
\cellcolor{table_white} \textcolor{table_blue}{ The ML-based approach leveraged by DL techniques has no need for CSI for direct mapping }  
\\ \hline
\cellcolor{table_white} \textcolor{table_blue}{\cite{jiang2020learning} }                                                         & 
\cellcolor{table_white} \textcolor{table_blue}{{Bypassing Channel Prediction }}                                                         & 
\cellcolor{table_white} \textcolor{table_blue}{ Unsupervised Learning-DNN }     & 
\cellcolor{table_white} \textcolor{table_blue}{ Proposed algorithm bypassing channel prediction process, requires fewer pilots compared to prior studies with the channel estimation }  
\\ \hline
\cellcolor{table_white} \textcolor{table_blue}{\cite{ge2021beamforming} }                                                         & 
\cellcolor{table_white} \textcolor{table_blue}{{Reduced Hardware Complexity }}                                                         & 
\cellcolor{table_white} \textcolor{table_blue}{Unsupervised Learning-DNN}     & 
\cellcolor{table_white} \textcolor{table_blue}{ Proposed deep-transfer learning-based algorithm requesting less sampled data for training process resulting in reduced hardware complexity and training load} 
\\ \hline
\cellcolor{table_white} \textcolor{table_blue}{\cite{liu2021deep}}                                                         & 
\cellcolor{table_white} \textcolor{table_blue}{{Learning Channel Behaviour}}                                                         &
\cellcolor{table_white} \textcolor{table_blue}{Deep Reinforcement Learning}     & 
\cellcolor{table_white} \textcolor{table_blue}{Deep Reinforcement Learning-based novel architecture capable of learning channel behaviour} 
\\ \hline
\end{tabular}
\end{table*}

\subsection{Passive Beamforming Design with DL for System Optimization}

Passive beamforming design is among the most critical problems for the RIS technology to be included in {the} next generation of communication systems. The success of phase reconfiguration not only increases the quality of signal transformation but also increases the adaptability of RISs to different communication environments such as indoor/outdoor applications. Especially in highly dynamic channels, such as application scenarios with high mobility, the inadequacy of existing control systems has revealed the necessity of a new perspective. In preserving the fully passive nature of RISs with reduced costs and system complexity, truly intelligent approaches have a bright future for next-generation communications. Therefore, DL-based passive beamforming techniques have extensively been addressed by the literature as presented in Table~\ref{tab:DL-based passive beamforming}. 
Current DL-based studies on RIS-aided communication systems are composed of various NN structures depending on their advantages for target applications. \acrshort{mlp}, a member of the feed-forward artificial-NN and one of the most basic types of DNN, can be given as an instance.~\cite{taha2021enabling} offers an \acrshort{mlp} architecture for beamforming design on the RIS side. The considered RIS scheme, which has a hybrid structure in which active and passive elements are present simultaneously, performs pilot training with randomly distributed active components. The trained data set, in which the beamforming on the RIS side is established as an input-output pair, is built with the supervised learning scheme. The use of active elements can be considered as a drawback of this study. The possible advantages of the supervised learning-based MLP algorithm with fully passive RIS have been inspected in~\cite{alexandropoulos2020phase}  to overcome the storage and computational performance limitations in the dynamic positioning of multiple RISs. The MLP-based NN has fed by dynamic channel information as well as RIS positioning information. The proposed position-trained NN results are superior to benchmark results retrieved by {an} exhaustive study.

\cite{ma2020distributed} offers beamforming optimization on the RIS side using the federated learning  approach. After the training phase of the \acrshort{mlp}  architecture, model updates are calculated for each user using local data sets. Model updates are received from the parameter server to which the RIS is connected. The results of the study indicate that the transfer overhead is reduced thanks to the federated architecture. However, because of the passive nature of RISs, the scenario in which it is constantly connected to the parameter server, makes this architecture inefficient. At that point,~\cite{gaoUnsupervised2020unsupervised} sought a solution to cure unit modulus constraints of passive reflecting surfaces from previous studies and conducted an approach that also reduces model complexity during the training phase. This study proposes an unsupervised learning architecture, in which the direct and reflected channels are the inputs, and the output is the phase values of the response beamforming of the RIS while the mapping is unveiled. This architecture is capable of making online predictions thanks to the trained DNN in the offline phase. However, the output parameters are unique for each learning phase. This contradiction will act as a layer. Therefore, each learning data will instantly output different values. However, different methods have been proposed in the literature to prevent this stratification and can be treated as further study on enhancement of this study. 

Deep reinforcement learning, appears as a promising method among the DL-based techniques with its high learning capability and other recent advantages {in} the decision-making process. As a result, it also appears as a strong candidate for various fields in RIS-assisted next-generation communication systems. The importance of deep reinforcement learning comes from its remarkable ability to make inferences from limited knowledge. The literature covers a wide range of topics, such as providing energy-efficient and robust processes or extending coverage, but it can primarily be evaluated at a higher level as passive beamforming applications.

\cite{fengDRL2020deep} proposes the deep deterministic policy gradient (\acrshort{ddpg}) technique for a DL-assisted RIS scheme, which combines the deep-Q network and policy gradient (PG), where the continuous action space is in use to accelerate the training phase. The presented model can adapt {quickly} to changing channel data and environmental conditions thanks to the continuous action space of DDPG algorithm. Yet, using multiple NNs increases the number of parameters in the learning phase. Therefore, the training overhead will cause high system load and storage requirements as well as hardware costs. To overcome the computational complexity and hardware limitations introduced by multiple NNs,~\cite{ge2021beamforming} offers a model combining deep-transfer learning algorithm and unsupervised learning which requests less sampled data for the training process. Therefore, results demonstrated that hardware complexity and training load problems in previous studies are addressed by this scheme. It is also inspected that a less transmit power request at the BS side is achieved to tackle hardware cost and limitations as well as process time.

Here, the DDPG algorithm has also been used by~\cite{Lin2020gong2020optimization} to optimize and accelerate the learning process. The DDPG algorithm is responsible for searching the optimum action for each decision during the network learning phase. Thus, the learning phase will be accelerated by optimizing the search space. However, this somewhat contradicts the model-free nature, which is the main philosophy of the DL approach, as the learning process will require an optimization toolbox and signal models for {the} learning process. At this point, instead of optimization-based techniques, it would be more appropriate to follow methods using completely deep model-independent learning approaches. To compensate model-dependent structure of current deep reinforcement learning-DDPG applications,~\cite{feriani2021robustness},~\cite{Zhang2020DRLtaha2020deep}, and~\cite{huang2020reconfigurable} proposed their advanced deep reinforcement learning techniques making the systems robust in changing environmental conditions. By these advanced deep reinforcement learning techniques, not only the channel information but also channel {behavior} can be learned similar to~\cite{liu2021deep}.~\cite{feriani2021robustness} has achieved a higher SNR for noisy channels and mobile scenarios {compared} to conventional deep reinforcement learning methods. The authors of~\cite{huang2020reconfigurable} demonstrated that their algorithm could enhance the BER performance in every cycle thanks to the evaluation of previous rewards for actions, so that enhance optimization of {the} phase matrix. The channel {behavior} can be interpreted by evaluating historical line-of-sight path channels to determine optimal phase configuration for further actions.

\cite{Zhang2020DRLtaha2020deep} developed an active learning scheme that adapts rapidly to changing environmental conditions while maintaining high performance. However, since reinforcement learning works with a reward mechanism, it has a longer operation time and lower performance compared to supervised learning. The fact that the considered scheme does not have a layered structure causes it to lag behind supervised learning performance. However,~\cite{mehmood2021throughput} has demonstrated that a design outperforming supervised learning in terms of both energy and time consumption can be created without the use of a layered structure. The proposed algorithm with trained NNs by the approach of deep unsupervised learning optimizes both energy expenditure and phase configuration to outperform even the Genetic algorithm, which is considered a groundbreaking development in optimization. That being said, this improvement comes with a slight loss in throughput performance compared to the genetic algorithm. However, that loss can be compensated with an increased number of antennas. 

Unsupervised learning has also been used to optimize computational complexity. In~\cite{nguyen2021machine}, efficient relay selection optimization is performed with low system complexity by the proposed algorithm. On the other hand,\cite{huangChen2021deep} outperforms previous deep reinforcement learning-based techniques in terms of computational complexity for complicated phase shift optimization scenarios by creating direct relation between optimization parameters and throughput. The relay selection method is also conducted to reduce propagation loss over increasing distance by the proposed deep reinforcement learning model that can learn from the environment.

Propagation loss is also the subject of terahertz communication which has a bright future for next-generation communication systems.~\cite{huangTerahertz2020hybrid} seeks a method that addresses the challenge in path loss optimization for RIS-aided terahertz communication which has high molecular absorption and attenuation. The proposed method implementing hybrid beamforming provides 50\% coverage extension compared to ZF and alternating beamforming.

In addition to prior applications, joint optimization of transmission power with phase shift configuration by deep reinforcement learning in RIS-assisted systems provides major advantages for optimizing achievable rates and achieving an energy-efficient system.~\cite{nguyenD2D2021deep} performs resource allocation optimization for device-to-device (\acrshort{d2d}) networks. The proposed scheme conducts a proximal policy optimization algorithm to define the policy gradient of deep reinforcement learning that seeks the maximized reward. The presented model outperforms the random phase shift matrix selection algorithm with random relay selection and maximum power transmission algorithm that also performs phase-shift configuration optimization in terms of achievable rate and computational time performance. The great reward achieved by~\cite{nguyenD2D2021deep} in which the deep reinforcement learning was involved, was enhanced by~\cite{leeJung2020deep} in an energy-efficiency manner.~\cite{leeJung2020deep} has aimed to present a fully energy-efficient method by optimizing the ON/OFF state of RIS elements besides transmit power. In this scheme, the access point receives available energy information from CSI in order to make the best decision for the current state via a backhaul link that connects the RIS to the BS.~\cite{Zhang2020DRLtaha2020deep} terminates the necessity of controlling RIS from any infrastructure in order to introduce standalone operation. The proposed study shows that RIS is capable of learning and adjusting optimal relay selection using the proposed deep reinforcement learning approach with almost no training overhead, thanks to the online learning phase. 

Advanced DL applications also promise intelligent environments in the case of mobile user scenarios for RIS-assisted communication systems.~\cite{huang2019indoor} used the DL approach in phase configuration on the RIS side and developed a process that yields user location-dependent results. An NN using the location information as input is trained for many reference points to give optimized phase outputs. In other words, this study offers an efficient method for configuring an RIS in the indoor communication environment. The fingerprinting database in the training of weights and biases forms the concept of location-based learning, which unveils a mapping between measured user location and RIS unit cells to maximize the received signal strength. Nevertheless, both the inefficiency of RIS positioning and the inefficient result in the throughput must be investigated.

\cite{ozdougan2020deep} used the DL approach to optimize the phase configuration on the RIS side, trained feedforward NN with reflected symbol signals and aimed to obtain the optimum phase matrix. The facts that the system cannot superior the least square estimation methods for high symbol powers and causes high signaling overhead are the points that need to be addressed. Also, the system should be analyzed under the assumption of imperfect channel estimation. However, channel acquisition is a significant challenge in RIS-assisted schemes due to complicated channel information caused by a large number of reflective elements. Therefore,~\cite{sheen2021deep} and~\cite{jiang2020learning} proposed approaches unveiling a direct mapping between optimal phase configuration and achievable rate.\cite{jiang2020learning} uses the pilot signals in the DNN trained by {an} unsupervised learning approach to determine the optimal phase configuration instead of using it in the channel acquisition. The proposed algorithm bypassing the channel prediction process requires fewer pilots compared to prior studies with channel estimation. In~\cite{sheen2021deep}, the considered ML-based approach leveraged by DL techniques has no need for initial CSI for direct mapping. The proposed algorithm is capable of learning the relation between RIS configuration and achievable rate correlated with dynamic receiver positioning. Since~\cite{sheen2021deep} is one of the studies that {give} the closest results to the benchmarks with the perfect channel assumption, the proposed method gleams as a bright future direction so that RIS shift optimization can be executed without the need for perfect CSI from the infrastructure.



{Following} a different perspective from prior studies,~\cite{song2021truly} concentrates on the advantages of the RIS architecture for the field of PHY security. This study intends to optimize the secrecy rate at the receiver by tuning meta elements on the RIS side. Therebeside, real-time tuned reflecting elements allow phase configuration optimization on the RIS side using the DL approach and this proposed a technique that yields similar outcomes to traditional methods but at the same time {reduces} the computational complexity considerably. 

Considering these studies and the rest of the literature, we conclude that the results obtained in combining RIS architectures with DL approaches do not adequately meet the flexibility and required performance for 6G and beyond wireless communication systems. In particular, the limitations are presented by both the transmitter and the user perspectives, such as the high computational complexity, the high training overhead, and the perfect \acrshort{csi} assumption in the transmitter. Besides, the use of active elements on smart surfaces in optimizing the phase configuration does not comply with the fully passive nature of RISs. The urgent need to achieve high energy efficiencies at the lowest possible cost should be considered. It is clear that the literature still contains many gaps that require work in light of these problems. Examination of different RIS positioning scenarios, optimization of system performances over computer simulation{,} and application scenarios under different environmental conditions, in which the DL approach enhances channel estimation and phase configuration optimization, is vital in designing promising perspectives on related research problems for 6G and beyond.

Another technical limitation that makes current system designs less viable is the assumption of a static environment. Though, in dynamic scenarios, the mobility of users includes some additional parameters such as speed, acceleration, and position that need to be considered in the process. Therefore, the learning process for the data traffic of a mobile user becomes more complicated due to dynamic channels. Moreover, the system parameters cannot be specified as constants because of the infinite amount of mobility. This situation makes it challenging to employ supervised techniques in which input-output pairs are labeled and learned from properly arranged data. Considering the benefits that ML techniques provide in the face of complex data in various fields, methods such as computer vision or neural language processing can also be regarded from the perspective of communication engineering. At this point, structures such as convolutional neural networks may be preferred instead of more superficial DNN structures.

RISs are comprised of complex formed subwavelength elements. As a consequence, a fundamental restriction in continuing RIS research is the lack of accurate and controllable models that define reconfigurable metasurfaces as a function of their EM features. As the majority of studies assume, the passive reflection behavior of RISs is not applicable in practice. Components such as material composition, polarization, and angles influence the RISs response to radio waves. In other words, RISs do not just reflect the waves but also re-design them. For RIS topologies, practical EM-based circuit designs should be used, which take mutual linking and unit cell configurations into account, similar to \cite{abeywickrama2020intelligent}.







\subsection{DL Techniques for Resource Allocation in RIS-aided NOMA Systems}

\begin{figure}[t]
   \centering
   \includegraphics[width=\linewidth,height=8cm]{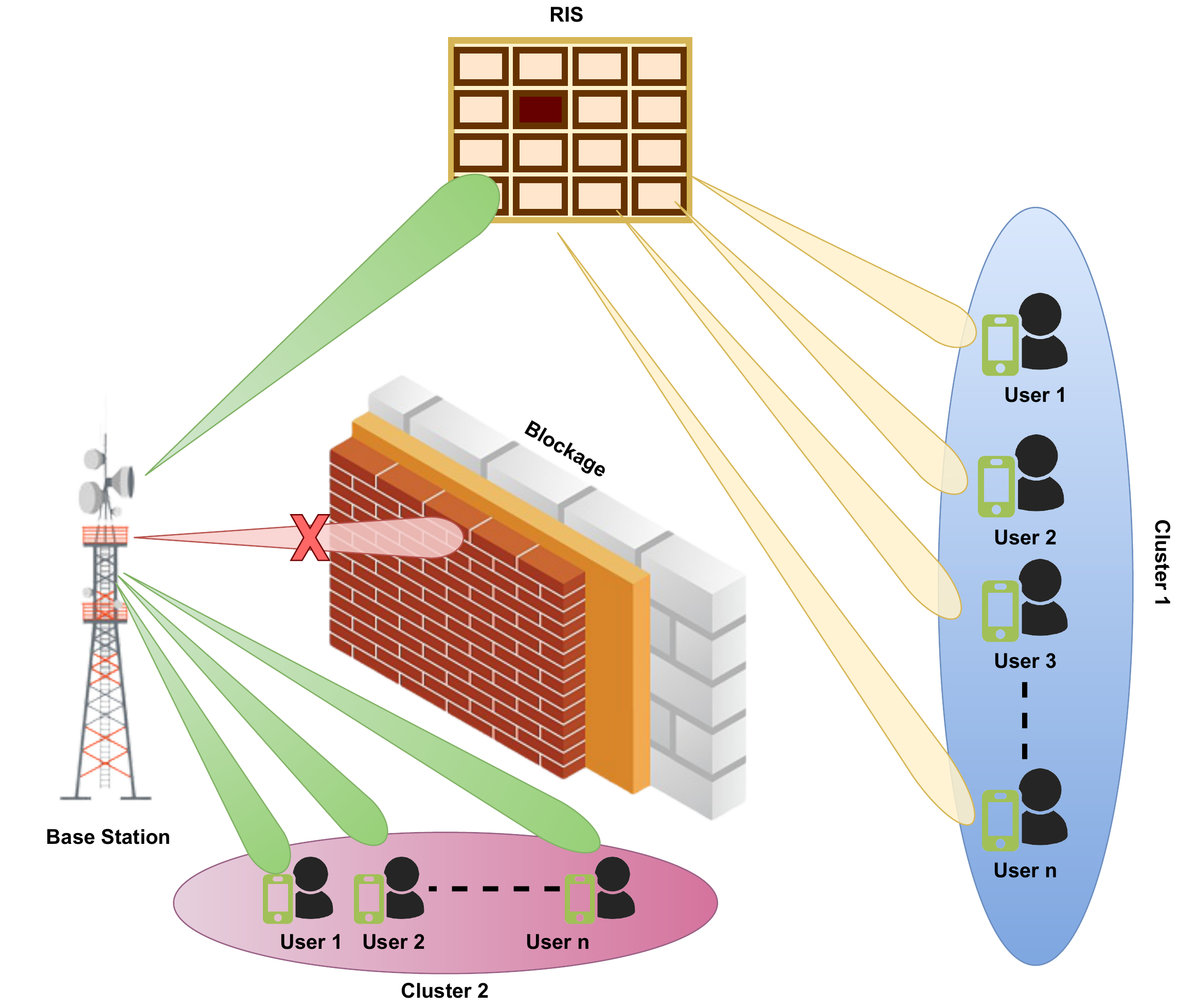}
   \caption{RIS-assisted downlink-NOMA with multiple clusters.}
   \label{fig:FIG_NOMA}
\end{figure}

NOMA stands out as another promising technique for increasing the efficiency of future massive MIMO systems. Therefore, NOMA-MIMO techniques have been studied extensively in the literature to examine how to improve the spectrum and energy efficiency required by 6G and beyond technologies and to increase the total capacity of multiple users~\cite{ding2015application}. Thus, NOMA sparkles as a promising solution for challenging fields of 6G communication systems such as massive IoT that {require} high data capacity and lower latency with high spectral efficiency. Fig.~\ref{fig:FIG_NOMA} illustrates an RIS-assisted downlink-NOMA scenario where the users having correlated channels are clustered together. Effective clustering schemes with individuals having minimum correlation with other clusters are provided to conduct efficient resource allocation and enhanced throughput while decreasing interference.

Besides all the aforementioned advantages of NOMA, DL approaches  will be critical in improving existing systems to meet the high{-}performance system requirements of next-generation perspectives. In the most general sense, the studies aimed at improving the performance of existing NOMA systems using the DL approach can be categorized under the detection of channel characteristics and SIC. Power optimization, user clustering, signal detection, decoding techniques, and phase configuration can be cited as {the} main areas where the DL approaches are employed to improve existing NOMA schemes. Since DL approaches are vital in increasing the performance of traditional RIS-assisted NOMA systems, numerous studies have been carried out in relevant areas as presented in Table~\ref{tab:RIS-aided NOMA}.


In the current NOMA literature, the intelligent system approach with DL has been mainly seen in user clustering scenarios that have already been studied extensively due to their critical role, as mentioned earlier.~\cite{cui2018unsupervised} provided a new idea about possible contributions of intelligent approaches in next-generation communication systems by presenting the effect of the DL approach integrated into a conventional RIS-{assisted} NOMA architecture. The proposed study presents a user clustering algorithm that reduces computational complexity and creates the optimum power allocation with sequential decoding. The study, which designed a \emph{K}-means-based ML algorithm for user clustering, benefited from the correlation in the user channels of mmWave systems. So the proposed \emph{K}-means-based algorithm also has the ability to do online user clustering. This characteristic allows it to outperform traditional \textit{K}-means-based algorithms in terms of performance in lower system complexity. Therebeside, a conventional \textit{K}-means-based Gaussian mixture model is conducted as a clustering technique used to interference alignment to cope with growing network traffic by~\cite{gao2021machine}. The proposed method aims to perform passive beamforming optimization on the RIS side with the ML algorithm. By this approach, they sought an optimized solution for signal decoding, power allocation{,} and user clustering under QoS constraints while the MISO-NOMA downlink scheme is considered for the proposed network. A novel Deep Q-network algorithm is adopted in the regulation of the power allocation process and the phase configuration on the RIS side.

The advantages of handling clustering techniques with intelligent approaches are also studied and discussed in RIS-assisted scenarios by~\cite{gao2020resource},~\cite{ni2021integrating}, and~\cite{yang2021machine}. In an RIS-assisted NOMA communication scheme,~\cite{gao2020resource} optimized the phase configuration using the deep reinforcement learning approach, thus aiming to increase the spectrum efficiency by developing a deep reinforcement learning approach solution for resource allocation. The authors of this work also propose multiple clustering schemes providing by the BS under the dynamic number of users assumption for \acrshort{miso} scheme. A novel Deep-Q network{-}based algorithm is introduced for the phase matrix configuration and power allocation. The proposed algorithm is superior to the benchmarks such as traditional RIS-assisted orthogonal multiple access (\acrshort{oma}) systems.
 
\begin{table*}
\caption{An overview of DL-based applications in RIS-aided NOMA studies.}
    \label{tab:RIS-aided NOMA}
\begin{tabular}{|p{1.3cm}|p{3cm}|p{12.388cm}|}
\hline
\cellcolor{table_blue} \textbf{\textcolor{table_white}{Reference}}&
\cellcolor{table_blue} \textbf{\textcolor{table_white}{Type of   Algorithm}}& 
\cellcolor{table_blue} \textbf{\textcolor{table_white}{Application Scnearios}}

\\ \hline

\cellcolor{table_white} \textcolor{table_blue}{\cite{gao2020resource} }                                                            & 
\cellcolor{table_white} \textcolor{table_blue}{Deep Reinforcement Learning-Deep Q}                                            & 
\cellcolor{table_white} \textcolor{table_blue}{A novel Deep-Q network-based algorithm for the phase matrix configuration and power allocation for a dynamic number of multiple users in RIS assisted downlink NOMA scheme}    
\\ \hline
\cellcolor{table_white} \textcolor{table_blue}{\cite{ni2021integrating} }                                                           & 
\cellcolor{table_white} \textcolor{table_blue}{Federated Learning}                    & 
\cellcolor{table_white} \textcolor{table_blue}{Over-the-air federated learning algorithm for resource allocation with RIS-assisted hybrid network}        
\\ \hline
\cellcolor{table_white} \textcolor{table_blue}{\cite{yang2021machine} }                                                         &
\cellcolor{table_white} \textcolor{table_blue}{Deep Reinforcement Learning-DDPG}     & 
\cellcolor{table_white} \textcolor{table_blue}{DDPG-based optimization algorithm capable to learn long-term policy for configuring phase shifting as well as user clustering under dynamic states}                
\\ \hline

\cellcolor{table_white} \textcolor{table_blue}{\cite{gao2021machine}}                                                            & 
\cellcolor{table_white} \textcolor{table_blue}{Deep Reinforcement Learning- Deep Q}                                            & 
\cellcolor{table_white} \textcolor{table_blue}{A novel deep Q-network algorithm adopted to regulate the power allocation process and \textbf{K}-means-based Gaussian mixture model for user clustering  }             
\\ \hline
\cellcolor{table_white} \textcolor{table_blue}{\cite{shehab2021deep}}                                                           & 
\cellcolor{table_white} \textcolor{table_blue}{Deep Reinforcement Learning}                    & 
\cellcolor{table_white} \textcolor{table_blue}{Deep reinforcement learning approach  involved in phase prediction and tuning, aiming the lowest possible training data load}                  

\\ \hline
\end{tabular}
\end{table*}

Thanks to the reward mechanism in Deep reinforcement learning,~\cite{yang2021machine} used a structure that can adapt to the dynamic number of users presented by~\cite{gao2020resource} to learn the channel behaviour.~\cite{yang2021machine} proposed a long-term stochastic optimization technique aiming to maximize the sum rate for multiple user NOMA downlink communication. They aimed at joint optimization of phase configuration on RIS side and user clustering. The proposed DDPG-based optimization is also capable of taking optimal action for dynamic states by various scenarios. This makes it possible to learn and apply long-term policy for configuring phase shifting as well as user clustering. The presented algorithm has optimized the sum data rate of mobile users, outperforming the results of traditional OMA technologies.
However, compared to conventional OMA solutions, the high system complexity and high training load created by NOMA schemes {are} an issue that needs to be addressed.~\cite{ni2021integrating} proposed the relaxation-then-quantization method that addresses the problem of optimizing the balance between system complexity and performance. This study combined the over-the-air federated learning algorithm with a RIS-assisted hybrid network and developed a flexible method that can adapt to different channel conditions. The presented study performs resource allocation by optimizing the transmitter power and phase configuration on the RIS side while meeting different QoS requirements. It is aimed to maximize the potential hybrid rate by performing these joint optimizations. 

High training overhead created by NOMA schemes is also addressed by~\cite{shehab2021deep} aiming to optimize total system capacity for the absence of CSI, using an approach similar to other downstream NOMA schemes. It is intended to learn the optimum phase configuration of the RIS, where the Deep reinforcement learning approach is involved in phase prediction and tuning, with the lowest possible training data load. In the same breath, as mentioned before in this survey, the performance degradation of imperfect SIC on the user data rate was studied. As a result of the parametric study, it was determined that the SIC imperfection rate grew ten times {causing} the decline of the system performance by 10


When the current RIS-aided NOMA literature is considered, it is evident that the existing studies are insufficient in terms of system complexity, energy efficiency and adaptation capability to different channel conditions required by 6G and beyond communication technologies. In particular, the high complexity of the systems and optimization problems has pushed the current practical studies to single antenna system assumption. Therefore, performance improvement of the RIS-assisted NOMA in MIMO systems emerges as a subject that needs to be addressed. In RIS-NOMA scenarios, imperfect SIC adoption is the main factor that degrades system performance and limits the number of users. Considering the current literature, it is evident that SIC techniques using conventional optimization algorithms are insufficient to obtain adequate performance. For this reason, novel grouping techniques have been developed and applied such as~\cite{zhangGaogrouping2021deep}. However, the development of advanced interference cancellation processes seem{s} as a bright future for next-generation NOMA architecture. Advanced DL approaches for SIC optimization will undoubtedly increase the performance of RIS-assisted NOMA systems.


\section{PHY Security Meets DL}   \label{sec:security}

In this section, we present an overview of \acrshort{dl}-based PHY security studies. For this purpose, firstly, the PHY security concept will be briefly introduced and \acrshort{dl}-based PHY security methods will be categorized for their attack type: spoofing, jamming, and eavesdropping. In each subsection, our investigation results will be presented with explanations and comparisons. 

Wireless communication is highly vulnerable to malicious attacks at \acrshort{phy} as a result of the open nature of the wireless channel. Since the channel is accessible by any party, an adversary can easily interfere with legitimate communication at \acrshort{phy}. 
Traditionally, a countermeasure for PHY attacks is designed at upper layers, e.g., network or transport layers. However, traditional methods are often resource-inefficient or might have limited protection in 6G networks or emerging \acrshort{iot} applications. Therefore, providing effective countermeasures against PHY attacks is an open problem. 

In the past years, many researchers focused on PHY security methods to tackle this problem. These methods are essentially based on exploiting \acrshort{phy} attributes for security purposes. 
Although PHY security shows great {promise}, it also holds significant challenges for practical implementations. With the emergence of \acrshort{dl} networks, researchers sought \acrshort{dl}-based PHY solutions to overcome existing challenges or further improve security performance. Readers are referred to Table~\ref{tab:secrelatedsurveys} for a list of related survey papers in this field.


\begin{figure} 
    \centering  
  \subfloat[\label{fig:spoofing}]{ 
       \includegraphics[width=0.6\linewidth]{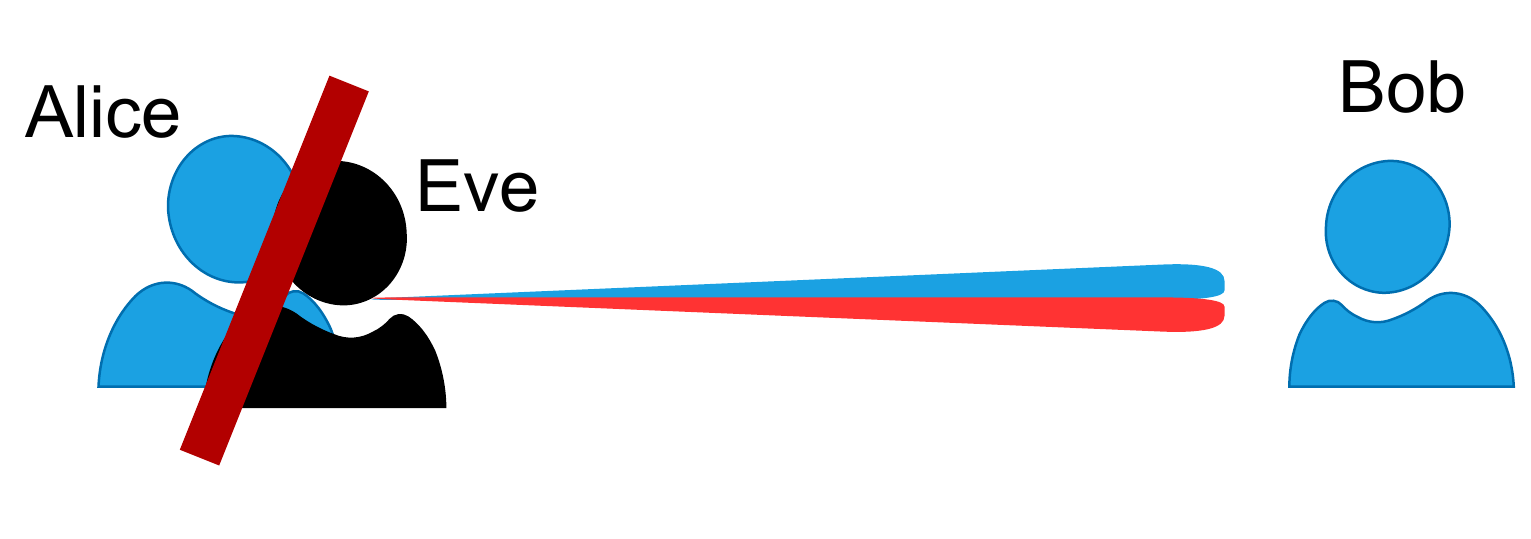}}
     \hfill 
  \subfloat[\label{fig:jamming}]{%
   \includegraphics[width=0.55\linewidth]{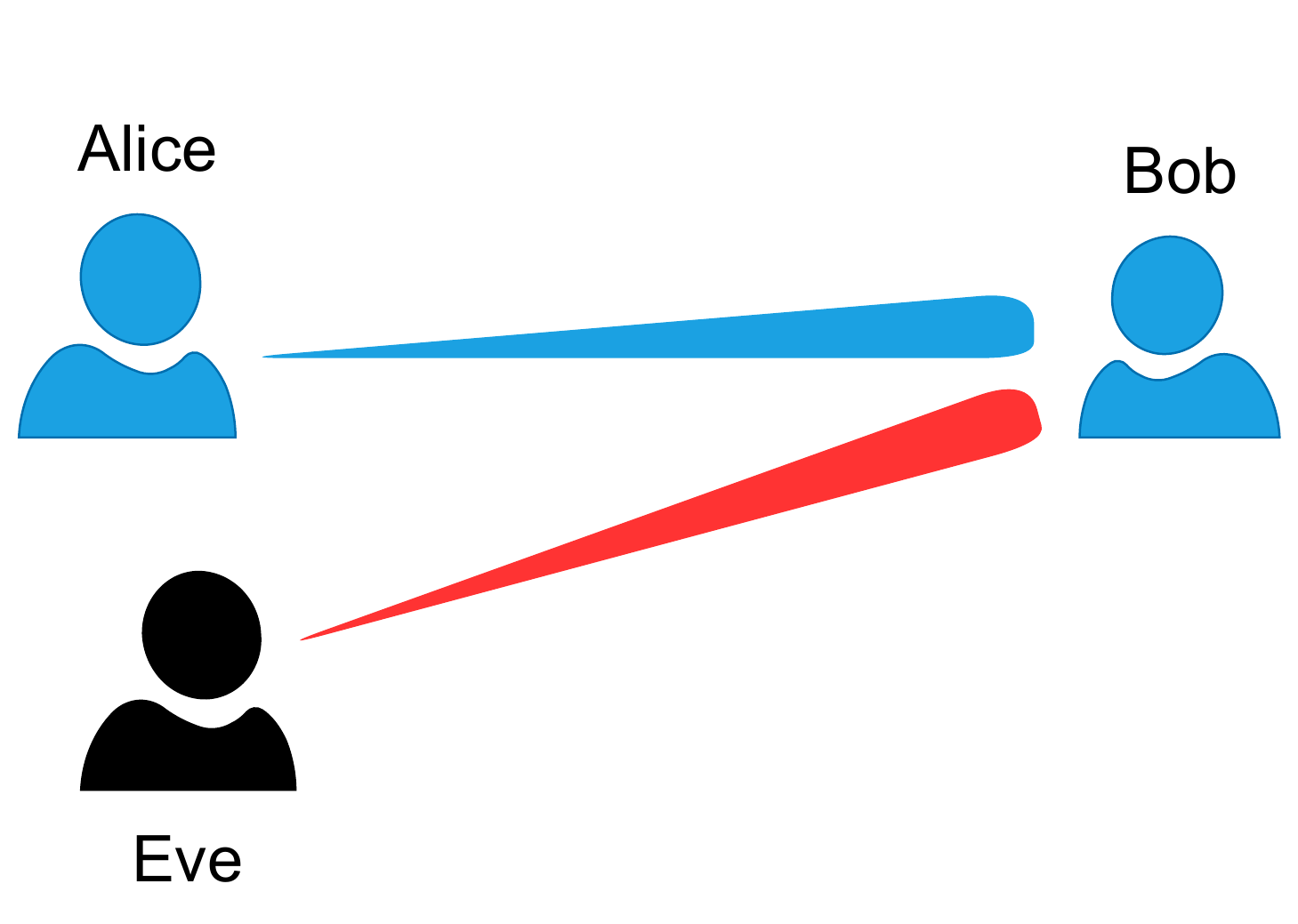}}
    \hfill 
  \subfloat[\label{fig:eavesdropping}]{%
        \includegraphics[width=0.55\linewidth]{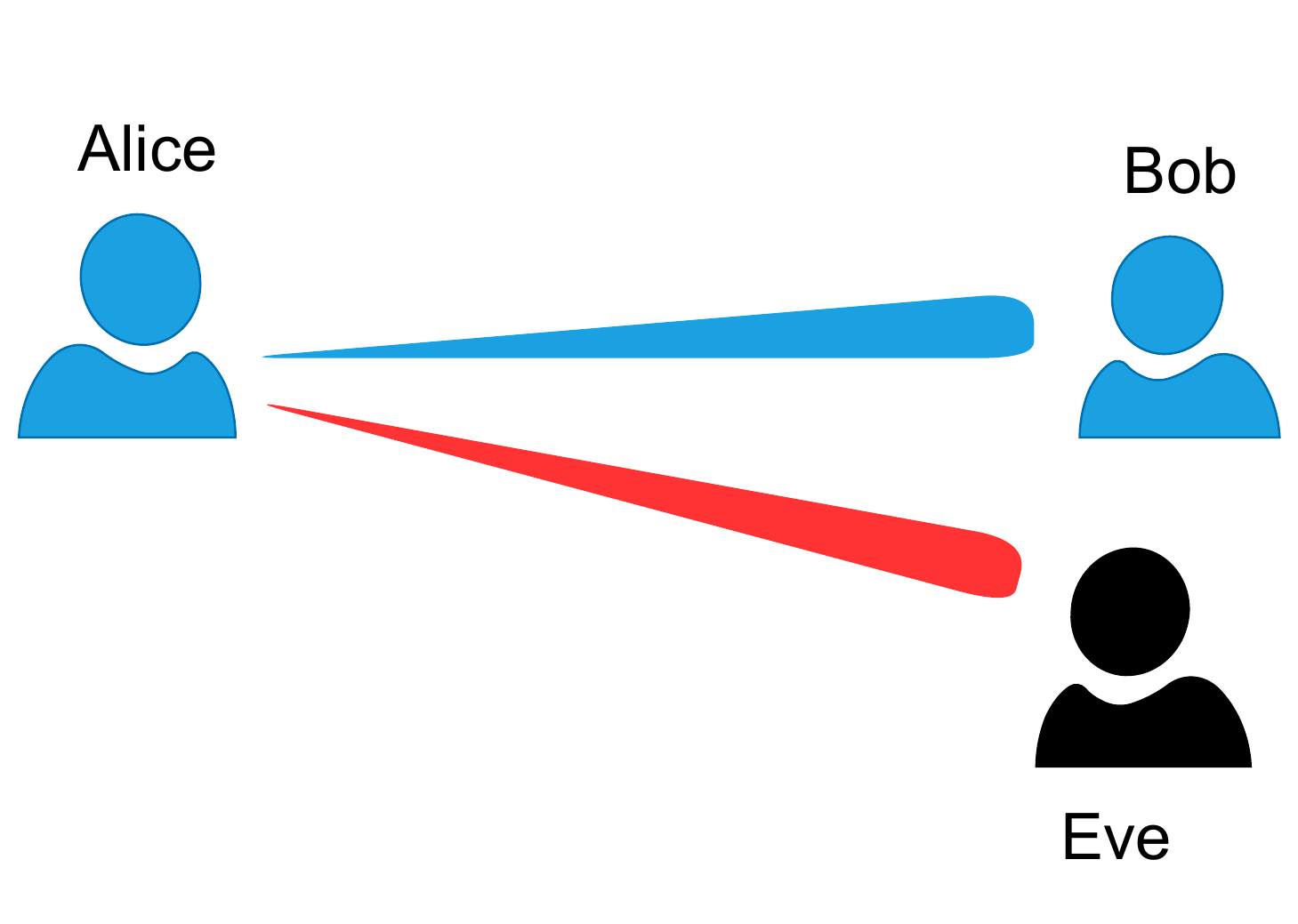}}
      \caption{Three different PHY attack types: (a) Spoofing: Eve aims to impersonate Alice. (b) Jamming: Eve aims to block the communication between Alice and Bob. (c) Eavesdropping: Eve aims to obtain the messages that are sent from Alice.}
  \label{fig:attacktypes} 
\end{figure}

\begin{table}
    \caption{Existing survey papers related to \acrshort{dl}-based security or \acrshort{phy} security.}

\centering
\begin{tabular}{|M{1.3cm}|M{0.5cm}|M{5cm}|}


\hline
 \cellcolor{table_blue} \textbf{\textcolor{table_white}{Survey}} &  \cellcolor{table_blue} \textbf{\textcolor{table_white}{Year}} &  \cellcolor{table_blue} \textbf{\textcolor{table_white}{Title}}  \\ \hline
\cellcolor{table_white} \textcolor{table_blue}{OShea \textit{et al.}~\cite{Burak_18_2017_AnIntroDLPHY}}  & \cellcolor{table_white} \textcolor{table_blue}{2017}         
& \cellcolor{table_white} \textcolor{table_blue}{An introduction to deep learning for the physical layer}       \\ \hline

\cellcolor{table_white} \textcolor{table_blue}{Wu \textit{et al.}~\cite{Wu2018}}  & \cellcolor{table_white} \textcolor{table_blue}{2018}         
& \cellcolor{table_white} \textcolor{table_blue}{A survey of physical layer security techniques for 5G wireless networks and challenges ahead}       \\ \hline

\cellcolor{table_white} \textcolor{table_blue}{Suomalainen \textit{et al.}~\cite{Suomalainen2020}}  & \cellcolor{table_white} \textcolor{table_blue}{2020}         
& \cellcolor{table_white} \textcolor{table_blue}{\acrshort{ml} threatens 5G security}       \\ \hline

\cellcolor{table_white} \textcolor{table_blue}{Benzaid \textit{et al.}~\cite{Benzaid2020}}  & \cellcolor{table_white} \textcolor{table_blue}{2020}         
& \cellcolor{table_white} \textcolor{table_blue}{\acrshort{ai} for beyond 5G Networks: A cyber-security defense or offense enabler?}       \\ \hline

\cellcolor{table_white} \textcolor{table_blue}{Jiang \textit{et al.}~\cite{Jiang2020}}  & \cellcolor{table_white} \textcolor{table_blue}{2020}         
& \cellcolor{table_white} \textcolor{table_blue}{Short survey on physical layer authentication by machine-learning for 5G-based internet of things}       \\ \hline

\cellcolor{table_white} \textcolor{table_blue}{Nguyen \textit{et al.}~\cite{Nguyen2021}}  & \cellcolor{table_white} \textcolor{table_blue}{2021}         
& \cellcolor{table_white} \textcolor{table_blue}{Enabling AI in future wireless networks: A data life cycle perspective}       \\ \hline

\cellcolor{table_white} \textcolor{table_blue}{Wang \textit{et al.}~\cite{Wang2020}}  & \cellcolor{table_white} \textcolor{table_blue}{2020}         
& \cellcolor{table_white} \textcolor{table_blue}{Physical layer authentication for 5G communications: Opportunities and road ahead}       \\ \hline

\cellcolor{table_white} \textcolor{table_blue}{Al-garadi \textit{et al.}~\cite{Al-garadi2020}}  & \cellcolor{table_white} \textcolor{table_blue}{2020}         
& \cellcolor{table_white} \textcolor{table_blue}{A survey of machine and deep learning methods for internet of things (IoT) security}       \\ \hline

\end{tabular}
    \label{tab:secrelatedsurveys}
\end{table}

We believe that an effective categorization for \acrshort{dl}-based security studies would be based on the attack types. The most common three attack types (spoofing, jamming, and eavesdropping) are illustrated in Fig.~\ref{fig:attacktypes} can be defined as follows.

\begin{itemize}
    \item \textit{Spoofing:} In spoofing attacks, the adversary (Eve) aims to imitate the identity of the legitimate transmitter (Alice) such that the receiver (Bob) thinks that the message is sent from Alice.
    \item \textit{Jamming:} In jamming attacks, Eve aims to block legitimate communication by introducing jamming signals to the channel such that Bob obtains highly distorted signals.
    \item \textit{Eavesdropping:} In eavesdropping attacks, Eve seeks to obtain the message sent from Alice by simply eavesdropping the channel.
\end{itemize}

   In the rest of this section, \acrshort{dl}-based PHY solutions against these three fundamental attacks are presented with comprehensive comparisons.


\subsection{Anti-Spoofing Solutions}


\begin{table*}
    \caption{An overview of DL-based anti-spoofing literature (Red hues indicate no USRP/Real-world data and green hues indicate USRP/Real-world data).}
    \centering
    \begin{tabular}{|M{2cm}|M{1.4cm}|M{1.4cm}|M{3cm}|M{4.1cm}|M{3cm}|}
        \hline
        
        \cellcolor{table_blue} \textbf{\textcolor{table_white}{Paper}} & 
        \cellcolor{table_blue} \textbf{\textcolor{table_white}{DL Structure}} & 
        \cellcolor{table_blue} \textbf{\textcolor{table_white}{USRP/ Real-world data}} & 
        \cellcolor{table_blue} \textbf{\textcolor{table_white}{{Model Input}}} & 
        \cellcolor{table_blue} \textbf{\textcolor{table_white}{Pros}} & 
        \cellcolor{table_blue} \textbf{\textcolor{table_white}{Cons}} \\
        \hline

        \cellcolor{table_white} \textcolor{table_blue}{Liao \textit{et al.}~\cite{Liao2019}} & 
        \cellcolor{table_white} \textcolor{table_blue}{\acrshort{cnn}} &
        \cellcolor{table_green} &
        \cellcolor{table_white} \textcolor{table_blue}{{CSI vector}} & 
        \cellcolor{table_white} \textcolor{table_blue}{MU, comprehensive analysis, good results with small dataset} &
        \cellcolor{table_white} \textcolor{table_blue}{Require training after coherence time, CSI requirement} \\
        \hline
        
        \cellcolor{table_white} \textcolor{table_blue}{Qiu \textit{et al.}~\cite{Qiu2019a}} & 
        \cellcolor{table_white} \textcolor{table_blue}{\acrshort{cnn}} &
        \cellcolor{table_red} &
        \cellcolor{table_white} \textcolor{table_blue}{{Channel estimation matrix}} &
        \cellcolor{table_white} \textcolor{table_blue}{MU, feature extraction} &
        \cellcolor{table_white} \textcolor{table_blue}{CSI requirement} \\
        \hline
        
        \cellcolor{table_white} \textcolor{table_blue}{Liao \textit{et al.}~\cite{Liao2019a}} & 
        \cellcolor{table_white} \textcolor{table_blue}{\acrshort{dnn} \acrshort{cnn} PCNN} &
        \cellcolor{table_green} &
        \cellcolor{table_white} \textcolor{table_blue}{{CSI matrix}} &
        \cellcolor{table_white} \textcolor{table_blue}{MU, comprehensive analysis} &
        \cellcolor{table_white} \textcolor{table_blue}{CSI requirement} \\
        \hline
        
        \cellcolor{table_white} \textcolor{table_blue}{Wang \textit{et al.}~\cite{Wang2019}} & 
        \cellcolor{table_white} \textcolor{table_blue}{\acrshort{cnn} \acrshort{rnn} CRNN} &
        \cellcolor{table_red} &
        \cellcolor{table_white} \textcolor{table_blue}{{CSI vector}} &
        \cellcolor{table_white} \textcolor{table_blue}{Can capture spectral dependencies, comprehensive analysis} &
        \cellcolor{table_white} \textcolor{table_blue}{Single-transmitter, CSI requirement} \\
        \hline
        
        \cellcolor{table_white} \textcolor{table_blue}{Qiu \textit{et al.}~\cite{Qiu2020}} & 
        \cellcolor{table_white} \textcolor{table_blue}{\acrshort{cnn}} &
        \cellcolor{table_green} &
        \cellcolor{table_white} \textcolor{table_blue}{{Augmented data-adaptive matrix (formed from CSI)}} &
        \cellcolor{table_white} \textcolor{table_blue}{Non-line-of-sight investigation} &
        \cellcolor{table_white} \textcolor{table_blue}{Single-transmitter, CSI requirement} \\
        \hline
        
        \cellcolor{table_white} \textcolor{table_blue}{Li \textit{et al.}~\cite{Li2021}} & 
        \cellcolor{table_white} \textcolor{table_blue}{Feed-forward NN} &
        \cellcolor{table_green} &
        \cellcolor{table_white} \textcolor{table_blue}{{Virtual channel vector}} &
        \cellcolor{table_white} \textcolor{table_blue}{mmWave MIMO, feature selection, channel correlation investigation} &
        \cellcolor{table_white} \textcolor{table_blue}{Single-transmitter, CSI requirement} \\
        \hline
        
        \cellcolor{table_white} \textcolor{table_blue}{Liao \textit{et al.}~\cite{Liao2020}} & 
        \cellcolor{table_white} \textcolor{table_blue}{\acrshort{dnn}} &
        \cellcolor{table_green} &
        \cellcolor{table_white} \textcolor{table_blue}{{Channel response vector}} &
        \cellcolor{table_white} \textcolor{table_blue}{Data augmentation, good results with extremely small dataset} &
        \cellcolor{table_white} \textcolor{table_blue}{CSI requirement} \\
        \hline
        
        
    \end{tabular}
    \label{tab:spoofing}
\end{table*}

In traditional communication systems, authentication algorithms are widely employed as a countermeasure against spoofing attacks. The authentication algorithm relies on the fact that Alice owns a prior information that indicates her authenticity. {On} the receiving side, Bob uses an algorithm (a detection/classification process) to check if Alice has this information. Although traditional methods are very effective at classifying non-authorized users, delivering prior information to Alice securely or keeping it secure is a highly vulnerable step. In PHY security methods, the role of prior information is transferred to channel or device characteristics {that} are unique to {the} location, time{,} or device. As a result, PHY authentication methods are able to catch attackers even {if} Alice's credentials are leaked. However, selecting practical features from the channel and designing the detection process is a challenging task. Recently, \acrshort{dl} networks are implemented to overcome these challenges and improve detection accuracy. A list of the \acrshort{dl}-based PHY authentication methods is given in Table~\ref{tab:spoofing} with their pros and cons.

\acrshort{ml}-aided PHY authentication is first introduced by Xiao \textit{et al.} in~\cite{Xiao2015}. The study considers a common PHY authentication scenario where the receiver applies a binary hypothesis test based on the received signal strength indicator (\acrshort{rssi}) values. The decision mechanism first calculates the test statistics, i.e., the normalized distance between the received \acrshort{rssi} value and a reference one. Secondly, test statistics are compared with a threshold to decide if the signal is spoofed or not. Here, the selection of the threshold is a critical step that directly affects the detection performance. As the main contribution, the authors propose a  Q-learning-based threshold selection algorithm. The algorithm basically searches the action and state planes to find the threshold that maximizes the Q-function\footnote{Q-function calculates the expected reward of an action in Q-learning algorithms. Q-learning or other reinforcement learning algorithms optimize their objective by maximizing a reward value. The reward values are calculated by trial-and-error since the system is unsupervised (i.e., labeled data does not exist). In larger problems, Q-function is often approximated with \acrshort{dl} algorithms.}. The authors also investigate the performance of the proposed algorithm using \acrshort{usrp} modules. Test results indicate that a dynamic threshold selection with Q-learning outperforms fixed threshold selection. In~\cite{Xiao2016}, the authors extend their previous work in~\cite{Xiao2015} to include Dyna-Q learning method. In addition to the results of~\cite{Xiao2015}, experiments reveal that Dyna-Q model can improve both learning speed and authentication performance compared to Q-learning. The authors also consider the \acrshort{mimo} scenario in~\cite{Xiao2017} by extending the algorithms given in~\cite{Xiao2016}. The results show that both using reinforcement learning and increasing {the} number of antennas improve the detection performance.

The most significant drawback of~\cite{Xiao2015,Xiao2016,Xiao2017} is that the receiver is assumed to know reference \acrshort{rssi} values for the detection mechanism. This assumption is dropped in~\cite{Xiao2018} by generating a decision result with logistic regression. Specifically, the receiver uses previous RSSI information to train a logistic regression model and estimates the authenticity of the current RSSI value. As a result, the proposed method is able to {detect spoofers successfully} without prior RSSI information, even under dynamic and unknown channel models. On the other hand, the study considers a network where the receiver obtains RSSI information from multiple spatially distributed nodes (landmarks) with multiple antennas to improve detection performance. In~\cite{Xiao2018a}, the authors extend their work by reducing the computational overhead of the regression model. \acrshort{usrp} experiments verify that the logistic regression-based decision mechanism can successfully detect spoofers under unknown channel models.

\acrshort{dl} is introduced to PHY authentication by the authors of~\cite{Liao2019, Qiu2019a, Liao2019a, Wang2019} simultaneously and independently in 2019. Also, another independent study~\cite{Qiu2020} {was} published in 2020. In~\cite{Liao2019}, a MU mobile edge network where a node aims to authenticate packets coming from multiple users is considered. The authentication system exploits past \acrshort{csi} values that are obtained with pilot transmission to discriminate unauthorized nodes. Instead of applying a traditional hypothesis test, the decision task is given to a \acrshort{cnn}. As a challenging task, the receiver requires that the authentication of a node should be within the coherence time of the last \acrshort{csi} that is used in the training process. When the coherence window expires, the receiver updates its \acrshort{cnn} parameters with a new training process. The authors present a comprehensive investigation of the proposed method where various algorithms are tested for their training speed. Simulation results show successful detection rates even with very small training data sizes. Also, simulation results are verified with \acrshort{usrp} modules in a real-world test environment.  

Similar to~\cite{Liao2019}, the authors of~\cite{Qiu2019a} and~\cite{Liao2019a} consider a PHY authentication method where the receiver node obtains packets coming from multiple nodes. These studies are mainly based on the same idea and differ from each other in terms of \acrshort{cnn} structure and performance analysis. In~\cite{Qiu2019a}, staying in a coherence window is not considered. Instead, the study focuses on the impact of the PHY attributes that are used to extract features. Also, the study includes \acrshort{snr} investigation and lacks practical implementation. In~\cite{Liao2019a}, the authors implement a \acrshort{dnn}, a \acrshort{cnn} and a preprocessing added \acrshort{cnn} model. The study gives a complete investigation of these three models in terms of computational complexity, training time, transmitter number, hidden layer number, antenna number, and authentication accuracy. Although \acrshort{dnn} model gives the best authentication rates, the proposed preprocessing added \acrshort{cnn} model achieves close results with less computational complexity. Also, the study presents \acrshort{usrp} implementations to verify the practical feasibility of the proposed models.

The authors of~\cite{Wang2019} consider a three-node network that includes a transmitter, a receiver, and an eavesdropper. Similar to~\cite{Liao2019} and~\cite{Qiu2019a}, this study exploits wireless channel features as input of a \acrshort{dnn} in order to identify spoofers. This study distinguishes from~\cite{Liao2019} and~\cite{Qiu2019a} by thoroughly implementing and investigating \acrshort{cnn}, \acrshort{rnn} and convolutional RNN models. The \acrshort{rnn} model is shown to be able to capture the spectral dependencies and improve detection performance. Moreover, the convolutional RNN model can further improve the detection performance by cascading \acrshort{cnn} and \acrshort{rnn} models. This work also proves the superiority of \acrshort{dl} on PHY authentication with numerical comparisons to traditional Neyman-Pearson hypothesis testing. 

In~\cite{Qiu2020}, another \acrshort{dl}-based PHY authentication method is proposed. Similar to~\cite{Wang2019}, this model consists of a single legitimate transmitter and exploits channel characteristics. Also, this study considers adding newly obtained \acrshort{csi} data to the training dataset as used in~\cite{Wang2019}. In simulations, this adaptive approach is compared with a non-adaptive scenario (i.e., new data is not added to the training set). The results prove the superiority of the adaptive approach. This work differs from~\cite{Wang2019} by its \acrshort{dnn} structure and its investigation of non-line-of-sight scenarios. 

In~\cite{Li2021}, the authors consider a mmWave \acrshort{mimo} scenario where channel characteristics are highly sensitive to spatial location. For this purpose, the authors propose a novel channel feature to improve the detection accuracy of mmWave \acrshort{mimo} networks. The proposed channel feature is used in two spoofing detection mechanisms: i) traditional Neyman-Pearson hypothesis testing, ii) feed-forward NN. Hypothesis testing is considered for static channels that have constant channel correlation. The feed-forward NN is considered for dynamic channels with varying channel correlations. It has been shown that the proposed new channel feature can improve detection performance for both static and dynamic channel models. This work differs from previous \acrshort{dl}-based approaches by its novel channel feature and its focus on the channel correlation level.  

Although we can not compare the models of~\cite{Liao2019, Qiu2019a, Liao2019a, Wang2019, Qiu2020, Li2021} with the same system parameters, the authors of these works state that their methods can achieve $97\%$ detection accuracy rates at their peak training time.

One of the biggest drawbacks of \acrshort{dl}-based models is the requirement of large datasets. It is a known fact that the performance of \acrshort{dl} models can significantly drop with limited data. In order to overcome this problem, data augmentation methods have been proposed in \acrshort{dl} literature. Data augmentation techniques are mainly based on generating artificial data from the existing dataset. The limited data problem of PHY security \acrshort{dl} networks has been considered by Liao \textit{et al.} in~\cite{Liao2020}. Liao \textit{et al.} previously worked on \acrshort{dl}-assisted PHY authentication in~\cite{Liao2019a}. In~\cite{Liao2020}, the authors argue that wireless networks are especially vulnerable to limited data as a result of channel coherence time. Hence, the authors propose three data augmentation methods to improve the training speed and authentication accuracy. Also, the authors perform experiments on real-world datasets to verify their results.

The majority of studies in the literature consider the application of \acrshort{dl} at the legitimate side (i.e., to improve security). In~\cite{Nooraiepour2021}, the authors take the opposite view and investigate \acrshort{dl}-assisted spoofing attacks. In particular, this study focuses on \acrshort{ofdm} systems that are deemed secure against spoofing attacks. Simulation results show that data-driven spoofing attacks are able to effectively disrupt \acrshort{ofdm} systems that are traditionally considered secure.

DL-based anti-spoofing methods are proved to be highly beneficial when compared to traditional threshold-based detection mechanisms. DL models learn the best decision mechanism from data when traditional methods suffer from the problem of threshold selection. An inaccurate threshold selection in a mathematical detection model can easily lead to high miss detection or false alarm rates. Moreover, DL-based methods in the literature are able to deal with large networks where authentication of multiple nodes is required. Creating a DL model that classifies multiple nodes simultaneously is possible. On the other hand, the greatest challenge of DL-based anti-spoofing methods is to obtain information on the spoofer. DL-based methods require training data to generate a model that can distinguish between a spoofer and a legitimate user. However, including any information from {the} spoofer to the training data is not practical. Unfortunately{,} current studies {include} the spoofer's channel in the training process to train their models. This assumption is a critical drawback on implementing DL-based methods in real scenarios.

\subsection{Anti-Jamming Solutions}

\begin{table*}
    \caption{An overview of DL-based anti-jamming literature.}
    \centering
    \begin{tabular}{|M{1.8cm}|M{2cm}|M{3cm}|M{4.8cm}|M{3.6cm}|}
        \hline
        
        \cellcolor{table_blue} \textbf{\textcolor{table_white}{Paper}} & 
        \cellcolor{table_blue} \textbf{\textcolor{table_white}{DL Structure}} & 
        \cellcolor{table_blue}
        \textbf{\textcolor{table_white}{{Model Input}}} & 
        \cellcolor{table_blue} \textbf{\textcolor{table_white}{Pros}} & 
        \cellcolor{table_blue} \textbf{\textcolor{table_white}{Cons}} \\
        \hline

        
        \cellcolor{table_white} \textcolor{table_blue}{Han \textit{et al.}~\cite{Han2017}} &         \cellcolor{table_white} \textcolor{table_blue}{\acrshort{cnn}-based DQN} &
        \cellcolor{table_white} \textcolor{table_blue}{{State matrix}} &
        \cellcolor{table_white} \textcolor{table_blue}{Applicable to any channel model} &
        \cellcolor{table_white} \textcolor{table_blue}{Limited jammer model, no real-world data} \\
        \hline
        
        \cellcolor{table_white} \textcolor{table_blue}{Liu \textit{et al.}~\cite{Liu2018}} &         \cellcolor{table_white} \textcolor{table_blue}{Recursive \acrshort{cnn}-based \acrshort{dqn}} &
        \cellcolor{table_white} \textcolor{table_blue}{{Spectrum sensing matrix}} &
        \cellcolor{table_white} \textcolor{table_blue}{Improved jammer model, raw data as feature, cost of frequency hops are included} &
        \cellcolor{table_white} \textcolor{table_blue}{More training time as a result of raw data, no real-world data} \\
        \hline
        
        \cellcolor{table_white} \textcolor{table_blue}{Bi \textit{et al.}~\cite{Bi2019}} &         \cellcolor{table_white} \textcolor{table_blue}{CNN and LSTM-based Double DQN} &
        \cellcolor{table_white} \textcolor{table_blue}{{User ID, position, channel information}} &
        \cellcolor{table_white} \textcolor{table_blue}{More stable than DQN} &
        \cellcolor{table_white} \textcolor{table_blue}{No real-world data} \\
        \hline
        
        \cellcolor{table_white} \textcolor{table_blue}{Xu \textit{et al.}~\cite{Xu2020}} & 
        \cellcolor{table_white} \textcolor{table_blue}{Transformer encoder-based Double DQN} &
        \cellcolor{table_white} \textcolor{table_blue}{{Spectrum sensing matrix}} &
        \cellcolor{table_white} \textcolor{table_blue}{More stable than DQN, improved throughput} &
        \cellcolor{table_white} \textcolor{table_blue}{No real-world data} \\
        \hline
        
        
        \cellcolor{table_white} \textcolor{table_blue}{Tingpeng \textit{et al.}~\cite{Tingpeng2018}} & 
        \cellcolor{table_white} \textcolor{table_blue}{\acrshort{cnn}} &
        \cellcolor{table_white} \textcolor{table_blue}{{Signal matrix from electronic information system}} &
        \cellcolor{table_white} \textcolor{table_blue}{Can identify three jammer types} &
        \cellcolor{table_white} \textcolor{table_blue}{Limited analysis, no real-world data} \\
        \hline
        
        \cellcolor{table_white} \textcolor{table_blue}{Cai \textit{et al.}~\cite{Cai2019}} & 
        \cellcolor{table_white} \textcolor{table_blue}{Le-Net5 \acrshort{cnn}} &
        \cellcolor{table_white} \textcolor{table_blue}{{Spectrum sensing matrix}} &
        \cellcolor{table_white} \textcolor{table_blue}{Can identify multi-tone jammers} &
        \cellcolor{table_white} \textcolor{table_blue}{Limited analysis, ideal jammer model, no real-world data} \\
\hline

        \cellcolor{table_white} \textcolor{table_blue}{Liu \textit{et al.}~\cite{Liu2019}} & 
        \cellcolor{table_white} \textcolor{table_blue}{\acrshort{cnn}} &
        \cellcolor{table_white} \textcolor{table_blue}{{Spectrum sensing matrix}} &
        \cellcolor{table_white} \textcolor{table_blue}{Both jammer identification and frequency hopping strategies are considered, cost of frequency hops are included, comprehensive numerical analysis} &
        \cellcolor{table_white} \textcolor{table_blue}{No real-world data} \\
\hline
        
    \end{tabular}
    \label{tab:jamming}
\end{table*}

Jamming attacks can dramatically disrupt the communication between Alice and Bob since Alice's signals easily become unrecognizable when superimposed with the jamming signal. Traditionally, spread spectrum techniques are implemented to combat jammers. Frequency hopping schemes aim to dodge jammers by continually changing the communication frequency. Direct sequence spread spectrum methods spread the communication to a larger bandwidth to reduce the impact of the jamming signal. Designing an efficient frequency hopping pattern is an open issue and {is} considered by many researchers. In the last decade, various efforts have been made to leave the design step to a \acrshort{dl} network. Since continuously dodging a jammer over a spectrum is a sequential game where Alice makes its hops depending on new observation{s}, reinforcement learning is a powerful candidate to improve performance. Table~\ref{tab:jamming} contains several \acrshort{dl}-based anti-jamming studies that exist in the literature with their pros and cons.

One of the first examples of learning-based anti-jamming approaches can be seen in~\cite{Gwon2013}. The study considers a wireless network that consists of two competing teams. Each team consists of a jammer and a receiver node, where each jammer node tries to interrupt the communication of the opposing team's receiver while each receiver node aims to avoid the jammed frequencies. The authors propose various Q-learning approaches to select the frequency hops of a receiver node and the target frequencies of a jammer node. The proposed approach assumes that each node has spectrum sensing capabilities and each team has control channels for receiver-friendly jammer communication. The authors investigate the performance of Q-learning methods against conventional methods and each other.

The accuracy of the Q-learning-based detection results can be insufficient in high{-}dimensional data, and the learning speed can increase dramatically. In order to capture non-linear relations in the data and increase both learning speed and accuracy, \acrshort{dl}-based jamming detection models have been proposed in~\cite{Han2017, Liu2018, Bi2019}, and~\cite{Xu2020}. These studies essentially focus on \acrshort{dl} techniques to find an efficient frequency hopping strategy. In~\cite{Han2017}, a cognitive radio network is considered where the secondary user applies {a} deep Q-learning network (\acrshort{dqn}) to avoid jammed frequencies while not interfering with the primary users. The proposed model uses Q-learning to decide on the jammed frequencies and select the appropriate frequency band. The Q-function is calculated with a \acrshort{cnn} to improve accuracy and time complexity. Q-learning states are calculated based on the received SNIR values (features extracted to avoid jammed frequencies) and the primary user occupation information. The proposed method does not require knowledge of the jammer or channel model. As a result, it is applicable to any environment. Computer simulation results show that \acrshort{cnn}-based Q-learning yields higher learning rates than traditional Q-learning methods. Also, the receiver attains higher SNIR values. 

A similar anti-jamming approach is considered in~\cite{Liu2018}. Contrary to~\cite{Han2017}, the method does not assume that the jammer follows the same transmission slot structure of the legitimate communication. Also, this method uses raw spectrum information to feed \acrshort{cnn} instead of using extracted features. In~\cite{Bi2019}, Q-function is approximated with double deep Q-learning.~\cite{Bi2019} can be distinguished from~\cite{Han2017, Liu2018} by their double \acrshort{dqn} choice instead of \acrshort{dqn}. Although double \acrshort{dqn} is shown to be more stable than \acrshort{dqn} in the literature,~\cite{Bi2019} only compares their method with traditional Q-learning and lacks any comparison with \acrshort{dqn}. As a strong side, the authors implement three types of networks: fully connected network, \acrshort{cnn} and long{-}short{-}term{-}memory. The authors of~\cite{Xu2020} improve the results of~\cite{Bi2019} by implementing double \acrshort{dqn} with a more efficient network model. Specifically, the authors exploit transformer encoder to implement double \acrshort{dqn} and obtain improved results compared to \acrshort{cnn}-based double \acrshort{dqn}. Unlike previous studies, in~\cite{Qiu2019}, \acrshort{dl} is used for receiver design. Specifically, a \acrshort{dl} algorithm is implemented at the receiver of a continuous phase modulation scheme in order to improve the prediction of the received signals under single or multi-tone jamming attacks. Numerical results showed that \acrshort{dl} assisted receiver can improve \acrshort{ber} performance of the system by $3-5$ dB under single-tone jamming attacks.

Detection of the jammer type is an important step for anti-jamming communications. Instead of focusing on avoiding the jammed frequencies (which primarily considers Q-learning), the detection of jammer type focuses on identifying jammer features. Conventional model-based methods first convert raw information into features using mathematical models and solve a classification problem to identify jammer type{s}. Applying \acrshort{dl} in this process can significantly overcome the inaccuracies of mathematical models. In~\cite{Tingpeng2018}, the authors proposed a \acrshort{cnn}-based jammer identification method which imitates the image processing applications of \acrshort{dl}. This method {first} converts one-dimensional data into image format in order to train the network. Although simulation results indicate $92\%$ accuracy for identifying three jammer types (single-tone jammers that use Gaussian noise with different mean and variance), this study lacks comparison with any benchmark model. An improved detection model is proposed in~\cite{Cai2019} where a simplified Le-Net5 \acrshort{cnn} model is used. In the performance evaluation, the authors include multi-tone jammers. The method achieves $92\%$ accuracy (which is similar to~\cite{Tingpeng2018}), yet the authors remark that the method attains a higher learning speed. As major drawbacks, this study lacks comparison with a benchmark model and assumes an ideal jamming pattern, i.e., the jammer can instantaneously shift frequencies.

The above studies mainly focus on either avoiding the jammer with frequency hopping or identifying the jammer type. In~\cite{Liu2019}, the authors focus on both concepts by firstly identifying the jammer type and then proposing an efficient frequency hopping pattern. The proposed method firstly sweeps the time domain and captures multiple frames of spectral energy (i.e., two-dimensional spectral energy information in a time interval). The frames are fed into a \acrshort{cnn} to identify the jammer type. In the next step, traditional Q-learning is used to provide a frequency hopping pattern. As an important contribution, the authors also consider the cost of frequency hops {in} the system model. 

An overview of the \acrshort{dl}-based anti-jamming literature is presented in Table~\ref{tab:jamming}. These studies mainly focus on improving the reinforcement learning model with \acrshort{dl} assistance. Based on our investigations, we conclude that \acrshort{dl} can improve the frequency hopping accuracy of anti-jamming systems and bring robustness against unbalanced environments. Although most of these studies include extensive computer analyses, {the} literature heavily lacks proof of concepts or testbed implementations with real-world datasets.

\subsection{Anti-Eavesdropping Solutions}

\begin{table*}
    \caption{An overview of DL-based anti-eavesdropping literature.}
    \centering
    \begin{tabular}{|M{2.3cm}|M{2cm}|M{2.5cm}|M{5.3cm}|M{3.3cm}|}
        \hline
        \cellcolor{table_blue} \textbf{\textcolor{table_white}{Paper}} & 
        \cellcolor{table_blue} \textbf{\textcolor{table_white}{DL Structure}} & 
        \cellcolor{table_blue} \textbf{\textcolor{table_white}{{Model Input}}} & 
        \cellcolor{table_blue} \textbf{\textcolor{table_white}{Pros}} & 
        \cellcolor{table_blue} \textbf{\textcolor{table_white}{Cons}} \\
        \hline

        \cellcolor{table_white} \textcolor{table_blue}{Fritschek \textit{et al.}~\cite{Fritschek2019}} & 
        \cellcolor{table_white} \textcolor{table_blue}{\acrshort{ae}} &
        \cellcolor{table_white} \textcolor{table_blue}{{One-hot-encoded message vector}} &
        \cellcolor{table_white} \textcolor{table_blue}{Provides a trade-off between secrecy and \acrshort{ber}, Eve has \acrshort{dl} abilities} &
        \cellcolor{table_white} \textcolor{table_blue}{Bob has a better channel than Eve, signal alphabet is infinite, no real-world data} \\
        \hline
        
        \cellcolor{table_white} \textcolor{table_blue}{Sun \textit{et al.}~\cite{Sun2020}} & 
        \cellcolor{table_white} \textcolor{table_blue}{\acrshort{cnn}-based \acrshort{ae}} &
        \cellcolor{table_white} \textcolor{table_blue}{{One-hot-encoded message vector}} &
        \cellcolor{table_white} \textcolor{table_blue}{Include an authentication model} &
        \cellcolor{table_white} \textcolor{table_blue}{Limited analysis, no real-world data} \\
        \hline
        
        \cellcolor{table_white} \textcolor{table_blue}{Besser \textit{et al.}~\cite{Besser2020}} & 
        \cellcolor{table_white} \textcolor{table_blue}{\acrshort{ae}} &
        \cellcolor{table_white} \textcolor{table_blue}{{Message vector}} &
        \cellcolor{table_white} \textcolor{table_blue}{Provides a trade-off between secrecy and \acrshort{ber}, BPSK signals are more practical, extensive theoretical and numerical analysis} &
        \cellcolor{table_white} \textcolor{table_blue}{No real-world data} \\
        \hline
        
        \cellcolor{table_white} \textcolor{table_blue}{Zhang \textit{et al.}~\cite{Zhang2019}} & 
        \cellcolor{table_white} \textcolor{table_blue}{Fully-connected \acrshort{nn} } &
        \cellcolor{table_white} \textcolor{table_blue}{{Channel matrices of Bob and Eve}} &
        \cellcolor{table_white} \textcolor{table_blue}{\acrshort{mimo} network} &
        \cellcolor{table_white} \textcolor{table_blue}{No real-world data} \\
        \hline
        
        \cellcolor{table_white} \textcolor{table_blue}{Zhang \textit{et al.}~\cite{Zhang2021}} & 
        \cellcolor{table_white} \textcolor{table_blue}{\acrshort{nn}} &
        \cellcolor{table_white} \textcolor{table_blue}{{Channel matrices and error bounds}} &
        \cellcolor{table_white} \textcolor{table_blue}{Applicable to cognitive radio networks (can be sensitive to primary users), can work without \acrshort{csi}, thorough investigation} &
        \cellcolor{table_white} \textcolor{table_blue}{No real-world data} \\
        \hline
        
        \cellcolor{table_white} \textcolor{table_blue}{Li \textit{et al.}~\cite{Li2020}} & 
        \cellcolor{table_white} \textcolor{table_blue}{\acrshort{lstm}, echo state network} &
        \cellcolor{table_white} \textcolor{table_blue}{{Position vectors and transmit antenna vectors}} &
        \cellcolor{table_white} \textcolor{table_blue}{Applicable to D2D communication, energy efficient} &
        \cellcolor{table_white} \textcolor{table_blue}{No real-world data} \\
        \hline
        
    \end{tabular}
    \label{tab:eavesdropping}
\end{table*}

Encryption/decryption methods are the conventional countermeasures against eavesdropping attacks. An encryption method requires secret information (key) shared by only Alice and Bob. The security of an encryption method is based on the assumption that Eve is unable to decrypt the message without the key. However, this assumption can be falsified with high computational power. Moreover, distributing the key to Alice and Bob without leaking it to Eve is a major drawback of the conventional encryption methods. Instead of hiding the message inside a cipher, PHY security methods aim to prevent Eve from correctly obtaining the messages at \acrshort{phy}. For this purpose, PHY security methods focus on various techniques such as coding, BF, or artificial noise to nullify Eve's channel. However, designing an effective model for these techniques is a challenging task and still draws the attention of many researchers. In the past years, various \acrshort{dl}-assisted models have been proposed in the literature to improve the secrecy rate. A list of the reviewed \acrshort{dl}-based anti-eavesdropping methods is given in Table~\ref{tab:eavesdropping} with their pros and cons.     

\acrshort{e2e} learning with \acrshort{ae}s is an emerging concept that can significantly improve communication rates under unpredictable environments. The success of \acrshort{ae}s on communication drew the attention of wireless security researchers in the past years. A direct implementation of \acrshort{ae}s for the secure communication purpose is presented in~\cite{Fritschek2019}. The main idea of the study is to include a security objective to the loss function of an \acrshort{ae}. In other words, the loss function does not maximize only the legitimate channel; it also aims to minimize the eavesdropper's channel. However, this objective as the difference {between} Bob and Eve's channel capacity is hard to compute. For this reason, the authors implement this objective as a modified version of the difference {in} cross-entropy losses. The method uses clusters that contain redundant symbols to confuse Eve. The proposed loss function enables the users to choose the level of the information loss in the legitimate channel (and the leaked information to Eve as a trade-off). As a strong side, Eve is considered to have \acrshort{nn} abilities. The main drawback of the study is that Eve is assumed to have more noise than Bob. Another secure \acrshort{ae} model is proposed in~\cite{Sun2020}. Similar to~\cite{Fritschek2019}, the authors consider the design of a secure loss function based on cross-entropy. As an addition,~\cite{Sun2020} also implements an authentication method into the \acrshort{ae}.

In~\cite{Besser2020}, the authors use \acrshort{dl} to generate wiretap codes. The main idea behind the study is to include the error rates of the eavesdropper into the loss function of the \acrshort{ae}. The study successfully analyzes the problems of implementing this idea and presents a thorough mathematical basis. Also, the design allows a trade-off between information leakage to the eavesdropper and \acrshort{ber} of the legitimate receiver. The authors also verify their theoretical results with computer simulations by comparing their scheme with polar wiretap codes. A similar approach is also considered in~\cite{Nooraiepour2021a}. Contrary to~\cite{Besser2020}, this study constrains their alphabet to \acrshort{bpsk} and uses a different loss function.   

A secure communication model for MIMO networks is proposed in~\cite{Zhang2019}. Instead of focusing on \acrshort{e2e} communication, the proposed \acrshort{dnn} aims to find the optimum covariance matrix of the input signal. The \acrshort{dnn} is fed with a large dataset containing input channels and covariance matrices in offline training. Compared to conventional analytical solutions, \acrshort{dnn}-based covariance matrix approximation reduces the time-complexity. Also, it is a practical solution to power and time-restricted applications since the computational load is at the offline training stage.

\acrshort{dl}-aided secure communication for cognitive radio networks is first considered in~\cite{Zhang2021}. In addition to the traditional secure communication objective function, the proposed method includes two additional constraints: i) transmit power of the secondary user should be under a threshold, ii) leaked interference to the primary user should be under a threshold. The authors exploit a \acrshort{dl} network to approximate the best transmit power allocation scenario for secure transmission with given cognitive radio constraints. Also, the method is able to find a solution to the optimization problem even without \acrshort{csi}. This study thoroughly compares the proposed method with conventional optimization techniques for their secrecy rate performance, leaked information to the primary user, and computation time. Their results show that \acrshort{dl} can heavily reduce the computational time and complexity without any significant performance loss. 

In~\cite{Li2020}, D2D communication networks are considered. The authors propose a \acrshort{dl}-assisted algorithm, which selects the transmit antennas and the device pairs in order to prevent pairing with eavesdroppers.

\begin{figure*}
   \centering
   \includegraphics[width=13cm]{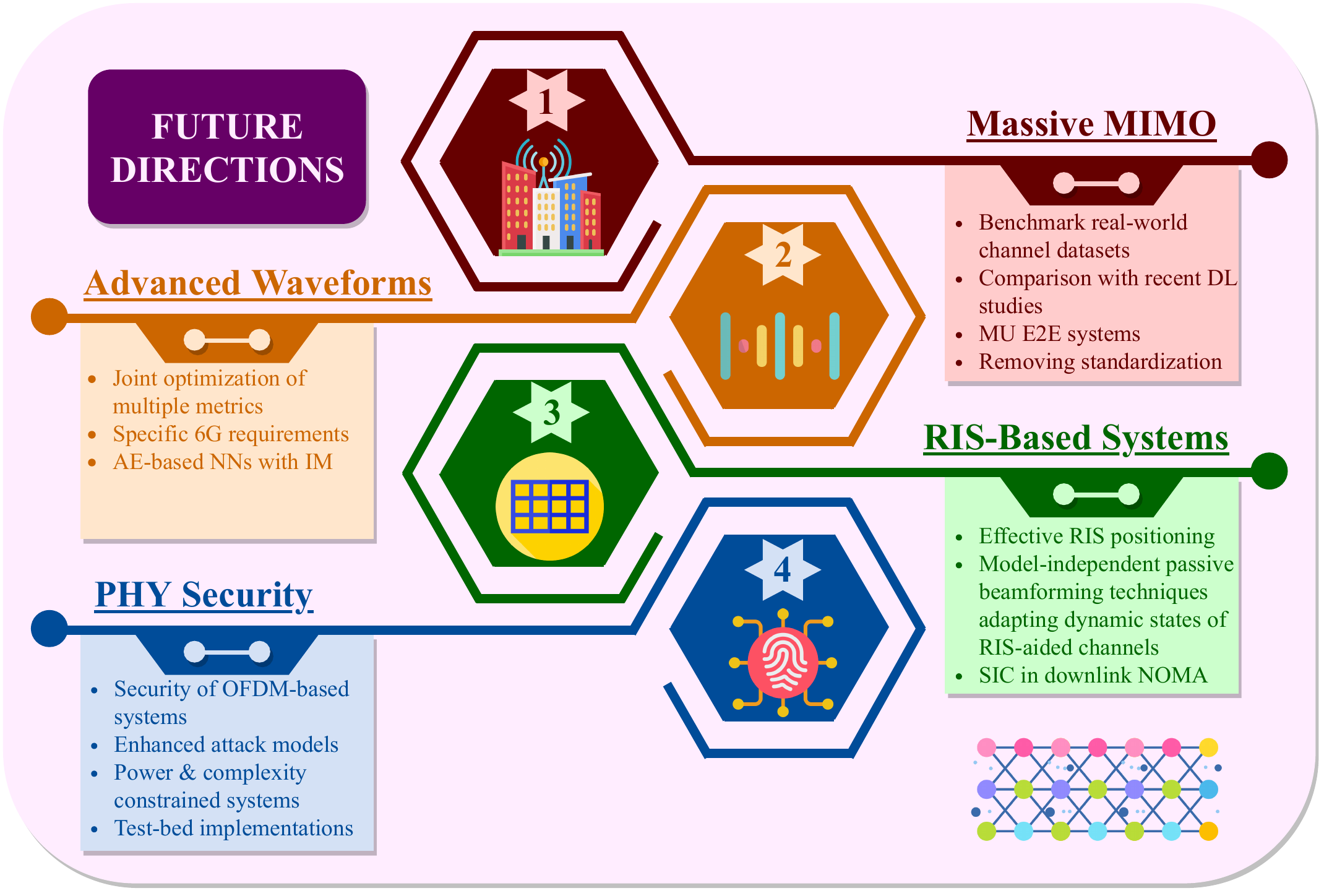}
   \caption{Future directions for DL-based PHY techniques in four primary research items: massive MIMO, advanced waveform designs, RIS-empowered systems, and PHY security.}
   \label{fig:FutureDirections}
\end{figure*}


\acrshort{dl}-based anti-eavesdropping studies are listed in Table~\ref{tab:eavesdropping}. Most of these works are based on utilizing \acrshort{ae}s to find secure encoding or BF schemes. Since traditional methods are model-based, they are highly vulnerable to unexpected variations in the environment. On the other hand, \acrshort{dl}-based anti-eavesdropping literature proves that the data-driven nature of \acrshort{dl} networks can significantly improve the robustness against unexpected variations. 

One of the challenges that DL literature faces is to include {the} eavesdropper's channel capacity in the loss function. Contrary to the spoofing or jamming scenarios, DL-based anti-eavesdropping methods do not face a classification problem. Essentially, the DL models are expected to optimize {their} variables for the secrecy rate metric which includes {the} eavesdropper's channel capacity. However, known DL models and loss functions are not fully applicable for this task. As a result, unique DL models are still needed to improve the performance of DL-based methods. Similar to anti-jamming literature, we should remark that anti-eavesdropping studies lack proof of concepts and testbed implementations with real-world datasets. Moreover, \acrshort{dl}-assistance is applied to a limited number of communication scenarios, i.e., \acrshort{ris} or \acrshort{ofdm} scenarios are missing. The following section dwells on the future aspects of our four leading research directions and presents our comments.

\section{Conclusions and Future Directions}
As we have often stated, we are on the verge of a potential revolution in wireless communications. The deployment process of the 5G technology has come a long way all over the world these days and the active research on the 6G technology has also gained tremendous momentum. It is an undeniable fact that 6G technology will usher in radical paradigm shifts, new thrilling applications, and groundbreaking technologies. AI will undoubtedly play a huge role in this irresistible revolution emerging with 6G. Specifically, DL approaches have been building a concrete ground and will most likely continue to overtake other AI branches for the bright future of wireless communication technologies. In this article, we have provided a comprehensive overview of the existing DL-based PHY techniques and shed light on the huge potential of emerging DL approaches under four main research directions {toward} 6G, namely massive MIMO systems, MC waveform designs, RIS-aided communications, and PHY security. We have analyzed the progress made so far in detail and determined major future research directions in each {of} these fields, as demonstrated in Fig.~\ref{fig:FutureDirections}. Please note that it is possible to add other \acrshort{phy} research directions to the framework that we have covered in this article. In fact, the usage of DL grows gradually on other \acrshort{phy} technologies. For instance, the literature also includes \acrshort{dl}-powered studies on \acrshort{mmwave} or terahertz communications. It has been shown that \acrshort{dl} can be employed to efficiently solve various challenges of high{-}frequency communications. Moreover, \acrshort{dl} is also applied for the challenges of drone or unmanned aerial vehicle communication networks. However, the \acrshort{dl} literature on these technologies {is} still developing{,} and including them {in} our \acrshort{6g} framework is a future direction.

\subsection{{Future Directions on Massive MIMO Systems}}

We have investigated the major developments of DL applications for MIMO systems from various aspects and identified the missing points that should attract more attention. The first issue from an intelligent receiver perspective is that most detectors available in the literature require perfect CSI to operate successfully. Since it is challenging to estimate channels in practice and even harder in massive MIMO systems due to their plenty of transmit antennas, developing a system model with a perfect CSI assumption remains a pipe dream. Therefore, receiver frameworks based on DL techniques composing channel estimators and symbol detectors should become more common in the future to support practicability. Secondly, evaluating the proposed models in a computer simulation environment using generic channel model assumptions may be deceptive since there are many more impairments in real-life scenarios, and we do not know how they impact the performance of DL models. Thus, we suggest creating benchmark channel datasets that include real-life impairments and encapsulate diverse settings for various wireless communication standards, such as 5G and Wi-Fi 7. These benchmark datasets will allow for fair comparisons of DL models. Apart from these, the third vital topic for rapid progress in this field is the development of benchmark system models that researchers can compare their DL models. The number of novel DL frameworks has been increasing in tandem with the advancement of the literature. Hence, new DL-based frameworks should be compared to existing novel data-driven methods to prevent the conglomeration of similar models.

In terms of intelligent transmitter designs and E2E perspectives, our initial advice is to concentrate solely on MU system models that require MU interference cancellation since almost all communication systems have been serving MU cases for a while. Thus, DL-based precoders should not only eliminate the effect of the channel on each user but also prevent a user's signals from interfering {with} others. The second point is practicability, as in the case of DL-based receiver designs. We believe that experiments using SDR testbeds in various environments will help us examine the feasibility of DL-based models in dynamically changing scenarios. It is also possible to understand the hardware requirements for efficient DL operations. The last research direction is the actual enabler of the PHY revolution. Considering DL approaches' ability to optimize a communication system as a whole in an E2E fashion, we question if the next-generation wireless communication technologies can arise without the restrictive standardization process that requires continuous regulations. The ultimate objective is to delegate defining parameters, signaling, and other issues that must be resolved throughout the standardization process to DL networks, therefore shortening the time between generations.

\subsection{{Future Directions on Multi-Carrier Waveform Designs}}

Various MC waveforms are proposed by researchers and OFDM has emerged as the most popular one among them. As noted earlier, OFDM has been used in modern wireless communication systems including 4G and 5G. Although designed on top of OFDM, other MC waveforms such as FBMC and GFDM could not seize the throne. Despite OFDM being the most powerful one  due to its simplicity and flexibility, other OFDM-based waveforms can also be seen as candidates for 6G wireless networks. Nonetheless, these waveforms are still needed to be improved to meet the high demands of 6G. At this point, DL comes into the picture as a great tool in order to enhance performance and reduce the complexity of MC waveforms.

\begin{figure*}[t]
   \centering
    
   \includegraphics[width=14.5cm,height=3.5cm]{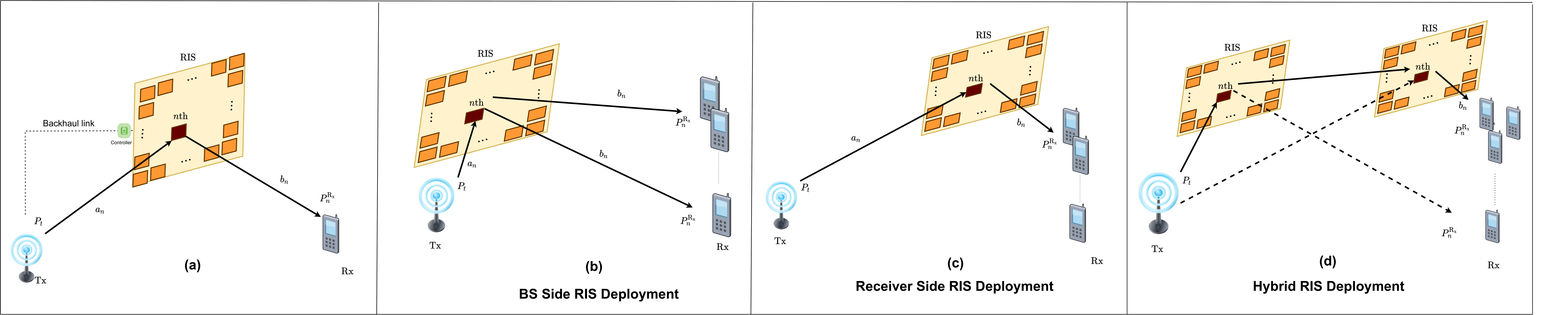}
    
   \caption{Effective RIS deployment scenarios (a) non-location specific positioning, (b) BS-side positioning, (c) user-side positioning and (d) hybrid-side positioning.}
  
   \label{fig:deployment}
\end{figure*}


{In} the second step, we have reviewed emerging DL-based solutions for various MC waveforms. With a comprehensive assessment, we conclude that the existing DL-based methods have great potential to improve the performance of the aforementioned waveforms for 6G and may outperform traditional algorithms. However, there are still unaddressed issues related to the application of DL-based solutions for MC waveforms. Firstly, most studies focus on optimizing only one performance metric such as BER and PAPR. However, in MC systems, there are multiple key performance indicators, which should be considered while designing the system. Therefore, we believe that it is important to design DL-based techniques that optimize multiple performance metrics jointly to improve the overall performance. Secondly, while designing a target DL method, specific requirements of different 6G applications, such as eMBB, URLLC, mMTC, and their possible combinations, need to be considered. Thirdly, most studies in {the} literature focus on enhancing the performance of classical OFDM. Designing DL-based techniques for other MC waveforms is still an open topic. Fourthly, the combination of AE-based NNs with IM to increase {the} data rate or improve error performance would be an interesting design problem. Lastly, we observe that most of the proposed DL-based solutions have not considered real-world data sets. In order to reveal the true potential of DL-based techniques, they need to be implemented in practical scenarios by employing wireless communication testbeds.

\subsection{{Future Directions on RIS-Aided Physical Layer Communications}}

Considering the needs of 6G and beyond wireless communication technologies, it is vital to reduce the overall system complexity by preserving the passive structure of RISs and achieving high performance at a low cost. Thus, it may be crucial to consider the advantages of optimal RIS positioning.
In light of the aforementioned approaches, the question of how effective RIS configuration should be in DL assistance in order to get the best performance will arise. In line with these approaches, applicable RIS positioning scenarios can be evaluated for effective signaling and channel estimation at the user side, reducing signaling overhead and optimizing the transmit power.

\begin{figure}[t]
   \centering
   \includegraphics[width=\linewidth,height=8cm]{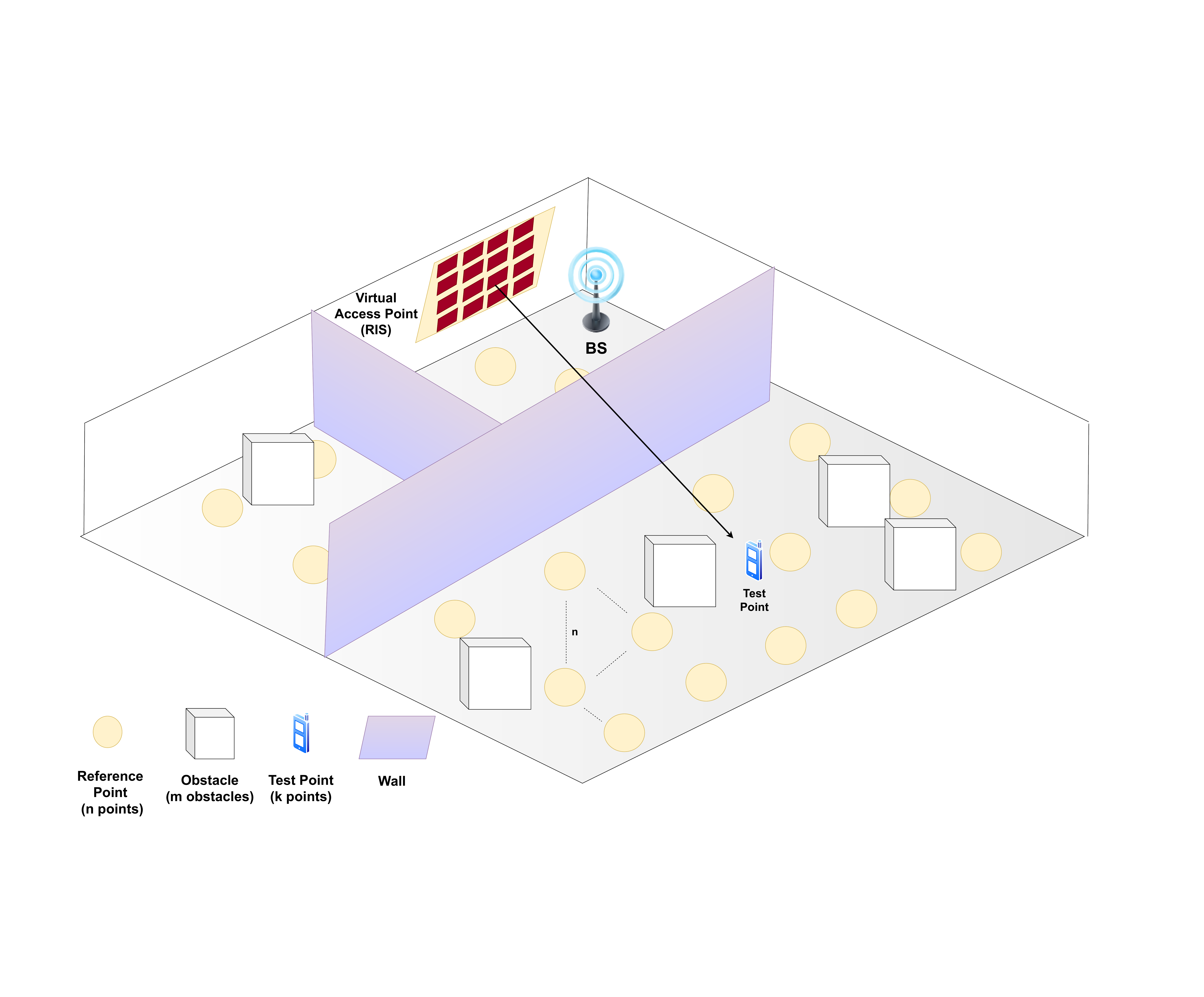}
   \caption{A location-based trained indoor application.}
   \label{fig:indoor}
\end{figure} 


In Fig.~\ref{fig:deployment}, we provide four major RIS deployment scenarios. As shown in Fig.~\ref{fig:deployment}(b)-(c), an RIS can be deployed on the BS side or the receiver side, or it can be positioned in a hybrid scheme as in Fig.~\ref{fig:deployment}(d) to combine its advantages in certain aspects. It appears as the most advantageous option in terms of signaling overhead due to the close distance between the RIS and the BS so that the channel in between is almost static. This also eliminates the coordination burden when compared to distributed RIS deployments on the user side. The BS-side positioning has the ability {to adapt} to instant changes in user channels and different QoS requirements. On the flip side, compared to the BS-side deployment, it provides a more advantageous solution in providing a direct LOS in user-side RIS positioning. This is shown as the only aspect where user-side deployment is superior to the BS-side in~\cite{you2020deploy}. However, this situation can be treated as an issue that can be improved with the DL approach. Then, BS-side deployment scenarios might gleam as  bright solutions for optimizing DL-assisted RIS applications.
Similarly, appropriate positioning by using intelligent surfaces as virtual access points in which BS-side positioning, might match the power of an active access point by using fully passive elements. These virtual access points increase environmental scattering and provide spatial multiplexing gain to the MIMO system as in~\cite{dunna2020scattermimo}. However, the assumption of {the} presence of initial CSI can be considered as a drawback in these systems. At this point, the development of a DL-based approach eliminating the need for initial CSI will provide significant gains in performance.

BS-side RIS positioning, as seen in Fig.~\ref{fig:deployment}(b), and {the} creation of virtual APs on the BS side may play an essential role in reducing signaling overhead and system complexity on the transmitter side. Furthermore, as shown in Fig.~\ref{fig:deployment}(d), the hybrid positioning scenario should be investigated to see if it yields more appealing results than BS-side positioning due to the optimization burden caused by increasing system parameters. At this juncture, a novel DNN architecture can be used in the learning process of a high number of parameters for the estimation of the complex channel information brought by the hybrid positioning scenario. These scenarios can be combined with a DL approach similar to~\cite{ozdougan2020deep} and phase optimization on the RIS-side can be placed with a model-independent and trained DNN. In order to observe the variation of system performance according to changing environmental conditions and to determine the optimum working policy, it is vital to create dynamic {user} scenarios for the BS-side RIS deployment. Thereby, a long-term policy with the ability to learn for dynamic scenarios can be constituted. By an approach similar to~\cite{huang2019indoor}, in a scenario{,} as given in Fig.~\ref{fig:indoor}, a DNN that performs the RIS phase configuration with various position information can be considered. Thus the signal level at different user positions can be increased effectively with DL.

DL approaches also have great significance in enhancing the performance {of} downlink NOMA-MIMO systems and increasing the overall system performance in terms of {user} sum-rate. In this way, lower system complexity and power consumption can be addressed, which will enable beyond 5G wireless communication technologies. Positioning scenarios while preserving the passive nature of RISs can also be considered here first. In DL-based studies on RIS-aided downlink-NOMA schemes, a common predicament arises with the assumption of having {an} initial CSI. However, obtaining CSI in a NOMA system scheme is another challenge due to the complexity of the underlying communication model. These schemes might be integrated with \acrshort{rnn}-like DL approaches that estimate channel state using channel statistics, aiming to reduce latency in scenarios without initial CSI. Furthermore, the assumption of imperfect SIC at the decoding stage makes preventing MUI difficult with {the} increasing number of users. This is one of the most critical factors degrading the total user performance due to the inefficiency of current SIC techniques and restricting user capacity. Inspired by~\cite{sim2020deep}, the combination of a novel DL approach seeking the perfect SIC with DL-assisted RIS phase optimization can address the aforementioned challenges in scenarios using the downlink-NOMA scheme. Novel DL algorithms suitable for these scenarios can be tested with DNN or CNN architectures and appropriate loss functions can be developed.

{As another direction to leverage novel DL methods, simultaneously transmitting and reflecting RISs (STAR-RIS) can be considered as a field that has a gleaming future. By simultaneously transmitting and reflecting the incident signals, STAR-RISs may serve both sides of users placed at their front and back, in contrast to RISs, which are often distinguished by their reflecting-only property.The concept of STAR-RISs has been extensively elaborated by researchers in utilizing the unique communication framework to produce smart radio settings, which has been motivated by its appealing advantages \cite{mu2021simultaneously}, \cite{xu2021star}, \cite{liu2021star}. Nevertheless, the fractional shape of the objective function and non-convex restrictions make it difficult to solve the energy efficiency maximization problem by locating the global optimal solution. DL can be considered as a potential solution to overcome the difficulties posed by current optimization methods.}


\subsection{{Future Directions on Physical Layer Security}}
{In} the final step, we have put our emphasis on emerging PHY security systems. The existing studies in the literature prove the positive impact of DL on PHY security. Although various network or attack models {have been} considered in the past, there are still scenarios where the impact of DL has not been tested. For example, {the} secrecy capacity of OFDM-IM/OFDM systems or RIS-assisted networks {is} widely considered in the literature. However, exploration of the impact of DL on these scenarios is still an open area. Moreover, the existing literature is very immature such that mostly raw ideas are considered. Their extensions to different DL networks and various system models (e.g., power/complexity constraints, improved attacker models) {are} missing and, we believe, is an interesting future direction to unlock the potential of DL.

To sum up, we conclude that DL architectures might have a huge potential to shape the PHY design of future radios{,} and challenging open problems exist to unlock the true potential of AI-based approaches for future wireless systems.

\section*{Acknowledgment}
The authors would like to thank Vestel Electronics and Dr. Basak Ozbakis for their financial support of this article under Vestel Electronics-Koc University Industry Cooperation Project No. OS.00170

\bibliographystyle{IEEEtran}
\bibliography{ref}	


\begin{IEEEbiography}[{\includegraphics[width=1in,height=1.25in,clip,keepaspectratio]{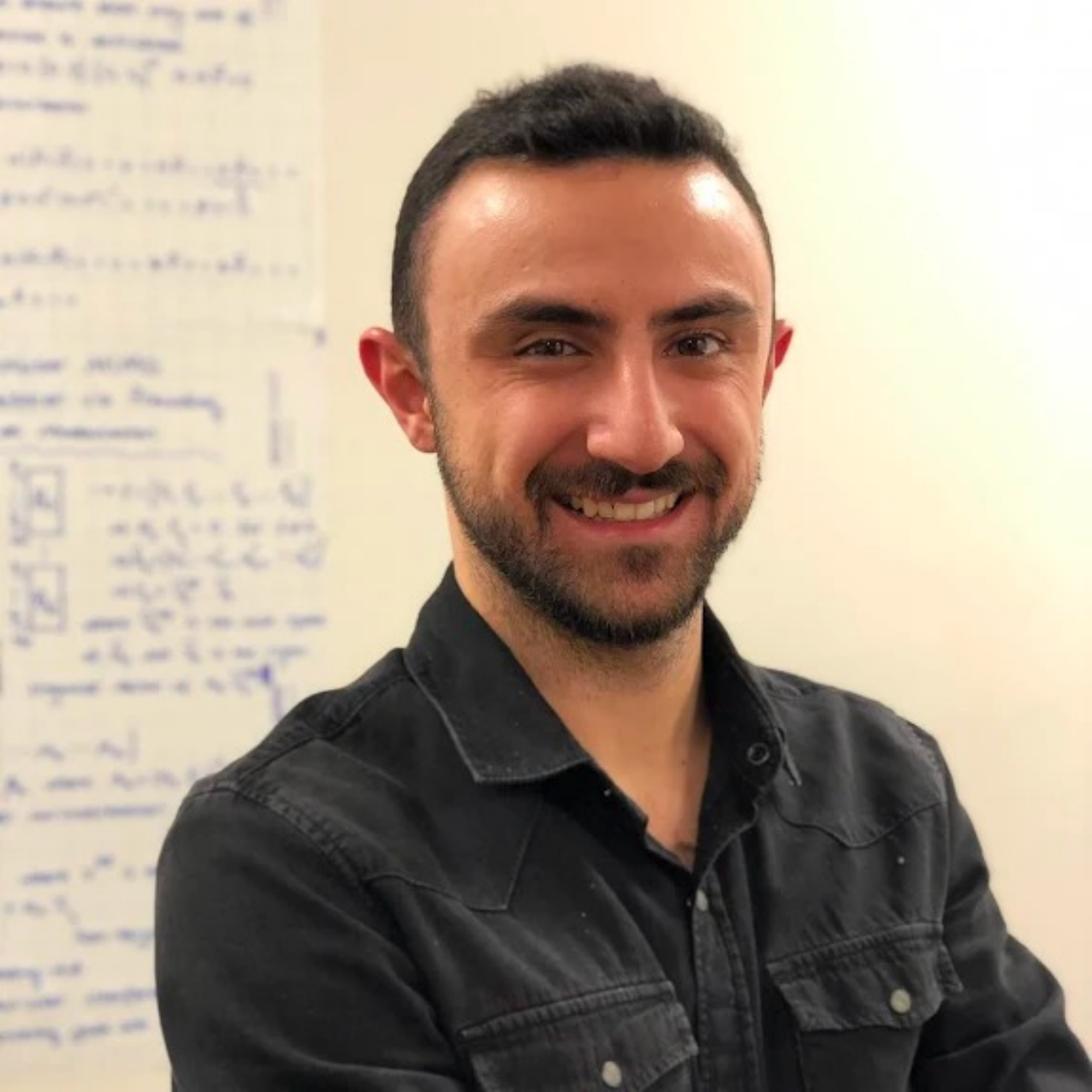}}]{Burak Ozpoyraz } received his B.Sc. degree from Electronics and Communication Engineering Department of Istanbul Technical University, Istanbul, Turkey, in 2019, and M.Sc. degree from Electrical and Electronics Engineering Department of Koc University, Istanbul, Turkey, in 2022. He is currently a Ph.D. student, and working as a research and teaching assistant in the Faculty of Electrical Engineering and Information Technology in RWTH Aachen University. His research interests include physical layer security, index modulation, massive multiple-input multiple-output systems, deep learning applications in the physical layer of wireless communications, and universal software defined peripherals.
\end{IEEEbiography}

\begin{IEEEbiography}[{\includegraphics[width=1in,height=1.25in,clip,keepaspectratio]{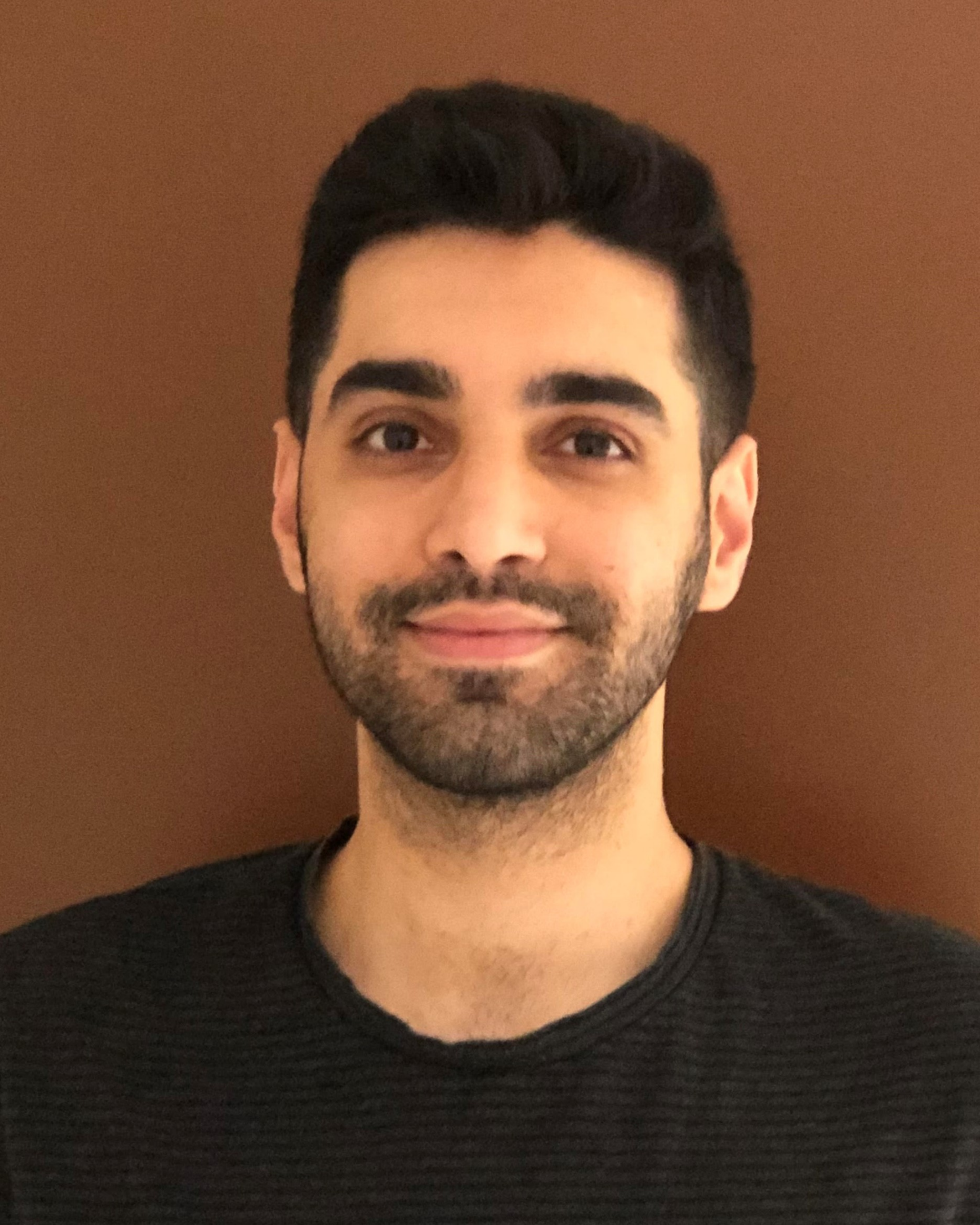}}]{Ali Tugberk Dogukan } received the B.S. degree from Istanbul Technical University, Istanbul, Turkey, in 2018, and the M.S. degree from Koç University, Istanbul, Turkey, in 2020. He is currently a Research and Teaching Assistant at Koç University while pursuing his Ph.D. degree at the same university. His research interests include waveform design, signal processing for communications, deep-learning for wireless communication, index modulation, and software defined radio-based practical implementation. He served as a Reviewer for several IEEE letters and journals.
\end{IEEEbiography}

\begin{IEEEbiography}[{\includegraphics[width=1in,height=1.45in,clip,keepaspectratio]{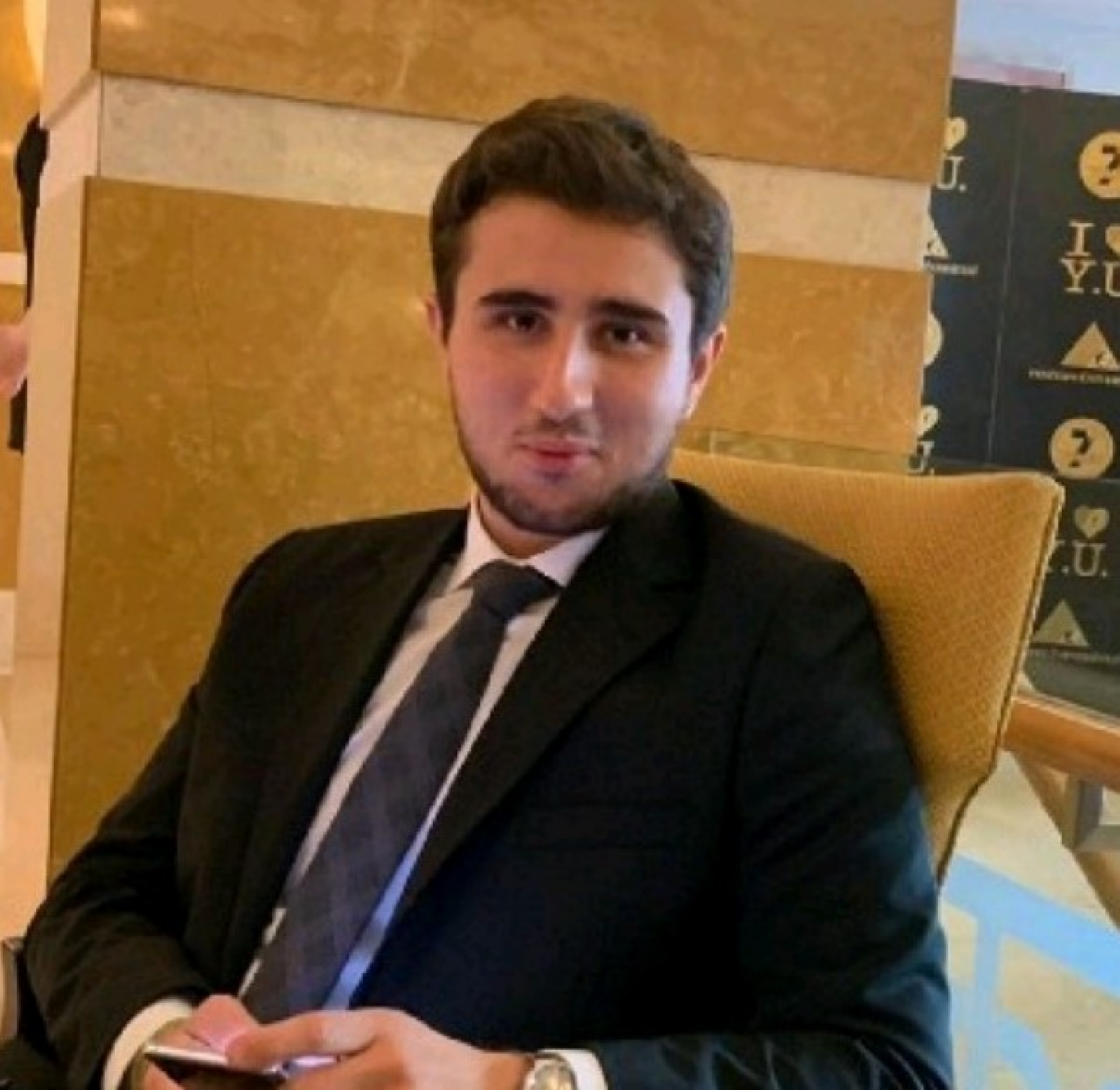}}]{YARKIN GEVEZ } received the B.S. degree in electrical and electronics engineering from the Yeditepe University, Istanbul, Turkey, in 2019, and the M.S degree from the University of Ontario, Institue of Technology Toronto, Canada, in 2020. He is currently pursuing the Ph.D. degree in electrical and electronics engineering at Koc University, Istanbul, Turkey. His research interests include Wireless Communications, Reconfigurable Intelligent Surfaces and Deep Learning Techniques for 5G and beyond Wireless Networks.
\end{IEEEbiography}

\begin{IEEEbiography}[{\includegraphics[width=1in,height=1.25in,clip,keepaspectratio]{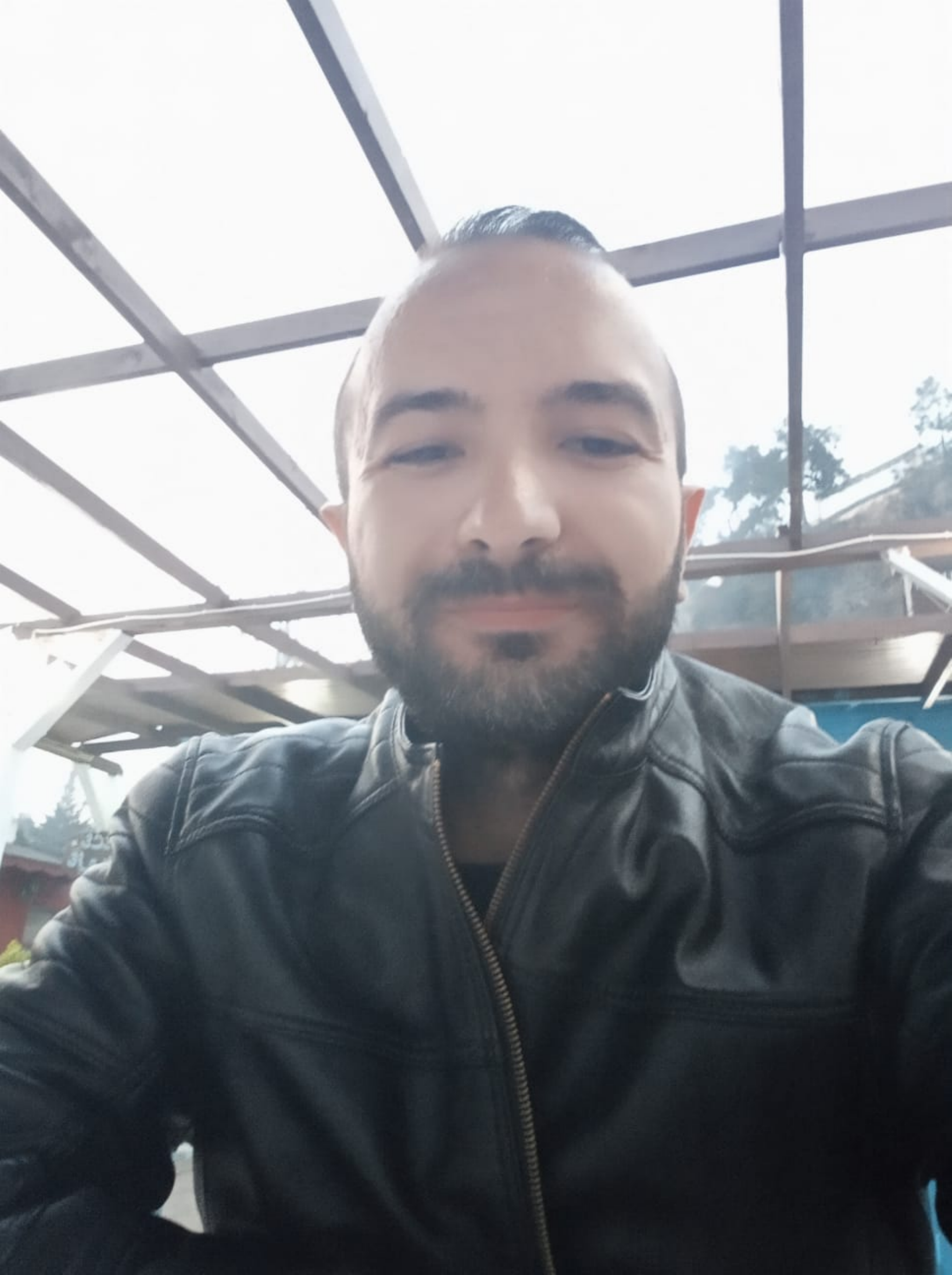}}]{Ufuk Altun } was born in Isparta, Turkey, in 1993. He received the B.Sc. and M.Sc. degrees, in electronics and communication engineering, from the Istanbul Technical University, Istanbul, Turkey, in 2017 and 2020, respectively. He is currently a Research Assistant at Trakya University, Edirne, Turkey and a Ph.D. student at Koc University, Istanbul, Turkey. His research interests include physical layer security (currently focuses on its possible interaction with autoencoders, machine learning and deep learning algorithms), reconfigurable intelligence surfaces, OFDM-IM and indoor localization.
\end{IEEEbiography}

\begin{IEEEbiography}[{\includegraphics[width=1in,height=1.25in,clip,keepaspectratio]{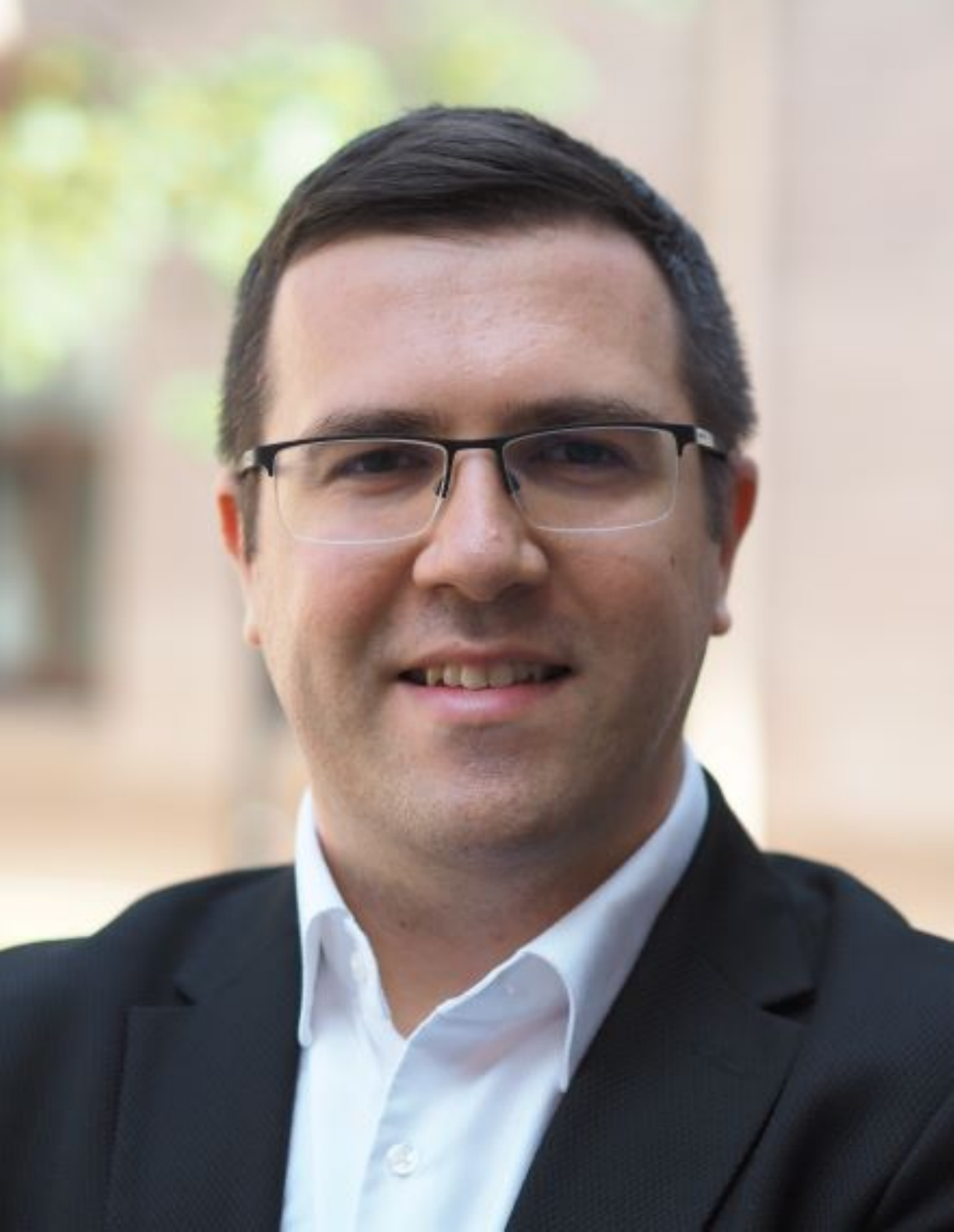}}]{Ertugrul Basar } received his Ph.D. degree from Istanbul Technical University in 2013. He is currently an Associate Professor with the Department of Electrical and Electronics Engineering, Ko\c{c} University, Istanbul, Turkey and the director of Communications Research and Innovation Laboratory (CoreLab). His primary research interests include beyond 5G systems, index modulation, intelligent surfaces, waveform design, and signal processing for communications. Dr. Basar currently serves as an Editor of \textit{IEEE Transactions on Communications} and \textit{Frontiers in Communications and Networks}. He is a Young Member of Turkish Academy of Sciences and a Senior Member of IEEE.
\end{IEEEbiography}

\end{document}